\documentclass[aps,prb,showpacs,onecolumn,notitlepage,superscriptaddress,nofootinbib]{revtex4-2}

\usepackage{bm,color,amsmath,graphicx,psfrag,accents,float,amssymb}
\usepackage{multirow}
\usepackage[colorlinks=true,linkcolor=blue,citecolor=blue]{hyperref}
\usepackage{dcolumn}
\usepackage{xcolor}
\usepackage{comment}
\usepackage{tcolorbox,mathtools,ulem}
\usepackage{mathrsfs}

\newcolumntype{L}[1]{>{\raggedright\arraybackslash}p{#1}}
\newcolumntype{C}[1]{>{\centering\arraybackslash}p{#1}}
\newcolumntype{R}[1]{>{\raggedleft\arraybackslash}p{#1}}

        \definecolor{AAcolor}{rgb}{0.7,0.1,0.4}

			\newcommand{\e}[1]{\begin{align}{#1}\end{align}}	
			\newcommand{\lin}{\notag \\}
		
		\newcommand{\f}[2]{\frac{#1}{#2}}
		\newcommand{\tf}[2]{\tfrac{#1}{#2}}
		
		\newcommand{\la}[1]{\label{#1}}

		\newcommand{\q}[1]{Eq.\ (\ref{#1})}
		\newcommand{\qq}[2]{Eqs.\ (\ref{#1})-(\ref{#2})}
		\newcommand{\s}[1]{Sec.\ \ref{#1}}
		\newcommand{\fig}[1]{Fig.\ \ref{#1}}		
		\newcommand{\app}[1]{App.\ \ref{#1}}				
		
		\newcommand{\ocite}[1]{Ref.\ \onlinecite{#1}}

		\newcommand{\ri}{\rightarrow}
		\newcommand{\leri}{\leftrightarrow}
		\newcommand{\lea}{\leftarrow}
				
		\newcommand{\iwith}{\ins{with}}

		\newcommand{\sgn}{\text{sgn}}

		\newcommand{\imp}{\;\;\Rightarrow\;\;}
		\newcommand{\eq}{=&\;}
		\newcommand{\condeq}[1]{\as \substack{\sma{#1}\\=}\as}

		\newcommand{\limit}[1]{\substack{\text{lim}\\#1}\;}		
		 
		\newcommand{\appr}{\approx &\;}

		\newcommand{\R}{\mathbb{R}}
		\newcommand{\Z}{\mathbb{Z}}

\newcommand{\mathsout}[1]
{\bgroup\mathchoice
  {\sbox0{$\displaystyle{#1}$}%
    \usebox0\hspace{-\wd0}%
    \rule[0.5\ht0-0.5\dp0-.5pt]{\wd0}{1pt}}%
  {\sbox0{$\textstyle{#1}$}%
    \usebox0\hspace{-\wd0}%
    \rule[0.5\ht0-0.5\dp0-.5pt]{\wd0}{1pt}}%
  {\sbox0{$\scriptstyle{#1}$}%
    \usebox0\hspace{-\wd0}%
    \rule[0.5\ht0-0.5\dp0-.5pt]{\wd0}{1pt}}%
  {\sbox0{$\scriptscriptstyle{#1}$}%
    \usebox0\hspace{-\wd0}%
    \rule[0.5\ht0-0.5\dp0-.5pt]{\wd0}{1pt}}%
\egroup}

\newcommand{\tr}{\text{Tr}}

	\newcommand{\eikr}{e^{i\bk \cdot \br}}

\newcommand{\diverge}{\nabla \cdot}

\newcommand{\nabk}{\nabla_{\boldsymbol{k}}}
\newcommand{\nabkp}{\nabla_{\boldsymbol{k}'}}

\newcommand{\per}{\perp}

\newcommand{\krec}{\bk_{rec}}
\newcommand{\kexc}{\bk_{exc}}
\newcommand{\kext}{\bk_{ext}}
\newcommand{\kave}{\bk_{ave}}
\newcommand{\bkp}{\bk'}
\newcommand{\bqp}{\bq'}
\newcommand{\delbk}{\delta \bk}

\newcommand{\var}{\varepsilon}

\newcommand\as{\;\;\;\;}

\newcommand{\hbp}{\hat{\bp}}
\newcommand{\hbq}{\hat{\bq}}
\newcommand{\hbr}{\hat{\br}}
\newcommand{\hbv}{\hat{\bv}}

\newcommand{\bd}{\boldsymbol{d}}

\newcommand{\bj}{\boldsymbol{j}}
\newcommand{\bk}{\boldsymbol{k}}

\newcommand{\bp}{\boldsymbol{p}}
\newcommand{\bq}{\boldsymbol{q}}
\newcommand{\br}{\boldsymbol{r}}

\newcommand{\bv}{\boldsymbol{v}}

\newcommand{\bA}{\boldsymbol{A}}

\newcommand{\bG}{\boldsymbol{G}}

\newcommand{\bP}{\boldsymbol{P}}
\newcommand{\bQ}{\boldsymbol{Q}}
\newcommand{\bR}{\boldsymbol{R}}
\newcommand{\bS}{\boldsymbol{S}}

\newcommand{\bze}{\boldsymbol{0}}

\newcommand{\beps}{\boldsymbol{\epsilon}}

\newcommand{\bOmega}{\boldsymbol{\Omega}}
\newcommand{\bsigma}{\boldsymbol{\sigma}}
\newcommand{\bSigma}{\boldsymbol{\Sigma}}
\newcommand{\btau}{\boldsymbol{\tau}}

\newcommand{\fraks}{\mathfrak{s}}

\newcommand{\W}{{\cal W}}

\newcommand{\vectwo}[2]{\begin{pmatrix} {#1}\\{#2} \end{pmatrix}}

\newcommand{\sy}{\sigma_{\sma{2}}}
\newcommand{\sz}{\sigma_{\sma{3}}}

\newcommand{\tx}{\tau_{\sma{1}}}

\newcommand{\tz}{\tau_{\sma{3}}}

\newcommand{\ins}[1]{\;\;\;\;\text{#1}\;\;\;\;}

\newcommand{\ang}{\mbox{\normalfont\AA}}

\newcommand{\imag}{{\text{Im}\;}}
\newcommand{\real}{{\text{Re}\;}}

\newcommand{\cala}{{\cal A}}

\newcommand{\cale}{{\cal E}}

\newcommand{\cali}{{\cal I}}

\newcommand{\calp}{{\cal P}}

\newcommand{\calv}{{\cal V}}

\newcommand{\noi}[1]{\noindent (#1)}

\newcommand{\braket}[2]{\big\langle #1 \,|\, #2 \big\rangle}
\newcommand{\ketbra}[2]{|\,  #1  \big\rangle \big\langle #2 \,| }
\newcommand{\braopket}[3]{\big\langle #1 \,|\, #2 \,|\, #3 \big\rangle}
\newcommand{\bra}[1]{\langle\,#1\,|}
\newcommand{\ket}[1]{|\,#1\,\rangle}

\newcommand{\half}{\frac{1}{2} }

\newcommand{\expect}[1]{\left\langle#1\right\rangle}

\newcommand{\pdg}[1]{{#1}^{\phantom{\dagger}}}

\newcommand{\bpm}{\begin{pmatrix}}
\newcommand{\epm}{\end{pmatrix}}

\newcommand{\bal}{\begin{align}}

\newcommand{\eps}{\epsilon}

\newcommand{\dg}[1]{#1^{\scriptstyle{\dagger}}}

\newcommand{\sma}[1]{\scriptscriptstyle{#1}}

\newcommand{\bgt}{B'}
\newcommand{\blt}{B}
\newcommand{\link}{(B'{>} B)}
\newcommand{\tran}{B'\lea B}

\newcommand{\jexc}{\bj_{\text{exc}} }
\newcommand{\jintra}{\bj_{\text{intra}} }
\newcommand{\jrec}{\bj_{\text{rec}} }
\newcommand{\jtran}{\bj_{\text{tran}} }
\newcommand{\jloop}{\bj_{\text{loop}} }

\newcommand{\sexc}{shift$_{\text{exc}}$ }
\newcommand{\sintra}{shift$_{\text{intra}}$ }
\newcommand{\srec}{shift$_{\text{rec}}$ }

\newcommand{\sloop}{\bS_{\text{loop}} }

\newcommand{\jdosup}{JDOS_{\sma{\uparrow}}}
\newcommand{\jdosupdo}{JDOS_{\sma{\uparrow\downarrow}}}
\newcommand{\twoupdo}{2_{\sma{\uparrow\downarrow}}}

\newcommand{\aven}[2]{\langle \langle\, #1\, \rangle \rangle_{#2}}

\begin{document}

\title{Anomalous shift and optical vorticity in the steady photovoltaic current}

\author{Penghao Zhu}\affiliation{Department of Physics and Institute for Condensed Matter Theory, University of Illinois at Urbana-Champaign, Urbana, Illinois 61801, USA}
\affiliation{Department of Physics, The Ohio State University, Columbus, OH 43210, USA}
\author{A. Alexandradinata} \affiliation{Department of Physics and Santa Cruz Materials Center, University of California Santa Cruz, Santa Cruz, CA 95064, USA}


\begin{abstract}
Steady illumination of a non-centrosymmetric semiconductor results in a bulk photovoltaic current, which is contributed by real-space displacements (`shifts') of charged quasiparticles as they transit between Bloch states. The shift induced by  interband excitation via absorption of photons has received the prevailing attention. However, this excitation-induced shift can be far outweighed  ($\ll$) by the shift induced by intraband relaxation, or by the shift induced by radiative recombination of electron-hole pairs. This outweighing ($\ll$)  is attributed to (i) time-reversal-symmetric, intraband Berry curvature, which results in an anomalous shift of quasiparticles as they scatter with phonons, as well as to (ii) topological singularities in the interband Berry phase (`optical vortices'), which makes the photovoltaic current extraordinarily sensitive to the linear polarization vector of the light source. Both (i-ii) potentially lead to  nonlinear conductivities  of order $mAV^{-2}$, without finetuning of the incident radiation frequency, band gap, or joint density of states. A case study of BiTeI showcases the anomalous shift and optical vorticity in a realistic material. 
\end{abstract}
\date{\today}

\maketitle

{\tableofcontents \par}

\section{Introduction}

\begin{figure}[h]
\centering
\includegraphics[width=13 cm]{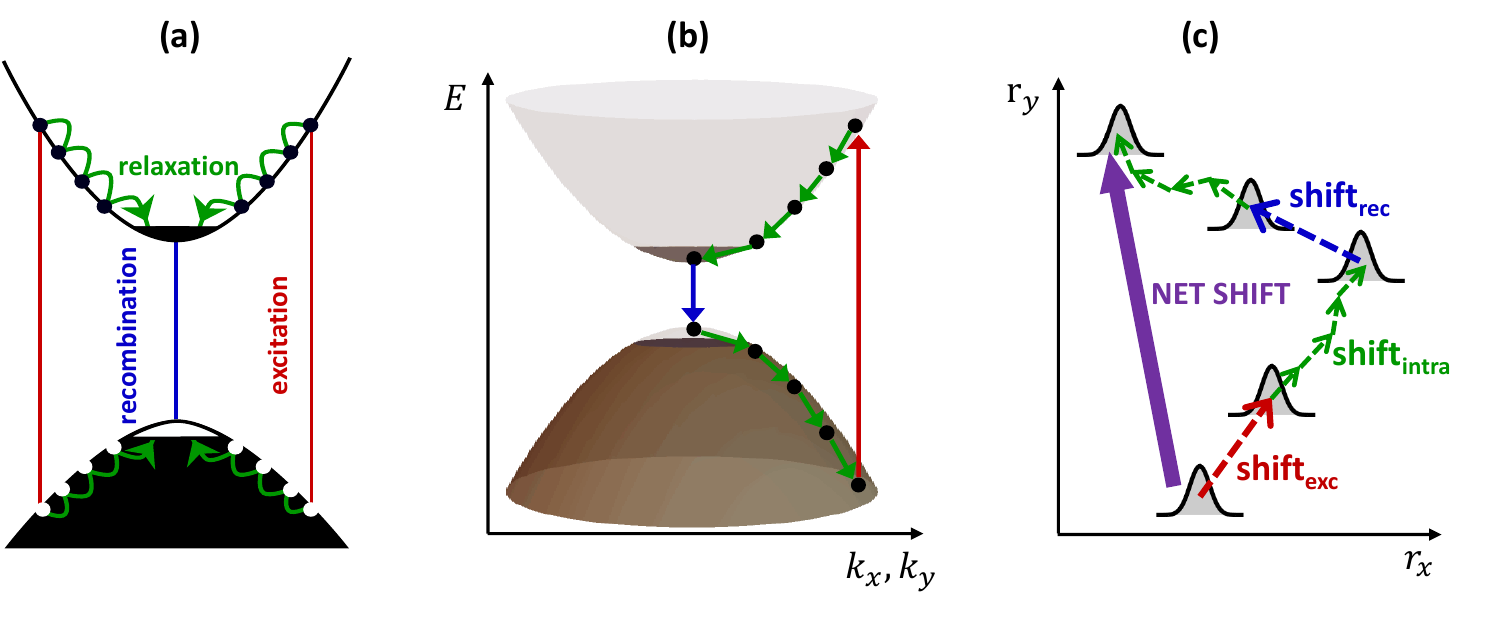}
\caption{(a) The kinetic processes of excitation, relaxation and recombination in a steady state of a homogeneously illuminated semiconductor. (b) These kinetic processes can be recast as loops in energy-momentum space, if one views a hole as an electron going backward in time. To each transition between Bloch states we associate a shift of a wavepacket in real space, as illustrated representatively by the dashed arrows in panel (c); the net shift over all possible loops results in the steady shift current.   
}
\label{fig:coverpicture}
\end{figure}

Light that is harvested for large-scale power transmission needs to be rectified, i.e.,  converted from electromagnetic waves at solar frequencies to a direct or low-frequency current. Rectification in a non-centrosymmetric, non-magnetic semiconductor results in a bulk direct current that is proportional to the radiation intensity, in the lowest order response. This \textit{bulk photovoltaic current} has a  contribution attributed to an asymmetry in the fermionic quasiparticle distribution\cite{danishevskii_dragging,grinberg_lightpressure,belinicher_ballistic} and a second contribution attributed to the real-space displacements (or `shifts') of quasiparticles as they transit between Bloch states; cf.\ \fig{fig:coverpicture}.\cite{baltz_bulkPV,belinicher_kinetictheory} \\

The shift induced by interband \textit{exc}itation via absorption of photons -- in short, shift$_{\text{exc}}$ -- has received the prevailing attention. However, the steady photovoltaic current is also contributed by a shift induced by \textit{\text{rec}}ombination of electron-hole pairs, as well as a shift induced by \textit{\text{intra}}band relaxation via scattering with phonons or impurities. Both \srec and \sintra have been emphasized by Belinicher-Ivchenko-Sturman (henceforth referred to as BIS) in their 1982 kinetic theory of the shift current, which accounts for the steady, non-equilibrium quasiparticle distribution.\cite{belinicher_kinetictheory} In contrast, \srec and \sintra have been ignored in all recent literature, which  either (a) disregarded relaxation completely,\cite{sipe_secondorderoptical,parker_diagrammatic,youngrappe_firstprinciples,LiangTan_review,chong_firstprinciplesnonlinear,julen_wannierinterpolation,ahn_riemanniangeometry} or (b) were agnostic about the nature of the relaxation mechanism, e.g., by naive relaxation-time approximations,\cite{kraut_anomalousbulkPV,matsyshyn_acceleration,holder_trsbreaking,hornung_magnetophotogalvanic} or (c) adopted relaxation mechanisms that are unrealizable in experiments, e.g., by scattering with a `fermionic bath'\cite{morimoto_nonlinearoptic,morimoto_excitonic,matsyshyn_rabiregime}.\footnote{(d) Barik and Sau have considered  electron-phonon scattering as a relaxation mechanism for the shift current.\cite{barik_nonequilibriumnature} However, they
assumed without justification that phonon-mediated scattering does not result in a shift.  A detailed criticism of the recent literature is provided in \app{sec:krautbaltz}.}  The one-sided interpretation of shift currents as a dissipation-less, `hot carrier effect'\cite{LiangTan_review} cannot explain the vanishing photocurrent in the low-temperature polar phase of organic charge-transfer complexes.\cite{nakamura_chargetransfercomplex}  To recapitulate, excitation, recombination and relaxation induce shifts which may counteract or synergize, and a complete model of the kinetic processes is required to quantitatively predict the steady photovoltaic current.\cite{sturman_ballisticandshift} \\

\begin{figure}[H]
\centering
\includegraphics[width=11 cm]{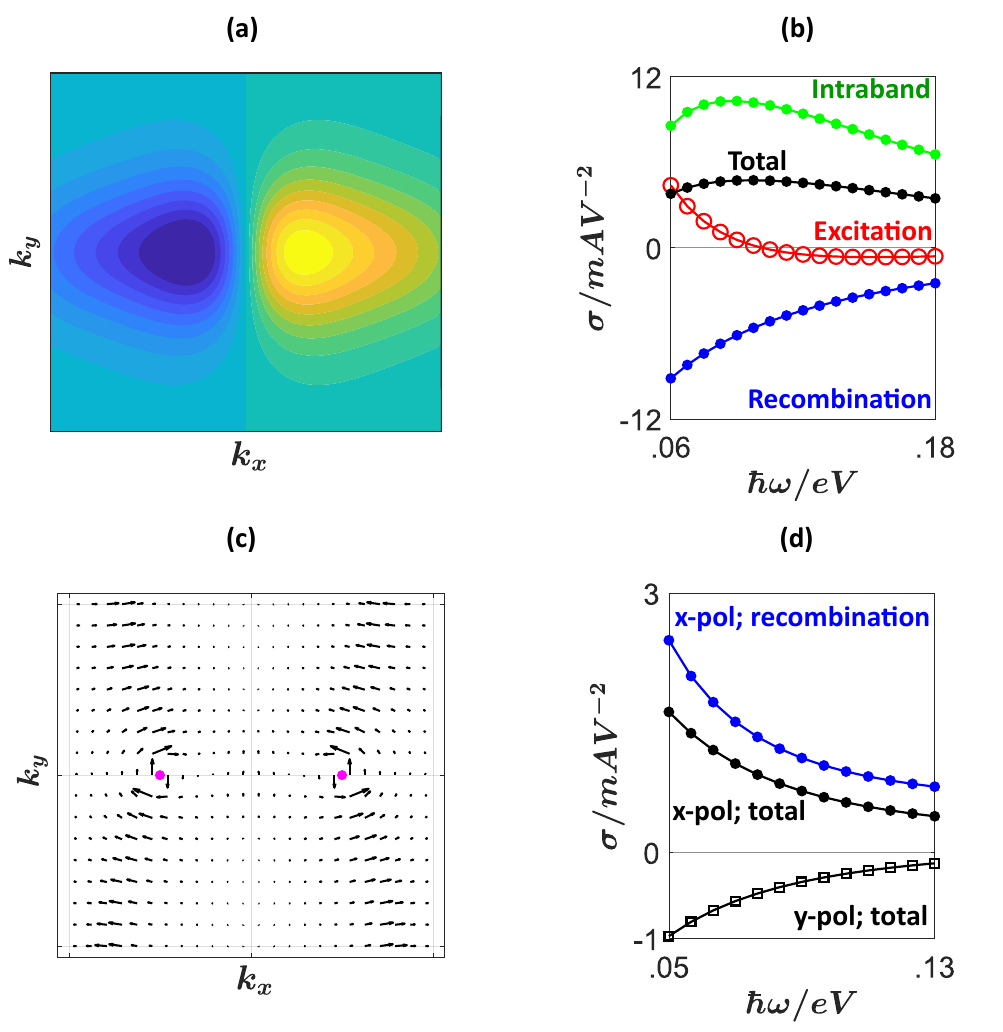}
\caption{ Panels (a-b) derive from a model Hamiltonian with intraband Berry curvature, and (c-d) from a different model Hamiltonian with optical vortices. In panel (a), a time-reversal-symmetric distribution of  intraband Berry curvature is plotted as a color field, with yellow (blue) representing positive (resp. negative) curvature. Panel (b) plots the nonlinear conductivity which is defined though $j_a=\bsigma^a_{\omega}|\cale_{\omega}|^2$, with $\cale_{\omega}$ the amplitude of an incident electric wave of frequency $\omega$, assuming that the light source is unpolarized. The black curve represents the net conductivity $\bsigma^y_{\omega}$, while the red, green, blue curves represent the components of $\bsigma^y_{\omega}$ contributed by interband excitation, intraband relaxation, and interband recombination, respectively. The intraband component is manifestly dominant. (c) When a Bloch quasiparticle (with wavevector $\bk$) is optically excited from the valence to the conduction band, the quasiparticle is displaced by a $\bk$-dependent \sexc vector [\q{photonicshift}]; the circulation of this  vector field has a quantized contribution attributed to optical vortices.
(d)  The black curves represent $\bsigma^y_{\vec{x},\omega}$ for an $\vec{x}$-polarized light source vs  $\bsigma^y_{\vec{y},\omega}$ for a $\vec{y}$-polarized light source; the blue curve represents the recombination component of $\bsigma^y_{\vec{x},\omega}$, which dominates the total current.}
\label{fig:coverpic2}
\end{figure}

While the \sintra/\srec currents have been explored for simplified models of piezoelectrics and pyroelectrics,\cite{belinicher_kinetictheory} there has not been an attempt to relate the \sintra/\srec currents to notions of   quantum wave function geometry which have revitalized the condensed matter field. Here, we identify scenarios (unimagined by BIS) in which the   \sintra  or \srec current dominates over the shift$_{\text{exc}}$ current by an order of magnitude. Such dominance is attributed to two quantum geometric properties of the Bloch wave function, namely (i) time-reversal-symmetric, \textit{intra}band Berry curvature [cf.\ \fig{fig:coverpic2}(a)], which results in an anomalous shift of quasiparticles as they are scatter with phonons, and (ii) topological singularities in the \textit{inter}band Berry phase known as `optical vortices' [cf. \fig{fig:coverpic2}(c)], which makes the photovoltaic current extraordinarily sensitive to the linear polarization vector of the light source.

 Both effects (i-ii) will be demonstrated in model Hamiltonians with generic values of the joint density of states and without assuming a semimetallic band gap. The nonlinear conductivities  in our models are of order $mAV^{-2}$ without finetuning of the incident radiation frequency,\footnote{In contrast,  a number of proposals for large shift currents (at low frequencies) have relied on  small\cite{LiangTan_topoins} or vanishing\cite{ahn_lowfrequencydivergence,chingkit_photocurrentinweyl,xuyang_photovoltaicweyl} band gaps, which makes for a questionable application to solar cells.} as illustrated in \fig{fig:coverpic2}(b) and (d). For comparison with a prototypical ferroelectric, the nonlinear conductivity of PbTiO$_3$ has a maximum (over frequency) of $0.05mAV^{-2}$ when only \sexc is accounted for.\cite{youngrappe_firstprinciples}

\subsection{Outline of paper}

\begin{itemize}
    \item As a preliminary step to substantiating these results, \s{sec:kinetictheory} first reviews BIS's kinetic theory for the shift current\cite{belinicher_kinetictheory} and several salient properties of the non-equilibrium distribution of photo-excited carriers.\cite{esipov_temperatureenergy} In addition, we will formalize an underappreciated distinction between the transient and steady shift currents; in particular, the transient shift current in intrinsic semiconductors will be shown to be identical to the current calculated by Kubo-type perturbation theories (e.g., by Kraut and Baltz,\cite{kraut_anomalousbulkPV,baltz_bulkPV} and by Sipe and Shkrebtii\cite{sipe_secondorderoptical}), which assume  a weak perturbation from thermal equilibirium. The difference between the steady and transient shift currents will turn out to be the sum of the \sintra  and \srec currents.

\item \s{sec:anomalous} demonstrates  the relevance of  \sintra in the presence of time-reversal-symmetric intraband Berry curvature.

\item \s{sec:vortex} demonstrates the relevance of \srec in the presence of optical vortices.  Because \s{sec:anomalous} also introduces our method of calculating the shift current via loops, we recommend that \s{sec:anomalous} be read before \s{sec:vortex}. 

\item Sec.~\ref{sec:casestudy} showcases the importance of both  \sintra and \srec in the 3D polar semiconductor BiTeI, which has an appreciable Berry curvature as well as optical vorticity. 


\item Finally, \s{sec:discussion} summarizes our results, gives directions to finding photovoltaic material with the desired wave function geometry, comments on experimental discrepancies between the transient and steady photovoltaic current,   elaborates on the notion of loop currents, and discusses the potential of shift-current materials for solar cell applications.
\end{itemize}

\section{Kinetic theory of the shift current}\la{sec:kinetictheory}

\begin{figure}[H]
\centering
\includegraphics[width=10 cm]{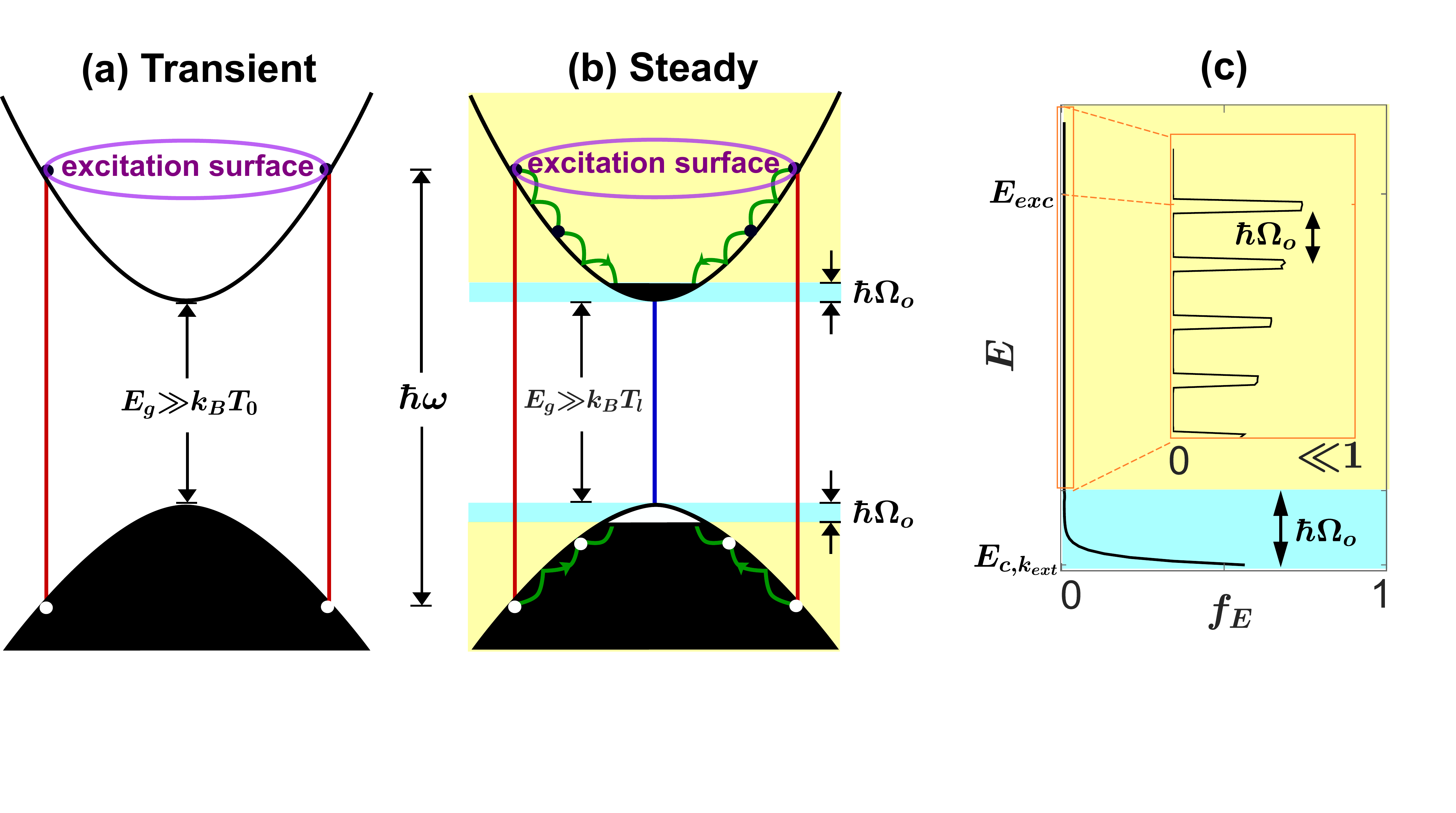}
\caption{(a) The excitation process in a transient state of a homogeneously illuminated semiconductor. White dots against a black background represent hole carriers. The purple ellipse should be understood as a cross-section of an ellipsoidal excitation surface. Panel (b) represents the steady state; yellow and blue energy intervals indicate the active and passive regions respectively. (c) A representative illustration of the quasiparticle distribution $f_{c\bk}$ in the conduction band, with $f_{E}$ being the average of $f_{c\bk}$ for all $\bk$ satisfying $E_{c\bk}=E$. Inset displays the same function $f_E$ in the active region, but with a much finer scale for the horizontal axis. The shape of this distribution is supported by theoretical models [cf.\ \app{app:kineticmodel}] and hot-carrier photoluminescence spectroscopy [e.g., Fig 25 in \ocite{zakharchenya_photoluminescence}].}
\label{fig:activeregion}
\end{figure}

The BIS kinetic theory presupposes that carrier-optical-phonon scattering (rather than carrier-carrier scattering) is the dominant mechanism of energy relaxation for photo-excited carriers in the `active region'. A carrier is said to be in the active region if its energy (defined with respect to the conduction/valence band extremum for an electron/hole) exceeds the optical phonon threshold: $E>\hbar\Omega_{o}$, as illustrated by the yellow energy intervals in \fig{fig:activeregion}(b). The dominance of carrier-optical-phonon scattering over carrier-carrier scattering occurs  for not-too-high carrier densities, which is typical of most continuous-wave laser experiments.\cite{esipov_temperatureenergy,zakharchenya_photoluminescence}\\

The BIS formula for the shift current can be compactly expressed as:
\e{\bj \eq -\f{|e|}{\calv}\sum_{B,B',m } \bS^{m}_{B' \lea B} \bigg( \cala^{m}_{B' \lea B}-\cale^{m}_{B \lea B'} \bigg); \as B=(b\bk),\as B'=(b'\bkp) \as m=(\bq p), \la{bismine}}
with $|e|$ the absolute value of the electron charge, $\calv$ the volume of the medium, and $B=(b\bk)$ a collective label for a Bloch state in band $b$ with wavevector $\bk$. $\sum_{B,B',m}$ sums over all possible  quasiparticle transitions ($B'\lea B$) mediated by a boson of mode $m$; $m=(\bq p)$ is specified by a bosonic wavevector $\bq$ and a bosonic branch/band $p$. $\bS$ and $\cala-\cale$, which appear in the summand of $\sum_{B,B',m}$, will be explained in turn.\\

The shift vector $\bS^{m}_{B' \lea B}$ is the real-space displacement of a Bloch quasiparticle as it transits from $B$ to  $B'$, by way of absorbing/emitting a boson of mode $m$. For phonons,
\e{Phonon:\as \bS^{m}_{B'\leftarrow B}= -(\nabk+\nabkp)\arg V^m_{B',B} +\bA_{b'b'\bkp}-\bA_{bb\bk}= -\bS^{-m}_{B\leftarrow B'},\la{phononicshift}}
with $V^m_{B'B}$ being the electron-phonon matrix element [cf.\ \q{defineUem}], $\bA_{bb'\bk}=\braket{u_{b\bk}}{i\nabk u_{b'\bk}}_{\sma{\text{cell}}}$ the Berry connection\footnote{$u_{b\bk}$ denotes the cell-periodic component of the Bloch function; the inner product $\braket{x}{y}_{\sma{\text{cell}}}$ involves an integral over the intracell coordinate; cf.\ \q{intracell}.}, and $-m=(-\bq p)$ being the momentum-reversed partner of $m=(\bq p)$.\footnote{In the BIS paper, all mode indices were omitted from their phononic shift, and their electron-phonon-matrix element was never explicitly defined, but one may guess that $\bS^{\bkp-\bk,p}_{b\bkp \lea b\bk}$ in our notation corresponds to $\bR_b(\bkp,\bk)$ in their notation.} 
For a photonic mode $m$ with linear polarization vector $\beps_m$,
\e{Photon:\as \bS^{m}_{b'\bk \lea b\bk}\eq -\nabk \arg \beps_m \cdot\bA_{b'b\bk}+\bA_{b'b'\bk}-\bA_{bb\bk}=-\bS^{-m}_{b\bk \lea b'\bk}. \la{photonicshift}}
We have assumed that the photon wavelength greatly exceeds the lattice period; within the dipole approximation, photon-mediated transitions are \textit{vertical} $\equiv \, \bk$-preserving [cf.\ red and blue lines in \fig{fig:activeregion}(b)], and the shift vector  depends on $m=(\bq p)$ only through $\beps_m$. For this reason, we often use $\bS^{m}_{b'\bk \lea b\bk}\equiv \bS^{\beps_m}_{b'\bk \lea b\bk}$ synonymously, when $m$ is photonic.   \q{photonicshift} is henceforth referred to as the \textit{photonic shift vector}, and \q{phononicshift} as the \textit{phononic shift vector}. In either case, the sign of the shift vector is inverted  if  the Bloch labels are interchanged and the bosonic wavevector simultaneously inverted: $\bq \ri -\bq$,  reflecting that forward and backward transitions (between the same pair of Bloch states) result in opposite shifts. \\

For either type of boson, $\cala^{m}_{B' \lea B}$ (resp. $\cale^{m}_{B \lea B'}$) is the transition probability rate for absorbing (resp. emitting) a boson of mode $m$. As explicitly written in \qq{phononrate}{photonrate}, both $\cala$ and $\cale$ have the golden-rule forms that are familiar from Dirac's perturbation theory.\cite{dirac_quantumradiation} In particular, $\cala^m\propto N_m$ and $\cale^m \propto (N_m+1)$, with $N_m$ the average occupancy of the boson $m$. For phonons, $N_m$ is assumed to follow the Planck distribution with lattice temperature $T_l$; for photons, $N_m$ is a sum of  thermal and non-thermal contributions, with the latter being generated by the light source. Additionally, both $\cala$ and $\cale$ depend on the quasiparticle distribution functions $f_{B}$ in a manner consistent with Pauli's exclusion principle: $\cala^{m}_{B' \lea B}\propto (1-f_{B'})f_{B}$ and $\cale^{m}_{B \lea B'}\propto (1-f_{B})f_{B'}$. Consequently, $\bj=\bj[f_{B},N_m]$ depends on $f_{B}$ and $N_m$ through $\cala$ and $\cale$; however, the dependence on $N_m$ will subsequently be made implicit: $\bj[f_{B},\mathsout{N_m}]$, to simplify notation.  \\

Our expression for the shift current is derived in \app{app:derivebis} and is slightly more general than the expression presented in the BIS paper,\cite{belinicher_kinetictheory} in that ours allows for interband, phonon-mediated transitions while theirs do not.\\

Let us consider three scenarios for the quasiparticle and bosonic distributions:\\

\noi{I-no source} Without photo-excitation by a source, quasiparticles, phonons and photons are all thermalized with an equilibrium temperature $T_0$, and \q{bismine} manifestly vanishes owing to   detailed balance: $\cala^m_{B'\lea B}=\cale^m_{B\lea B'}$; cf.\ \q{detailedbalance}. \\

\noi{II-transient state} This balancing is disrupted when the light source is switched on.  At the onset of radiation, the quasiparticle distribution retains its equilibrium value (the Fermi-Dirac function $f^{T_0}_{B}$) but the non-thermal photons drive a transient current $\jtran=\bj[f^{T_0}_{B}]$, which is purely attributed to vertical, interband transitions throughout the \textit{excitation surface}, as illustrated in \fig{fig:activeregion}(a).  The excitation surface ($ES$) is defined as the surface in the Brillouin zone where the difference in conduction- and valence-band energies equals the source photon energy: $E_{c\bk}-E_{v\bk}=\hbar\omega$. The meaning of being `purely attributed' is that the sum over all quasiparticle transitions [in \q{bismine}] is contributed nontrivially only by vertical transitions throughout the excitation surface, i.e., the value of $\bj[f^{T_0}_{B}]$ does not change  if the summation  $\sum_{\bk,\bkp}$ is restricted such that $\bk=\bkp$ lies on the excitation surface:\footnote{This is proven more elaborately in \app{app:jtranjexc}.}
\e{\jtran=\bj[f^{T_0}_{B}]= \bj[f^{T_0}_{B}]_{\bk=\bkp\in ES}. \la{jtran}}
If the non-thermal photons are well approximated by a classical electromagnetic wave, then $\bj[f^{T_0}_{B}]$ reduces to the Kraut-Baltz-Sipe-Shkrebtii formula\cite{kraut_anomalousbulkPV,baltz_bulkPV,sipe_secondorderoptical} calculated by Kubo-type perturbation theory, as demonstrated in \app{sec:krautbaltz}. \\

\noi{III-steady state} For $t\gg \tau_{\text{rec}} \sim 1ns$ (a typical time scale for radiative interband recombination\cite{esipov_temperatureenergy,sturmanfridkin_book}),  the transient current evolves to a steady current: $\bj[f_{B}]$, with the non-equilibrium distribution $f_{B}$ being the steady solution to a kinetic equation encoding all the processes in \fig{fig:coverpicture}(a); cf.\  \app{app:kineticmodel}. The difference between the equilibrium $f_B^{T_0}$ and the non-equilibrium $f_B$ is caricatured in \fig{fig:activeregion}(a) vs (b).
Henceforth, $\bj[f]$ and $\bj[f^{T_0}]$ will be our shorthand for the steady and transient currents, respectively.\\

While in principle the BIS formula [\q{bismine}] for $\bj[f]$ sums over all possible quasiparticle transitions, it is worth in practice to identify the \textit{predominant} transitions that make an outsize contribution to the summation; throughout this work, our use of `predominant' should be understood as significantly contributing to the \textbf{steady}, non-equilibrium shift current. \\

It is simplest to consider the predominant transitions in an intrinsic, direct-gap semiconductor, with $E_{c\bk}-E_{v\bk}$ minimized at a single wavevector: $E_{c\kext}-E_{v\kext}=E_g$. The band gap is assumed to exceed the optical phonon energy ($E_g\gg\hbar\Omega_{o}$), such that phonon-mediated transitions  are intraband;  cf.\ green curves in \fig{fig:activeregion}(b). A typical electron-optical-phonon scattering time is $\tau^o\sim 100 fs$.\cite{lundstrom_book,na_damascelli_electronphonon}
The lattice temperature is assumed to be small ($k_BT_l\ll E_g,\hbar\Omega_o)$, so that the  emission of optical phonons outweighs the absorption. Supposing that carriers are optically excited into the active region with energy $E_{\text{exc}}$,  then the transitions illustrated in \fig{fig:activeregion}(b) predominate. Indeed, the vast difference in relaxation time scales: $\tau^o\ll \ll\tau_{\text{rec}}$,\cite{esipov_temperatureenergy,zakharchenya_photoluminescence}  favors fast, intraband   transitions by emission of optical phonons with energy $\geq \hbar\Omega_{o}$; carriers quickly relax into a \textit{passive region}, defined as the energy interval near a band extremum where the carrier energy $E<\hbar\Omega_o$ [cf.\ blue interval in \fig{fig:activeregion}(a)]; carriers in the passive region can no longer relax via optical phonons, and remain in the passive region till they are annihilated in the slower process of radiative recombination.\footnote{ Auger recombination empirically occurs at much higher photo-excited carrier densities than the present consideration.\cite{esipov_temperatureenergy}} The majority of photo-excited carriers are thus contained within the passive region, with a steady distribution that depends on whether  electron-electron scattering or electron-acoustic-phonon scattering is the dominant mechanism for energy relaxation in the passive region [\fig{fig:activeregion}(c)].  However, fine-grained details about the carrier distribution within the passive region do not matter to  estimating the shift current, because the optical phonon threshold $\hbar\Omega_o$ is typically a small fraction of the band gap $E_g$, and $E_g$ is the energy scale  for significant variations of the energy-dependent shift vectors.\footnote{The majority of recombination transitions occur at $\bk$ points close to $\kext$ and contained within the passive region. Each recombination transition is  associated with a photonic shift $\bS^{\beps}_{v\bk \lea c\bk }$ [\q{photonicshift}], which may as well be approximated as $\bS^{\beps}_{v\kext \lea c\kext }$, because the variation of the photonic shift vector within the passive region is small. Likewise, the current induced by phonon-mediated transitions within the passive region is outweighed by the current induced by phonon-mediated transitions outside the passive region, assuming that the active region is much bigger than the passive region. This assumption holds for most radiation frequencies, because $\hbar\Omega_{o}$ is a tiny fraction of the band width. To formalize this discussion, one may split the line integral in \q{sintra} to a short-line integral within the passive region and a long-line integral within the active region; the long-line integral dominates, because the Berry curvature typically varies on the scale of $E_g \gg \hbar\Omega_o$.} For additional details on the predominant relaxation mechanisms in a direct-gap semiconductor, we refer the reader to App. A and B.\\

 By decomposing the BIS formula [\q{bismine}] according to the three classes of transitions sketched in \fig{fig:coverpicture}(a), one obtains $\bj[f]=\jexc+\jintra+\jrec$, which is the precise meaning of the shift$_{\text{exc}}$, shift$_{\text{intra}}$ and \srec currents  mentioned colloquially in our introduction. To clarify, the \textit{intraband current} is extracted from \q{bismine} by restricting the band summations $\sum_{b,b'}$ by the condition $b=b'$: $\jintra =\bj[f]_{b=b'}$; the \textit{excitation-induced current}  is extracted  by restricting the wavevector summations $\sum_{\bk,\bkp}$ with the condition that $\bk=\bkp$ lies on the excitation surface: $\jexc=\bj[f]_{\bk=\bkp\in ES}$; the \textit{recombination-induced current} $\jrec$ restricts $\bk=\bkp$ to lie outside the excitation surface.  Explicit expressions for the threefold decomposition of $\bj$ are given in  \app{app:threefold}. \\

Let us argue that $\jexc$ in a steady state is well approximated by $\jtran$ in a transient state. Our argument  relies on the following property of
 the non-equilibrium quasiparticle distributions: for all
 $\bk \in ES, f_{c\bk}\ll 1$ and $(1-f_{v\bk})\ll 1$, as illustrated by the inset in \fig{fig:activeregion}(c).
The smallness of $f_{c\bk}$ and $(1-f_{v\bk})$ originates from the slowness in optical excitations compared to the fastness of inelastic collisions  by carrier-carrier and carrier-phonon scatterings. In other words, despite the continuous generation of electron-hole pairs by photon absorption, inelastic scattering processes are so efficient that the non-equilibrium carrier distribution (over the excitation surface) never builds up to significance; this statement is derived  rigorously in \app{app:jtranjexc}. Thus for the purpose of computing the excitation-induced current $(\jexc=\bj[f]_{\bk=\bkp\in ES})$ in a non-equilibrium state, one may as well input the equilibrium distribution: $\jexc\approx \bj[f^{T_0}]_{\bk=\bkp\in ES}$, since it also holds that $f^{T_0}_{c\bk}\ll 1$ and $(1-f^{T_0}_{v\bk})\ll 1$, assuming $k_BT_0 \ll E_g$. Recalling a similar expression for the transient current in \q{jtran}, we deduce that  $\jtran \approx \jexc,$  implying that $\jintra+\jrec$ is precisely what is missed from previous Kubo-type theories\cite{kraut_anomalousbulkPV,baltz_bulkPV,sipe_secondorderoptical} that purport to calculate a steady shift current.

\section{Anomalous shift} \la{sec:anomalous}

In connection to $\jintra$, our first main result is  that the phononic shift induced by  small-angle, intraband  scattering is  expressible in terms of the intraband Berry curvature:
\e{Phonon:\as \bS^{m}_{b\bkp\leftarrow b\bk}=\bS^{ano}_{b;\bkp\leftarrow \bk} + O(\delta k^3), \as \bS^{ano}_{b;\bkp\leftarrow\bk} = \bOmega_{b\bk_{ave}}\times \delbk,\la{anomalousshift}}
with the curvature defined as $\bOmega_{b\bk}=\nabla \times \bA_{bb\bk}$; $\kave =({\bk+\bkp})/{2}$ and $\delbk = \bkp-\bk$  are the average and difference in quasiparticle wavevectors, and $\delta k=||\delbk||$. The \textit{anomalous shift} ($\bS^{ano}$) is purely a geometric property of the quasiparticle wave function, and is insensitive to the nature of the electron-phonon coupling; such coupling affects the shift current only through the transition rate; cf.\ \q{phononrate}.\\

Our use of `anomalous' evokes a comparison with the anomalous velocity correction in the semiclassical equation of motion,\cite{chang_niu_hyperorbit,sundaram1999} which  gives an anomalous displacement:  $\delta{\br}_{ano}=\Omega_{b\bk}\times \delta{\bk}$ for a wave packet of Bloch states in band $b$. In the photovoltaic context, $\delta{\bk}$ is driven by a phonon-induced electric field rather than an externally applied field. Indeed,   phonons in non-centrosymmetric semiconductors induce macroscopic electric fields, which cause the electron-phonon matrix element  $V^m_{b\bkp,b\bk}$   to diverge as $\delbk\ri \bze$.\cite{gantmakherlevinson_book}  In the self-consistent-field approximation,\cite{vogl_electronphonon}   $V^m_{b\bkp,b\bk}=f^m_{\delbk}\braket{u_{b\bkp}}{u_{b\bk}}_{\sma{\text{cell}}}$ plus asymptotically irrelevant terms; $f^m_{\delbk}$ diverges as $1/\delta k$ for `polarization scattering'\cite{frohlich_polarscattering,frohlichmott_meanfreepath} with optical phonons, and as $1/\delta k^{1/2}$ for `piezo-acoustic scattering'\cite{meijer_piezoacoustic} with acoustic phonons.\footnote{The explicit expressions for $f^m$ can be found in equation (3.12) of \ocite{vogl_electronphonon} in the case of `polarization scattering', and in the sum of (3.15) and (3.16) for the case of `piezo-acoustic scattering'. In the general case, $f^m$ may have an anisotropic dependence on $\delbk$, but this does not affect the power exponent of the divergence. In the case of optical phonons, the divergence is cut off by a minimal $\delta k_{cut}$ which is determined by the minimal optical phonon energy $\hbar\Omega_{o}$; because $\hbar\Omega_o$ is much smaller than typical band widths, $\delta k_{cut}$ is much smaller  the Brillouin-zone period. The author of \ocite{vogl_electronphonon}, P. Vogl, dropped the factor $\braket{u_{b\bkp}}{u_{b\bk}}_{\sma{\text{cell}}}$ from all their long-wave-length expressions for the  electron-phonon matrix element [including (3.12), (3.15) and (3.16)], based on the fallacious belief that $\braket{u_{b\bkp}}{u_{b\bk}}_{\sma{\text{cell}}}=1+O(\delta k^2)$ can be chosen as a gauge choice for the wave function.  The error in this belief is explained in \app{app:deceptive}.} 
(With the possible exception of small-gap semiconductors, the polarization and piezo-acoustic scatterings typically dominate\cite{gantmakherlevinson_book} over the deformation scattering.\cite{bardeen_deformationpotential}) Let us substitute $V^m_{b\bkp,b\bk}$ in \q{phononicshift} with its  asymptotically dominant contribution. Since the symmetrized derivative of any function of $\delbk$ vanishes, we are led to evaluate
$(\nabk+\nabkp)\arg \braket{u_{b\bkp}}{u_{b\bk}}_{\sma{\text{cell}}} = \nabla_{\kave}(\bA_{bb\kave}\cdot\delbk)+O(\delta k^3).$ Subsequently applying the identities $(\nabla_{\kave}\times \bA)\times \delbk=(\delbk\cdot \nabla_{\kave})\bA-\nabla_{\kave}(\bA\cdot \delbk)$ and $\bA_{\bkp}-\bA_{\bk}=(\delbk\cdot \nabla_{\kave})\bA +O(\delta k^3)$, one obtains the anomalous shift in \q{anomalousshift}.\footnote{\app{app:alternative}	describes an alternative derivation of the anomalous shift by identifying $-\nabla_{\kave}(\bA_{\kave}\cdot\delbk)+\bA_{\bkp}-\bA_{\bk}$ as a line integral of the Berry connection over an infinitesimally thin parallelogram centered at $\kave$. This somewhat demystifies the appearance of the Berry curvature. Just as the intraband anomalous shift vector is expressible in terms of  geometric quantities over an intraband loop in momentum space, so is the excitation shift vector (generalized to non-vertical transitions) expressible in terms of geometric quantities over an interband loop~\cite{shi_geometric,wang_generalized}.}\\

\begin{figure}[H]
\centering
\includegraphics[width=13 cm]{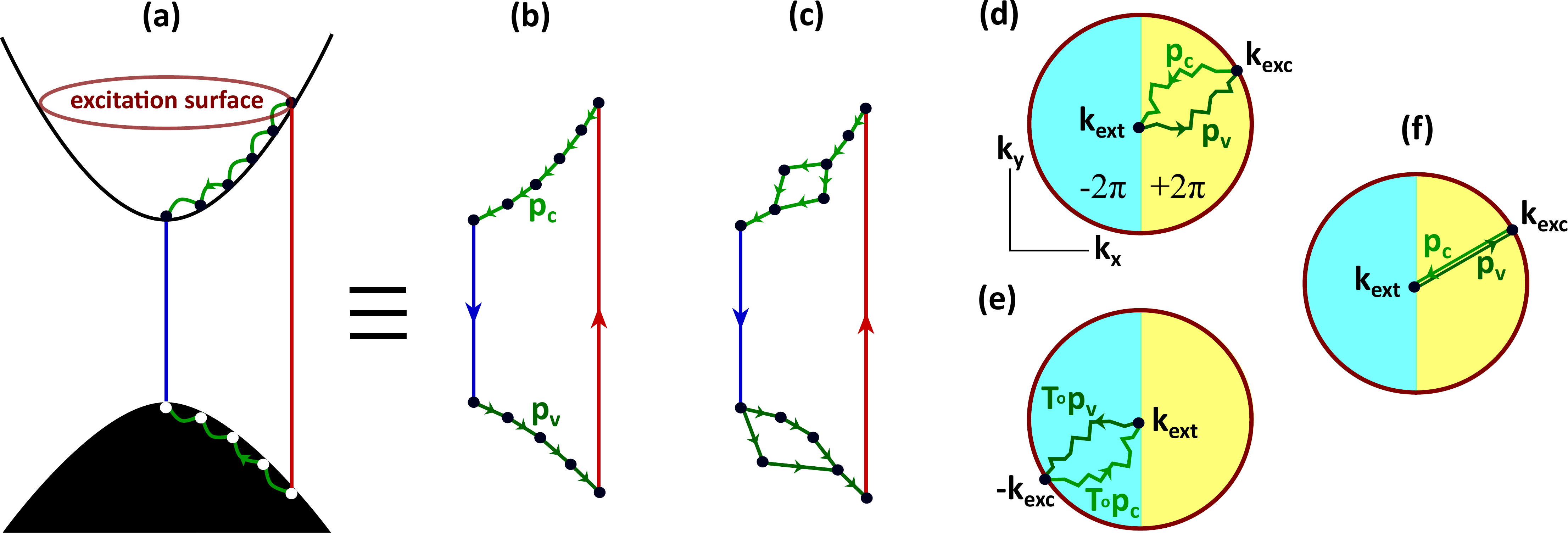}
\caption{Viewing a closed electron-hole trajectory [panel (a)] as an electron loop [panel (b)]. (c) An electron loop with a nonunique relaxation pathway.  (d) Top-down view of the same electron loop in panel (b). The excitation surface encloses a $2\pi$ flux of Berry curvature ($\Omega_{c\bk}^z$) in the yellow region, and a negative $2\pi$ flux in the cyan region.  (e) The time-reversed counterpart to the loop in panel (d). (f) Approximating the relaxation pathways as geodesic paths.  }
\label{fig:loops}
\end{figure}

The anomalous shift induces a large $\jintra$ if the {excitation surface} encloses a  time-reversal-symmetric distribution of Berry curvature. The minimal model to demonstrate this effect is quasi-two-dimensional, meaning that the band energies and cell-periodic wave functions $\ket{u_B}_{\sma{\text{cell}}}$ are approximately independent of one component of $\bk$, say, $k_z$. Let us consider a quasi-2D excitation surface that encircles a $2\pi$-quantum of Berry flux ($2\pi=\int\int \Omega^z_{c\bk}dk_xdk_y$) in the positive-$k_x$ half plane [yellow region in \fig{fig:loops}(d)]; by time-reversal symmetry, the same excitation surface must encircle a $(-2\pi)$-quantum of Berry flux in the negative-$k_x$ half plane [cyan region in \fig{fig:loops}(d)]. In short, we simply say that the excitation surface encloses a \textit{time-reversal-symmetric Berry flux} of $2\pi$.\footnote{In the quasi-2D context, the time-reversal-symmetric Berry flux is defined as $\int\int\Omega_{c}^z\Theta(k_x>0)dk_xdk_y$, with $\Theta(k_x>0)$ a projector to positive values of $k_x.$}\\

How do $\jintra, \jexc$ and $\jrec$ compare in this minimal model? Before getting too quantitative, one may gain some qualitative insight from comparatively evaluating \sintra, \sexc and \srec for  the representative electron-hole trajectory in \fig{fig:loops}(a), which describes the photoexcitation of an electron-hole pair at $\kexc$ on the excitation surface, the relaxation of the excited electron (hole) in the conduction (resp.\ valence) band,  and recombination at $\kext$.  Viewing a forward-moving hole as a backward-moving electron, this electron-hole trajectory becomes an oriented electron loop, which we denote as $loop[\kexc]$. This loop concatenates two interband links with two intraband pathways, $p_c(\kexc)$ and $p_v(\kexc)$, which correspond respectively to the conduction and valence band; cf.\ \fig{fig:loops}(b). The net shift associated to this loop, which we call the \textit{shift loop} $\equiv \bS_{loop[\kexc]}$,  is the summation of shift vectors over all one-electron transitions that make up the loop:\footnote{A general definition of the shift loop is given in \q{sloop} which applies beyond direct-gap semiconductors.} 
\e{\bS^{\beps}_{loop[\bk]}\eq  \bS^{\beps}_{\text{exc},\bk}+\bS_{\text{rec}}+\bS_{\text{intra},\bk} \la{decomposeshiftloop}\\
\bS^{\beps}_{\text{exc},{\bk}} \eq \bS^{\beps}_{c\bk \lea v\bk}, \la{sexc} \\ 
\bS_{\text{rec}}\eq \f{\int d\lambda_{\hbq} \sum_{p=1}^2|\beps_{\bq p}\cdot \bA_{cv\bk}|^2\;\bS^{\beps_{\bq p}}_{v\bk\lea c\bk}}{\int d\lambda_{\hbq} \sum_{p=1}^2|\beps_{\bq p}\cdot \bA_{cv\bk}|^2}\bigg|_{\bk=\kext},\la{srec}\\
\bS_{\text{intra},\bk} \eq \int_{p_c(\bk)}\bOmega_c \times d\bk + \int_{p_v(\bk)}\bOmega_v \times d\bk. \la{sintra}}
The first line [\q{decomposeshiftloop}] represents the threefold decomposition of  the shift loop into its excitation, recombination and intraband components. Assuming the light source is  linearly polarized with polarization vector $\beps_s$, the shift loop depends on  $\beps_s$  through the excitation-induced $\bS_{\text{exc}}$; cf.\ \q{sexc} with \q{photonicshift}.  The recombination shift $\bS_{\text{rec}}$ [\q{srec}] is an average of the photonic shift vector over all possible modes of the spontaneously-emitted photon: that is to say, fixing the photon energy by $\hbar c|\bq|=E_{c\kext}-E_{v\kext}$, one averages over all directions for $\bq$ (parametrized by the solid angle $\lambda_{\hbq}$) and over all transverse polarizations $\beps_{\bq p}$; 
this average is weighted by the transition rate, which is proportional to the square of the interband Berry connection  by the golden rule; cf.\ \q{photonrate}. Finally, we have taken the liberty of approximating the summation (over small momentum jumps) as line integrals over $p_c$ and $p_v$. \\

Let us argue for our minimal model that no symmetry enforces $\bS_{\text{intra}}=0$. Being quasi-two-dimensional implies that only the z component of $\bOmega_{c/v,\bk}$ is nonzero. As illustrated in \fig{fig:loops}(d), both relaxation pathways ($p_c$ and $p_v$) lie in the $\bk$-region with positive Berry curvature ($\Omega^z_c$) for the conduction-band states. In the two-band approximation, the Berry curvature of conduction- and valence-band states sum to zero: $\Omega^z_{c\bk}=-\Omega^z_{v\bk}$,\cite{spinorbitfree_AAchen} but this does not imply a cancellation in \q{sintra}, because $p_c$ and $p_v$ are oppositely oriented. There is also no cancellation with the time-reversed loop, which is indicated by $T\circ p_c$ and $T\circ p_v$ in \fig{fig:loops}(e): under time reversal, $\Omega^z_c \ri -\Omega^z_c$, but the orientation of $T\circ p_c$ is opposite to that of $p_c$. In the absence of symmetry-enforced cancellations, one expects that an anomalous shift current is a generic consequence of enclosed, time-reversal-symmetric Berry flux. This does not violate any symmetry principle, because optical excitation creates a non-equilibrium state with an arrow of time; this arrow manifests in the orientation of our loops. Our argument for the anomalous shift is widely generalizable: one may imagine a greater variety of enclosed time-reversal-symmetric Berry flux for which the line integrals in \q{sintra} are nonvanishing, and such imagination need not be restricted to semiconductors.\\

Moving beyond qualitative arguments, we would like to   quantify  the current for our quasi-2D semiconducting model. However, a brief diversion  is required to explain the calculational method we invented. Our method  introduces the \textit{affinity shift loop} as a figure of merit for the shift current:
\e{ASL_{\beps_s,\omega}\; = \big\langle\; f_{vc\bk} |\beps_s\cdot \bA_{cv\bk}|^2\bS^{\beps_s}_{loop[\bk]}\;\big\rangle_{\omega}; \as f_{vc\bk}=f_{v\bk}-f_{c\bk},\la{aveshiftloop}}
with $\langle \ldots \rangle_{\omega}$ meaning to average over all $\bk$ on the excitation surface:
\e{\langle \Xi(\bk)\rangle_{\omega} \eq \int \frac{d^3k}{(2\pi)^3}\;\f{\delta(E_{cv\bk}-\hbar\omega)}{\jdosup}\;\Xi(\bk); \as E_{cv\bk}=E_{c\bk}-E_{v\bk}. \la{averageES}}
$\jdosup$ stands for the  joint density of states for quasiparticles of one spin orientation:
\e{\jdosup \eq \int \f{d^3k}{(2\pi)^3} \,\delta(E_{cv\bk}-\hbar\omega). \la{definejdos}}

What is being averaged in \q{aveshiftloop} is  the shift loop weighted by the rate of optical excitations,   with $\beps_s$ the polarization vector of the light source; by Fermi's golden rule, this rate is proportional to $|\beps_s\cdot \bA_{cv\bk}|^2$, which we will refer to as the \textit{optical affinity}. The shift loop  is defined in \qq{decomposeshiftloop}{sintra}, with $p_c(\kexc)$  (resp.\ $p_v(\kexc)$) chosen to be the unique oriented path that (i) connects $\kexc \ri \kext$ (resp. $\kext\ri \kexc$), and  (ii) is tangential to $\bv_{c\bk}=\nabk E_{c\bk}$  (resp. $\bv_{v\bk}=\nabk E_{v\bk}$) at all points along the path.   We refer to $p_b$ (with $b=c$ or $v$) as a \textit{geodesic path}.\footnote{Construct the four-momentum $(P^0,P^1,P^2,P^3)=(E_{c\bk}/||\bv_{c\bk}||,\hbar k_x,\hbar k_y,\hbar k_z)$ and introduce  the Lorentzian metric $g_{\mu\nu}$, with $g_{11}=g_{22}=g_{33}=-g_{00}=-1$. Then $p_c(\kexc)$ can be viewed as a path $P^{\mu}(\lambda)$ that minimizes the action $\int_0^1 \surd [-g_{\mu\nu} \tfrac{dP^{\mu}}{d\lambda}\tfrac{dP^{\nu}}{d\lambda}] d\lambda $, given that the end points of the path are fixed to $\kexc$ and $\kext$.} If $E_{B}$ is an isotropic function of $\bk$, then  the geodesic path is simply the Euclidean-straight path connecting $\kexc$ to $\kext$; cf.\ \fig{fig:loops}(f). The motivation for geodesic paths is that the predominant relaxation pathways {[\fig{fig:loops}(d)]} do not deviate far from being geodesic [\fig{fig:loops}(f)]: each time a quasiparticle in the conduction band emits an optical phonon, the likeliest transition involves the smallest wavenumber change $\delta k$ [\fig{fig:loops}(a)],  since the electron-phonon matrix element diverges as $1/\delta k^2$;\cite{vogl_electronphonon} minimizing $\delta k=||\bkp-\bk||$ with the constraint  $E_{c\bk}-E_{c\bkp}=\hbar\Omega_o$ is approximately equivalent to   $\bk-\bkp$ being parallel to $\bv_{c\bk}$, given that the optical phonon energy $\hbar\Omega_o$ is small compared to typical band widths.\\

Defining the \textit{shift conductivity} through  $\bj[f]=\bsigma_{\eps_s,\omega}|\cale_{\omega}|^2$, with $\bj[f]$ the steady shift current given by the BIS formula [\q{bismine}], and [$\boldsymbol{\cale}(\br,t)=\beps_s\cale_{\omega}e^{i(\bq\cdot\br-i\omega t)} +$ complex conjugate]  being the incident electric wave, the shift conductivity relates to our figure of merit through
\e{\bsigma_{\beps_s,\omega} \approx - 1.53 mAV^{-2} \; \f{ASL_{\beps_s,\omega}}{\calv_{\sma{\text{cell}}}}\; \f{2\jdosup}{(\calv_{\sma{\text{cell}}}eV)^{-1}},\la{shiftconduct}}
with $1.53 mAV^{-2}=2\pi |e|^2/\hbar V$ in SI units, and $\calv_{\sma{\text{cell}}}$ being the volume of the primitive unit cell.\footnote{\q{shiftconduct} applies to direct-gap semiconductors in which  a single conduction and a single valence band are optically excited in the vicinity of a single extremal wavevector $\kext$.
In direct-gap semiconductors with multiple valleys/pockets (indexed by $v$), such as transition metal dichalcogenides, the total shift conductivity is proportional to $\sum_{v} ASL_{v} JDOS_{v}$.}    Because the shift loop  is threefold decomposable according to  \q{decomposeshiftloop}, one may likewise decompose 
\e{
\bsigma_{\beps,\omega}=\bsigma^{\text{exc}}_{\beps,\omega}+\bsigma^{\text{rec}}_{\beps,\omega}+\bsigma^{\text{intra}}_{\beps,\omega}, \la{decomposesigma}
}
with  $\bsigma^{\text{exc}}_{\beps,\omega}$ matching the Kraut-Baltz-Sipe-Shkrebtii formula\cite{kraut_anomalousbulkPV,baltz_bulkPV,sipe_secondorderoptical} from Kubo-type perturbation theory; cf.\  \app{sec:krautbaltz}. \\

Our relation between the shift conductivity and the affinity shift loop [\q{shiftconduct}] holds  at low temperature ($k_BT_l\ll \hbar\Omega_{o},E_g$) and for small optical phonon energy (compared to the band gap and the largest energy of  photo-excited carriers).\footnote{ The largest energy of photo-excited electrons is represented as $E_{\text{exc}}-E_{c,\kext}$ in \fig{fig:activeregion}(c).} The right-hand side of  \q{shiftconduct} should be understood as an approximation to the BIS formula [\q{bismine}]; the major error in this approximation originates from fixing $p_b$ to be a geodesic path, hence we refer to \q{shiftconduct} as the \textit{geodesic approximation} to the shift conductivity. In reality, an electron excited at $\kexc$ follows multiple relaxation pathways [as caricatured in \fig{fig:loops}(c)] which deviate from being geodesic and narrow. The geodesic approximation is therefore justified to the extent that small-angle scattering dominates over large-angle scattering.\footnote{The BIS formula [\q{bismine}] reduces asymptotically to the geodesic approximation [\q{shiftconduct}] in the limit of vanishing scattering angle, as proven in \app{app:smallanglelimit}. } A benchmarking of the approximation will shortly be presented.\\

While other groups have attempted to optimize the shift conductivity by maximizing the JDOS,\cite{cook_designprinciples} we adopt a wave-function-centric approach in maximizing the affinity shift loop. Assuming a generic value for  $\jdosup\approx (\calv_{\sma{\text{cell}}}eV)^{-1}$,  $||ASL_{\beps_s,\omega} || \sim \calv_{\sma{\text{cell}}}$ implies a conductivity of order $mAV^{-2}$.  \\

Returning to our quasi-2D model, we now demonstrate that  the intraband component [cf.\ \q{sintra}] of the affinity shift loop is indeed comparable in magnitude to $\calv_{\sma{\text{cell}}}$:
\e{ \bigg|\bigg|\;\big\langle (f_{v\bk}-f_{c\bk}) |\beps_s\cdot \bA_{cv\bk}|^2\bS_{\text{intra}, \bk}\big\rangle_{\omega}\bigg|\bigg| \sim \calv_{\sma{\text{cell}}}. \la{indeedcompare}}  
We adopt two heuristic approximations for a back-of-the-envelop calculation, namely that (i) the excitation surface is circular with radius $k_r=\pi/2a$ (assuming $\calv_{\sma{\text{cell}}}= a^3$), and that (ii) in the absence of unusually small band gaps,\footnote{ The Berry curvature only exhibits significant variations over a length scale that is comparable to the inverse of the band gap.} the $2\pi$ Berry flux is roughly homogeneous over the yellow semicircle enclosed by the excitation surface; cf.\ \fig{fig:loops}(d). Because $\Omega^z_{c\bk}=-\Omega^z_{v\bk}$ (in the two-band approximation) and $p_c=-p_v$ (presuming an electron-hole symmetry $E_{c\bk}= -E_{v\bk}$), the anomalous contribution to the shift loop simplifies to $\bS_{\text{intra},\kexc}=2\int_{p_c(\kexc)}\Omega^z_{c\bk}\vec{z}\times d\bk$, with $p_c(\kexc)$ a straight path connecting $\kexc$ to $\kext=(0,0,0)$. $\bS_{\text{intra}}(\kexc)=-16a \vec{y}/\pi$ if $\kexc=(k_r,0,0)$ and vanishes if $\kexc=(0,k_r,0)$; from this one deduces that the average of $\bS_{\text{intra},\kexc}$ over all $\kexc$ on the excitation surface is comparable to $-a\vec{y}$.
Assuming that $f_{v\kexc}-f_{c\kexc}\approx 1$  [cf.\ \s{sec:kinetictheory}], and that the interband connection is  generic-valued: $|\beps_s\cdot \bA_{cv\kexc}|^2\sim a^2$,  one finds   $\langle |\beps_s\cdot \bA_{cv}|^2\bS_{\text{intra}}\rangle_{\omega}\sim -\calv_{\sma{\text{cell}}}\vec{y}$, leading to \q{indeedcompare}. \\

It is of interest to demonstrate that our crudely-derived conclusion holds true in a precise calculation for a model Hamiltonian:\footnote{This $\bk\cdot\bp$ model can be extended to a tight-binding model by recognizing certain terms as Taylor-series coefficients for trigonometric functions.}
\e{H(\bk)=-E_o(\dg{z}\bsigma z) \cdot \bsigma, \as z=\vectwo{\tilde{k}_x-\tilde{k}_x^3/6}{\tilde{k}_y+i(\tilde{Q}-\tilde{k}_x^2-\tilde{k}_y^2)/2},  \as {E_o}=\f1{\tilde{P}}\frac{\hbar^2}{{m_f}a^2}. \la{modelham}}
 $\bsigma=(\bsigma_1,\bsigma_2,\bsigma_3)$ is the vector of Pauli matrices; $\tilde{k}_j=k_ja$ (for $j=x,y,z$) is a dimensionless wavenumber, with $a =5\ang$  a generic value for the lattice period; $m_f$ is the free-electron mass;  $\tilde{P}$ and $\tilde{Q}$ are dimensionless Hamiltonian parameters. Assuming that $\tilde{Q}<2$, \q{modelham} is the Hamiltonian of a direct-gap, quasi-2D semiconductor with band gap $E_g=E_o\tilde{Q}^2/2$, effective masses $m_x=m_y=\tilde{P}m_f/(2-\tilde{Q})$, $m_z=\infty$ (for both electrons and holes), and  $\jdosup \approx \tilde{P}/(8-4\tilde{Q})$ in units of $(\calv_{\sma{\text{cell}}}eV)^{-1}$. We choose $\tilde{Q}=1$ and $\tilde{P}=4$, such that $\jdosup \approx 1$ is generic-valued.\\
 
\begin{figure}[H]
\centering
\includegraphics[width=17 cm]{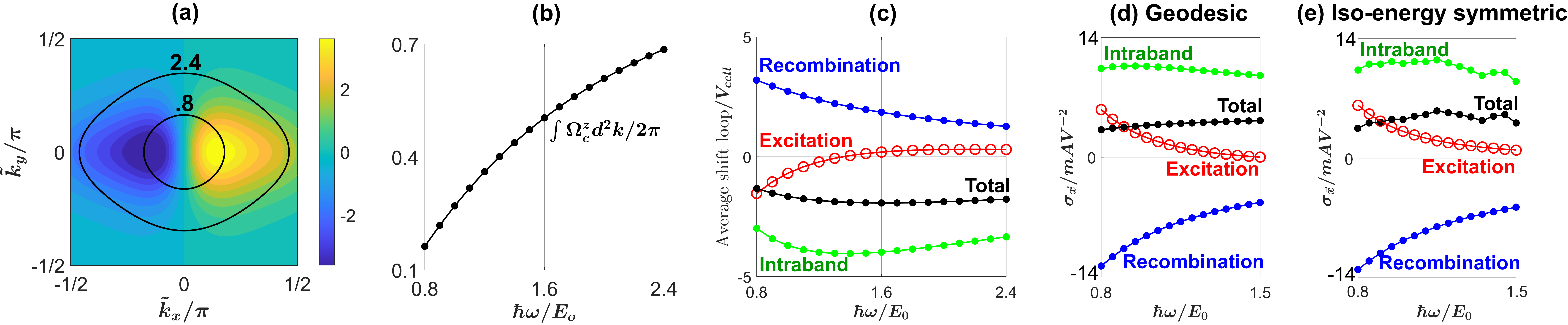}
\caption{Characterization of a quasi-2D model [\q{modelham} with $\tilde{Q}=1$] with a time-reversal-symmetric Berry flux. Panel (a) plots the  Berry curvature  as a colored background, with a color legend (on the right) specifying the value of $\Omega_{c\bk}^z/(\calv_{\sma{\text{cell}}})^{2/3}$; the black ellipses represent excitation surfaces for two source photon energies: $\hbar\omega=2.4E_0$ and $0.8E_0$, with  $E_o\approx 76 meV$. Panel (b) shows the $\omega$-dependence of the time-reversal-symmetric Berry flux enclosed by the excitation surface. The black curve in panel (c) represents the polarization-averaged affinity shift loop [$(1/2)\sum_{\beps_s\in \{\vec{x},\vec{y}\}}ASL^y_{\beps_s,\omega}/\calv_{\sma{\text{cell}}}$] vs $\omega$; the non-black curves represent  the three-fold decomposition of the polarization-averaged affinity shift loop: excitation (red), intraband relaxation (green), and recombination (blue). The shift conductivity $\bsigma^y_{\vec{x},\omega}$ (and its threefold decomposition) is calculated in the geodesic approximation in panel (d), and in the iso-energy symmetric approximation in panel (e).}
\label{fig:Q05_maintext}
\end{figure}

The model Hamiltonian has been chosen because it realizes a time-reversal-symmetric Berry flux of $2\pi=\int_{-\infty}^{\infty}\int_0^{\infty} \Omega_{c}^z dk_xdk_y$,\cite{AA_teleportation,nelsonAA_multicellularity,nelsonAA_delicatetopology} as illustrated in \fig{fig:Q05_maintext}(a). The time-reversal-symmetric Berry flux  enclosed by the excitation surface  varies from roughly $0.7(2\pi)$ [for 
$\hbar\omega=2.4E_0$] to $0.2(2\pi)$ [for 
 $\hbar\omega=0.8E_0$]; cf.\  \fig{fig:Q05_maintext}(b). The case of  $\hbar\omega=2.4E_0$ is not unlike the caricature we drew in \fig{fig:loops}(d).\\

The quasi-two-dimensionality and reflection symmetry ($x \ri -x$) of our model imply that only the y-component of the shift current can be nontrivial. We have numerically computed the y-component of the affinity shift loop  via \qq{decomposeshiftloop}{aveshiftloop} for the chosen model parameters.\footnote{The computation was simplified by setting $f_{v\bk}-f_{c\bk}=1$ in \q{aveshiftloop}, for reasons explained at the end of \s{sec:kinetictheory}. We have also approximated the geodesic paths as straight [cf.\ \fig{fig:loops}(f)], since the energy-momentum dispersion is roughly isotropic.}
\fig{fig:Q05_maintext}(c) plots the polarization-averaged affinity shift loop $[(1/2)\sum_{\beps_s\in \{\vec{x},\vec{y}\}}ASL^y_{\beps_s,\omega}]$   with respect to $\hbar\omega$ in the interval $[0.8E_0,2.4E_0]$; the threefold decomposition of 
$(1/2)\sum_{\beps_s}ASL^y$ is also illustrated.
For a broad range of frequencies ($\hbar\omega>1.3E_0$) where the encircled time-reversal-symmetric Berry flux exceeds $0.4(2\pi)$, the intraband component not only exceeds the excitation component by an order of magnitude, but also carries an opposite sign. These values for the affinity shift loop translates [via \q{shiftconduct}] to a shift conductivity  $\approx 4 mAV^{-2}$ for an unpolarized light source, as illustrated in \fig{fig:coverpic2}(b) for the same frequency range.\\

To benchmark the geodesic approximation that has been used in all conductivity calculations thus far, we also computed $\bsigma^y_{\vec{x},\omega}$ via the more traditional method of numerically simulating a steady quasiparticle distribution $f_B$ that sets the collisional integral to zero, and then inputting $f_B$ into \q{bismine}; the detailed procedure is described in \app{app:benchmark}. Because this procedure is numerically intensive, we resorted to approximating $f_{B}$ as an \textit{iso-energy symmetric} function of $\bk$, meaning $f_{b\bk}$ is constant over iso-energy $\bk$-surfaces of $E_{b\bk}$. \fig{fig:Q05_maintext}(d) and (e) show the same quantity $\bsigma^y_{\vec{x},\omega}$ calculated in the geodesic and iso-energy symmetric approximations, respectively. It is reassuring to see semi-quantitative consistency in the values of the shift conductivity and all its components, especially at lower photon frequencies where the iso-energy symmetric approximation is better justified.\footnote{The $\bk$-dependent transition rate for optical excitation [cf.\ \q{photonrate}] becomes increasingly iso-energy asymmetric at higher frequencies: the standard deviation of $|A^x_{cv}|^2$ (over the excitation surface at frequency $\omega$) increases from 8.6 percent  (of $\langle |A^x_{cv}|^2\rangle_{\omega}$) at $\hbar\omega=0.8E_0$ to 23.8 percent at $\hbar\omega=1.5E_0$. }

\section{Optical vorticity}\la{sec:vortex}

Having demonstrated the dominance of the intraband current $\jintra$  in the presence of time-reversal-symmetric \textit{\text{intra}}band Berry curvature, this section will demonstrate  the dominance of the recombination-induced current $\jrec$  in the presence of optical vortices -- topological singularities in the \textit{inter}band Berry phase. \\

Before discussing vortices properly, we  first consider a vortex-less scenario where the 
photon-mediated current components cancel out: $\jexc+\jrec \approx 0$. In understanding how this cancellation happens, it will become apparent that vorticity is one route to prevent such a cancellation. Let us then hypothesize a scenario where the photonic shift vector is roughly independent of the light polarization, and roughly homogeneous in the $\bk$-region enclosed by the excitation surface. This would imply that the shift vector at excitation [\q{sexc}] is opposite to the shift vector at recombination [\q{srec}], leading to a cancellation of the excitation- and recombination-induced currents; cf.\ \qq{aveshiftloop}{decomposesigma}.\\

\begin{figure}[H]
\centering
\includegraphics[width=12 cm]{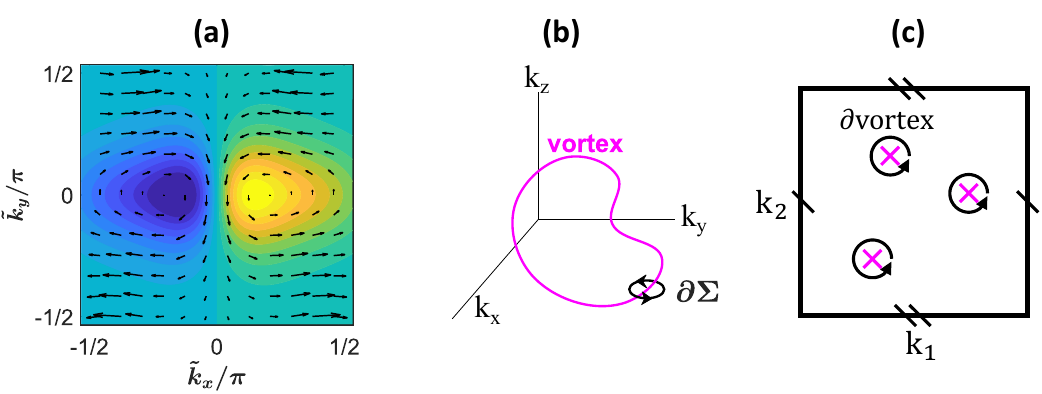}
\caption{ (a) Circulation of the photonic shift vector $\bS^{\vec{x}}_{c\bk\lea v\bk}$ associated to the intraband Berry curvature for our model Hamiltonian in \q{modelham}, with $\tilde{Q}=1$. The curvature is represented with the color legend  in \fig{fig:Q05_maintext}(a). (b) The magenta line represents a singularity in the interband Berry phase, i.e., an optical vortex. (c) Optical vortices appear as points (represented by magenta crosses) in a 2D closed $\bk$-manifold parametrized by $k_1$ and $k_2$.  }
\label{fig:circulation}
\end{figure}

The contrapositive implication of this thought experiment is that for $\jexc+\jrec$ to be significant,  the photonic shift at excitation must differ from the photonic shift at recombination. With great circulation comes great differences! The circulation of the photonic shift vector is defined by integrating \q{photonicshift} over a loop $\partial \bSigma$ in $\bk$-space:
\e{\oint_{\partial \bSigma} \bS^{\beps}_{c\bk\lea v\bk}\cdot \f{d\bk}{2\pi} =-\oint_{\partial \bSigma}  \nabk \arg[\beps\cdot \bA_{cv\bk}]\cdot \f{d\bk}{2\pi} + \int_{\bSigma}\big( \,\bOmega_{c\bk}-\bOmega_{v\bk}\,\big)\cdot \f{d^2\bSigma}{2\pi}. \label{eq:circ}}
The last term is derived by Stokes' theorem, and corresponds to  a generically nonquantized\footnote{With the inclusion of crystallographic symmetry and for a symmetric choice of $\partial \bSigma$, it is possible for the Berry-flux term to be integer-quantized, which makes the circulation of the shift vector a topological invariant.\cite{aa_topologicalprinciple} }  circulation associated to the intraband Berry phase, as illustrated in \fig{fig:circulation}(a) for our model Hamiltonian in \q{modelham}, with $\tilde{Q}=1$. Beyond our model Hamiltonian,  the possibility exists for an integer-quantized circulation stemming from the preceding term which involves the interband Berry connection.\cite{aa_topologicalprinciple,fregoso_opticalzero} One is led to consider an optical vortex -- a line in three-dimensional $\bk$-space where (i) $\beps\cdot \bA_{cv\bk}=0$, and (ii) the argument/phase of $\beps\cdot \bA_{cv\bk}$ winds nontrivially as $\bk$ is varied along any infinitesimal loop linked to  the vortex line, as representatively illustrated by the black curve in \fig{fig:circulation}(b). It is worth distinguishing between  $\vec{x}$-vortices (where $A_{cv\bk}^x=0$) and $\vec{y}$-vortices (where $A_{cv\bk}^y=0$). Because $A_{cv\bk}^x=0$ does not generally imply $A_{cv\bk}^y=0$,  one type of vortex may occur independently of the other.\\

\begin{figure}[h]
\centering
\includegraphics[width=10 cm]{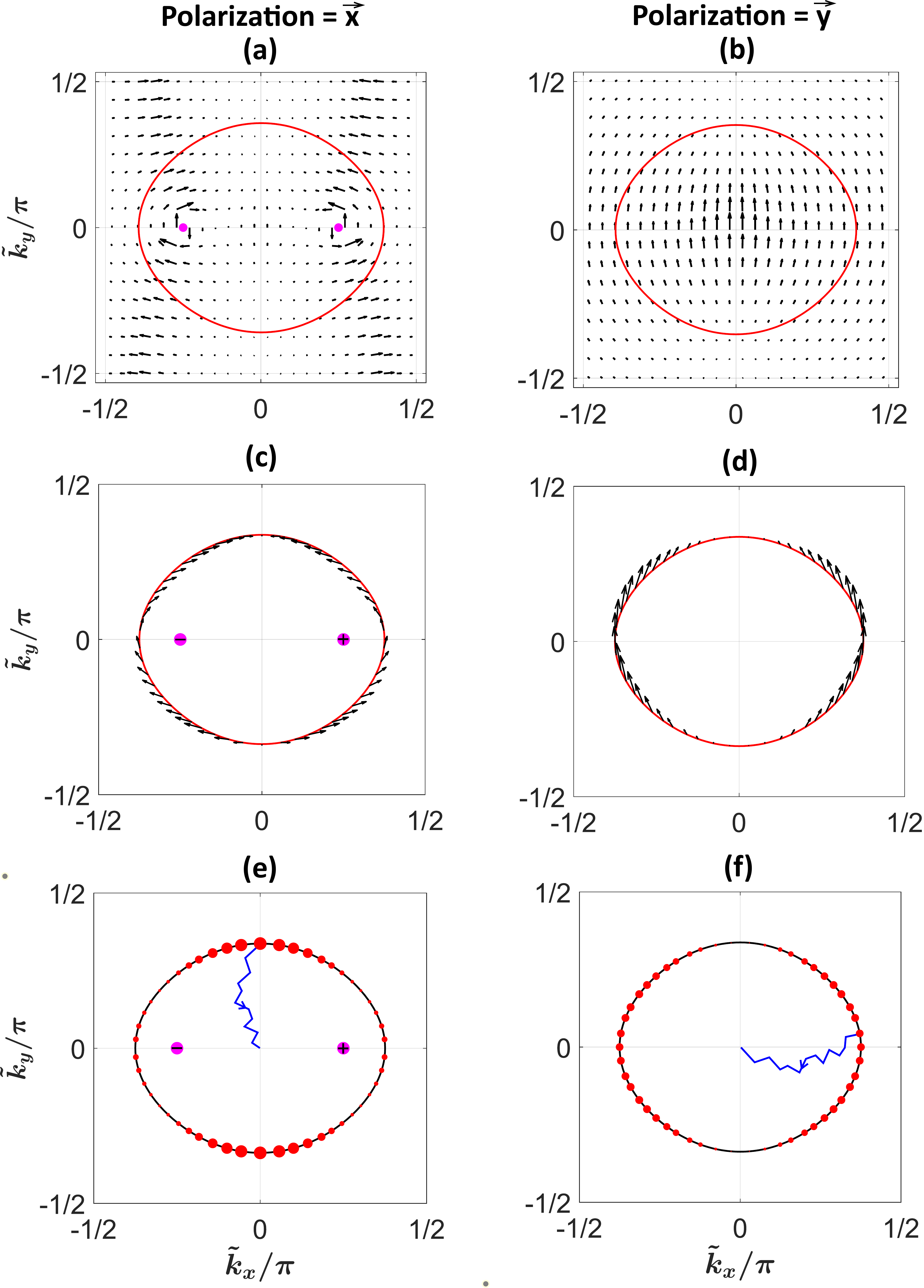}
\caption{Characterization of a quasi-2D model with a pair of $\vec{x}$-vortices, whose $\bk$-locations are indicated by pink dots in panels (a,c,e). Panels (a) and (b) depict the photonic shift vector field $\bS^{\beps_s}_{c\bk\lea v\bk}$, with $\beps_s=\vec{x}$ and $\vec{y}$ respectively. In panels (c) and (d), the red ellipse represents  the same   excitation surface; the arrows on the ellipse represent the vectors $|\beps_s\cdot \bA_{cv\bk}|^2\bS^{\beps_s}_{c\bk\lea v\bk}$ for $\bk$ on the excitation surface. In panels (e) and (f), the size of the red dots indicates the magnitude of $|\beps_s\cdot \bA_{cv\bk}|^2$ for $\bk$ on the excitation surface; the jagged blue line represents a predominant relaxation pathway from the excitation surface to the extremal wavevector: $\kext=\bze$.     }
\label{fig:Qm05_maintext}
\end{figure}

Consider the shift current in a quasi-2D model where $\vec{y}$-vortices are absent [\fig{fig:Qm05_maintext}(b)], but the excitation surface encircles  a pair of $\vec{x}$-vortices which are mutually related by time reversal and mirror reflection [\fig{fig:Qm05_maintext}(a)].\footnote{That optical vortices come in time-reversed pairs was proven in \ocite{aa_topologicalprinciple}.} \fig{fig:Qm05_maintext} is derived from the model Hamiltonian in \q{modelham} with a different set of parameters: $\tilde{Q}=-1$ and $\tilde{P}=12$, but fixed $\jdosup \approx  (eV\calv_{\sma{\text{cell}}})^{-1}$.\\

 Just as for the model studied in \s{sec:anomalous},  only the y-component of the shift conductivity ($\bsigma^y_{\vec{x},\omega}$) is symmetry-allowed to be nonzero.
$\bsigma^y_{\vec{x},\omega}$ for an $\vec{x}$-polarized light source is calculated via \qq{decomposeshiftloop}{shiftconduct}, and is shown in \fig{fig:coverpic2}(d) to be dominated by its recombination component. The same figure shows  the difference between $\bsigma^y_{\vec{x},\omega}$ and  $\bsigma^y_{\vec{y},\omega}$ to be of order $mAV^{-2}$. This represents an extraordinary sensitivity of the shift current to the source polarization, and motivates  $(\bsigma_{\vec{x},\omega}-\bsigma_{\vec{y},\omega}$) as an experimental indicator of $\vec{x}$-vorticity. More generally, for a quantity $B(\beps_s)$ that depends on the source polarization, we refer to $B(\vec{x})\neq B(\vec{y})$ as a \textit{linear disparity} in $B$.\footnote{In contrast, linear birefringence results from a linear disparity in the first-order-in-$\cale_{\omega}$ conductivity, which is associated to an alternating current.} The remainder of this section aims to demystify the large linear disparity in $\bsigma^y_{\beps_s,\omega}$, and the dominance of the recombination-induced current when $\beps_s=\vec{x}$.\\

Let us first understand the linear disparity in the excitation component of the conductivity; cf.\ \q{decomposesigma}. Recall that $\bsigma_{\beps_s,\omega}^{\text{exc}}$ is proportional to a weighted average of the photonic shift vector over the excitation surface, which we have denoted as $\langle\,|\beps_s\cdot \bA_{cv\bk}|^2\bS^{\beps_s}_{c\bk\lea v\bk}\,\rangle_{\omega}$; cf.\ \qq{decomposeshiftloop}{decomposesigma}. A major effect of $\vec{x}$-vorticity without $\vec{y}$-vorticity is that the orientation of $|\bA_{cv}^x|^2\bS^{\vec{x}}_{c\lea v}$, viewed as a vector field over $\bk$-space, tends to be more disordered than the vector field $|\bA_{cv}^y|^2\bS^{\vec{y}}_{c\lea v}$.\footnote{The vector norm of  $|\bA_{cv}^x|^2\bS^{\vec{x}}_{c\lea v}$ is not singular:  as $\bk$ approaches the vortex center, the quantized circulation implies $\bS^{\vec{x}}_{c\bk\lea v\bk}\ri \infty$,  but this divergence is compensated by $|\bA_{cv\bk}^x|^2\ri 0$.\cite{aa_topologicalprinciple} } In particular, along the excitation surface, the orientation of $|\bA_{cv}^x|^2\bS^{\vec{x}}_{c\lea v}$ exhibits rotations which are more pronounced than that of  $|\bA_{cv}^y|^2\bS^{\vec{y}}_{c\lea v}$, as comparatively illustrated in \fig{fig:Qm05_maintext}(c-d). Ceteris paribus, the average of a rotational vector field is smaller than the average of an irrotational vector field, hence  $||\langle |\bA_{cv}^x|^2\bS^{\vec{x}}_{c\lea v}\rangle_{\omega}|| < || \langle |\bA_{cv}^y|^2\bS^{\vec{y}}_{c\lea v}\rangle_{\omega}||$ and $|\vec{y}\cdot\bsigma^{\text{exc}}_{\vec{x},\omega}|<|\vec{y}\cdot\bsigma^{\text{exc}}_{\vec{y},\omega}|$.\\

A different argument is needed to understand  the linear disparity of the intraband conductivity: $\bsigma^{\text{intra},y}_{\beps_s,\omega}$ in  \q{decomposesigma}. For an $\vec{x}$-polarized source, the photon-mediated transition rate depends anisotropically on the orientation of $\kexc$; in particular,    $|A_{cv}^x|^2$ is suppressed  on segments of the excitation surface that are closer to the vortex, where $A_{cv}^x=0$; cf.\ \fig{fig:Qm05_maintext}(e). For a   $\vec{y}$-polarized source, the photon-mediated transition rate is also anisotropic but in the opposite sense: $|A_{cv}^y|^2$ is suppressed near the mirror-invariant line ($k_x=0$), where $A_{cv}^y=0$ by a dipole selection rule; cf.\ \fig{fig:Qm05_maintext}(f).\footnote{The conduction-band states transform in a different representation of mirror than the valence-band states, as detailed in \app{app:vortexmodel}. } The opposite senses of anisotropy imply that the predominant intraband relaxation pathways are roughly  parallel to the $k_y$ axis for an $\vec{x}$-polarized source [\fig{fig:Qm05_maintext}(e)], and parallel to the $k_x$ axis for a $\vec{y}$-polarized source [\fig{fig:Qm05_maintext}(f)]. Ceteris paribus, this implies a larger $\vec{y}\cdot \bsigma^{\text{intra}}_{\vec{y},\omega}$ for a $\vec{y}$-polarized source, because the y-component of the anomalous shift [\q{anomalousshift}] is proportional to the x-component of the momentum transfer: $\vec{y}\cdot \bS^{ano}_{b\bkp\lea b\bk}=\Omega^z_{b\kave}\delta k^x$.\\

A final argument explains the dominance of the recombination-induced current $\jrec$ over the excitation-induced current $\jexc$, for an $\vec{x}$-polarized source. Unlike $\jexc$, $\jrec$ is insensitive to the vortex-induced disordering in the orientation of  $\bS^{\vec{x}}_{c\lea v}$. To understand why, recall that the majority of recombination transitions occur at $\bk$ points close to the extremal wavevector $\kext$ and contained within the passive region; cf.\ \s{sec:kinetictheory} and \q{approxjrec}. Each recombination transition is  associated with a photonic shift $\bS^{\beps_m}_{v\bk \lea c\bk }$, which may as well be approximated as $\bS^{\beps_m}_{v\kext \lea c\kext}$, because the passive region typically occupies a tiny fraction of the Brillouin-zone volume. For the same reason, it is presumed that the optical vortex does not intersect the passive region. It follows that $\jrec$ depends on  $\bS^{\beps_m}_{v\kext \lea c\kext}$ but not on the vortex-induced disorder in $\bS^{\vec{x}}_{c\bk\lea v\bk}$. 
To wrap up the argument, the vortex-induced orientational disorder in $\bS^{\vec{x}}_{c\lea v}$ diminishes $\bsigma^{\text{exc}}_{\vec{x},\omega}$ but not $\bsigma^{\text{rec}}_{\vec{x},\omega}$; ceteris paribus, $||\bsigma^{\text{rec}}_{\vec{x},\omega}|| > ||\bsigma^{\text{exc}}_{\vec{x},\omega}||.$ This explains how `great differences' (between the excitation and recombination shifts) result from a `great circulation' (induced by a vortex).\\  

To recapitulate, we have qualitatively argued that  $\vec{x}$-vorticity leads to  $\bsigma^{\text{rec}}_{\vec{x},\omega}$ dominating over $\bsigma^{\text{exc}}_{\vec{x},\omega}$, as well as brings about  a linear disparity of both $\jexc$ and $\jintra$.\footnote{ Vorticity also results in a linear disparity of $\jrec$ due to the absorption coefficient being proportional to $\langle |\beps_s\cdot \bA_{cv}|^2\rangle_{\omega}$, but this is not a large effect in our model.} These arguments are quantitatively supported by model calculations detailed in \app{app:vortexmodel}; here, we will just summarize the salient conclusions: $\bsigma_{\vec{x},\omega}$ is dominated by the recombination-induced current; $\bsigma_{\vec{y},\omega}$ is dominated by the excitation-induced and intraband currents; the signs of $\bsigma^y_{\vec{x},\omega}$ and $\bsigma^y_{\vec{y},\omega}$ differ over a broad range of frequencies; the linear disparity in the conductivity is large:  $|\bsigma^y_{\vec{x},\omega}-\bsigma^y_{\vec{y},\omega}| \sim mAV^{-2}$ [cf.\ \fig{fig:coverpic2}(d)], assuming a generic value for the JDOS; the current response to unpolarized light is slightly smaller: $|\bsigma^y_{\vec{x},\omega}+\bsigma^y_{\vec{y},\omega}|/2 \sim 0.1 mAV^{-2}$; all these results hold without finetuning of the incident radiation frequency.\\

To find optical vortices in model Hamiltonians and realistic materials, let us develop the close relationship between vorticity and Berry curvature that has been suggested by
Eq.~\eqref{eq:circ}: for any  closed 2D $\bk$-manifold $\bSigma$ (which can be a two-toroidal or two-spherical cut of the 3D Brillouin zone), we establish a general theorem relating  the Chern numbers ($C_v,C_c$) of the  valence and conduction bands to the net optical vorticity $(Vort_{\beps})$:
\e{\text{Chern-vorticity theorem:}\as C_c-C_v= Vort_{\beps} =\sum_{\text{vortex}}\oint_{\partial \text{vortex}}\nabk  \arg[\beps\cdot \bA_{cv\bk}]\cdot \f{d\bk}{2\pi}. \label{eq:chernvortex}}
$Vort_{\beps}$ is the net circulation of the interband Berry phase over all $\beps$-vortex points in $\bSigma$, and $\partial \text{vortex}$ is an infinitesimal loop surrounding each $\beps$-vortex point as illustrated in Fig.~\ref{fig:circulation} (c).\footnote{An equivalent and manifestly gauge-invariant expression is $Vort_{\beps}=\sum_{vortex}\oint_{\partial \text{vortex}} \bS^{\beps}_{c\bk\lea v\bk}\cdot {d\bk}/{2\pi}$, with the photonic shift vector defined in  \q{photonicshift}. This expression differs  from Eq.~\eqref{eq:chernvortex} only in the line integral of $\bA_{cc}-\bA_{vv}$ over $\partial vortex$; this integral vanishes because $\partial vortex$ is an infinitesimal loop and $\bA_{cc}-\bA_{vv}$ is smoothly defined at the vortex point; cf.\ \fig{fig:chernvortex} in \app{app:chernvorticity}.} This theorem is derived by setting $\bSigma$ to be a closed manifold in \q{eq:circ}, such that the area integral of $\bOmega_c$ simplifies to the Chern number $C_c$ of the conduction band, and that of $\bOmega_v$ to $C_v$; the line integral of the  shift vector over $\partial \bSigma$ vanishes, but the line integral of the  interband Berry phase is contributed by the circulation around each vortex point, as elaborated in \app{app:chernvorticity}. The next section employs the Chern-vorticity theorem to identify vortices in $\rm{BiTeI}$.\\

\section{Case study of $\rm{BiTeI}$}
\label{sec:casestudy}
To demonstrate the effects of the anomalous shift and optical vorticity in a realistic material, we present a case study of $\rm{BiTeI}$, a 3D polar, layered semiconductor with  P3m1 space group symmetry~\cite{shevelkov_crystal,ishizaka_giant}. The large atomic number of $\rm{Bi}$ correlates with a large Rashba-type spin-orbit coupling~\cite{ishizaka_giant} and the promixity of $\rm{BiTeI}$ to a $\Z_2$ topological insulator~\cite{das_engineering}. A previous study~\cite{LiangTan_topoins} of $\rm{BiTeI}$ by Tan and Rappe exhibited the enhancement of $\vec{z}\cdot \bsigma^{\text{exc}}_{\vec{z}}$ for $\vec{z}$ parallel to the polar axis, assuming the band gap $|E_g|$ were made small (by hydrostatic pressure), and further assuming the photon frequency were finetuned to be comparably small: $\omega \approx |E_g|/\hbar$. This enhancement of $\vec{z}\cdot \bsigma^{\text{exc}}_{\vec{z}}$ originates from the divergence of the band-edge optical affinity at the phase transition ($E_g=0$) between the trivial and topological insulator; across this transition, $\vec{z}\cdot \bsigma^{\text{exc}}_{\vec{z}}$ changes sign. \\

Our case study demonstrates that:\\

\begin{figure}[h]
\centering
\includegraphics[width=0.8 \textwidth]{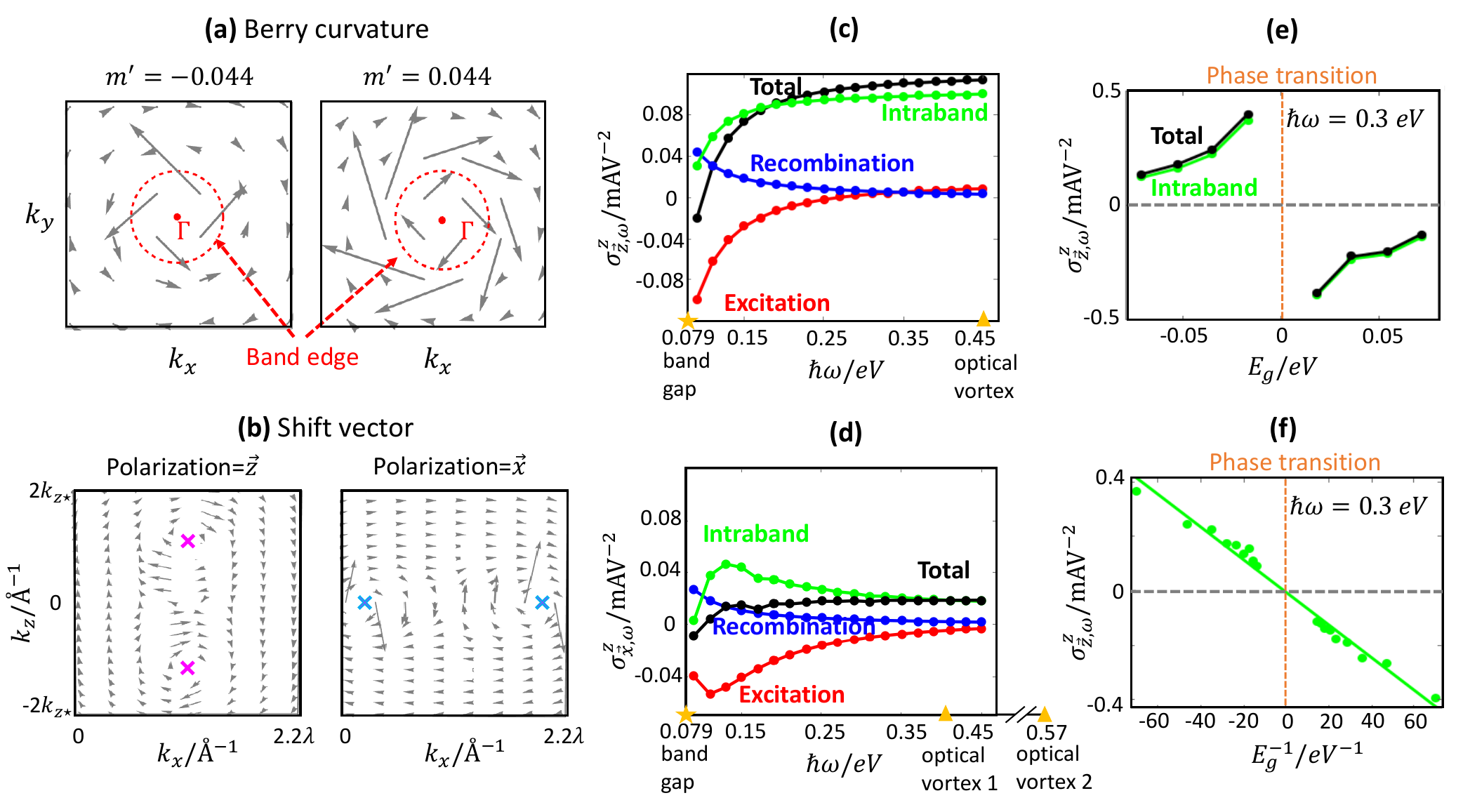}
\caption{ Panel (a) shows the Berry curvature vector field in the $k_{z}=0$ plane, for the model Hamiltonian in \q{eq: Hamiltonian} with  Hamiltonian parameter $m^{\prime}=m-A\lambda^2=\pm 0.044$ eV. Panel (b) plots the photonic shift vector fields ($\bS^{\vec{z}}_{c\lea v}$ and $\bS^{\vec{x}}_{c\lea v}$) in the $k_{x^{+}}-k_{z}$ half-plane for $m^{\prime}=- 0.044$. The shift vectors circulate around optical $\vec{z}$-vortices ($\vec{x}$-vortices), which are represented by magenta (navy) crosses. (c,d) For $m^{\prime}=- 0.044$ eV, we plot the shift conductivities ($\vec{z}\cdot \bsigma_{\vec{z}}$ and $\vec{z}\cdot \bsigma_{\vec{x}}$) and their threefold decomposition. Panel (e) plots the $\vec{z}\cdot\bsigma_{\vec{z}}$ and its intraband component versus $E_g$. Panel (f) plots  $\vec{z}\cdot\bsigma^{\text{intra}}_{\vec{z}}$ against $1/E_g$, with negative $E_g$ corresponding to the trivial side of the topological phase transition. }
\label{fig:BiTeI}
\end{figure}

\noi{i} The just-mentioned topological phase transition guarantees the existence of large Berry curvature [cf.\ \fig{fig:BiTeI}(a)] 
and optical vortices [\fig{fig:BiTeI}(b)], as per the Chern-vorticity theorem in Eq.~\eqref{eq:chernvortex}.  \\ 

\noi{ii} For photon frequencies such that the excitation surface is close to the optical vortex, the phonon-mediated $\vec{z}\cdot \bsigma^{\text{intra}}_{\beps}$ dominates over the photon-mediated $\vec{z}\cdot (\bsigma^{\text{rec}}_{\beps}+ \bsigma^{\text{exc}}_{\beps})$ [\fig{fig:BiTeI}(c)], owing to the excitation surface enclosing a larger volume of one-quasiparticle Bloch states with nontrivial the Berry curvature. \\

\noi{iii} Conversely for smaller frequencies ($\omega \approx |E_g|/\hbar$) such that  the excitation surface encloses a negligible amount of Berry curvature,  it is $\vec{z}\cdot (\bsigma^{\text{rec}}_{\beps}+ \bsigma^{\text{exc}}_{\beps})$  which dominates over $\vec{z}\cdot \bsigma^{\text{intra}}_{\beps}$. 
The net current is nonvanishing despite $\vec{z}\cdot \bsigma^{\text{rec}}_{\beps}$ and $\vec{z}\cdot \bsigma^{\text{intra}}_{\beps}$ opposing each other [\fig{fig:BiTeI}(d)], owing to an asymmetry of  the photon polarizations  in the  excitation and recombination processes.\\ 

\noi{iv} The recombination shift strongly depends on the symmetry of the Hamiltonian at the wavevectors of recombination. For BiTeI, a chiral symmetry reduces $|\vec{z}\cdot\bS_{rec}|$ to about a third of the lattice period, which makes the recombination shift current smaller than the other two components. \\

\noi{v} Because the $\bk$-locations of optical vortices depend on the light polarization [Fig.~\ref{fig:BiTeI}(b)], we find that $\vec{x}$-vortices suppress the anomalous shift more effectively than would $\vec{z}$-vortices, resulting in a linear disparity of the shift conductivity ($|\vec{z}\cdot \bsigma_{\vec{z}}-\vec{z}\cdot \bsigma_{\vec{x}}|\sim 0.1 mA/V^2$) at higher frequencies.\\

\noi{vi} If the topological phase transition is induced by tuning the band gap $|E_g|$ to zero at a fixed photon frequency, the discrete change in wave function topology manifests as a sign change of the {steady} shift current $\bj[f]$. The reason for this sign change is that $\bj[f]$ is dominated by the phonon-mediated $\jintra$  which changes sign [Fig.~\ref{fig:BiTeI}(e)]; the previously-calculated sign change of $\jexc$ (by Tan and Rappe\cite{LiangTan_topoins}) is irrelevant. The sign change of $\jintra$ is concomitant with a divergence of the band-edge intraband Berry curvature at the phase transition, which results in an approximate  $(1/E_g)$-divergence of the low-temperature $\vec{z}\cdot \bsigma_{\beps}$ [Fig.~\ref{fig:BiTeI}(f)], with negative (resp. positive) $E_g$ referring to a semiconductor (with band gap $|E_g|$) on the topologically trivial (resp. nontrivial) side of the phase transition. The $(1/E_g)$-divergence of $\vec{z}\cdot \bsigma_{\beps}$ is cut off when $E_g$ becomes comparable to the thermal energy ($k_BT_l$) or to an energy scale representing trigonal warping; the latter scale is estimated\cite{LiangTan_topoins} to be about $10meV$.\\

\noindent It may be seen from (ii-iii) that the winner in the competition (between shift$_{\text{intra}}$, \srec and shift$_{\text{exc}}$) depends sensitively on the photon frequency, to the extent that the net shift conductivity changes sign in the transition from a photon-dominated shift current (at low frequency) to a phonon-dominated shift current (at high frequency); cf. black curve in \fig{fig:BiTeI}(c-d). This exemplifies a general principle: because band wave functions can strong depend on energy, so can the shift current sensitively depend on the photon frequency. 
Point (v) exemplifies a general principle that optical vorticity makes the shift current sensitive to changes in light polarization. Point (vi) suggests  the bulk photovoltaic effect can provide smoking-gun evidence of the  
topological phase transition in BiTeI. Such  evidence is presently lacking: though it has been alleged that BiTeI is pressure-tunable to a phase transition, the experimental corroborations of this allegation (namely, a mininum of the resistivity~\cite{yanpeng_pressureBiTeI} or variations of the quantum  oscillation frequency~\cite{ideue_pressureBiTeI}) cannot be directly interpreted as a change in wave function topology. \\

\begin{figure}[h]
\centering
\includegraphics[width=0.8 \textwidth]{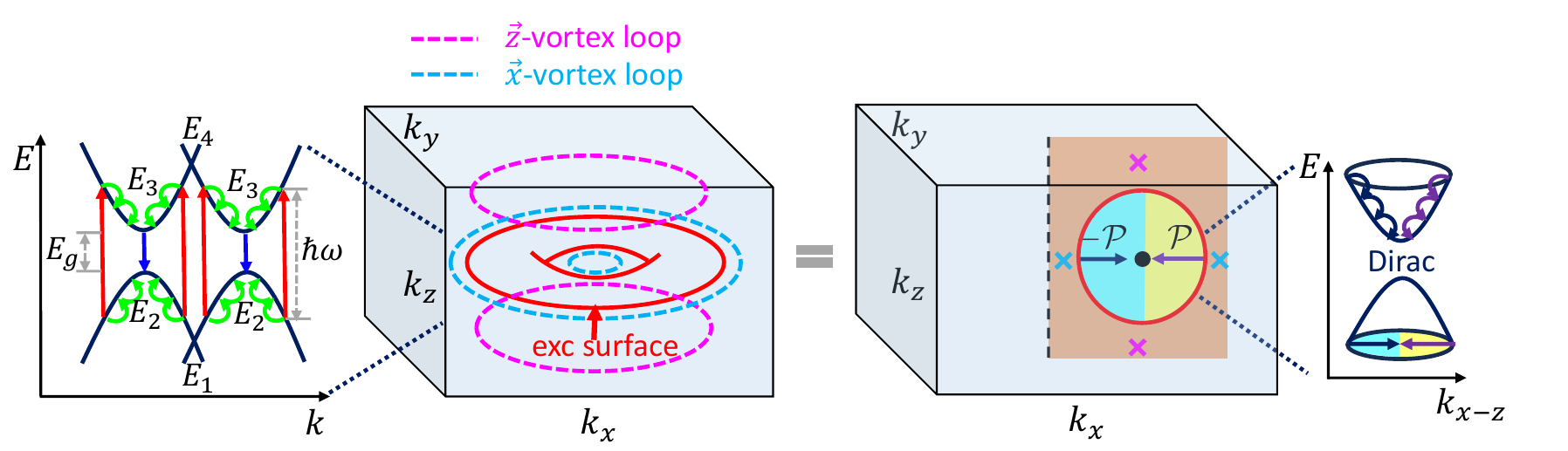}
\caption{ On the left: the box is a 3D Brillouin zone containing a toroidal excitation surface (outlined in red), a pair of $\vec{z}-$optical vortex loops (magneta dotted lines) and a pair of $\vec{x}-$optical vortex loops (navy dotted lines); the left-most $E$-vs-$k$ plot shows the band structure of $\rm{BiTeI}$ with the kinetic processes of excitation (red arrows), intraband relaxation (green) and recombination (blue).  On the right: an O(2) rotation symmetry allows us to focus on the $k_{x^{+}}-k_{z}$ half-plane, which contains a massive Dirac fermion in two momentum dimensions; the cross-section of the toroidal excitation surface is a circle (colored red) in the half-plane; the cross section of vortex loops are vortex points (indicated by magenta and navy crosses). The dark blue and purple arrows represent diametrically opposed geodesic paths for intraband relaxation.}
\label{fig:BiTeI_illustration}
\end{figure}

To substantiate our results, we employ an effective Hamiltonian for the four low-energy, spin-split bands of $\rm{BiTeI}$:
\e{
\label{eq: Hamiltonian}
H_{\text{BiTeI}}\eq \hbar v \big(\;\lambda\tau_{1}\sigma_{3}  -\tau_2 \bk\cdot \bsigma\;\big)+M\tau_{3}\sigma_{0};\as M=(m-Ak^2); \as k^2=k_x^2+k_y^2+k_z^2,
}
with $\tau_{1,2,3}$ and $\sigma_{1,2,3}$ being Pauli matrices for the orbital and spin degrees of freedom respectively; $\tau_0$ and $\sigma_0$ are identity matrices, and $A=0.5 eV\mathring{A}^{2}, \ \hbar v=0.7 eV\mathring{A}, \ \lambda=0.25/(\hbar v)=0.357 \mathring{A}^{-1}$ are ab-initio-derived~\cite{LiangTan_topoins} parameters. The spectrum of this model is given by
\begin{equation}
    E=\pm\sqrt{M^2+(\hbar v)^2\big[ \;k_z^2+ (k_{\perp}\pm\lambda)^2\;\big]}; \as k_{\perp}\equiv\sqrt{k_{x}^2+k_{y}^2}, 
\end{equation}
with each choice of either $\pm$ determining four energy levels: $E_{1}\leq E_{2}<E_{3}\leq E_{4}$ as illustrated in Fig.~\ref{fig:BiTeI_illustration}, with corresponding eigenstates: $\ket{u_{1\mathbf{k}}}$, $\ket{u_{2\mathbf{k}}}$, $\ket{u_{3\mathbf{k}}}$, and $\ket{u_{4\mathbf{k}}}$. When $m$ is tuned to the critical value $m_c=A\lambda^2$ (possibly by hydrostatic pressure~\cite{LiangTan_topoins}), the band gap vanishes with a concomitant  energy  degeneracy ($E_2=E_3$) along a loop defined by $k_{\perp}=\lambda$ and $k_{z}=0$. The circular shape of this loop reflects the O(2) rotational symmetry of the Hamiltonian.\footnote{The SO(2) rotational symmetry of the Hamiltonian manifests as $\hat{R}_{\theta}H_{\text{BiTeI}}(\bk)\hat{R}_{\theta}^{-1} = H_{\text{BiTeI}}(g_{\theta}\bk)$, with $\hat{R}_{\theta}=e^{i\theta\tau_{0}\sigma_{3}}$, and with $g_{\theta}\bk$ obtained from $\bk$ by a rotation of angle $\theta$ around the z axis. This SO(2) rotational symmetry is approximate; we neglect a trigonal warping whose energy scale is estimated\cite{LiangTan_topoins} to be about $10meV$. Time reversal symmetry is represented by $\hat{T}=i\tau_{3}\sigma_{2}K$, with $K$ implementing complex conjugation.  The Hamiltonian term proportional to $\tx \sz$ breaks both mirror   $[z\ri -z;\hat{M}_z=i\tz\sz]$ and  $[(x,y,z)\ri (-x,-y,-z); \hat{P}=\tau_3]$ parity symmetries.} \\

Away from the topological phase transition ($m\neq m_c$), the  Berry curvature of $\ket{u_{2\mathbf{k}}}$   is a circulating vector field illustrated in \fig{fig:BiTeI}(a), with the circulation flipping in orientation when $ m^{\prime}=m-m_c$ changes sign.  Because this circulation is $O(2)$-rotation symmetric, it may be understood by focusing on a single 2D slice of the Brillouin zone, say, the slice with $k_y=0$ and $k_x>0$, which we henceforth call the $k_{x^+}-k_{z}$ half-plane [see Fig.~\ref{fig:BiTeI_illustration}]. This half-plane is intersected by the energy-degenerate loop at $\mathbf{k}_{0}=(\lambda,0,0)$; by Taylor expanding $H_{\text{BiTeI}}$    around $\bk_0$  and projecting onto  bands $2$ and $3$, we obtain an effective Hamiltonian for a massive Dirac fermion in two momentum dimensions:
\begin{equation}
\label{eq:massivedirac}
H^{\prime}= (m'-\hbar v' q_x)\gamma_3 +\hbar v(q_x\gamma_1+q_z\gamma_2); \as \hbar v'= 2A\lambda; \as q_{x}=k_x-\lambda,\;q_z=k_z,
\end{equation}
with $\gamma_{1,2,3}$ being Pauli matrices of the reduced Hilbert space. Detailed derivations of $H^{\prime}$ can be found in App~\ref{app:sbandedge}.
It is known that the massive Dirac fermion is characterized by a large Berry curvature $\Omega^y_{c\bk}=-\Omega^y_{v\bk}$, for $\bk$ in a `hot spot' of width $\text{(band gap)}/\hbar v$;  assuming this width is small compared to the linear dimension of the Brillouin zone, the  Berry flux through the half-plane  changes by $2\pi$ when $m'$ changes sign: $\int_{half-plane} \Omega^y_{c\bk} dk_xdk_z=\pi \sgn{m'}$ 
\cite{bernevig_book}. \\

This jump of the Berry flux indicates the presence of optical vorticity on at least one side of the phase transition, meaning for either $m^{\prime}>0$ or $m^{\prime}<0$. This is because as $m^{\prime}$ is tuned from $0^-$ (negative infinitesimal) to $0^+$ (positive infinitesimal), the net vorticity $Vort_{\beps}$ of the half-plane must change discontinuously by $+2$ for any $\beps$, according to the Chern-vorticity theorem in \q{eq:chernvortex}.\footnote{Strictly speaking, the half-plane is not a closed 2D $\bk$-manifold, which precludes a direct application of the Chern-vorticity theorem. However, nearly the same logical considerations apply:  the eigenstates $\ket{u_{b\mathbf{k}}}$ continuously depends on $m^{\prime}$ except for $\bk=\bk_0$ (the band-touching point), thus  when  $m^{\prime}$ is tuned from $0^-$ to $0^+$, the photonic shift vector $\bS^{\beps}_{c\bk \leftarrow v\bk}$  is invariant for $\bk$ on the boundary of the half-plane. This implies that the change in  $2\int_{half-plane} \Omega^y_{c\bk} dk_xdk_z/(2\pi)$ equals the change in $Vort_{\beps}$, across the topological phase transition. }   This topological argument is verified by our numerical calculations: As shown in Fig.~\ref{fig:BiTeI}(b), for $m^{\prime}<0$, two $\vec{z}$-vortices ($\vec{x}$-vortices) are observed on the $k_{x^{+}}-k_{z}$ plane, as highlighted by magenta (navy) crosses. This indicates the presence of a pair of $\vec{z}$-vortex ($\vec{x}$-vortex) loops as illustrated in Fig.~\ref{fig:BiTeI_illustration} when $m^{\prime}<0$.\footnote{Due to the SO(2) symmetry, $\vec{y}$-vortex loops are related to $\vec{x}$-vortex loops because $|\vec{y}\cdot \bA_{cv}(R_{4}\bk)|=|\vec{x}\cdot \bA_{cv}(\bk)|$ where $R_{4}$ is the four-fold rotation in momentum space} While the locations of $\vec{x}$-vortex loops can only be determined numerically, direct calculations reveal that the $\vec{z}$-vortex loops reside at 
\begin{equation}
\label{eq:opticalzero}
\lambda=k_{\perp}, \ k_{z}=\pm k_{z\star}, \ \operatorname{with} \ k_{z\star}=\sqrt{\lambda^2-m/A}.
\end{equation}
In contrast, there is no vorticity for $m^{\prime}>0$. The circulation of the photonic shift vectors in Fig.~\ref{fig:BiTeI} (b) imply that both $\vec{z}$-vortex  ($\vec{x}$-vortex) loops have vorticity $-1$, which is consistent with the theorem's prediction that $Vort_{\beps}$ changes by $+2$.\\ 

The combination of Berry curvature and optical vorticity results in the three-fold decomposition of the shift current being highly sensitive to the photon frequency and polarization.  { Focusing first on $z$-polarized light,}  the only nonzero component of the shift current (allowed by $O(2)$ symmetry) is the z component.   \fig{fig:BiTeI}(c) illustrates the threefold decomposition of the shift conductivity $\vec{z}\cdot \bsigma_{\vec{z}}$ for a range of frequencies, including low frequencies that are comparable to  $|E_g|/\hbar$,   as well as  higher frequencies   where  the excitation surface approaches the pair of $\vec{z}$-vortex loops, as illustrated in \fig{fig:BiTeI_illustration}.  \\

We focus first on the high-frequency regime, where the optical affinity $|\vec{z} \cdot \mathbf{A}_{cv}|^2$  is reduced in the vicinity of the zeroes of $\vec{z} \cdot \mathbf{A}_{cv}$. Ceteris paribus, a reduction of the affinity would depress each of $\vec{z}\cdot \bsigma^{\text{exc}}_{\vec{z}},\vec{z}\cdot \bsigma^{\text{rec}}_{\vec{z}}$ and $\vec{z}\cdot \bsigma^{\text{intra}}_{\vec{z}}$, according to \q{aveshiftloop}. 
This depression is observed in  Fig.~\ref{fig:BiTeI}(c) for
 both $\vec{z}\cdot \bsigma^{\text{exc}}_{\vec{z}}$ and $\vec{z}\cdot \bsigma^{\text{rec}}_{\vec{z}}$; in contrast,  $\vec{z}\cdot \bsigma^{\text{intra}}_{\vec{z}}$ is enhanced rather than depressed, for two reasons:\\

\noi{a} The non-uniformity of the optical affinity (over the excitation surface) favors  $\vec{z}\cdot \bsigma^{\text{intra}}_{\vec{z}}$. On one hand, \fig{fig:BiTeI_illustration} shows that the optical affinity is more greatly reduced for large $|k_z|$ (closer to the vortex) than it is for small $|k_z|$ (further from the vortex), implying that the predominant relaxation paths are proximate  to the $k_x$ axis.
On the other hand, only those relaxation paths (for which $d\bk$ is perpendicular to $\vec{z}$) result in a large anomalous shift: $\vec{z}\cdot \bOmega \times d\bk$. Bringing both hands together, $\vec{z}$-vorticity preserves the horizontal relaxation paths which have a large anomalous shift, and deactivates the vertical relaxation paths which anyway have a negligible anomalous shift.\\

\noi{b} A higher photon frequency implies that the excitation surface encloses a larger volume of quasiparticle Bloch states with nontrivial Berry curvature, and this results in a larger anomalous shift.  For an intuitive understanding, consider reformulating the intra-conduction-band\footnote{The total interband contribution to the affinity shift loop is twice of \q{contributeASL}, owing to $\bOmega_{c\bk}\approx -\bOmega_{v\bk}$} contribution to the affinity shift loop [\q{aveshiftloop})]
\e{
\sum_{\mathcal{P}}\bigg(|\vec{z}\cdot\bA_{cv}|^2_{-\mathcal{P}}\int_{-\mathcal{P}}\;+\;|\vec{z}\cdot\bA_{cv}|^2_{\mathcal{P}}\int_{\calp}\bigg)\vec{z}\cdot\bOmega_{c\bk}\times d\bk, \la{contributeASL}
}
with $\calp$ and $-\calp$ representing diametrically opposed geodesic paths in a cross section of the torus enclosed by the excitation surface  (as  representatively illustrated by arrows in the blue and yellow semicircles of Fig.~\ref{fig:BiTeI_illustration});  $|\vec{z}\cdot\bA_{cv}|^2_{\pm\mathcal{P}}$ is the optical affinity evaluated at the intersection of $\pm\mathcal{P}$ with the excitation surface, and $\sum_{\calp}$ integrates over all pairs of $\pm \calp$ such as to entirely fill the torus. From \q{contributeASL}, one deduces that the anomalous interband contribution (to the affinity shift loop) increases with increasing photon frequency, because one integrates the Berry curvature over increasingly wider paths.\footnote{The increase of $\vec{z}\cdot \bsigma^{\text{intra}}$ (with respect to frequency) saturates when the excitation energy (measured from the conduction-band minimum) becomes comparable to the band gap: $E_{exc}-E_{c,\bk_{ext}}\sim |E_g|$, as illustrated in \fig{fig:BiTeI}(c-d). After all, this energy interval contains the Berry curvature `hot spot'.}  \\

Conversely, for  smaller frequencies ($\omega \approx |E_g|/\hbar$), the excitation surface lies closer to the extremal wavevectors $\kext$ but further away from the vortex loops; then   it is the photon-mediated $\vec{z}\cdot (\bsigma^{\text{rec}}_{\vec{z}}+ \bsigma^{\text{exc}}_{\vec{z}})$ which dominates over the phonon-mediated $\vec{z}\cdot \bsigma^{\text{intra}}_{\vec{z}}$ [cf.\ the trends in Fig.~\ref{fig:BiTeI}(c)], owing  to the Berry dipole moment  vanishing as the volume (enclosed by the excitation surface) shrinks.  
$\vec{z}\cdot \bsigma^{\text{exc}}_{\vec{z}}$ and $\vec{z}\cdot \bsigma^{\text{rec}}_{\vec{z}}$ oppose each other but do not cancel out, because the excitation shift $\vec{z}\cdot \bS_{c\leftarrow v}^{\vec{z}}$ is larger in magnitude than 
the recombination shift $\vec{z}\cdot\bS_{rec}$.\footnote{This is explained by  $\bS_{rec}$ being a weighted average of $\bS_{v\kext\leftarrow c\kext}^{{\beps}}$ over all possible polarization vectors ${\beps}$ of the spontaneously emitted photons [cf.\ \q{srec}]. Moreover, $\vec{z}\cdot \bS_{v\kext\leftarrow c\kext}^{\vec{x}}=\vec{z}\cdot \bS_{v\kext\leftarrow c\kext}^{\vec{y}}=0$ owing to a chiral symmetry [$\tau_{2}\sigma_{3} H_{\text{BiTeI}}(k_{x},k_{y},0)\tau_{2}\sigma_{3}=-H_{\text{BiTeI}}(k_{x},k_{y},0)$] which is elaborated in App~\ref{app:sbandedge}. This exemplifies a general principle that the recombination shift strongly depends on the symmetry of the Hamiltonian at the wavevectors of recombination. For BiTeI, chiral symmetry reduces $|\vec{z}\cdot\bS_{rec}|$ to about a third of the lattice period, whereas for the two-band model in \q{modelham}, the symmetries of reflection and quasi-two-dimensionality enhance $|\vec{y}\cdot\bS_{rec}|$  by precluding an orientational-disordered average, as explained in \app{app:vortexmodel}.}\\


In comparing the shift conductivites for $\vec{z}$- vs $\vec{x}$-polarized   light [Fig.~\ref{fig:BiTeI}(c) vs Fig.~\ref{fig:BiTeI}(d)], the starkest difference is that $\vec{z}\cdot \bsigma^{\text{intra}}_{\vec{x}}$ is a non-monotone function which is suppressed at higher photon frequency.\footnote{A minor difference between Fig.~\ref{fig:BiTeI}(c) and Fig.~\ref{fig:BiTeI}(d) is that $\vec{z}\cdot \bsigma^{\text{exc}}_{\vec{x}}$ is also non-monotonic and remains small at low frequencies. This occurs because $\vec{z}\cdot \bS_{v\kext\leftarrow c\kext}^{\vec{x}}=0,$ owing to an emergent chiral symmetry at the band edge, as elaborated in App~\ref{app:sbandedge}. } This is because, in contrast to $\vec{z}$-vortices and the above-mentioned point (a), $\vec{x}$-vortices lie on the $k_{x}$ axis [c.f. Fig.~\ref{fig:BiTeI_illustration}] and reduces the optical affinity $|\vec{x}\cdot \bA_{cv}|^2$ for small $|k_{z}|$. Thus for high frequencies, $\vec{x}$-vortices deactivate the horizontal relaxation paths which  have the largest anomalous shift ($\vec{z}\cdot \bOmega \times d\bk$); this effectively suppresses the intraband shift current and leads to a linear disparity ($|\vec{z}\cdot \bsigma_{\vec{x}}-\vec{z}\cdot \bsigma_{\vec{z}}|$) of order $0.1 mA V^{-2}$.\\

Let us close this section by explaining the  $(1/E_{g})$-divergence (and concomitant sign change) of $\vec{z}\cdot\bsigma_{\beps}^{\text{intra}}$ across the topological phase transition, as illustrated in Fig.~\ref{fig:BiTeI}(e,f). It suffices to show that the affinity shift loop also has a $(1/E_{g})$-divergence, according to the proportionality relation in \q{shiftconduct}. The contribution to the affinity shift loop [cf.\ Eq.~\eqref{contributeASL}] by  a  pair of diametrically-opposed geodesic paths ($\pm\mathcal{P}$)  can be further decomposed as
\begin{equation}
\label{eq:affweightedas}
\text{Aff}_{\text{ave}}\left(\int_{-\mathcal{P}}+\int_{\mathcal{P}}\right)\vec{z}\cdot \bOmega_{c\bk}\times d\bk+\delta \text{Aff}\left(\int_{-\mathcal{P}}-\int_{\mathcal{P}}\right)\vec{z}\cdot \bOmega_{c\bk}\times d\bk, 
\end{equation}
with $\text{Aff}_{\text{ave}}\equiv (|\vec{z}\cdot\bA_{cv}|^2_{-\calp}+|\vec{z}\cdot\bA_{cv}|^2_{\calp})/2$ and $\delta \text{Aff}\equiv (|\vec{z}\cdot\bA_{cv}|^2_{-\calp}-|\vec{z}\cdot\bA_{cv}|^2_{\calp})/2$. The term proportional to $\text{Aff}_{\text{ave}}$ is asymptotically irrelevant as $|E_{g}|$ approaches zero, owing to an emergent left-right symmetry of the massive Dirac fermion [\q{eq:massivedirac}] about the extremal wavevector.\footnote{This left-right symmetry is explained in \app{app:anomalousshiftintegrals}. } What remains is to evaluate the asymptotic behavior of the term proportional to $\delta \text{Aff}$: the integral ($\int_{-\mathcal{P}}-\int_{\mathcal{P}}$) of the anomalous shift vector diverges as $1/E_{g}$, because: (i) $\vec{z}\cdot \bOmega_{c\bk}\times d\bk/|d\bk|$ diverges as $E_g/|E_{g}|^3$  at the band extremum, which is a well-known type of divergence for massive Dirac fermions  [cf. \q{omegaeg}], and (ii) the width of the Berry curvature hot spot is of order $|m'|/\hbar v\propto |E_g|$. Combining both (i) and (ii), the second integral in Eq.~\eqref{eq:affweightedas} is estimated as (extremal value of curvature) $\times$ (hot-spot width), which is proportional to   $ E_g/|E_{g}|^3\times |E_g|= 1/E_{g}$. Because this divergence applies to any pair of diametrically opposed geodesic paths, the affinity shift loop must likewise diverge as $1/E_g$, and thus also $\vec{z}\cdot\bsigma_{\beps}^{\text{intra}} \approx \vec{z}\cdot\bsigma_{\beps}$.\\

There are two reasons why this divergence will be cut off in a more realistic model of BiTeI, meaning that the $1/E_g$ behavior breaks down in a narrow energy interval: $|E_g|<E_g^{cut}$: \\

\noi{i} The first reason is that not all photo-excited quasiparticles will relax  all the way down to the conduction-band bottom (where the Berry curvature diverges), but instead they will relaxe to a Maxwell-Boltzmann distribution with a characteristic thermal energy $k_BT_l$.\footnote{The preceding calculation of the $1/E_g$ divergence assumed   that as $|E_g|\ri 0$, $k_BT_l$ must likewise $\ri 0$; indeed, the geodesic approximation relied on  $k_BT_l \ll |E_g|$, as was explained in \s{sec:kinetictheory}. } \\

\noi{ii} The second reason is that the O(2) symmetry of our effective model of BiTeI is only approximate; in real BiTeI, the topological phase transition (between two topologically distinct semiconductors) is not intermediated by an $O(2)$-symmetric nodal-loop band touching, but by a $C_{3v}$-symmetric Weyl-semimetallic phase; the energy scale of the $C_{3v}$-symmetric trigonal warping is estimated to be about $10 meV$.\cite{LiangTan_topoins}\\

\noindent Both reasons suggest the $1/E_g$ behavior of   $\vec{z}\cdot\bsigma_{\beps}$ to be precluded with a cutoff $E_g^{cut}$ that is comparable to either $k_BT_l$ or $10meV$, whichever is larger.


\section{Discussion and outlook}\la{sec:discussion}

\subsection{The three-fold way}

The steady shift current density  in a direct-gap semiconductor has a three-fold decomposition: $\bj=\jexc+\jrec+\jintra$, corresponding respectively to current contributions  by interband \textit{{{exc}}}itation [cf.\ \q{sexc}],  interband  \textit{\text{rec}}ombination [\q{srec}], and \textit{\text{intra}}band relaxation [\q{sintra}]. While this threefold decomposition has been studied for simplified models of pyroelectrics and piezoelectrics,\cite{belinicher_kinetictheory} it is here that $\jrec+\jintra$ acquires a new dimension of understanding through the lens of wave function geometry. Geometrical notions  (such as the Berry phase) transcend  the traditional classification of piezoelectrics vs pyroelectrics, and provide overarching principles to guide our  interpretation of the out-of-equilibrium, many-body dynamics of photo-excited matter.\\

One of our main results is that the excitation-induced current density $\jexc$ can be outweighed by either of $\jintra$ and $\jrec$, especially in semiconductors characterized by large intraband Berry curvature or optical vortices (topological singularities in the interband Berry phase). Model semiconductors with large Berry curvature exhibit a shift-current conductivity that is of order  $mAV^{-2}$ without finetuning of the incident radiation frequency; in the presence of optical vortices, the conductivity can change by $\sim mAV^{-2}$ if the linear polarization vector flips by ninety degrees. 
These estimates of the conductivity  assumed a generic value of the joint density of states, but in principle the joint density of states can be further optimized\cite{cook_designprinciples} for a synergistic enhancement. To our knowledge, no measurement of the short-circuit conductivity in shift-current materials has reached the $mAV^{-2}$ range.

\subsection{Wave-function approach to photovoltaic materials}

Establishing the steady shift current in the broader framework of wave function geometry confers an advantage:  we acquire a Rosetta stone to  translate our vast body of knowledge (on topological materials) to concrete predictions of photovoltaic materials. Here are two cases in point:\\

\noi{i} Intraband relaxation due to electron-phonon scattering results in an anomalous shift that is proportional to the intraband Berry curvature; cf.\ \q{anomalousshift}.\footnote{Electron-phonon scattering is not the only mechanism for intraband relaxation in a direct-gap semiconductor; electron-impurity scattering also results in a shift,\cite{belinicher_kinetictheory} which may substantially contribute to $\jintra$ for dirtier samples. The impurity-mediated shift is closely analogous to the `side jump' in the anomalous Hall effect of magnetic metals.\cite{sinitsyn_coordinateshift} In their study of the `side jump',  Sinitsyn et al have argued that the impurity-mediated shift reduces to the anomalous shift [\q{anomalousshift}] under two assumptions: (i) dominance of small-angle scattering, and (ii) the cell-periodic component of the Bloch function is spatially homogeneous. The improbability of either assumption makes for a tenuous relation between the impurity-mediated shift and the anomalous shift.}  Let us juxtapose this anomalous shift with the nonlinear Hall effect predicted by Sodemann and Fu.\cite{sodemann_berrydipole} What matters to the anomalous shift current is the Berry curvature of all Bloch states enclosed by the excitation surface; this contrasts with the nonlinear Hall effect, which depends (at low temperature) on the Berry curvature of the Fermi surface. However, the two effects are not completely divorced:  a semiconductor with a large anomalous shift current is continuously tunable (e.g., by doping) to  a metal with a large nonlinear Hall effect. This is evident from \fig{fig:Q05_maintext}(a) if one imagines the excitation surface to be a Fermi surface. \\

\noi{ii} While the topological-matter community is well-versed in finding materials with large intraband Berry curvature, it is presently unclear which materials have optical vorticity. On one hand, a highthroughput ab-initio algorithm has been proposed in \ocite{aa_topologicalprinciple} to search for materials with optical vorticity. On the other hand, it would also be advantageous to identify general, topological principles which guarantee the existence of optical vorticity in certain classes of materials. One such principle is the Chern-vorticity theorem in \q{eq:chernvortex}, which relates the Chern number (of a 2D cross-section of a 3D Brillouin zone, or of a 2D Brillouin zone) to the net vorticity (of the same 2D cross-section or 2D Brillouin zone). This theorem has broad implications for the vorticity in topological semimetals  and topological insulators, one of which is the necessary existence of optical vorticity in BiTeI. \\

Our case study of the linear photogalvanic effect (LPGE) in semiconducting BiTeI [\s{sec:casestudy}] illustrates four principles: \\

\noi{a} Because the $\bk$-locations of optical vortices depend on the light polarization, the steady shift current sensitively depends on the light polarization. If the excitation surface is proximate to an optical vortex, the photonic shift vector is orientationally-disordered  over the excitation surface, which tends to reduce the excitation shift current.\\

\noi{b}  The recombination shift strongly depends on the symmetry of the Hamiltonian at the wavevectors of recombination. For BiTeI, chiral symmetry reduces the recombination shift vector to about a third of the lattice period.\\

\noi{c} Because band wave functions can strongly depend on energy, so can the steady shift current sensitively depend on the photon frequency. A rule of thumb is that the net shift conductivity changes sign in the transition from a photon-mediated shift current (at low frequencies corresponding to band-edge excitation) to a phonon-mediated anomalous shift current (at higher frequencies).\\

\noi{d} The steady shift current is sensitive to discrete changes of the wave function topology. In particular, the sign of the steady shift current changes sign across the $\Z_2$ topological phase transition, and the magnitude of said current is extraordinarily large in the vicinity of the transition.\\

\begin{figure}[h]
\centering
\includegraphics[width=0.5 \textwidth]{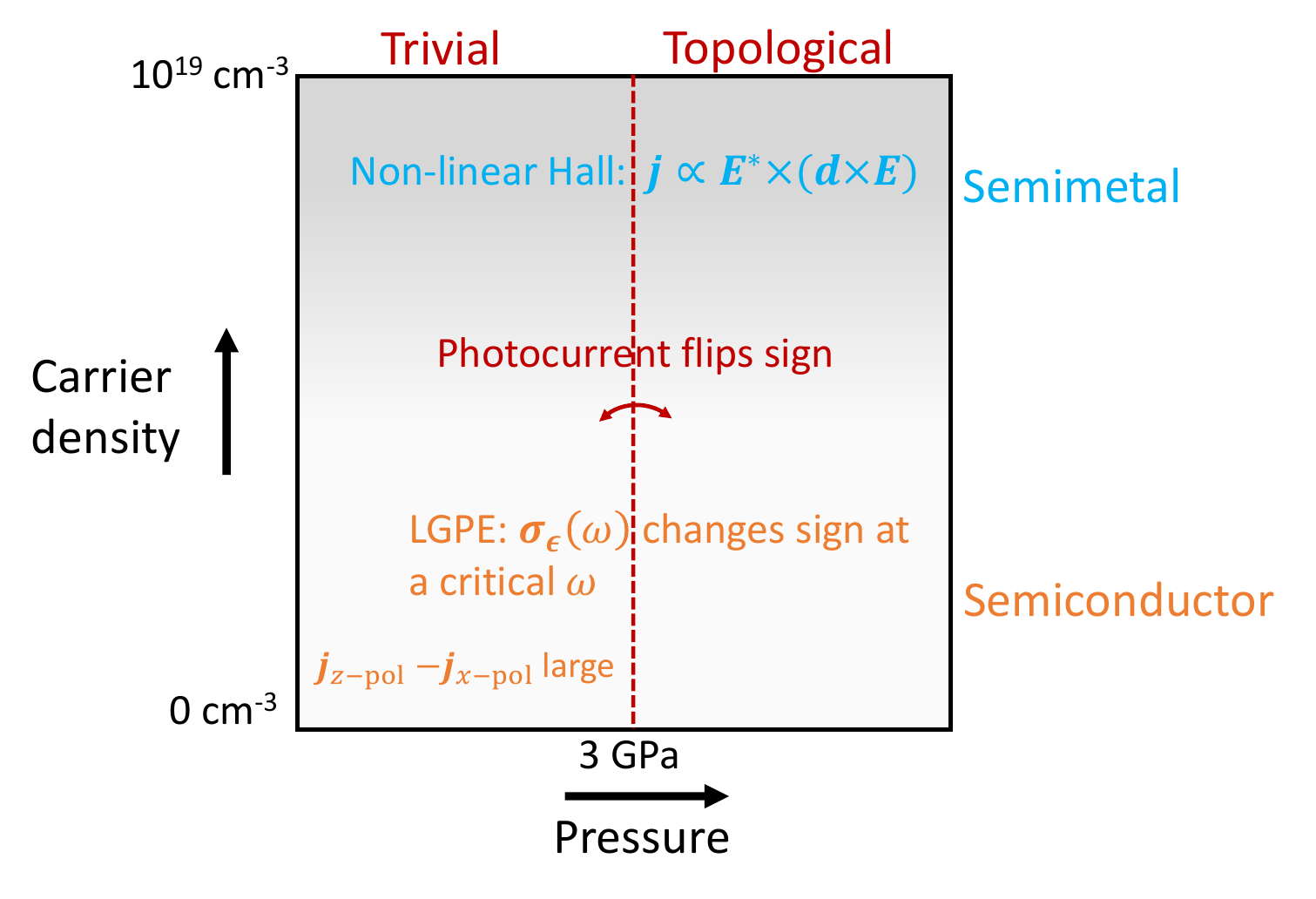}
\caption{A photovoltaic phase diagram of BiTeI. The carrier density can be tuned by varying the growth method\cite{tomokiyo_phasediagramBiTeI,lostak_opticalBiTeI} or by doping with Cu.\cite{changran_CudopedBiTeI} Pressure can be applied within a diamond anvil cell\cite{yanpeng_pressureBiTeI} or by chemical substitution.}
\label{fig:photovoltaicphasediagram}
\end{figure}

\noindent The experimental implications of (a-d) are summarized in Fig.~\ref{fig:photovoltaicphasediagram}.
The figure also illustrates the bulk photovoltaic current in semimetallic BiTeI: a previous theory~\cite{facio_berrydipoleBiTeI} has predicted a nonlinear Hall current that depends on the Berry curvature dipole $\bd$ of the Fermi surface and also changes sign across the topological phase transition. For light sources in the 100 THz regime (and higher), the bulk photovotaic current of semimetallic BiTeI is dominated by the circular photogalvanic effect (CPGE), \footnote{The CPGE is larger than the LPGE by a factor $\omega\tau\gg 1$, with $\omega$ being the photon frequency and $\tau\gtrsim 100$ fs being the momentum relaxation time.\cite{sodemann_berrydipole}} but this is not true for semiconducting BiTeI.\\

Our \textit{photovoltaic phase diagram} of BiTeI showcases the tight correlations between wave function geometry and the bulk photovoltaic effect over a wide range of carrier densities and on both sides of the topological phase transition. We hypothesize that similar correlations hold for other topological materials, suggesting  the bulk photovoltaic effect to be an unprecedented phenomenological framework to faithfully diagnose and sensitively characterize topological materials.




\subsection{Transient vs steady photovoltaic current}

\s{sec:kinetictheory} demonstrates that the steady, excitation-induced current $\jexc$ well approximates  the transient current  at the onset of radiation.  By substracting the transient  current (measured at early times) from the steady current (measured at late times), would one obtain $\jrec+\jintra$?\footnote{We consider an experimental geometry where the photon-dragged current vanishes,\cite{danishevskii_dragging,grinberg_lightpressure} and where the surface photovoltaic current is separable from the bulk photovoltaic current.\cite{alperovich_gaas}} Not quite, because the steady photovoltaic current includes not only the  shift current but also  the ballistic current.\footnote{The ballistic current results from a $(\bk \ri -\bk)$ asymmetry in the quasiparticle distribution;\cite{belinicher_ballistic,sturmanfridkin_book,alperovich_gaas,sturman_ballisticandshift} such asymmetry can be induced by electron-impurity, electron-phonon,\cite{zhenbang_phononballistic} and electron-hole interactions.\cite{zhenbang_electronholeballistic}. Does a large phonon-induced ballistic current  correlate with a large phonon-induced $\jintra$? We leave this open question for future investigations. }\\


It may be argued that the transient current is insensitive to the temperature of a photoexcited semiconductor,\footnote{$\jexc$ depends on temperature only through the photoexcitation transition rate, which is proportional to $(1-f_{c\bk})f_{v\bk}$, with $f$  the Fermi-Dirac distribution. Assuming the band gap greatly exceeds $k_BT_0$, $(1-f_{c\bk})f_{v\bk}\approx 1$ is insensitive to $T_0$. } while the steady current includes $\jintra$ which is sensitive to the temperature of the phonons. Suggestively, a substantial disparity in the temperature dependence (of the transient vs steady photocurrent) was observed for a ferroelectric charge-transfer complex, though the disparity was attributed by Nakamura et al to the formation of a Schottky barrier at the sample-electrode interface.\cite{nakamura_chargetransfercomplex}  
It would be interesting to see if this disparity persists for a different electrode whose work function is identical to that of the sample. \\

Conversely, it has been reported that the Kraut-Baltz-Sipe-Shrekbtii formula\cite{kraut_anomalousbulkPV,baltz_bulkPV,sipe_secondorderoptical} for the shift current adequately describes the photoconductivity measurements of n-GaP,\cite{hornung_nGaP} suggesting that $\jrec+\jintra$ is small for this material; this hypothesis can be tested by an ab-initio study of the intra/interband Berry phase of GaP, which we would love to see. 

\subsection{The loop approach to shift currents}

Our analysis of the direct-gap semiconductor relied on identifying a reduced set of quasiparticle transitions that concatenate into loops [cf.\ \fig{fig:loops}] and  predominantly contribute to the steady shift current [\q{bismine}].  \app{app:loop} shows how  to exactly reformulate the BIS formula [\q{bismine}] as a sum over \textit{loop currents}:
\e{ \bj=-\f{|e|}{\calv}\sum_{B,B',m } \bS^{m}_{B' \lea B} \bigg( \cala^{m}_{B' \lea B}-\cale^{m}_{B \lea B'} \bigg)=\sum_{\text{loop}}\jloop, \la{loopcurrentthm2}}
with $\jloop$ meaning the current contributed by  a closed flow line (in energy-momentum space) of one-electron probability. \\

The loop formulation holistically treats excitation, relaxation and recombination as inextricably linked processes; such linkage is  epitomized by the shift loop in  \qq{decomposeshiftloop}{shiftconduct}. Using loops allows to derive general properties of  the steady shift current that do not necessarily apply to the  transient shift current; in particular, a purported  relation between the shift current and interband polarization differences\cite{fregoso_opticalzero} is shown in \app{app:loopimp} to apply to the transient current but not the steady current. A related advantage of loops is calculational: approximating the steady shift current by a reduced family of predominant loops [e.g., via \q{shiftconduct}] requires far less computational resources than  simulating a quasiparticle distribution $f_B$ that sets the collisional integral to zero, and then inputting $f_B$ into the BIS formula; cf.\ \app{app:benchmark}.\\

Our loop current formulation is applicable beyond direct-gap semiconductors, with the caveat that the predominant loops may change depending on the context. For instance, recombination in indirect-gap semiconductors is intermediated by 
transitions between Bloch states and impurity-localized states; such transitions also contribute to the shift current.\cite{sturmanfridkin_book} In small-gap semiconductors or Dirac-Weyl semimetals, interband recombination may be contributed by electron-phonon scattering. It would be desirable to develop a theory of the steady shift current in Dirac-Weyl semimetals, for which the anomalous shift may potentially be large. It is hoped that photoconductivity measurements of TaAs\cite{osterhoudt_colossalBVPEweyl} would benefit from a re-interpretation of what exactly is causing the shift.


\subsection{The potential for solar cell applications}

A large short-circuit conductivity is not sufficient for solar cell applications;  also required is a large open-circuit photovoltage to generate sufficient electrical power.  Though shift-current materials can have open-circuit photovoltages that greatly exceed the band gap,\cite{brody_largephotovoltage,glass_highvoltage,koch_batio3expt}  the product of the short-circuit current and open-circuit photovoltage may be limited depending on the architecture of the shift-driven solar cell.\\

Let us first consider Pusch et al.'s model\cite{pusch_conversionefficiency}  of a shift-driven solar cell, in which a homogeneous shift-current-carrying intrinsic semiconductor is connected via leads to an external load; henceforth, we refer to this as the \textit{PRCE cell}. Assuming a few ideal conditions, namely that the contacts with the leads do not introduce additional resistance, and that temperature is sufficiently low ($k_BT\ll E_g$) to ignore the dark conductivity, the light-to-electrical energy conversion efficiency of a PRCE cell is calculated in  \app{app:efficiency} to be
\e{
\text{Eff} \eq \f1{4} \f{m_r\fraks^2/\tau_{tr}\tau_{rec}}{\hbar\omega},\la{efficiency1}
}
under monochromatic illumination with frequency $\omega$. Here, $m_r^{-1}=m_e^{-1}+m_h^{-1}$ is the reduced mass of an electron-hole pair in the parabolic-band approximation, $\tau_{tr}$ is the Drude-type transport lifetime for photo-excited carriers, and $\tau_{rec}$ is the recombination lifetime; cf.\ \s{sec:kinetictheory}.
$\fraks$ can be interpreted as the average shift per photo-excited electron-hole pair, and is expressible as a normalized affinity shift loop [cf.\ \q{aveshiftloop}]:
\e{ \fraks \eq \f{ASL_{\beps_s,\omega}}{\langle f_{vc\bk}|\beps_s\cdot \bA_{cv\bk}|^2\rangle_{\omega}}. \la{fraks}}
Our formula for the efficiency is essentially identical to Eq. (11) in Ref.\ \cite{pusch_conversionefficiency}, except that our $\fraks$ replaces their heuristically defined `average charge displacement $R$' with a precisely calculable formula; moreover, \qq{efficiency1}{fraks} with \qq{decomposeshiftloop}{aveshiftloop} clarify the oft ignored roles of $\jrec+\jintra$ in the operation of a shift-driven solar cell.\\

Even if $\fraks$ is of order the lattice period $a$, it may be seen that Eff $\ll \ll 1$ at solar frequencies and with typical values for $a\approx 5\ang$, $m_r\approx m_f$ (free-electron mass), $\tau_{tr} \approx 100fs$, and $\tau_{rec}\approx 1ns$ for radiative recombination. More appreciable efficiencies are expected for heavy-fermion materials dominated by faster non-radiative recombination,\cite{pankove_opticalprocesses} and with extraordinarily large lattice periods that characterize (Moir\'e) superlattices. Going beyond the PRCE model, inhomogeneous doping of the shift-current-carrying semiconductor would marry advantages of both the shift mechanism and conventional barrier layer photovoltaics; whether such a synergy is incremental or transformative remains to be seen.

\begin{center} \textbf{Acknowledgments} \end{center}

We thank Boris Sturman for patient explanations of the BIS theory, Benjamin Fregoso for illuminating discussions of third-order optical responses,  Michael Schuler for educating us on the electron-phonon interaction, and Pavlo Sukhachov for a detailed reading and commentary. An illuminating discussion with Andreas Pusch helped us formulate  a more nuanced discussion of shift-driven solar cells. 
This research was supported in part by the National Science Foundation under Grant No. NSF PHY-1748958. 
In the final stages of this work, PZ received support from the Center for Emergent Materials, an NSF MRSEC, under award number DMR-2011876.\\

\newpage

\appendix

\begin{center} \textbf{Appendix} \end{center}

The appendix contains several supplementary results for the specialized audience. Most of these results have been referenced and motivated in the main text. An organizational structure of the appendix is presented to help with navigation:\\

\noi{\app{app:glossary}} This  \textbf{glossary} collects many recurrent equations and symbols  for easy reference.  The equations include the Belinicher-Ivchenko-Sturman (BIS)  formula for the shift current [\app{app:glossarybis}] and its threefold decomposition into contributions by excitation, recombination and intraband transitions [\app{app:threefold}]. The BIS formula inputs the quasiparticle, photonic and phononic occupancies and outputs a current; certain assumptions about these occupancies are recorded in \app{app:occupancies}. All explicit calculations of the shift current in this work have been based on a two-band, direct-gap semiconducting model; the terminology that surrounds this model is collected in \app{app:smcmodel}.\\ 

\noi{\app{app:nonequil}} This appendix summarizes several salient aspects of the steady \textbf{non-equilibrium  distribution of photo-excited carriers}. Included is a  review of  the different relaxation mechanisms for a hot photo-excited carrier in a semiconductor [\app{app:relaxationmechanisms}] and a derivation of the associated kinetic model which applies in the regime of low carrier density [\app{app:kineticmodel}]. Simplified versions of this kinetic model are described if there is electron-hole symmetry [\app{app:ehsymmetry}] and if the collisional integral is constant along iso-energy surfaces [\app{app:eisotropy}].\\

\noi{\app{app:bisformula}} The \textbf{BIS formula of the shift current} is pedagogically derived in \app{app:derivebis}, numerically calculated in \app{app:benchmark}, and compared to other formulas of the shift current in \app{sec:krautbaltz}. This comparison elucidiates what is missing from the Kraut-Baltz-Sipe-Shkrebtii formula,\cite{kraut_anomalousbulkPV,baltz_bulkPV,sipe_secondorderoptical} as well as formulas derived from dissipative Floquet methods.\cite{morimoto_nonlinearoptic,barik_nonequilibriumnature,matsyshyn_rabiregime} \app{app:derivebis} also demonstrates that to describe the steady state perturbatively, the zeroth-order state is emphatically \underline{not} the thermal equilibrium state (in the absence of the light source).\\

\noi{\app{app:loop}} The steady shift current is \textbf{equivalently formulated in terms of loop currents}. 
The basic ingredients of the loop formulation are presented in  \app{app:flownetwork}, namely the loop decomposition of the probability flow network, the shift loop, and the loop current. A theorem derived in \app{app:loopthm} formalizes the equivalence between the BIS formula and a sum of loop currents. This theorem is applied  to revise  a purported  relation\cite{fregoso_opticalzero}  between the shift current and interband polarization differences [\app{app:loopimp}], and to derive the geodesic approximation of the shift conductivity for 3D semiconductors [\app{app:geodesic}] and quasi-2D semiconductors [\app{app:geodesicquasi2D}]. The geodesic approximation is rigorously justified as the small-angle-scattering limit of the BIS formula  in \app{app:smallanglelimit}.\\

\noi{\app{app:jtranjexc}} The \textbf{transient shift current} is shown here to be well approximated by the excitation-induced component of the steady shift current. This appendix rigorously elaborates an argument presented in \s{sec:kinetictheory}.\\

\noi{\app{app:vortexmodel}} The shift conductivity is calculated for a \textbf{model Hamiltonian with optical vortices}, to corroborate claims made in  \s{sec:vortex} about the vortex-induced linear disparity of the shift conductivity.\\

\noi{\app{app:chernvorticity}} A \textbf{theorem relating Chern numbers to optical vorticity}  [cf. Eq.~\eqref{eq:chernvortex}] is proven here.\\

\noi{\app{app:sbandedge}} A few facts which \textbf{support our case study on BiTeI} is presented here, including an effective Hamiltonian of a massive Dirac fermion that holds near the topological phase transition, as well as the  vanishing of the  shift current at the band edge, for $x$ and $y$-polarized light. \\

\noi{\app{app:alternative}}	An \textbf{alternative derivation of the anomalous shift vector}  is provided.\\

\noi{\app{app:deceptive}} A \textbf{misconception about the electron-phonon scattering rate} is exposed. The misconception traces back to a fallacious belief of a universally applicable gauge for the Bloch wave function.\\

\noi{\app{app:efficiency}} The \textbf{energy conversion efficiency} is calculated for a solar cell based on an intrinsic, shift-current-carrying semiconductor. \\

\section{Glossary}\la{app:glossary}

\subsection{The Belinicher-Ivchenko-Sturman  formula}\la{app:glossarybis}

We use `Belinicher-Ivchenko-Sturman (BIS)  formula' to refer to two sets of equations for the phonon-mediated and photon-mediated shift current. The former current is given by 
\e{\bj \eq -\f{|e|}{\calv}\sum_{B,B',m } \bS^{m}_{B' \lea B} \bigg( \cala^{m}_{B' \lea B}-\cale^{m}_{B \lea B'} \bigg); \as B=(b\bk),\as B'=(b'\bkp) \as m=(\bq p), \la{bismine2}} 
with the phononic shift vector given by
\e{Phonon:\as \bS^{m}_{B'\leftarrow B}= -(\nabk+\nabkp)\arg V^m_{B',B} +\bA_{b'b'\bkp}-\bA_{bb\bk}= -\bS^{-m}_{B\leftarrow B'},\la{phononicshift2}}
and with the difference in absorption and emission transition rates given by 
\e{
Phonon:\as \cala^{m}_{B'\lea B}-\cale^{m}_{B \lea B'}\eq \delta_{\bq,\bkp-\bk}\frac{2\pi}{\hbar} |V^{m}_{B'B}|^2\delta(E_{B'B}-\hbar\omega_m)\bigg\{\big(1-f_{B'}\big)f_{B}N_{m}-\big(1-f_{B}\big)f_{B'} (N_{m}+1)\bigg\},\la{phononrate}
	}
with $E_{B'B}=E_{B'}-E_B$.   Actually, the above equations are more general than those presented in \ocite{belinicher_kinetictheory}, in that the above equations allow for interband phonon-mediated transitions, while the formula in the BIS paper does not. This being a minor generalization, we will anyway refer to our final formula as the BIS formula.  \\

The BIS formula for the photon-mediated shift current combines \q{bismine2} with 
\e{
Photon:\as \bS^{m}_{b'\bk \lea b\bk}\eq -\nabk \arg \beps_m \cdot\bA_{b'b\bk}+\bA_{b'b'\bk}-\bA_{bb\bk}=-\bS^{-m}_{b\bk \lea b'\bk}; \la{photonicshift2} \\
\cala^{m}_{B'\lea B}-\cale^{m}_{B \lea B'}\eq \delta_{\bk,\bkp}\f{( 2\pi e)^2\omega_{m}}{\calv} \;\big|\beps_m\cdot \bA_{b'b\bk}\big|^2\delta(E_{B'B}- \hbar \omega_{m})\bigg\{\big(1-f_{B'}\big)f_{B}N_{m}-\big(1-f_{B}\big)f_{B'} (N_{m}+1)\bigg\},\la{photonrate}
	}
in Gaussian units.  In \ocite{belinicher_kinetictheory}, $\cala^{m}_{B'\lea B}-\cale^{m}_{B \lea B'}$ appears with an additional multiplicative factor of $1/n^2$, with $n$ the refractive index; this factor should not be there, according to our derivation in \app{app:derivebis}.\footnote{The $1/n^2$ factor is absent in the Sturman-Fridkin monograph\cite{sturmanfridkin_book} which followed after the BIS paper. A factor of $1/n$ appears only in the conversion of photon occupancies to the radiation intensity; cf.\ \app{sec:krautbaltz} . }\\ 

In the steady state, the  quasiparticle distribution $f_B$ satisfies a non-detailed balance condition that represents an invariance under simultaneous collisions with all bosons:
\e{
\text{For all $B$,}\as 0=\sum_{m}\sum_{B'}\big( \cala^m_{B\lea B'}+\cale^m_{B\lea B'} -\cala^m_{B'\lea B}-\cale^m_{B'\lea B} \big), \la{completebis}
}
with $\sum_m$ summing over all photonic and phononic modes. The right-hand side of the above equality may be viewed as the collisional integral evaluated to second order in the electron-boson coupling, i.e., the integral has a form expected from Fermi's golden rule. \qq{bismine2}{completebis} represents a closed set of equations to determine the shift current: one first determines $f_B$ from \q{completebis} then inputs $f_B$ into \q{bismine2}.

\subsection{Quasiparticle, photon and phonon occupancies}\la{app:occupancies}

The above equations show that the BIS current is a functional of the quasiparticle, photon and phonon occupancies: 
\e{\bj= \bj[f_{B},N_m^{phot},N_m^{phon}], \la{definejsteady}} 
with the dependence on occupancies given solely by the transition rates $\cala=\cala[f_B,N_m^{phot},N_m^{phon}]$ and $\cale=\cale[f_B,N_m^{phot},N_m^{phon}]$.\\

Throughout this work, $f$ symbolizes the occupancy of charged, fermionic quasiparticles that are long-lived in an insulator,\cite{kohn_effectivemass} though strictly speaking we do not account for the renormalization of the wave function.\cite{sham_shallowimpurity}
When there is no $T$ (for temperature) superscript on $f_B$,  $f_B$ should be understood as the non-equilibrium, steady distribution determined in a kinetic model [cf.\  \s{app:kineticmodel}], hence $f_B$ generically deviates from the thermal
\e{\text{Fermi-Dirac distribution}:\as f_B^T= \f1{e^{(E_B-\mu)/k_BT}+1}. \la{fermidirac}}

The phonons are assumed to thermalize with a lattice temperature $T_l$, meaning  that the phonon occupancy is a Planck distribution:
\e{\text{Phonon}: \as N_m=N_m^{T_l}=\f1{e^{\hbar\omega_m/k_BT_l}-1}. \la{planck}}
We will not always have the $phon$ or $phot$ superscript on $N_m$, so the meaning of $m$ should be deduced from the local context. The minimal frequency for optical phonons is defined to be the optical phonon threshold:
\e{  
\text{Optical phonon}: \as min\{\omega_m\} =\Omega_o;\as k_BT_l \ll \hbar\Omega_o \imp N_m^{T_l} \ll 1, \la{opticalphononocc}
}
$\hbar\Omega_o$ typically comparable to $k_B$ times room temperature. We assume in calculations of the shift conductivity that the lattice temperature is small compared to the optical phonon energy, hence the thermal occupancy of optical phonons is negligible. \\

The photon occupancy is 
assumed to be a sum of thermal and source-generated contributions: \e{\text{Photon}: \as N_m=N_m^{T_l}+\Delta N_s \delta_{m,m_s}; \as \Delta N_s \gg 1. \la{photonoccupancy}}
We have assumed that the source is bright ($\Delta N_s \gg 1$) and produces  photons of a single, linearly-polarized mode:
\e{ \text{Source:}\as \text{mode}\;=m_s;\as \text{frequency}\;=\omega_s=c||\bq_s||>E_g/\hbar; \as \text{polarization vector}\;=\beps_s\in \R^3. \la{sourceparameters}}
In the classical approximation to the radiation field, $\Delta N_s$ can be be expressed in terms of the electric-wave amplitude according to:\footnote{This may be derived from the standard relation\cite{heitler_quantumradiation,sakurai_advanced} between the classical electromagnetic vector potential and the photon number: \e{ \boldsymbol{\cale} = -\f1{c} \f{\partial \bA^s_{em}}{\partial t}; \as \bA^s_{em}=2c\sqrt{\f{h\,\Delta N_s}{\omega_s \,\calv}}\beps_s \,\cos(\bq_s\cdot\br-\omega_s t).\notag}  }
\e{ \boldsymbol{\cale} = \cale_{\omega}\beps_s e^{i(\bq_s\cdot \br-\omega_s t)} + c.c.;\as |\cale_{\omega}|^2 =2\pi \f{\hbar\omega \Delta N_s}{\calv}. \la{elecfield}}
The vector-valued \textit{shift conductivity} is defined by the nonlinear current response:
\e{  \bj=\bsigma_{\beps,\omega}|\cale_{\omega}|^2, \la{shiftconductivity} }
with $\bj$ the shift current in \q{bismine2}.\\

A basic property of the quasiparticle transition rates is that the absorption and emission rates cancel out if the fermions and bosons are thermalized with the same temperature:
\e{
\text{Detailed balance}: \as \big(\cala^{m}_{B'\lea B}-\cale^{m}_{B \lea B'}\big)_{f^{T}_{B},f^{T}_{B'},N^{T}_{m}}=0. \la{detailedbalance}
}
This holds for both phonons and phonons, as one may verify by substituting the Fermi-Dirac and Planck distributions into \qq{phononrate}{photonrate}.

\subsection{Direct-gap semiconducting model} \la{app:smcmodel}

All explicit calculations of the shift conductivity in this work are based on a model of a direct-gap intrinsic semiconductor with two bands (not counting spin):
\e{ \text{Band indices}:\as & b=c\; \text{(conduction)}; \as b= v \; \text{(valence)}; \lin
  \text{Bloch labels:}\as  &C=(c\bk); \as V=(v\bk).}
$E_{C}-E_{V}$  is assumed to be minimized at a single wavevector, which we call the \textit{extremal wavevector}: 
\e{ \text{min}\{E_{cv\bk}\} =E_{cv\kext}=E_g; \as E_{cv\bk}=E_{C}-E_{V}. \la{kext}}
The band gap $E_g$ is assumed to be large compared to the lattice temperature $T_l$ (with source turned on) and the equilibrium temperature $T_0$ (with source turned off), hence the equilibrium Fermi-Dirac occupancies are close to being binary:
\e{ k_BT_0 \ll E_g \imp  f^{T_0}_{C}\approx 0;  \as f^{T_0}_{V}\approx 1. \la{binaryfermidirac} }
Iso-energy surfaces of a band are defined to $\bk$-surfaces in which the band dispersion is constant:
\e{ \text{Iso-energy surface of band $b$ and energy $E$}\; \equiv \; \text{all}\; \bk \; \text{satisfying}\as E_{B}= E.}

The source photon frequency is assumed large enough that resonant absorption can occur across the  band gap, and the excitation surface is defined as the $\bk$-surface where resonant optical absorption can occur:
\e{ES\equiv \;\text{excitation surface}\;\equiv \;\text{all}\; \bk \; \text{satisfying}\as E_{cv\bk}=\hbar\omega_s > E_g.\la{excsurf}} 
We define $E_{b,exc}$ as the energies of $b$-band states on the excitation surface:
\e{ \text{Excitation energy}: \as E_{c,exc}=\{E_{C}\;|\; \bk \in ES\}; \as E_{v,exc}=\{E_{V}\;|\; \bk \in ES\}. \la{excenergy}} 
We will encounter symmetric models where $E_{c,exc}$ is degenerate for all conduction-band states on the excitation surface, meaning the excitation surface is an iso-energy surface of energy $E_{c,exc}$.\\

For photonic modes that mediate resonant interband transitions, their thermal occupancy is negligible:
\e{\text{Resonant photon}: N_m^{T_l}\delta(E_{cv\bk}-\hbar\omega_m) \ll \delta(E_{cv\bk}-\hbar\omega_m)  \as \Leftarrow \as k_BT_l \ll E_g.\la{resonantphotonocc}}
The passive and active regions of either band are defined with respect to the optical phonon threshold [\q{opticalphononocc}]:
\e{ \text{Conduction}: \as &E_{c\bk}-E_{c\kext} > \hbar\Omega_o \; \text{(active)}; \as  0<E_{c\bk}-E_{c\kext} < \hbar\Omega_o \; \text{(passive)};\lin
\text{Valence}: \as &-(E_{v\bk}-E_{v\kext}) > \hbar\Omega_o \; \text{(active)}; \as  0<-(E_{v\bk}-E_{v\kext}) < \hbar\Omega_o \; \text{(passive)}. \la{activepassive}
}

\subsection{Excitation, recombination and intraband components of the shift current}\la{app:threefold}

As discussed in \s{sec:kinetictheory}, the steady shift current can be decomposed into  contributions by excitation, recombination and intraband relaxation processes:
\e{ \text{Threefold decomposition}:\as  \bj=\jexc +\jintra +\jrec. \la{jthreefold}}
Here, we present the explicit expressions for each of the three components in \qq{definejexc}{definejrec}. \\

The \textit{excitation-induced current} is defined to be  the shift current contributed by interband, vertical transitions over the excitation surface [\q{excsurf}]:
\e{\text{Excitation-induced current:} \as \jexc\eq \bj[f]_{\bk=\bkp\in ES} \approx -\twoupdo\f{|e|}{\calv}\sum_{\bk} \bS^{\beps_s}_{c\bk \lea v\bk}I_{exc\bk}^{m_s}; \la{definejexc}\\ 
\text{Excitation rate:} \as I_{exc\bk}^{m_s}\eq  \f{( 2\pi e)^2\omega_s}{\calv} \;\big|\beps_s\cdot \bA_{cv\bk}\big|^2f_{vc\bk}\delta(E_{cv\bk}- \hbar \omega_s)\Delta N_s. \la{exciterate} } 
$\twoupdo=2$, with the additional subscript reminding us this two originates from the spin degree of freedom. $f_{vc\bk}$ is defined to be $f_{v\bk}-f_{c\bk}$. \q{definejexc} has been presented for the case where a pair of spinless bands (labelled $b=v$ and $c$) are optically excited; if there are more than a pair, simply sum the right-hand side of \q{definejexc} over all pairs.\\

\q{definejexc} is derived by restricting $\sum_{\bk\bkp}$ in \q{bismine2} with the condition that $\bk=\bkp$ lies on the excitation surface, and this is the meaning of $\bj[f]_{\bk=\bkp\in ES}$.  In principle,
$\sum_m$ in $\bj[f]_{\bk=\bkp\in ES}$ sums over all bosonic modes with the same frequency $\omega_m=\omega_s$ as the source-generated photons. This $\sum_m$ may be restricted to photonic modes, because the band gap is presumed to exceed the optical phonon energies.  Photon-mediated vertical transitions over the excitation surface can be divided into five classes, according to \q{photonrate} and \q{photonoccupancy}:\\

\noi{i} Absorption of thermal photons with a rate $\propto N_{m}^{T_l}f_v(1-f_c)$,\\

\noi{ii} Stimulated emission of thermal photons with a rate $\propto N_{m}^{T_l}f_c(1-f_v)$, \\ 

\noi{iii} Absorption of source-generated photons with a rate $\propto \Delta N_s f_v(1-f_c)$, \\

\noi{iv} Stimulated emission of source-generated photons with a rate $\propto \Delta N_s f_c(1-f_v)$, and\\

\noi{v} Spontaneous emission of photons with a rate $\propto f_c(1-f_v)$. \\ 

\noindent In practice, only (iii) and (iv) are significant. Here is why. Since the time scale  to spontaneously  emit  photons ($\sim 1\,ns$) greatly exceeds the time scale for scattering with phonons ($\sim 100\,fs$), the contribution of (v) to the shift current is negligible. By our assumptions that temperature is low and that carriers are resonantly excited, $k_BT_l \ll E_g \leq \hbar\omega_s=\hbar\omega_m$, hence the Planck occupancy $N_{m}^{T_l}\ll 1$. It follows that $\Delta N_s \gg 1 \gg N_{m}^{T_l}$, and we then assume  (i) $\ll$ (iii), and (ii) $\ll$ (iv). Keeping only (iii) and (iv) leads to \q{definejexc}.\\

The \textit{intraband current} is defined to be  the shift current contributed by intraband transitions:
\e{\text{Intraband current}:\as  \jintra =\sum_{b=v,c}\jintra^b; \as \jintra^b=\bj[f]_{b=b'}=-\twoupdo\f{|e|}{\calv|}\sum_m^{phon}\sum_{\bk,\bkp } \bS^{m}_{b\bkp \lea b\bk} \bigg( \cala^{m}_{b\bkp \lea b\bk}-\cale^{m}_{b\bk \lea b\bkp} \bigg), \la{definejintra}}
with $\cala-\cale$ given in \q{phononrate}. When bands do not overlap on the energy axis (as is true for our two-band semiconducting model), intraband transitions may be restricted to phononic modes, because the typical quasiparticle band velocity is much less than the speed of light. \\

It is useful to decompose the intraband current into contributions by acoustic and optical phonons: $\jintra^b = \jintra^{a,b}+\jintra^{o,b}$.  Assuming (a) $k_BT_l\ll \hbar \Omega_o$, (b) the active region is much bigger than the passive region ($|E_{b,exc}-E_{b,\kext}|\gg \hbar\Omega_o$), and that (c) small-angle-scattering predominates, the effect of acoustic phonons is substantially outweighed by that of optical phonons: $||\jintra^{a,b}||\ll ||\jintra^{o,b}||$. Here is why.
Assumption (c) allows us to employ the small-angle-limit of the phononic shift in \q{anomalousshift}; because this anomalous shift  is proportional to $\delta k=||\bkp-\bk||$, the net effect of transitions $\bkp \lea \bk$  within the passive region are ignorable compared to transitions within the much larger active region; cf.\ a similar argument made in \s{sec:kinetictheory}. Within the active region, transitions mediated by optical phonons are predominantly that of spontaneous emission, because the thermal occupancies of optical phonons  are small; cf.\ \q{opticalphononocc}. Transitions mediated by spontaneous emission of optical phonons predominantly result in a larger shift than  transitions mediated by acoustic phonons; this is because the time scales for individual collisions are comparable to $100fs$ for both types of phonons,\cite{lundstrom_book} but for optical phonons, $\delta k$ has a lower bound given by $\Omega_o$ divided by the carrier group velocity, while the only lower bound for electron-acoustic-phonon scattering is the trivial bound $\delta k>0$. Altogether, these considerations lead to the intraband shift current being dominated by:
\e{ \jintra^{o,b} \appr \twoupdo\f{|e|}{\calv|}\sum_m^{\text{optical phonons}}\sum_{\bk,\bkp } \bS^{m}_{b\bkp \lea b\bk} \cale^{sp,m}_{b\bk \lea b\bkp}; \la{jintraoptical}\\
\cale^{sp,m}_{b\bk \lea b\bkp} \eq \delta_{b,b'} \delta_{\bq,\bkp-\bk}\frac{2\pi}{\hbar} |V^{m}_{B'B}|^2\delta(E_{B'B}-\hbar\omega_m)\big(1-f_{B}\big)f_{B'}. \la{sponemirate1}}
\\

The \textit{recombination-induced current} is defined to be  the shift current contributed by vertical photon-mediated transitions ($c\bk \leri v\bk$) for $\bk$ outside the excitation surface:
\e{ \text{Recombination-induced current:}\as \jrec \eq \bj[f]_{\bk=\bk' \notin ES}=\twoupdo \f{|e|}{\calv}\sum_m^{phot}\sum_{\bk } \bS^{m}_{c\bk \lea v\bk}I_{rec\bk}^m;\la{definejrec}\\ 
\text{Recombination rate:}\as I_{rec\bk}^m \eq (1-\delta_{\bk,ES})\bigg( \cale^{m}_{v\bk \lea c\bk}-\cala^{m}_{c\bk \lea v\bk} \bigg), \la{recombrate}}
with $\cala-\cale$ given in \q{photonrate}.
We have introduced $\delta_{\bk,ES}$ as the projector to the excitation surface, and $1-\delta_{\bk,ES}$ as the complementary projector. \q{definejrec} may be simplified on the basis of two considerations:\\

\noi{a} The $\sum_{\bk}$ in \q{definejrec} may be further restricted to  a small $\bk$-volume corresponding to the passive region, according to arguments presented in \s{sec:kinetictheory} and \app{app:relaxationmechanisms}.\\

\noi{b} Because of the projection in \q{definejrec}, $\jrec$ depends on the thermal photon occupancy $N^{T_l}_m$ but not the source-generated occupancy $\Delta N_s$; since the thermal occupancies of resonant photons are small [cf.\ \q{resonantphotonocc}], one may as well retain only the transitions mediated by spontaneous emission. \\

\noindent Altogether, (a-b) imply
\e{
\jrec \appr \twoupdo \f{|e|}{\calv}\sum_m^{phot}\sum_{\bk}^{pass} \bS^{m}_{c\bk \lea v\bk} \cale^{sp,m}_{v\bk \lea c\bk}, \la{approxjrec} \\
\cale^{sp,m}_{v\bk \lea c\bk} \eq \delta_{\bk,\bkp}\f{( 2\pi e)^2\omega_{m}}{\calv} \;\big|\beps_m\cdot \bA_{cv\bk}\big|^2\delta(E_{cv}- \hbar \omega_{m})\big(1-f_{v}\big)f_{c}.\la{spontphotonrate}
}
We take $\sum_{\bk}^{pass}$ to mean an integral over the passive $\bk$-volume of either the conduction or valence band [cf.\ \q{activepassive}], whichever of the two volumes is smaller.\\

The threefold decomposition of the shift current in \q{jthreefold} imply a threefold decomposition of the shift conductivity defined in \q{shiftconductivity}:
\e{
\bsigma_{\beps,\omega}=\bsigma^{\text{exc}}_{\beps,\omega}+\bsigma^{\text{intra}}_{\beps,\omega}+\bsigma^{\text{rec}}_{\beps,\omega}; \as  \bsigma^{\text{exc}}_{\beps,\omega} =\f{\jexc}{|\cale_{\omega}|^2}; \as \bsigma^{\text{intra}}_{\beps,\omega} =\f{\jintra}{|\cale_{\omega}|^2}; \as \bsigma^{\text{rec}}_{\beps,\omega} =\f{\jrec}{|\cale_{\omega}|^2}. \la{sigmathreefold}
}


\section{The non-equilibrium distribution of photo-excited carriers}\la{app:nonequil}

\subsection{Relaxation mechanisms for photo-excited carriers}\la{app:relaxationmechanisms}

The steady shift current in a semiconductor cannot be calculated without understanding some basic aspects of the relaxation mechanisms and distribution of photo-excited carriers, which we review here. (We use `carrier' as a shorthand for hole and electron quasiparticles.) Much of this brief review derives from a more comprehensive review by Esipov and Levinson.\cite{esipov_temperatureenergy} \\

Which scattering process dominates the energy relaxation of carriers depends on (i) the radiation intensity $J$ generated by a source, and (ii) the energy $E$ of the carrier. \\

The dependence on $J$ is because the rate of carrier-carrier scattering via the instantaneous Coulomb interaction increases with the density $n$ of photo-excited carriers,\footnote{If the majority of photo-excited carriers follow a Maxwellian distribution, the rate of carrier-carrier scattering is simply proportional to $n$; cf.\ Eq. (2.3.6) in Ref.\ \cite{esipov_temperatureenergy}.} and $n$ is roughly proportional to  $J$.\\

The dependence on the carrier energy $E$ is because the matrix elements for scattering depend on the initial and final states. An especially strong dependence develops for $E$ near the optical phonon threshold $\hbar\Omega_o$, which is defined as the smallest optical phonon energy. Our convention is that $E$ for  an electron (resp. hole) carrier is set to zero at the conduction-band mimimum (resp. valence-band maximum). For $E<\hbar\Omega_o$ (the passive region), a carrier is forbidden by energy conservation against emitting optical phonons, and energy relaxation is substantially suppressed relative to $E>\hbar\Omega_o$ (the active region); cf.\ \q{activepassive} and \fig{fig:activeregion}(b-c).\\

We define an upper bound $n_h$ to the carrier density, such that if $n\lesssim n_{h}$ (meaning much less than or comparable in magnitude), scattering by optical phonons is the \textit{primary}/dominant mechanism of energy relaxation for photo-excited carriers \textit{in the active region}; if $n\gg n_h$, it would be carrier-carrier scattering that dominates energy relaxation in the active region. For instance,  $n_h\sim 10^{18}cm^{-3}$ for GaAs.\cite{esipov_temperatureenergy} We assume throughout this paper that optical phonons are the primary energy relaxers in the active region. Because
the typical carrier-optical-phonon scattering time  $\tau^o\sim 100fs$,\cite{lundstrom_book,na_damascelli_electronphonon} which is far smaller than the interband recombination time $(\tau_{\text{rec}}\sim 1ns$),\cite{sturmanfridkin_book,esipov_temperatureenergy} the majority of carriers would relax into the passive region where they await recombination.\footnote{Because $\hbar\Omega_o\sim 30 meV$, and a typical band width $\sim 1eV$, it takes at most thirty emissions of optical phonons for a hot carrier to relax into the active region. $30 \times 100fs$ is still much less than $\tau_{\text{rec}}$. } In other words, the steady electron (hole) distribution in the passive region accounts for most of the electrons in the conduction band (resp. holes in the valence band), as illustrated in \fig{fig:activeregion}(c). \\

It is also useful to identify the secondary/subdominant mechanism for energy relaxation in the active region; the two candidates are carrier-carrier scattering and carrier-acoustic-phonon scattering. We assume that the subdominant mechanism in the active region is also the dominant mechanism for energy relaxation in the passive region, where carrier-optical-phonon scattering `switches off' discontinuously.  
Let us  define a second density $n_l \ll n_h$, such that in the intermediate density range: $n_l \ll n \lesssim n_h$  (e.g., $10^{13}cm^{-3}\ll n \lesssim 10^{18}cm^{-3}$ for GaAs), carrier-carrier scattering is the subdominant relaxer in the active region; and in the low-density regime: $n\ll n_l$, carrier-acoustic-phonon scattering   is the subdominant relaxer in the active region.\\

For at least a number of semiconductors, steady-state measurements of hot-carrier photoluminescence spectra\footnote{ For instance, see \ocite{shah_photoexcitedhot} and \ocite{meneses_hotcarriers}; more experiments are reviewed in \ocite{zakharchenya_photoluminescence} and \ocite{esipov_temperatureenergy}.}  support the hypothesis that photo-excited electrons in the passive region  largely follow a nondegenerate Maxwellian distribution with a source-dependent chemical potential $\mu_e$ and electron temperature $T_e$; likewise, the majority of photo-excited holes in the passive region are Maxwellian with parameters $\mu_h$ and $T_h$. $\mu_e$ and $T_e$ are distinct from the equilibrium chemical potential and temperature: $T_e$ simply equals the non-equilibrium lattice temperature $T_l$  if electron-acoustic-phonon scattering is the dominant energy relaxer in the passive region ($n \ll n_l$); however, $T_e$ may  exceed $T_l$  if electron-electron scattering is the dominant energy relaxer in the passive region ($n_l \ll n \lesssim n_h$).  Typically, both $k_BT_e$ and $k_BT_h <\hbar \Omega_o$, so most of the photo-excited carriers  occupy only a smaller fraction of the passive region, and recombination transitions predominantly occur between electrons (with energy $\lesssim k_BT_e$)  and holes (with energy $\lesssim k_BT_h$). In large part, the theory that is presented in the main text is agnostic about fine-grained details of the carrier distribution within the passive region, meaning the theory is generally applicable whether or not a Maxwellian distribution develops in the passive region. However, if it does develop, then explicit kinetic models can be constructed that are based on the diffusive approximation for energy relaxation; cf.\ \app{app:eisotropy} and \app{app:jtranjexc}.

\subsection{The kinetic model in the low-density regime}\la{app:kineticmodel}

We will introduce a kinetic model that holds in the low-density regime ($n\ll n_l$) and forms the basis for numerical simulations of the BIS formula in \app{app:benchmark}.\\

The collisional integral for a quasiparticle  in a two-band semiconductor may decomposed into vertical photon-mediated transitions and intraband phonon-mediated transitions:
\e{
I_{c\bk} \eq I_{\bk}^{phot} + I_{c\bk}^{phon}; \as  I_{v\bk} = -I_{\bk}^{phot} +I_{v\bk}^{phon}. \la{ick}}
The photon-mediated component can be further decomposed into an excitation rate [\q{exciterate}] and recombination rate [\q{recombrate}], depending on whether $\bk$ lies on the excitation surface or not: 
\e{
I_{\bk}^{phot}\eq \delta_{\bk,ES}I^{m_s}_{exc\bk} -(1-\delta_{\bk,ES}) \sum_{m}^{phot}I^m_{rec\bk}.\la{iphot}
}
The phonon-mediated component can be decomposed into incoming transitions that increase the quasiparticle occupancy  and outgoing transitions that decrease  the quasiparticle occupancy:
\e{
I_{b\bk}^{phon} \eq \sum_{m}^{phon}\sum_{\bkp}\big( \cala^m_{b\bk\lea b\bkp}+\cale^m_{b\bk\lea b\bkp} -\cala^m_{b\bkp\lea b\bk}-\cale^m_{b\bkp\lea b\bk} \big),\la{iphon}
}
with $\cala$ and $\cale$ defined in \q{phononrate}.
Each of $I_{\text{exc}},I_{\text{rec}},\cala$ and $\cale$  depends on   the quasiparticle distribution $f_B$; this distribution is   is defined to be steady if it sets the collisional integral to zero:
\e{ \text{Steady distribution:}\as I_{c\bk}[f_B]=I_{v\bk}[f_B]=0  \as \text{for all}\as \bk. \la{steadydist}} \\

Let us first address the contribution to $I_{b\bk}^{phon}$ by carrier-optical-phonon scattering, which is assumed to be the dominant energy relaxation mechanism in the active region; cf.\ \app{app:relaxationmechanisms}. As justified in \app{app:threefold}, one may neglect the absorption and stimulated emission of optical phonons,  retaining only the transition rate for spontaneous emission:  $\cala^m_{b\bk\lea b\bkp}-\cale^m_{b\bkp\lea b\bk} \ri - \cale^{sp,m}_{b\bkp\lea b\bk}$, with $\cale^{sp}$ defined in \q{sponemirate1}. We assume that small-angle polarization scattering predominates over deformation scattering.\cite{gantmakherlevinson_book} For simplicity in modelling, we focus on polarization scattering by a single branch of longitudinal optical phonons, in which case the phonon mode $m$ is fully specified by a phonon wavevector $\bq$; for small $\bq$, the phonon  frequency is assumed to be approximately a constant  equal to $\Omega_o$.  The asymptotic expression for the collisional integral is then given by:\footnote{For general expressions, see Eq. (3.12) in \ocite{vogl_electronphonon} and the discussion in \s{sec:anomalous}. For the specific case of longitudinal optical phonons, Sec 1.3.E in \ocite{mahan_manyparticle} contains a concise derivation. }
\e{\text{Optical phonon:}\as &I^o_{b\bk} =\sum_{\bq\bkp}\big(\cale^{sp,\bq}_{b\bk\lea b\bkp} -\cale^{sp,\bq}_{b\bkp\lea b\bk} \big),\\
&\cale^{sp,\bq}_{b\bk \lea b\bkp}  = \delta_{\bkp-\bk,\bq}\frac{2\pi}{\hbar} |V^{\bq}_{b\bkp b\bk}|^2\delta(E_{b\bkp\bk}-\hbar\Omega_{o})\big(1-f_{b\bk}\big)f_{b\bkp} ; \\
& |V^{\bkp-\bk}_{b\bkp b\bk}|^2=|V^{\bk-\bkp}_{b\bk b\bkp}|^2\approx \frac{\hbar}{2\pi}\f{\zeta a}{\mathcal{V}} \f{\left|\braket{u_{b\bkp}}{u_{b\bk}}_{\sma{\text{cell}}}\right|^2}{|\bkp-\bk|^2}. \la{longi} }
$\zeta$ is a coupling parameter with dimensions of energy over time.
The inner product of cell-periodic Bloch functions is related to the quantum metric tensor\cite{provost_riemannianstructure,resta_insulatingmatter} as
\e{|\braket{u_{b\bkp}}{u_{b\bk}}_{\sma{\text{cell}}}|^2 \eq  1-\delta k_i\delta k_j g_{b\bk}^{ij}+O( \delta k^3 );\as  g_{b\bk}^{ij} = \real \braket{\nabk^iu_b}{\nabk^ju_b}_{\sma{\text{cell}}} -A_{bb\bk}^iA_{bb\bk}^j; \as \delta \bk=\bkp-\bk, \la{metric}}
with $\bA_{bb\bk}$ the intraband Berry connection.\footnote{Recently, the electron-phonon coupling has been related to an orbital-projected analog of the Fubini-Study metric.\cite{jiabin_nontrivialquantum}} Below room temperature ($k_BT_l\ll \hbar \Omega_o$), spontaneous emission of optical phonons dominates over stimulated emission and absorption, meaning we drop all terms in \q{longi} that are proportional to the Planck occupancy: $N_{\bq}^{T_l} \ll 1$.\\

Next we attend to the contribution to $I_{b\bk}^{phon}$ by carrier-acoustic-phonon scattering, which has been assumed to be the subdominant energy relaxation mechanism in the active region; cf.\ \app{app:relaxationmechanisms}.  Deformation scattering with acoustic phonons is typically outweighed by piezo-acoustic scattering.\cite{gantmakherlevinson_book} 
The precise expression of the transition rate/matrix element for piezo-acoustic scattering will not be required, and because we will eventually employ a diffusive Fokker-Planck approximation to the collisional integral. For now, it is worth knowing that the matrix element depends on the quasiparticle band index only through:\footnote{See Eq. (3.15) and (3.16) in Ref.\ \cite{vogl_electronphonon}, bearing in mind a remark made in \s{sec:anomalous} about a missing factor. } 
\e{|V^{\bkp-\bk,p}_{b\bkp, b\bk}|^2\propto|\braket{u_{b\bkp}}{u_{b\bk}}_{\sma{\text{cell}}}|^2,\la{bandind}}
 just as for polarization scattering with optical phonons in \q{longi}.

\subsubsection{Electron-hole symmetric kinetic model }\la{app:ehsymmetry}

Because it is numerically intensive to simulate  a steady distribution that satisfies $I_{c\bk}[f_B]=I_{v\bk}[f_B]=0$ for all $\bk$, we will resort to two model assumptions. The first is that  band energies and electron-phonon-scattering matrix elements are
\e{\text{Electron-hole symmetric}: \as E_{c\bk}=-E_{v\bk}; \as |V^m_{c\bkp,c\bk}|^2 = |V^{-m}_{v\bk,v\bkp}|^2, \la{ehsymmetry}}
with  $-m$ being the momentum-inverted counterpart of $m$.
This symmetry condition ensures for the phononic transition rates that
\e{
\cala^m_{c\bk\lea c\bkp}[1-f_v,1-f_c] =\cala^{-m}_{v\bkp\lea v\bk}[f_c,f_v]; \as \cale^m_{c\bk\lea c\bkp}[1-f_v,1-f_c] =\cale^{-m}_{v\bkp\lea v\bk}[f_c,f_v],
}
as may be verified by inspecting \q{phononrate}; the meaning of $\cala[1-f_v,1-f_c]$ is to replace $f_{c\bk} \ri 1-f_{v\bk}$ and $f_{v\bk}\ri 1-f_{c\bk}$ for all terms in $\cala$, and for all $\bk$. The photonic transition rate satisfies: $I_{\bk}^{phot}[1-f_v,1-f_c]=I_{\bk}^{phot}[f_c,f_v]$, even without assuming electron-hole symmetry; cf.\ \q{ehsymmetry}. Altogether,
\e{
I_{c\bk}[1-f_v,1-f_c] \eq I_{\bk}^{phot}[f_c,f_v] +\sum_{m}^{phon}\sum_{\bkp}\big( \cala^{-m}_{v\bkp\lea v\bk}+\cale^{-m}_{v\bkp\lea v\bk} -\cala^{-m}_{v\bk\lea v\bkp}-\cale^{-m}_{v\bk\lea v\bkp} \big)_{f_c,f_v} = -I_{v\bk}[f_c,f_v].
}

Thus the steady-state condition $I_{c\bk}=I_{v\bk}=0$ is solved by an electron-hole-symmetric distribution: $f_{c\bk}=1-f_{v\bk}$. In particular, if $f_{c\bk}$ is found such that $I_{c\bk}[f_c,1-f_c]=0$, then it is guaranteed that $I_{v\bk}[f_c,1-f_c]=0$. \\

Let us check that our model for the quasiparticle Hamiltonian [$H(\bk)$ in \q{modelham}] and carrier-phonon scattering [\qq{iphon}{longi}], is electron-hole-symmetric in the sense of \q{ehsymmetry}. Suppose that the conduction- and valence-band wave functions are related by an anti-unitary operation: $\ket{u_{c\bk}}_{\sma{\text{cell}}}=\hat{C}\ket{u_{v\bk}}$, which implies $\braket{u_{c\bk}}{u_{c\bkp}}=\braket{u_{v\bkp}}{u_{v\bk}}$. Because the electron-phonon matrix element (for both polarization and piezo-acoustic scatterings) only depends on the band index through  $|\braket{u_{b\bk}}{u_{b\bkp}}|^2$ [cf.\ \q{bandind}], $|V^m_{c\bk,c\bkp}|^2 =|V^m_{v\bk,v\bkp}|^2$. Applying a  general property of electon-phonon matrix elements: $V^m_{b\bk,b\bkp}=\overline{V^{-m}_{b\bkp,b\bk}}$ [cf.\ \q{Vnotselfadj}], one obtains the second equation in \q{ehsymmetry}. For the specific model Hamiltonian in \q{modelham}, the anti-unitary operation is simply $\hat{C}=\sy K$, with $\sy$ the second Pauli matrix and $K$ implementing complex conjugation. More generally, any $H(\bk)$ that is a sum of Pauli matrices satisfies
\e{
\sy \overline{H(\bk)} \sy =-H(\bk); \as E_{c\bk}=-E_{v\bk}.
}
The last condition  further implies that the excitation surface is an iso-energy surface:
\e{ 0=E_{cv\bk}-\hbar\omega_s = 2E_{c\bk}-\hbar\omega_s, \la{exciso}}
for any source radiation frequency $\omega_s$. In other words, the set of excitation energies $E_{c,exc}$  defined in \q{excenergy} is just a single energy.

\subsubsection{Iso-energy symmetric kinetic model }\la{app:eisotropy}

To recapitulate, we want to numerically simulate an electron-hole-symmetric distribution $f_{c\bk}$ such that $I_{c\bk}[f_c,1-f_c]=0$. Having reduced the problem to a single band by electron-hole symmetry, one may as well drop the band index on $f_{c\bk}\ri f_{\bk}$, $E_{c\bk}\ri E_{\bk}$ and $E_{c,exc}\ri E_{\text{exc}}$. We further redefine $E=0$ to be the minimal energy for the conduction band.  \\

To simplify the simulation of $f_{\bk}$, we further assume that  $f_{\bk}$  is approximately \textit{iso-energy symmetric}, meaning that $f_{\bk}$ is approximately constant under variation of $\bk$ within an iso-energy surface for $E_{\bk}$:
	\e{f_{\bk}\appr f_E; \as f_E =  \aven{f_{\bk}}{\bk E}\lin
 \aven{\Xi(\bk)}{\bk E}\eq \sum_{\bk}\,\delta_{\bk E}\,\Xi_{\bk};\as \delta_{\bk E}= \f{\delta(E_{\bk}-E)}{\calv g_E}, \la{isoenergy}}
 with 	$g_E$ meaning the density of conduction-band states per unit volume ($\calv$) and per spin orientation.
We refer to $\aven{\ldots}{\bk E}$ as \textit{iso-energy averaging}, and
$f_E$ as the \textit{iso-energy-averaged distribution}. For the purpose of computing the shift current, $f_{\bk}\approx f_E$ is justified to the extent that the collisional integral in \q{ick} is  iso-energy symmetric: $I_{\bk}\approx I_E=\aven{I_{\bk}}{\bk E}$, which constrains the model Hamiltonians that we allow ourselves to numerically simulate.\footnote{In general, it should be expected that the non-equilibrium distribution is iso-energy asymmetric with  respect to inverting $\bk$: $f_{\bk}\neq f_{-\bk}$. This is possible because the continuous absorption of photons creates a non-equilibrium state with a direction for time. Consequently, the asymmetry ($f_{c\bk}- f_{c,-\bk}$) is proportional to the light intensity\cite{sturmanfridkin_book} and contributes to a `ballistic current'\cite{belinicher_ballistic} but not the shift current.}  \\

By  averaging the kinetic equation $I_{\bk}[f_c,1-f_c]=0$ over an iso-energy surface, one obtains:\footnote{The following kinetic equation is very similar to one studied in \ocite{esipov_oscillatoryeffects}; however, we would rather not presume they adopted the same premises as we have adopted.
} 
	\e{E>E_{cut}:\as  g_EI_E[f]=G_{\sma{\uparrow}}[f]\,\delta(E-E_{\text{exc}})-\f{g_Ef_E}{\tau^o_{E}} +\f{g_{E_+}f_{E_+}}{\tau^o_{E_+}} -\f{g_Ef_E}{\tau_{\text{rec}}}-\partial_E j_E^s =0; \as E_{\pm}=E\pm \hbar\Omega_o. \la{esipovmodel2}}
We will explain the terms on the right-hand side in turn:\\

\noi{i} Recalling the excitation energy $E_{\text{exc}}$  [\q{excenergy}] to be the energy of conduction-band states on the excitation surface, $G_{\sma{\uparrow}}\,\delta(E-E_{\text{exc}})$ is the rate of increase in the quasiparticle number density $g_Ef_E$ due to the absorption of source-generated photons.\footnote{$G_{\sma{\uparrow}}\,\delta(E-E_{\text{exc}})$ is derived by applying $\sum_{\bk}\delta(E_{\bk}-E_{\text{exc}})/\calv$ to $I^{phot}_{\bk}$ [\q{iphot}] and retaining terms which are proportional to the source photon number $\Delta N_s$; cf.\ \q{photonoccupancy}.} In other words, $G_{\sma{\uparrow}}[f]$ is the rate at which source-generated photons are absorbed per unit volume and per spin orientation; 
\e{G_{\sma{\uparrow}}[f]=\f{\alpha_{\sma{\uparrow}}[f]\cali_{rad}}{\hbar\omega} \la{defineGup}}
can be expressed as a product of the single-spin absorption coefficient and the radiation intensity, divided by the source photon energy.\\

\noi{ii} $-gf/\tau^o\big|_{E}$  (resp. $+gf/\tau^o\big|_{E_+}$) in \q{esipovmodel2} represents an outflow (resp. inflow) of electrons due to spontaneous emission of optical phonons. $\tau^o_E$ is the average time  for a quasiparticle with energy $E$ to spontaneously emit an optical phonon; in the passive region ( $E<\hbar\Omega_{o}$), spontaneous emission is forbidden by energy conservation, hence we set $\tau^{o}_{E}=\infty$. One may relate $\tau^o_E$ to the collisional integral $I_{\bk}^{phon}$ by
\begin{equation}
\label{zeta}
\begin{aligned}
& \f{g_{E_+}f_{E_+}}{\tau^o_{E_+}}-\f{g_Ef_E}{\tau^o_{E}}=\frac{\zeta a}{\calv^2}\sum_{\bk\bkp}^{cut}\delta(E_{\bk}-E)\frac{\left|\braket{u_{\bkp}}{u_{\bk}}_{\sma{\text{cell}}}\right|^2}{|\bk-\bkp|^2}\bigg\{f_{\bkp}\delta(E_{\bkp}-E_+)
-f_{\bk}\delta(E_{\bkp}-E_-)
\bigg\}.
\end{aligned}
\end{equation}
The right-hand side is obtained by applying $\sum_{\bk}\delta(E_{\bk}-E)/\calv$ to the component of $I_{\bk}^{phon}$ corresponding to spontaneous emission of optical phonons [cf.\ \q{longi}] and dropping all terms which are  nonlinear in the quasiparticle distribution: $f_{\bk}f_{\bkp}\ll f_{\bk}$ and $f_{\bkp}$; bear in mind that nondegenerate fermion statistics ($f_{\bk}\ll 1$) apply to a wide range of continuous-wave laser experiments.\cite{esipov_temperatureenergy} Because we are employing an asymptotic expression that is valid for small-angle scattering, we have introduced a cutoff in $\sum^{cut}_{\bk\bkp}$, so that $\delta k=|\bk-\bkp|$ is much less than the linear dimension of the Brillouin zone.  \\

\noi{iii} $-{g_Ef_E}/{\tau_{\text{rec}}}$ represents the quasiparticle loss rate due to interband recombination by spontaneous emission of photons.   The effects of absorption and stimulated emission of thermal photons are negligible, as was explained in \app{app:threefold}. In numerical simulations, we just take  ${\tau_{\text{rec}}}\sim 1ns$  to be 
 a typical, energy-averaged  time scale for interband recombination.\cite{sturmanfridkin_book,esipov_temperatureenergy}. In principle, one could refine the model by replacing  ${\tau_{\text{rec}}}\ri{\tau_E^{\text{rec}}}$, with $\tau^{\text{rec}}$  depending on $E$ through the energy dependences of the dipole matrix element  and $f_E$.\footnote{${g_Ef_E}/{\tau_E^{\text{rec}}}=\sum_{\bk}\delta(E_{\bk}-E_{\text{exc}})\cale^{sp,m}_{v\bk \lea c\bk}/\calv$, with the spontaneous emission rate defined in \q{spontphotonrate}.}
 In practice, what matters to the  
 shift current is the order-of-magnitude difference: $\tau^{\text{rec}}_E \gg \gg \tau^o_E$, which guarantees that recombination transitions predominantly occur in the passive region, independent of  the precise energy dependence of $\tau^{\text{rec}}$; cf.\ \s{sec:kinetictheory} and \q{approxjrec}.  \\
 
\noi{iv} $-\partial_Ej_E^s$ is the rate of change of $g_Ef_E$ induced by electron-acoustic-phonon scattering. A negative $j_E^s$ represents a scattering-induced relaxation of  the number density $g_Ef_E$ toward decreasing energies, so we refer to $j_E^s$ as the \textit{energy-axis current}. In principle, this current should be an integral of $f_E$; however, the smallness of acoustic-phonon energies relative to typical electron energies allows to employ the   
diffusive Fokker-Planck approximation:\cite{gantmakherlevinson_book,physicalkinetics_lifshitzpitaev}
 \e{j_E^s= -\f{g_EE}{\tau^{s}_E}\bigg(1 + k_BT_l\partial_E\bigg)f_E. \la{jEs}}
$E/\tau^s_E$ is the dynamic friction coefficient,\cite{gantmakherlevinson_book} which  is interpretable as minus the  `drift speed' of a number-density-valued wavepacket on the energy axis. The form of $(1 + k_BT_l\partial_E)$ encodes an Einstein relation between the dynamic friction coefficient and the diffusion coefficient.\footnote{For electron-acoustic-phonon scattering, the Einstein relation is derived most directly from simplified expressions in Section 4.5 of \ocite{gantmakherlevinson_book}, assuming that $k_BT_l\gg$ the acoustic phonon energy. In a subsequent discussion in \app{app:jtranjexc}, we will also need an analogous Einstein relation for electron-electron scattering, which has been derived in \ocite{esipov_temperatureenergy}.}\\

Our previous assumption  that energy relaxation is dominated by optical phonons can now be expressed as a mathematical inequality, namely that the dynamic friction coefficient is much less than the energy relaxation rate due to spontaneous emission of optical phonons: 
	\e{E>\hbar\Omega_o:\as\eta_E = \f{E}{\hbar\Omega_o}\f{\tau^o_E}{\tau^s_E}\ll 1. \la{defineeta}}\\
 
The diffusive approximation is valid on the conditions that the density of states is analytic  and the  collisions are quasi-elastic. The former condition rules out van Hove singularities.\cite{vanhove_singularities} The latter condition means precisely that the change in a quasiparticle's energy (due to a collision) is much less than the quasiparticle's initial energy~\cite{gantmakherlevinson_book}. This holds for most quasiparticle energies, since acoustic-phonon energies are a very small fraction of the quasiparticle band width. We introduce a cutoff energy $E_{cut}$ which is comparable to the typical acoustic phonon energy, such that the diffusive approximation holds for $E>E_{cut}$.\\

 For energies less than the cutoff, we adopt the following kinetic equation:
 \e{  
0<E<E_{cut}: \as g_EI_E[f]=\f{g_{E_+}f_{E_+}}{\tau^o_{E_+}}-\f{g_Ef_E}{\tau_{\text{rec}}} - \frac{j_{E_{cut}}^s}{E_{cut}}=0. \la{belowcut}
 }
$-j_{E_{cut}}^s$ represents a (downward $\equiv$ energy-relaxing) current of  the number density $g_Ef_E$
across the cutoff energy [cf.\ \q{jEs}]; any density that relaxes across the cutoff  is equally distributed between all conduction-band states  below the cutoff.\footnote{The $-j_{E_{cut}}^s/E_{cut}$ term can be viewed as a collisional term $g_EI^s_E[f]$ due to the secondary scattering mechanism. The crudeness in our approximation lies in assuming $g_EI^s_E[f]$ is independent of $E$, for $E$ below the cutoff. This amounts to assuming that $I^s_E$ is independent of $E$, because the density of states is energy-independent for a quasi-2D parabolic band. } This  crude modelling of scattering below the cutoff can in principle be improved upon, but we remind the reader that the steady shift current is insensitive to fine details of the quasiparticle distribution within the passive region, owing to arguments explained in \s{sec:kinetictheory}. Despite the crudeness of the model, the model ensures that all phonon-mediated collisions conserve the total number of quasiparticles within the conduction band. In other words, if all the collisional terms  in \qq{esipovmodel2}{belowcut}, with the exception of terms involving $G_{\sma{\uparrow}}$ and $\tau_{\text{rec}}$, are collectively denoted  as $g_E I^{\text{intra}}_E$, then $\int_{0}^{\infty} g_E I^{\text{intra}}_E dE=0$. 

\section{Belinicher-Ivchenko-Sturman  formula for the shift current}\la{app:bisformula}

\subsection{Derivation of the Belinicher-Ivchenko-Sturman formula}\la{app:derivebis}

It has been expressed to the authors that the Belinicher-Ivchenko-Sturman theory\cite{belinicher_kinetictheory} is difficult to penetrate. To our  knowledge, no explicit derivation of the BIS formula yet exists in the literature. We will therefore derive their main formulas for pedagogy. Precisely, we mean to derive the form of the phonon-mediated (resp.\ photon-mediated) shift current to be 
\q{bismine2}, with the phononic (resp.\ photonic) shift vector given in \q{phononicshift2} [resp.\ \q{photonicshift2}], and with the difference in absorption and emission transition rates given by \q{phononrate} [resp.\ \q{photonrate}]. \\

Since the BIS formula encodes the spontaneous emission of photons, the derivation requires to quantize the radiation field. If one were to quantize the radiation field but retain a first-quantized electron description, one would derive an analog of the BIS formula that is only applicable to nondegenerate Fermi statistics, i.e., one would miss a spontaneous-emission term that is nonlinear in the distribution function [cf.\ \q{photonrate} below]. To properly account for the Pauli exclusion principle in the presence of spontaneous emission, it is necessary to apply second quantization to the electron. It is fortuitous but misleading that terms which are nonlinear in the distribution function cancel out if one considers only photon absorption and stimulated emission [\q{photonrate} with $N_{m}+1 \approx N_{m}$]; thus it has been possible for theories (based on first quantization of the electron and a classical theory of radiation) to neglect the exclusion principle and yet derive correct  formulas for the transient shift current, as will be elaborated in \app{sec:krautbaltz}.\\

Our derivation also manifests how a perturbation theory of the steady state differs dramatically from a perturbation theory of the transient state. Most practitioners who calculate nonlinear optical responses are calculating the transient response, and their zeroth-order state is the thermal equilibrium state in the absence of the light source. In \textbf{steady-state perturbation theory}, the zeroth-order state is emphatically \underline{not} a thermal state; instead, \app{app:zerothrho} proves rigorously that if the state is steady, the zeroth-order quasiparticle distribution satisfies a non-detailed balance condition that represents an invariance under simultaneous collisions with all bosons. Moreover, we have no reason to believe that the zeroth-order state in steady perturbation theory is perturbatively connected to the thermal equilibrium state (in the absence of the light source).\\

 The outline of the derivation is:\\

\noi{i} \app{sec:defineshiftcurrent} sets up the problem and establishes the notation. We review salient properties of the independent-electron Hamiltonian, the crystal momentum representation, the independent-boson Hamiltonian, Fock space and the electron-boson interaction.  Finally, we  express the shift current in terms of stationary density matrices, and derive a perturbative expression for the stationary density matrix  in the Lippmann-Schwinger scattering formalism.  \\

\noi{ii} The Lippman-Schwinger formula for the stationary density matrix is expressed in terms of second-quantized matrix elements; these elements will be reduced to first-quantized matrix elements in  \app{sec:algebra}. The result of this reduction is an intermediate formula for the photonic and phononic shift current in \q{eq6bis} and \q{eq6bisphonon}, respectively.  These intermediate formulas are more formal than optimal: they are expressed in terms of an infinite number of band-off-diagonal matrix elements of the position operator. \\

\noi{iii} \app{app:sumrulephotonic} derives an optimal expression for  the photonic shift current, with help from  a sum rule derived from the first-quantized commutation relation between position and canonical momentum.\\

\noi{iv} \app{app:sumrulephononic} derives an optimal expression for  the phononic shift current, with help from  a sum rule derived from the first-quantized commutation relation between position and the phonon-induced potential-energy field.\\

\noi{v} \app{app:zerothrho} demonstrates that the zeroth-order density matrix is not thermal; instead, the zeroth-order quasiparticle distribution satisfies a non-detailed balance condition that represents an invariance under simultaneous collisions with all bosons.

\subsubsection{Preliminaries}\la{sec:defineshiftcurrent}

We decompose our Hamiltonian into two independendent-particle terms  and an electron-boson interaction:
\e{H=H_0+U, \as H_0=H_0^{ele} + H_0^{bos}.}
We will first explain the independent-particle terms:\\

\noindent \underline{Independent-electron Hamiltonian and the crystal momentum representation}\\

$H_0^{ele}$ is a mean-field Hamiltonian for independent electrons in a crystalline medium:
\e{H_0^{ele}= \sum_{B}E_B \dg{c}_B\pdg{c}_B, \as [\pdg{c}_B,\dg{c}_{B'}]=\delta_{B,B'}, \as B=(b,\bk), \as B'=(b',\bkp)\la{H0ele}}
where $[x,y]= xy- yx$ (the commutator) and $B$ is a collective index for both the band label and crystal wavevector. We assume throughout this work that spin-orbit coupling is negligible; to simplify notation, $b$ should be understood as a spinless band label, and $H$ as a Hamiltonian in one spin sector; only in the final steps will the current be multiplied by two to account for the spin degeneracy of bands.\\

$c_B$ annihilates an electronic state with a wave function of the Bloch form: $\eikr u_{b\bk}(\br)/\sqrt{\calv}$, with $u_{b\bk}(\br)=u_{b\bk}(\br+\bR)$ being periodic in Bravais-lattice translations and $\calv$ the volume of the medium. These cell-periodic functions are normalized as 
\e{\braket{u_{b\bk}}{u_{b'\bk}}_{\sma{\text{cell}}}=\delta_{b,b'}, \as \braket{X}{Y}_{\sma{\text{cell}}}=\int \f{d\btau}{\calv_{\sma{\text{cell}}}} \overline{X(\btau)}{Y(\btau)} \la{intracell}, }
with $\btau$ the intracell coordinate, $\delta_{b,b'}$ a Kronecker delta function for the band labels, and $\calv_{\sma{\text{cell}}}$ the real-space volume of the primitive unit cell. The orthonormality  and completeness of our basis of Bloch waves reads  as
\e{ \braket{B}{B'}_{1}=\delta_{B,B'}=\delta_{bb'}\delta_{\bk\bkp}, \as I_{1}=\sum_{B}\ketbra{B}{B}_{1}. \la{orthocom}}
$I$ is the identity operator, and the superscript $1$  in \q{orthocom} reminds us that we are dealing with a first-quantized, one-particle Hilbert space.  \\

Our notation for $\bk$ suggests misleadingly that $\bk$  is a discrete wavevector: $\sum_B=\sum_b\sum_{\bk}$ and $\delta_{B,B'}=\delta_{\bk,\bkp}\delta_{b,b'}$. However, for  the position operator  to have a well-defined action on periodic Bloch states, one must take $\calv$ to be infinite,\cite{blount_formalisms} hence $\delta_{\bk\bkp}= (2\pi)^3\delta(\bk-\bkp)/\calv$ should be understood as a shorthand for a Dirac delta function, and we will be applying certain identities that apply to Dirac delta functions but not Kronecker delta functions:
\e{ &\nabk \delta_{\bk\bkp}= -\nabkp \delta_{\bk\bkp}; \as f_{\bkp}\nabk\delta_{\bk\bkp} -f_{\bk}\nabk\delta_{\bk\bkp}=\delta_{\bk\bkp}\nabk f_{\bk}.\la{dirac2}
}
$\sum_{\bk}$ should also  be understood as an integral over the Brillouin zone: $\calv \int_{BZ} d^3k/(2\pi)^3$. With these caveats in mind, we present 
the first-quantized position, canonical momentum and velocity operators in the crystal momentum representation:\cite{blount_formalisms}
\e{  \br_{BB'}\eq   \braopket{B}{\hbr}{B'}_{1}  = i\delta_{bb'}\nabk\delta_{\bk\bkp} +\delta_{\bk\bkp}\bA_{bb'\bk}; \as \bA_{bb'\bk}=\braket{u_{b\bk}}{i\nabk u_{b'}}_{\sma{\text{cell}}}\la{rbb}\\
\bp_{BB'}\eq   \braopket{B}{\hbp}{B'}_{1} = \delta_{\bk\bkp}\bP_{bb'\bk}; \as \bP_{bb'\bk}=\braopket{u_{b\bk}}{\hbp}{u_{b'\bk}}_{\sma{\text{cell}}} =m_f\bv_{bb'\bk}, \la{pbb}\\
\bv_{BB'}\eq   \braopket{B}{\hat{\bv}}{B'}_{1} = \delta_{\bk\bkp}\bv_{bb'\bk};\as \bv_{bb'\bk}= \braopket{u_{b\bk}}{\hat{\bv}}{u_{b'\bk}}_{\sma{\text{cell}}}.\la{definevbb}}
We have assumed in the absence of spin-orbit coupling that $\hbv=\hbp/m_f$, with $m_f$ the free-electron mass. It is also worth defining the band-off-diagonal position operator as 
\e{  
\hbr_{\text{off}} \eq \sum_{BB'} \br^{\text{off}}_{BB'}\ketbra{B}{B'}_{1}; \as \br^{\text{off}}_{BB'}=\delta_{\bk\bkp}\bA^{\text{off}}_{bb'\bk}; \as \bA^{\text{off}}_{bb'\bk}=\bA_{bb'\bk}(1-\delta_{bb'}),\la{defineroff}
} 
 which is related to the band-off-diagonal elements of the velocity operator:\cite{blount_formalisms}
	\e{  \f{\bv^{\text{off}}_{b'b\bk}}{E_{bb'\bk}} = -\f{i}{\hbar}\bA^{\text{off}}_{b'b\bk}; \as E_{bb'\bk}=E_{b\bk}-E_{b'\bk}. \la{defineXk}}
\\

\noindent \underline{Independent-boson Hamiltonian}\\

$H_0^{bos}$ is the independent-boson Hamiltonian absent the zero-point energy:
\e{H_0^{bos}=\sum_{m}\hbar\omega_m \dg{a}_m\pdg{a}_{m'},\as \pdg{a}_m\dg{a}_{m'}+\dg{a}_{m'}\pdg{a}_m=\delta_{m,m'}, \as m=(\bq,p),\la{freeboson}}
where the index $m=(\bq p)$  runs over both photonic and phononic modes.\\

We follow E. Fermi's prescription\cite{fermi_quantumradiation} in quantizing the transverse/solenoidal component of the electromagnetic vector potential  in the Coulomb gauge.\cite{heitler_quantumradiation,sakurai_advanced} For photons,  $\bq$ is a wavevector in  $\R^3$ with a cutoff: $\hbar cq<E_{cut}$;  the cutoff energy may be taken as the largest energy difference between the Bloch bands which are excited by the light source.\footnote{This cutoff is imposed for self-consistency: our use of the dipole approximation requires that $q$ is much less than the linear dimension of the BZ.} $p\in \{1,2\}$ specifies one of the two possible transverse polarizations for a given $\bq$; we adopt a linearly polarized basis, meaning the polarization vector is real-valued: $\beps_m \equiv \beps_{\hbq}^{(p)}=\beps_{-\hbq}^{(p)}\in \R$. The photon frequency is polarization-independent: $\omega_m=cq$ with $q=||\bq||$.\\

For phonons,  $\bq$ is a wavevector in the BZ, and $p=3,4,\ldots,3 N_{nuc}+2$ a label for a nondegenerate phonon band, with $N_{nuc}$ being the number of nuclei per primitive unit cell. $\omega_{\bq p}=\omega_{-\bq p}$ is the renormalized phonon dispersion.\cite{shamziman_electronphonon,keating_selfconsistentphonon}\\

Altogether, $\sum_m=\sum_{p}\sum_{\bq}$ with $p$ running over $3N_{nuc}+2$ values, and $\sum_{\bq}=\calv \int d^3q/(2\pi)^3$ with the integration domain depending on $p$, and
$\delta_{m,m'}$ should be understood as $\delta_{\bq,\bqp}\delta_{p,p'}$. \\

\noindent \underline{Fock space}\\

Eigenstates of the independent-particle Hamiltonian are labelled by electronic occupancies $n_B\in \{0,1\}$ and bosonic occupancies $N_m\in \{0,1,2,\ldots\}$:
\e{  (H_0-E_{\mu})\ket{\mu}=0, \as \mu=\bigg(\{n^{\mu}_B\}_B,\{N_m^{\mu}\}_m\bigg), \as E_{\mu}= \sum_BE_Bn^{\mu}_B +\sum_m \hbar\omega_m N^{\mu}_m. }
Throughout this appendix, Greek symbols (like $\mu$) are used as a collective index for all electronic and bosonic occupancies.  $\{n_B\}_B$ means a set of occupancies for all Bloch states, but we will often use the shorthand:  $\{n_B\}_B\ri \{n\}.$ Likewise for $\{N_m\}_m\ri \{N\}.$ We will refer to $\ket{\mu}$ as an \textit{independent-particle state}. The set of independent-particle states forms an orthonormal basis  ($\braket{\mu}{\nu}=\delta_{\mu,\nu}$) for the combined-electron-boson Fock space. The resolution of identity is given by 
\e{I=\sum_{\mu}\ketbra{\mu}{\mu}=\sum_{\{n\}}\sum_{\{N\}}\ketbra{\{n\}\{N\}}{\{n\}\{N\}}, \as \sum_{\{n\}}=\prod_B\sum_{n_B=0}^1, \as \sum_{\{N\}}=\prod_m\sum_{N_m=0}^{\infty}.\la{resolution}}

 Any  operator $O^e$ with an $e$ superscript should be understood as acting only in the electronic Fock space, which is spanned by \textit{independent-electron states} denoted as  $\ket{\{n\}}_e$. (The existence or absence of subscripts distinguishes kets in  different Hilbert spaces.) We will focus on bilinear electronic operators 
 \e{O^e=\sum_{B,B'} O_{B,B'}\dg{c}_B\pdg{c}_{B'}; \as O^e_{\mu\nu} =   \braopket{\{n^{\mu}\}}{O^e}{\{n^{\nu}\}}_e,\la{defineOe}}
 with matrix elements denoted as $O^e_{\mu\nu}$; the commutator of two bilinear operators is expressible as:
 \e{[G^e,O^e]=\sum_{BB'} [G,O]_{BB'}\dg{c}_B\pdg{c}_{B'};\as  G^e=\sum_{B,B'} G_{B,B'}\dg{c}_B\pdg{c}_{B'}. \la{commutatorelecop}}

\noindent \underline{Electron-boson interaction}\\

We decompose $U$ into a tensor product of operators acting in the electronic and photonic Fock spaces:
\e{U=\sum_m U_m^e (\pdg{a}_m +\dg{a}_{-m});\as U^e_m=\sum_{B,B'}U^m_{B,B'}\dg{c}_{B}\pdg{c}_{B'}=\dg{(U^e_{-m})}, \la{defineU}}
with $-m=(-\bq,p)$ the momentum-reversed partner of $m=(\bq p)$. $U^e_m=\dg{(U^e_{-m})}$ ensures that $U$ is self-adjoint. For  $m$ that  is photonic (resp. phononic), $U^m_{B,B'}$ is defined as the \textit{electron-photon} (resp. \textit{electron-phonon}) \textit{matrix element}:  
\begin{align}
U^{\bq p}_{b\bk,b'\bkp}=\begin{cases}\delta_{\bk,\bkp}\breve{W}^{\bq p}_{bb'\bk}; \as \breve{W}^{\bq p}_{bb'\bk}=\sqrt{\tf{he^2}{\omega_{\bq p}\calv}}\beps_{\hbq}^{(p)} \cdot \bv_{bb'\bk};\as &(\text{electron-photon}) \\
V^{\bq p}_{b\bk,b'\bkp}=\delta_{\bk,\bkp+\bq}\breve{V}^{\bq p}_{bb'\bk}; \as  \breve{V}^{\bq p}_{bb'\bk}=\calv^{-1}\sum_{\bG}^{RL} \widetilde{PE}^{\bq p}_{\bq+\bG} \braopket{u_{b\bk}}{e^{i\bG\cdot \hbr}}{u_{b'\bk-\bq}}_{\sma{\text{cell}}}.\as &(\text{electron-phonon}) 
\end{cases}\la{defineUem}
\end{align}
We will describe each matrix element in turn.\\

The electron-photon matrix element is derived from the first-order term in the non-relativistic minimal coupling: $|e|\bA_{\perp}\cdot \bv/ c$, with $\bv$ the second-quantized electron velocity operator and $\bA_{\perp}$ the quantized electromagnetic vector potential satisfying $\diverge \bA_{\perp}=0$.\cite{heitler_quantumradiation} 
 The photonic expression in \q{defineUem} is valid in the dipole approximation. Within this approximation, $U^e_m=U^e_{-m}=\dg{(U^e_m)}$ is self-adjoint. Minimal coupling also results  in an electron-photon interaction proportional to $e^2$, but such a coupling does not contribute to the shift current because it cannot induce interband transitions within the dipole approximation [cf.\ \q{definephotonicshiftcurrent} below].\\

We adopt a simplified electron-phonon matrix element $\breve{V}^{\bq p}_{bb'\bk}$ which is derived in the adiabatic approximation (where phonons are frozen from the electron's perspective) and by applying the Hartree approximation to electron-electron interactions.\cite{vogl_electronphonon,shamziman_electronphonon} 
In the expression for $\breve{V}^{\bq p}_{bb'\bk}$, $\sum_{\bG}^{RL}$ sums over all reciprocal-lattice vectors, $\bq$  is a wavevector in the Brillouin zone $BZ$ and $\bq+\bG=\bQ$  a wavevector in $\R^3$. $\widetilde{PE}^{m}_{\bQ}=\int_{\R^3} e^{-i\bQ\cdot \br} PE^{m}_{\br}d\br$  is a Fourier transform of the one-electron potential energy $PE^{m}_{\br}$ induced by annihilating a phonon of mode $m$. $PE_{\br}^m$ is self-consistently\cite{shamziman_electronphonon} screened in a crystalline medium, and is linearly related to the bare potential energy $PE_{\br}^{m;0}$:
\e{\widetilde{PE}^{\bq p}_{\bq+\bG}\eq \sum_{\bG'}^{RL}\varepsilon^{-1}_{\bq+\bG,\bq+\bG'}\widetilde{PE}^{\bq p; 0}_{\bq+\bG'}= \overline{\widetilde{PE}^{-\bq p}_{-\bq-\bG}},}
with $\varepsilon^{-1}_{\bQ,\bQ'}= \overline{\varepsilon^{-1}_{-\bQ,-\bQ'}}$ the static, inverse dielectric function in the  Hartree approximation.\footnote{An explicit expression can be found in Eq. (12.16) of Ref.\ \cite{shamziman_electronphonon}.} The bare potential energy is expressible in terms of $\tilde{v}_{\bQ}=4\pi e^2/||\bQ||^2$, the Fourier transform of the Coulomb interaction:
\e{
\widetilde{PE}^{\bq p; 0}_{\bq+\bG} \eq i\tilde{v}_{\bq+\bG}\sum^{nuclei}_j \bigg(\tf{\hbar N_{\sma{\text{cell}}}Z_{j}^2}{2\omega_{\bq p}M_j}\bigg)^{\sma{1/2}}(\bq+\bG)\cdot \beps_{\bq p}^{j} e^{-i\bG\cdot  \br_j}=\overline{\widetilde{PE}^{-\bq p;0}_{-\bq-\bG}}, 
 } 
 with the caveat that $\tilde{v}_{\bze}=0$ to account for the electrical neutrality of the entire medium.\cite{bardeenpines_electronphonon}  
$N_{\sma{\text{cell}}}=\calv/\calv_{\sma{\text{cell}}}$ is the number of primitive unit cells; $j$ labels the nuclei in one primitive unit cell; a nucleus labelled $j$ has a charge $Z_j|e|$, mass $M_j$, and real-spatial coordinate $\br_j$; $\beps_{m}^{j}=\overline{\beps_{-m}^{j}}$ is the polarization vector of the $j$'th nucleus.\footnote{The above expressions are obtained from equations (2.9) to (2.11) in \ocite{vogl_electronphonon}.}   \\

It is worth defining a first-quantized operator whose matrix elements (with respect to Bloch waves) are identical to the electron-phonon matrix element [\q{defineUem}]:
\e{  \hat{V}^{\bq p} =\calv^{-1}\sum_{\bG}^{RL}{\widetilde{PE}^{\bq p}_{\bq+\bG}} {e^{i(\bq+\bG)\cdot \hbr}}; \as \braopket{B}{\hat{V}^{m}}{B'}_{1}=V^{m}_{BB'}.\la{eleconephononop}}
Because $PE^{m}_{\br}$ is the one-body potential induced by a complex-valued wave (rather than a standing wave), the potential is not real-valued but satisfies $PE^{m}_{\br}=\overline{PE^{-m}_{\br}}$; moreover, $\hat{V}^m$ is not  self-adjoint: 
\e{ \dg{(\hat{V}^{m})}= \hat{V}^{-m}=\hat{T} \hat{V}^{m}\hat{T}^{-1}; \as \overline{V^{m}_{BB'}}= V^{-m}_{B'B},\la{Vnotselfadj}}
with $\hat{T}$ being the first-quantized, time-reversal operator.\\

\noindent \underline{Shift current in terms of density matrices}\\

We adopt the Schr\"odinger representation in which $a_m,\dg{a}_m,U$ and $H$ are all time-independent, i.e., $a_m$ is not accompanied with the multiplicative factor $e^{-i\omega_mt}$. This allows to  solve for the stationary density matrix 
\e{\partial_t\rho = -\f{i}{\hbar}[H,\rho]=0, \as \rho=\rho^{(0)}+\rho^{(1)}+\rho^{(2)}+\ldots \la{rhostat}}
in time-independent perturbation theory, with $\rho^{(n)}$ proportional to the $n$'th power of the perturbation $U$. \\

Because $\rho$ is stationary, $-|e|\tr [\bv \rho]$ represents a direct current. $-|e|\tr [\bv \rho^{(0)}]$ represents the direct current in the absence of the light source, and vanishes by time-reversal symmetry.
We will see in \s{sec:algebra} that $\rho^{(1)}$ does not contribute to the direct current, but $\rho^{(2)}$ does. The shift current is the second-order direct  current contributed by band-off-diagonal  elements of the velocity matrix $\bv_{bb'\bk}$: 
\e{\bj= -\frac{|e|}{\calv} \tr \big[\bv_{\text{off}}\rho^{(2)}\big]; \as \bv_{\text{off}}=\sum_{b,b',\bk}\bv^{\text{off}}_{bb'\bk}\dg{c}_{b\bk}\pdg{c}_{b'\bk}; \as \bv^{\text{off}}_{bb'\bk}=\bv_{bb'\bk}(1-\delta_{b,b'}).\la{definephotonicshiftcurrent}}
Band-diagonal elements contribute to the `ballistic current',\cite{belinicher_ballistic} which we do not touch upon in this work.\\

\noindent \underline{Stationary density matrix from the Lippmann-Schwinger formalism}\\

We will derive $\rho^{(n)}$ based on the Lippmann-Schwinger scattering formalism,\cite{lippmann_schwinger,weinberg_foundations} which we briefly review.\\

For any independent-particle state $\ket{\mu}$ with energy $E_{\mu}$, one can construct an `in' state  $\ket{\mu_+}$ that is an eigenstate of the full Hamiltonian with the same energy:
\e{ (H-E_{\mu})\ket{\mu_+}=0; \as \ket{\mu_+}=\ket{\mu} +G^+_{E_{\mu}}U\ket{\mu}; \as G^+_E=\f1{E-H+i0^+},}
with $G^+$ the retarded Green's function and $0^+$ a positive infinitesimal. An `in' state has the same normalization as its independent-particle counterpart.\cite{weinberg_foundations} Since the set of independent-particle states forms an orthonormal basis, so then does the set of all `in' states: $\braket{\mu_+}{\nu_+}=\delta_{\mu,\nu}$.\\

Let us  motivate the imaginary infinitesimal by a wave packet interpretation proposed in \ocite{weinberg_foundations}. The above correspondence between $\ket{\mu}$ and $\ket{\mu_+}$ allows to parametrize $\ket{\mu_+}$ by the one-particle wavevectors $(\bk_1,\bk_2,\ldots,\bq_1,\bq_2,\ldots$) of electrons and bosons that make up $\ket{\mu}$. Thus it is possible to form a wave packet by smoothly linearly combining $\ket{\mu_+}$ with slightly different values for the one-particle wavevectors. The $i0^+$ guarantees that such a wave packet behaves essentially as a superposition of independent particles in the far past: $t\ri -\infty$.\footnote{One can construct `out' states by flipping the sign of $i0^+$, such that the wave packet becomes essentially non-interacting in the far future. This wave packet interpretation is elaborated in Chapter 3 of \ocite{weinberg_foundations}.  In other derivations of the conductivity,\cite{holder_trsbreaking,lingyuan_fermisurface} $i0^+$ appears as a result of  an adiabatic turn-on process in accordance with Kubo tradition,\cite{kubo_statisticalmechanical} yet no such adiabatic process exists in the typical experiment, e.g., with lasers.} The use of `in' states thus simulates a scattering process in which localized wavepackets of electrons and bosons are initially separated (in real space) but subsequently approach each other, and in so doing evolves into an entangled, polaritonic/polaronic state with a nontrivial current.\\

Let us construct a density matrix by summing over outer products of `in' states weighted by probability coefficients $F_{\mu}$:
\e{\rho = \sum_{\mu}F_{\mu}\ketbra{\mu_+}{\mu_+};\as 1\eq \sum_{\mu}F_{\mu}. \la{rhoperturb}}
Because the `in' state is an eigenstate of $H$, $\rho$ satisfies the stationary condition in \q{rhostat}.
By iteratively expanding the Green's function in a perturbative series 
\e{ G^+ = G^+_{0} + G^+_{0}UG^+_{0} + G^+_{0}UG^+_{0}UG^+_{0}+\ldots; \as G^+_{0;E}=\f1{E-H_0+i0^+},  }
one obtains a perturbative series for the density matrix:
\e{\rho = \sum_{\mu}F_{\mu}\ketbra{\mu_+}{\mu_+}=\rho^{(0)}+\rho^{(1)}+\rho^{(2)}+\ldots; \as \rho^{(0)}= \sum_{\mu}F_{\mu}\ketbra{\mu}{\mu}.}
Because the zeroth-order component  $\rho^{(0)}$ is stationary with respect to the non-interacting Hamiltonian $H_0$, one may as well
take $F_{\mu}$ to be a product of one-particle probabilities $p_{n_B}$ and $P_{N_m}$: 
\e{F_{\mu}=p_{\{n^{\mu}\}}P_{\{N^{\mu}\}}, \as p_{\{n\}}=\prod_B p_{n_B}, \as P_{\{N\}}=\prod_m P_{N_m}.\la{defineFmu}}
The sense in which $p$ and $P$ are one-particle probabilities is that
\e{1\eq \sum_{n_B=0}^1p_{n_B}=\sum_{N_m=0}^{\infty}P_{N_m};\as \expect{n_B}=\sum_{n_B=0}^1p_{n_B}n_B=\sum_{\mu}F_{\mu}n_B^{\mu}; \as \expect{N_m}=\sum_{N_m=0}^{\infty}P_{N_m}N_m=\sum_{\mu}F_{\mu}N_m^{\mu}, \la{definepP} }
with $\expect{n_B}$ and $\expect{N_m}$ being the average number of electrons and bosons with the one-particle labels $B$ and $m$, respectively. In a generic, non-equilibrium state,  $p_{n_B}$ does not have the Fermi-Dirac form, and instead satisfies a non-detailed balance condition that represents an invariance under simultaneous collisions with all bosons, as detailed in \app{app:zerothrho}.   \\

It is convenient to introduce the shorthand
\e{F_{\mu\nu}=F_{\mu}-F_{\nu}, \as E_{\mu\nu}=E_{\mu}-E_{\nu},}
and express $\rho^{(1)}$ and $\rho^{(2)}$ in terms of their matrix elements in the independent-particle basis:
\e{\rho^{(1)}_{\mu\nu}\eq \braopket{\mu}{\rho^{(1)}}{\nu} =\f{F_{\mu\nu}U_{\mu\nu}}{E_{\mu\nu}-i0^+} \la{rho1}\\
\rho^{(2)}_{\mu\nu}  \eq \sum_\lambda U_{\mu\lambda}U_{\lambda\nu}\bigg[ \f{F_\lambda}{(E_{\lambda\mu}+i0^+)(E_{\lambda\nu}-i0^+)} + \f{F_\nu}{(E_{\nu\mu}+i0^+)(E_{\nu\lambda}+i0^+)} +  \f{F_\mu}{(E_{\mu\lambda}-i0^+)(E_{\mu\nu}-i0^+)}\bigg]\lin
\eq \sum_\lambda \f{U_{\mu\lambda}U_{\lambda\nu}}{E_{\mu\nu}-i0^+}\bigg[  \f{F_{\mu\lambda}}{E_{\mu\lambda}-i0^+} + \f{F_{\nu\lambda}}{E_{\lambda\nu}-i0^+} \bigg]. \la{fflast}
}

\subsubsection{From second-quantized matrix elements to first-quantized matrix elements}\la{sec:algebra} 

We need only concern ourselves with matrix elements $\rho^{(n)}_{\mu\nu}$ with $\ket{\mu}$ and $\ket{\nu}$ having identical occupations numbers for all bosonic modes.
After all, for any operator $O=O^e\otimes (identity)$ that acts trivially in the bosonic Fock space, 
\e{\tr [O\rho]=\sum_{\mu\nu}O_{\nu\mu}\rho_{\mu\nu}\delta_{\{N^{\mu}\},\{N^{\nu}\}}.\la{electronicop}}
In particular, \q{electronicop} holds for $O$ being the electronic velocity operator $\bv$. An immediate implication is that $\rho^{(1)}$ does not contribute to the direct current: $\tr[\bv \rho^{(1)}]=0$, because $\rho^{(1)}_{\mu\nu} \propto U_{\mu\nu}$ [cf.\ \q{rho1}] and $U$ necessarily changes the boson number; cf.\ \q{defineU}.\\

Let us apply \q{electronicop} to the shift current [\q{definephotonicshiftcurrent}] with $\rho^{(2)}=\sum_\lambda U_{\mu\lambda}U_{\lambda\nu} \ldots $  given in \q{fflast}. If $U_{\lambda\nu}$ represents the creation (resp.\ annihilation) of a boson of mode $m$, then $U_{\mu\lambda}$ must represent the annihilation (resp.\ creation) of a boson of the same mode. Thus, $\rho^{(2)}_{\mu\nu}\delta_{\{N^{\mu}\},\{N^{\nu}\}}=$ 
\e{ \sum_{\lambda,m} \f{\braopket{\mu}{U_m^e a_m}{\lambda} \,\braopket{\lambda}{U^e_{-m} \dg{a}_m}{\nu} + \braopket{\mu}{U^e_{-m} \dg{a}_m}{\lambda} \,\braopket{\lambda}{U^e_m {a}_m}{\nu}}{E_{\mu\nu}-i0^+}\bigg[  \f{F_{\mu\lambda}}{E_{\mu\lambda}-i0^+} + \f{F_{\nu\lambda}}{E_{\lambda\nu}-i0^+} \bigg] \delta_{\{N^{\mu}\},\{N^{\nu}\}}.\la{interme}}
In particular, \q{electronicop} implies there are no
`cross terms'  proportional to $\langle \ldots U^e_m \ldots \rangle \langle \ldots U^e_{m'}\ldots \rangle$ with $m$ photonic and $m'$ phononic.\\

 \q{interme} manifests
two classes of intermediate states $\ket{\lambda}$ -- one with $\{N^{\lambda}\}$ differing from $\{N^{\mu}\}$ only in that $N^{\lambda}_m=N^{\mu}_m+1$, and another with $\{N^{\lambda}\}$ differing from $\{N^{\mu}\}$ only in that $N^{\lambda}_m=N^{\mu}_m-1$. We distinguish the two classes by the notation $\{N^{\lambda}\}=\{\ldots,N^{\mu}_m\pm 1,\ldots \}$, which allows to express $\rho^{(2)}_{\mu\nu} \delta_{\{N^{\mu}\},\{N^{\nu}\}}=$
\e{\sum_{\{n^\lambda \}}\sum_{m} \bigg\{ \f{(U_m^e)_{\mu\lambda} (U^e_{-m})_{\lambda\nu}}{E^e_{\mu\nu}-i0^+}(N_{m}^{\mu}+1)\bigg[  \f{p_{\{n^{\mu}\}}P_{\{N^\mu \}}- p_{\{n^{\lambda}\}}P_{\{\ldots,N_m^\mu +1,\ldots \}} }{E^e_{\mu\lambda}-\hbar\omega_m-i0^+} + \f{p_{\{n^{\nu}\}}P_{\{N^\mu \}}- p_{\{n^{\lambda}\}}P_{\{\ldots,N_m^\mu +1,\ldots \}}}{E^e_{\lambda\nu}+\hbar\omega_m-i0^+}\bigg]\lin
+\f{(U_{-m}^e)_{\mu\lambda} (U^e_{{m}})_{\lambda\nu}}{E^e_{\mu\nu}-i0^+} N_{m}^{\mu}\bigg[  \f{p_{\{n^{\mu}\}}P_{\{N^\mu \}}- p_{\{n^{\lambda}\}}P_{\{\ldots,N_m^\mu -1,\ldots \}}}{E^e_{\mu\lambda}+\hbar\omega_m-i0^+} + \f{p_{\{n^{\nu}\}}P_{\{N^\mu \}}- p_{\{n^{\lambda}\}}P_{\{\ldots,N_m^\mu -1,\ldots \}}}{E^e_{\lambda\nu}-\hbar\omega_m-i0^+}\bigg]\bigg\}\delta_{\{N^{\mu}\},\{N^{\nu}\}}.\la{saveink}}
The above expression utilizes the definition of $O^e$ in \q{defineOe} and a new definition for the electronic component of the total energy:
\e{E^e_{\mu} = \sum_B n^{\mu}_B E_B; \as E^e_{\mu\nu}=E^e_{\mu}-E^e_{\nu}.}
The factors of $N_{m}^{\mu}$ and $N_{m}^{\mu}+1$ in \q{saveink} are obtained from  the standard matrix elements for bosonic creation and annihilation.\\

In evaluating $\tr \big[\bv_{\text{off}}\rho^{(2)}\big]$, we first perform a partial trace by summing over the bosonic occupancies. In this manner, one converts expressions involving $N^{\mu}_m$ to expressions involving average occupancies:
\e{\tr \big[\bv_{\text{off}}\rho^{(2)}\big]\eq \sum_{\{n^\mu,n^\nu, n^\lambda \}}\sum_{m}  \f{(\bv^e_{\text{off}})_{\nu\mu}}{E^e_{\mu\nu}-i0^+}\bigg\{ (U_m^e)_{\mu\lambda} (U^e_{-m})_{\lambda\nu} \bigg[\f{p_{\{n^\mu \}}\expect{N_m+1} - p_{\{n^\lambda \}}\expect{N_m} }{E^e_{\mu\lambda}-\hbar\omega_m-i0^+} + \f{p_{\{n^\nu \}}\expect{N_m+1} - p_{\{n^\lambda \}}\expect{N_m}}{E^e_{\lambda\nu}+\hbar\omega_m-i0^+}\bigg] \lin
&\as + (U_{-m}^e)_{\mu\lambda} (U^e_m)_{\lambda\nu}\bigg[\f{p_{\{n^\mu \}}\expect{N_m} - p_{\{n^\lambda \}}\expect{N_m+1} }{E^e_{\mu\lambda}+\hbar\omega_m-i0^+} + \f{p_{\{n^\nu \}}\expect{N_m} - p_{\{n^\lambda \}}\expect{N_m+1}}{E^e_{\lambda\nu}-\hbar\omega_m-i0^+}\bigg]\bigg\}.\la{saveink2}}
	Let us apply the relation between band-off-diagonal elements of the velocity operator and band-off-diagonal elements of the position operator [\q{defineXk}], which translates to the following identity in second quantization:
	\e{\f{(\bv^e_{\text{off}})_{\nu\mu}}{E^e_{\mu\nu}-i0^+} =-\f{i}{\hbar}\sum_{\bk bb'}\bA^{\text{off}}_{b'b\bk}   \braopket{\{n^{\nu}\}}{\dg{c}_{b'\bk}\pdg{c}_{b\bk}}{\{n^{\mu}\}}_e \equiv -\f{i}{\hbar}(\bA^e_{\text{off}})_{\nu\mu}.\la{plugin}}
In dropping the $i0^+$, we have assumed that $E^e_{\mu\nu}=E_{bb'\bk}$ (for some $b\neq b'$) is nonzero for the bands and wavevectors of interest; it is worth recalling that $b$ does not include the spin label, hence one should not expect an energy degeneracy owing to spin. By plugging \q{plugin} into \q{saveink2} and recognizing that two of the four terms are complex conjugates of the other two, $\tr \big[\bv_{\text{off}}\rho^{(2)}\big]=$
\e{ -\frac{i}{\hbar} \sum_{\{n^\mu,n^\nu, n^\lambda \}}\sum_{m} (\bA^e_{\text{off}})_{\nu\mu}\bigg[  (U_m^e)_{\mu\lambda} (U^e_{-m})_{\lambda\nu} \f{p_{\{n^\mu \}}\expect{N_m+1} - p_{\{n^\lambda \}}\expect{N_m} }{E^e_{\mu\lambda}-\hbar\omega_m-i0^+} +  (U_{-m}^e)_{\mu\lambda} (U^e_m)_{\lambda\nu} \f{p_{\{n^\nu \}}\expect{N_m} - p_{\{n^\lambda \}}\expect{N_m+1}}{E^e_{\lambda\nu}-\hbar\omega_m-i0^+}\bigg]+c.c. .\la{saveink3}}
Let us interchange variables $\{n^{\mu}\}\leftrightarrow\{n^{\nu}\}$ for the first term and $\{n^{\lambda}\}\leftrightarrow\{n^{\nu}\}$ for the second, and then apply the resolution of identity within the electronic Fock space: $\sum_{\{n^\mu\}}\ketbra{\{n^{\mu}\}}{\{n^{\mu}\}}=I^e$.
 \e{\tr \big[\bv_{\text{off}}\rho^{(2)}\big]\eq \frac{i}{\hbar} \sum_{\{n^\nu, n^\lambda \}}\sum_{m} \f{(U_m^e)_{\nu\lambda}\big[\bA^e_{\text{off}},U^e_{-m}\big]_{\lambda\nu}}{E^e_{\nu\lambda}-\hbar\omega_m-i0^+}\bigg(\expect{N_m}\big(p_{\{n^\nu\}}-p_{\{n^\lambda \}}\big) + p_{\{n^\nu\}}\bigg) + c.c.,\la{saveink4}}
with $[\bA^e_{\text{off}},U^e_m]_{\lambda\nu}$ meaning a matrix element of the commutator of two electronic operators, as defined in \qq{defineOe}{commutatorelecop}. By splitting $\sum_m=\sum_m^{photon}+\sum_m^{phonon}$ in \q{saveink4}, one decomposes $\tr \big[\bv_{\text{off}}\rho^{(2)}\big]=\tr \big[\bv_{\text{off}}\rho_{phot}^{(2)}\big]+\tr \big[\bv_{\text{off}}\rho_{phon}^{(2)}\big]$, which we 
separately tackle.\\

\noindent \underline{Evaluating $\tr \big[\bv_{\text{off}}\rho_{phot}^{(2)}\big]$}\\

Recalling the definitions of $U^e_m$, $\bA^{\text{off}}_{\bk}$ and $\bA^e_{\text{off}}$ in \q{defineUem}, \q{defineXk} and \q{plugin}, and that $U^e_m=U^e_{-m}$ within the dipole approximation,
\e{(U_m^e)_{\nu\lambda}[\bA^e_{\text{off}},U^e_{-m}]_{\lambda\nu}= \tf{he^2}{\omega_m\calv}\sum_{\bkp\bk}\sum_{aa'bb'}\beps_m\cdot \bv_{bb'\bk}\bigg[\bA^{\text{off}}_{\bkp},\beps_m\cdot \bv_{\bkp}\bigg]_{aa'}\braopket{\{n^\nu \}}{\dg{c}_{b\bk}\pdg{c}_{b'\bk}}{\{n^\lambda \}}_e\braopket{\{n^\lambda \}}{\dg{c}_{a\bkp}\pdg{c}_{a'\bkp}}{\{n^\nu \}}_e,  }
with $a,a',b$ and $b'$ being band labels, and $[\bA^{\text{off}}_{\bkp},\beps_m\cdot \bv_{\bkp}]$ being a commutator of two matrices in the band indices.\\

The product $\expect{\dg{c}_{b\bk}\pdg{c}_{b'\bk}}_e \expect{\dg{c}_{a\bkp}\pdg{c}_{a'\bkp}}_e$ is given by \\

\e{
\text{(i)}\; \delta_{AA'}\delta_{BB_{2}}n_{b\bk}^{\nu}n^{\nu}_{a\bkp}\delta_{\{n^\nu \},\{n^\lambda \}}+\text{(ii)}\; (1-n_B^{\lambda})n^{\lambda}_{B_2}n^{\nu}_B(1-n^{\nu}_{B_2})\delta_{AB_{2}}\delta_{A'B}\delta_{\{n^\nu \},\{n^\lambda \} -B_2+B},\la{similarlogic}
}
with $B=(b\bk)$, $B_2=(b^{\prime}\bk)$, $A=(a\bk^{\prime})$, and $A^{\prime}=(a^{\prime}\bk^{\prime})$, and $\{n^{\lambda}\}-B_{2}+B$ labels an electronic Fock basis state that differs from $\{n^{\lambda}\}$ only in having one-particle state $B_2$ be unoccupied and $B$ be
occupied.
\q{similarlogic} implies two additive contributions to $(U_m^e)_{\nu\lambda}[\bA^e_{\text{off}},U^e_{-m}]_{\lambda\nu}=$(i')+(ii'), namely
\e{
  &\text{(i')}\;= \tf{he^2}{\omega_m\calv}\bigg\{\sum_{b\bk}\beps_m\cdot \bv_{bb\bk}n_{b\bk}^{\nu}\bigg\}\bigg\{\sum_{b'\bkp}\bigg[\bA^{\text{off}}_{\bkp},\beps_m\cdot \bv_{\bkp}\bigg]_{b'b'} n^{\nu}_{b'\bkp}\bigg\}\delta_{\{n^\nu \},\{n^\lambda \}};\lin
&\text{(ii')}\;= \tf{he^2}{\omega_m\calv}\sum_{\bk}\sum_{bb'}\beps_m\cdot\bv_{bb'\bk}\bigg[\bA^{\text{off}}_{\bk},\beps_m\cdot \bv_{\bk}\bigg]_{b'b}(1-n_B^{\lambda})n^{\lambda}_{B_2}n^{\nu}_B(1-n^{\nu}_{B_2})\delta_{\{n^\nu \},\{n^\lambda \}- B_2+B  },\la{similarlogic1}
}
where we replace the dummy index $a$ by $b^{\prime}$ in (i'). Plugging (i')+(ii') into \q{saveink4} leads to two additive contributions to $\tr \big[\bv_{\text{off}}\rho_{phot}^{(2)}\big] =$(i'')+(ii''). \\


It should be seen that (i'') is at least fourth order in the electron charge and therefore does not contribute to the second-order $\tr \big[\bv_{\text{off}}\rho_{phot}^{(2)}\big]$. This follows from 
\e{\sum_{\{n^\nu, n^\lambda \}}\f{\expect{N_m}\big(p_{\{n^\nu\}}-p_{\{n^\lambda \}}\big) + p_{\{n^\nu\}}}{E^e_{\mu\lambda}-\hbar\omega_m-i0^+}n_{b\bk}^{\nu}n^{\nu}_{b'\bkp}\delta_{\{n^\nu \},\{n^\lambda \}} =\sum_{\{n\}}\f{p_{\{n\}}}{-\hbar\omega_m}n_{b\bk}n_{b'\bkp}=-\f{\expect{n_{b\bk}}\expect{n_{b'\bkp}}}{\hbar\omega_m}, \la{oneway}}
and 
\e{\text{(i'')}\;\propto e^2\sum_{\bk}\beps_m\cdot \bv_{bb\bk}\expect{n_{b\bk}}=e^2\sum_{\bk}\beps_m\cdot \bv_{bb\bk}\f{\expect{n_{b\bk}}-\expect{n_{b,-\bk}}}{2} \propto e^4.\la{e4}
}
Due to time-reversal symmetry,  $\bv_{bb\bk}$ is odd under $\bk\ri -\bk$. The same symmetry would constrain $\expect{n_{b\bk}}$ to be an even function, if the average were taken in a state of thermal equilibrium. However, optical excitation creates a non-equilibrium state that breaks time-reversal symmetry, which is reflected in a nonzero    $(\expect{n_{b\bk}}-\expect{n_{b,-\bk}})$ that is proportional to the source intensity, i.e., to $e^2$.\footnote{The `ballistic current' is essentially $-|e|/2\calv \sum_{b\bk}\bv_{bb\bk} (\expect{n_{b\bk}}-\expect{n_{b,-\bk}}).$\cite{belinicher_ballistic}}\\

What remains of $\tr \big[\bv_{\text{off}}\rho_{phot}^{(2)}\big]$ is (ii''). To evaluate (ii''), we point out that the energy denominator in \q{saveink4} reduces to
\e{
\f{(1-n_B^{\lambda})n^{\lambda}_{B_2}n^{\nu}_B(1-n^{\nu}_{B_2})\delta_{\{n^\nu \},\{n^\lambda \} -B_2+B  }}{E^e_{\nu\lambda}-\hbar\omega_m-i0^+} = \f{(1-n_B^{\lambda})n^{\lambda}_{B_2}n^{\nu}_B(1-n^{\nu}_{B_2})\delta_{\{n^\nu \},\{n^\lambda \} -B_2 +B }}{E_{bb'\bk}-\hbar\omega_m-i0^+}. \la{point1}
}
We need two more identities which follow from $p_{\{n\}}$ being a probability function for independent particles [cf.\ \q{defineFmu}]:
\e{
\sum_{\{n^\nu, n^\lambda \}}p_{\{n^\nu\}}(1-n_B^{\lambda})n^{\lambda}_{B_2}n^{\nu}_B(1-n^{\nu}_{B_2})\delta_{\{n^\nu \},\{n^\lambda \} -B_2 +B } \eq \sum_{n^{\lambda}_B,n^{\lambda}_{B_2}} (1-n_B^{\lambda})n^{\lambda}_{B_2} \sum_{n^{\nu}_B}p_{n^{\nu}_B}{n^{\nu}_B} \sum_{n^{\nu}_{B_2}}p_{n^{\nu}_{B_2}}{(1-n^{\nu}_{B_2})}=\expect{n_B}\expect{1-n_{B_2}} \lin
\sum_{\{n^\nu, n^\lambda \}}p_{\{n^\lambda \}}(1-n_B^{\lambda})n^{\lambda}_{B_2}n^{\nu}_B(1-n^{\nu}_{B_2})\delta_{\{n^\nu \},\{n^\lambda \} -B_2 +B } \eq\expect{n_{B_2}}\expect{1-n_{B}}.\la{point2}
}  
Altogether, the photonic shift current is expressible as
\e{\bj_{phot}
\eq - \frac{2\pi i |e|^3}{\omega_{m}\calv^2}\sum_m^{\text{photon}} \sum_{bb'\bk}\bigg\{{N_m}{f_{bb'\bk}}-{f_{b'\bk}}\big(1-{f_{b\bk}}\big)\bigg\}\f{\beps_m\cdot \bv_{b'b\bk}\bigg[\bA^{\text{off}}_{\bk},\beps_m\cdot \bv_{\bk}\bigg]_{bb'}}{E_{bb'\bk}+\hbar\omega_m+i0^+} +c.c.; \as f_{bb'\bk}={f_{b\bk}}-{f_{b'\bk}}. \la{eq6bis}
}
In this last step, we interchanged $b\leri b'$ and simplified our notation as $\expect{N_m}\ri N_m$ and $\expect{n_B}\ri f_B$, to be consistent with the rest of the paper.\\

To go from \q{eq6bis} to the final expression for the photonic shift current [\q{bismine2}, \q{photonicshift2} and \q{photonrate}] involves a sum rule derived from the first-quantized commutation relation: $[\hat{r}^n,\hat{p}^{n'}]=i\hbar\delta_{n,n'}$, with $n$ and $n'$ denote the components of three-vectors. We follow this through in \app{app:sumrulephotonic}.\\

\noindent \underline{Evaluating $\tr \big[\bv_{\text{off}}\rho_{phon}^{(2)}\big]$}\\

Recalling the definitions of $U^e_m$, $\breve{V}^m$, $\bA^{\text{off}}_{\bk}$ and $\bA^e_{\text{off}}$ in \q{defineUem}, \q{defineXk} and \q{plugin},
\e{(U_{m}^e)_{\nu\lambda}[\bA^e_{\text{off}},U^e_{-m}]_{\lambda\nu}= \sum_{\bk\bkp}\sum_{aa'bb'}\breve{V}^{m}_{bb'\bk}\bigg(\bA^{\text{off}}_{\bkp}\breve{V}^{-m}_{\bkp}-\breve{V}^{-m}_{\bkp} \bA^{\text{off}}_{\bkp+\bq}\bigg)_{aa'}\braopket{\{n^\nu \}}{\dg{c}_{b\bk}\pdg{c}_{b'\bk-\bq}}{\{n^\lambda \}}_e\braopket{\{n^\lambda \}}{\dg{c}_{a\bkp}\pdg{c}_{a'\bkp+\bq}}{\{n^\nu \}}_e,  }
with $m=(\bq,p)$, $a,a',b$ and $b'$ being band labels, and $\bA^{\text{off}}_{\bkp}\breve{V}^{-m}_{\bkp}$ being a product of two matrices indexed by band labels. \\

Imitating \q{similarlogic} and \q{similarlogic1}, we find 
two additive contributions to $(U_m^e)_{\nu\lambda}[\bA^e_{\text{off}},U^e_{-m}]_{\lambda\nu}=$(i')+(ii'), the first of which is nontrivial only if the phonon wavevector vanishes:
\e{
  &\text{(i')}\;= \delta_{\bq,\bze}\bigg\{\sum_{b\bk}\breve{V}^m_{bb\bk}n_{b\bk}^{\nu}\bigg\}\bigg\{\sum_{b'\bkp}\bigg[\bA^{\text{off}}_{\bkp},\breve{V}^{-m}_{\bkp}\bigg]_{b'b'} n^{\nu}_{b'\bkp}\bigg\}\delta_{\{n^\nu \},\{n^\lambda \}};\lin
 &\text{(ii')}\;= \sum_{\bk}\sum_{bb'}\breve{V}^m_{bb'\bk}\bigg(\bA^{\text{off}}_{\bk-\bq}\breve{V}^{-m}_{\bk-\bq}-\breve{V}^{-m}_{\bk-\bq} \bA^{\text{off}}_{\bk}\bigg)_{b'b}(1-n_B^{\lambda})n^{\lambda}_{B_3}n^{\nu}_B(1-n^{\nu}_{B_3})\delta_{\{n^\nu \},\{n^\lambda \} - B_3+B  },
}
with $B=(b\bk)$ and $B_3=(b',\bk-\bq)$. The contribution to (i') is only by zero-wavevector optical phonons, since zero-wavevector acoustic phonons do not admit quantization.\footnote{One way to see this is that in the quantization of the displacement field, the prefactor in front of $a_m$ is inversely proportional to $\sqrt{\omega_m}$.\cite{gantmakherlevinson_book}} Plugging (i')+(ii') into \q{saveink4} leads to two additive contributions to $\tr \big[\bv_{\text{off}}\rho_{phon}^{(2)}\big] =$(i'')+(ii''). \\

It should be seen that (i'') is at least fourth order in the electron-boson coupling and therefore does not contribute to the second-order $\tr \big[\bv_{\text{off}}\rho_{phon}^{(2)}\big]$. To appreciate this, apply \q{oneway} once again, noting that the $1/\omega_m$ factor in \q{oneway} is well-defined for optical phonons as $\bq \ri \bze$. Then,
\e{\text{(i'')}\;\propto \bigg(\sum_{\ldots}\breve{V}^{m}_{\ldots}\ldots\bigg)\sum_{b\bk}\bigg[\bA^{\text{off}}_{\bk},\breve{V}^{-m}_{\bk}\bigg]_{b,b}\expect{n_{b\bk}}= \bigg(\sum_{\ldots}\breve{V}^{m}_{\ldots}\ldots\bigg)\sum_{b\bk}\big[\bA^{\text{off}}_{\bk},\breve{V}^{-m}_{\bk}\big]_{b,b}\f{\expect{n_{b\bk}}-\expect{n_{b,-\bk}}}{2}.\la{e2v2}
}
Because  $(\expect{n_{b\bk}}-\expect{n_{b,-\bk}})$ is proportional to $e^2$, altogether (i'') is quadratic in both the electron-photon and electron-phonon couplings.\\

To arrive at the last line in \q{e2v2}, we had applied that $\big[\bA^{\text{off}}_{\bk},\breve{V}^{\bze p}_{\bk}\big]_{b,b}$ is odd under $\bk$-inversion, owing to time-reversal symmetry. Indeed, the anti-unitary nature of time reversal:
\e{\hat{T}\ket{u_{b,-\bk}}_{\sma{\text{cell}}}=e^{i\phi_{b\bk}}\ket{u_{b\bk}}_{\sma{\text{cell}}}; \as\braket{u_B}{\hat{T}u_{B'}}_{\sma{\text{cell}}}=\braket{u_{B'}}{\hat{T}^{-1}u_B}_{\sma{\text{cell}}}, \la{timereversecell}}
 results in a transposition of the band labels for matrix elements:
\e{\bA^{\text{off}}_{bb',-\bk}=e^{i(\phi_{b\bk}-\phi_{b'\bk})}\bA^{\text{off}}_{b'b\bk}; \as \breve{V}^{\bze p}_{b'b,-\bk}=e^{i(\phi_{b'\bk}-\phi_{b\bk})}\breve{V}^{\bze p}_{bb',\bk} \imp \as \big(\bA^{\text{off}}_{-\bk}\breve{V}^{\bze p}_{-\bk}\big)_{b,b}=\big(\breve{V}^{\bze p}_{\bk}\bA^{\text{off}}_{\bk}\big)_{b,b}. }
To elaborate on the middle equality, we utilize our general expression for the self-consistently-screened electron-phonon matrix element [\q{defineUem}] and massage the matrix element as:
\e{\braopket{u_{b',-\bk}}{e^{i\bG\cdot \hbr}}{u_{b,-\bk}}\eq \braopket{u_{b',-\bk}}{\hat{T}^{-1}e^{-i\bG\cdot \hbr}\hat{T}}{u_{b,-\bk}}=\overline{\braopket{\hat{T}u_{b',-\bk}}{e^{-i\bG\cdot \hbr}}{\hat{T}u_{b,-\bk}}}=e^{i(\phi_{b'\bk}-\phi_{b\bk})}\braopket{u_{b\bk}}{e^{i\bG\cdot \hbr}}{u_{b'\bk}},  }
omitting the $cell$ superscript in the above equation.\\

What remains of $\tr \big[\bv_{\text{off}}\rho_{phon}^{(2)}\big]$ is (ii''). To evaluate (ii''), we follow steps closely analogous to \qq{point1}{point2}, replacing the Bloch label $B_2\ri B_3$. This leads to the following expression for the phononic shift current:
\e{\bj_{phon}
\eq - \frac{ i |e|}{\hbar \calv}\sum^{\text{phonon}}_m \sum_{bb'\bk}\bigg\{{N_m}{(f_{B_3}-f_{B})}-{f_{B}}\big(1-{f_{B_3}}\big)\bigg\}\f{\bigg(\bA^{\text{off}}_{\bk-\bq}\breve{V}^{-m}_{\bk-\bq}-\breve{V}^{-m}_{\bk-\bq} \bA^{\text{off}}_{\bk}\bigg)_{b'b}\breve{V}^m_{bb'\bk}}{E_{B_3}-E_{B}+\hbar\omega_m+i0^+} +c.c., \la{eq6bisphonon2}
}
with $m=(\bq,p), B=(b\bk)$ and $B_3=(b',\bk-\bq)$. Utilizing our definition of the band-off-diagonal position operator [\q{defineroff}] and the first-quantized electron-phonon operator [\q{eleconephononop}],
\e{[\hbr_{\text{off}},\hat{V}^{-m}]_{B'B}=\braopket{B'}{[\hbr_{\text{off}},\hat{V}^{-m}]}{B}_{1}; \as \sum_{\bkp}[\hbr_{\text{off}},\hat{V}^{-m}]_{B'B}V^m_{BB'}= \bigg(\bA^{\text{off}}_{\bk-\bq}\breve{V}^{-m}_{\bk-\bq}-\breve{V}^{-m}_{\bk-\bq} \bA^{\text{off}}_{\bk}\bigg)_{b'b}\breve{V}^m_{bb'\bk},
}
with $B'=(b'\bkp)$. This identity can be inserted into \q{eq6bisphonon2} to obtain an equivalent expression for the phononic shift current:
\e{\bj_{phon}
\eq - \frac{ i |e|}{\hbar \calv}\sum^{\text{phonon}}_m \sum_{BB'}\bigg\{{N_m}{(f_{B'}-f_{B})}-{f_{B}}\big(1-{f_{B'}}\big)\bigg\}\f{[\hbr_{\text{off}},\hat{V}^{-m}]_{B'B}V^m_{BB'}}{E_{B'}-E_{B}+\hbar\omega_m+i0^+} +c.c.. \la{eq6bisphonon}
}

 To go from \q{eq6bisphonon} to the final expression for the phononic shift current [\q{bismine2}, \q{phononicshift2} and \q{phononrate}] involves a sum rule derived from  $[\hbr,\hat{V}^{-m}]=0$. The zero is because $\hat{V}^{-m}$ is defined in terms of the position operator but not the momentum operator [\q{eleconephononop}]. We follow this through in \app{app:sumrulephononic}.

\subsubsection{Sum rule for the photonic shift current}\la{app:sumrulephotonic} 

The first-quantized commutation relation 
\e{
i\hbar\delta_{n,n'}\delta_{BB'}\eq \braopket{B'}{[\hat{r}^n,\hat{p}^{n'}]}{B}_{1}= \sum_{B''}\big( r^n_{BB''}p^{n'}_{B''B'}-p^{n'}_{BB''}r^n_{B''B'}\big). \la{sumrule}
}
will be used to prove:
\e{
\big[A^{\text{off}n},P^{n'}\big]_{bb'}=i\hbar\delta_{n,n'}\delta_{b,b'}+ \big[-i\nabk^n  + (A^n_{b'b'\bk}-A^n_{bb\bk})\big]P^{n'}_{bb'}, \la{idenbis}
}
with all $\bk$-dependent quantities evaluated at the same $\bk$. By inserting \qq{rbb}{pbb} into the right-hand side of \q{sumrule} and carrying out $\sum_{B''}$, 
\e{
i\hbar\delta_{n,n'}\delta_{BB'}\eq iP^{n'}_{bb'\bkp}\nabk^n\delta_{\bk\bkp} -iP^{n'}_{bb'\bk}\nabk^n\delta_{\bk\bkp}+\delta_{\bk\bkp}[A^n,P^{n'}]_{bb'}.
}
By applying the second Dirac-delta identity [\q{dirac2}] and separating diagonal and off diagonal components of $A^n$ one derives \q{idenbis}.\\

Let us plug \q{idenbis} into our expression for the shift current [\q{eq6bis}]. It should be remarked that the $i\hbar\delta_{n,n'}\delta_{b,b'}$ term in \q{idenbis} does not contribute to the current because the band-diagonal velocity $\bv_{bb\bk}$ is an odd function of $\bk$ and the rest of the integrand may be taken as even.\footnote{An argument can be constructed that is analogous to the one used in \q{e4}.} What remains is
\e{\bj_{phot}
\eq \imag \sum_m \frac{4\pi  |e|^3}{\omega_m\calv^2} \sum_{bb'\bk}\bigg\{{N_m}{f_{bb'\bk}}-{f_{b'\bk}}\big(1-{f_{b\bk}}\big)\bigg\}\f{\beps_m\cdot \bv_{b'b\bk}\big[-i\nabk  + (\bA_{b'b'\bk}-\bA_{bb\bk})\big]\beps_m\cdot \bv_{bb'\bk}}{E_{bb'\bk}+\hbar\omega_m+i0^+}, \la{eq6bis2}
}
with all $\bk$ subscripts omitted for simplicity. By applying the  Sokhotski–Plemelj theorem: $1/(x+i0^+)=CPV[1/x] -i\pi \delta(x)$, with $CPV$ meaning Cauchy's principal value, one can decompose $\bj=(a)+(b)$, with
\e{
(a) \propto&\; \sum_{\bk}\bigg\{{N_m}{f_{bb'\bk}}-{f_{b'\bk}}\big(1-{f_{b\bk}}\big)\bigg\}CPV\f{\imag \beps_m\cdot \bv_{b'b\bk}\big[-i\nabk  + (\bA_{b'b'\bk}-\bA_{bb\bk})\big]\beps_m\cdot \bv_{bb'\bk}}{E_{bb'\bk}+\hbar\omega_m},\lin
(b) \eq -\sum_m \frac{4\pi^2  |e|^3}{\omega_m\calv^2} \sum_{bb'\bk}\bigg\{{N_m}{f_{bb'\bk}}-{f_{b'\bk}}\big(1-{f_{b\bk}}\big)\bigg\}\delta({E_{b'b\bk}-\hbar\omega_m)} \real \beps_m\cdot \bv_{b'b\bk}\big[-i\nabk  + (\bA_{b'b'\bk}-\bA_{bb\bk})\big]\beps_m\cdot \bv_{bb'\bk}.\la{jb}
}

(a) vanishes by time-reversal symmetry, which imposes that
\e{\imag \beps_m\cdot \bv_{b'b\bk}\big[-i\nabk  + (\bA_{b'b'\bk}-\bA_{bb\bk})\big]\beps_m\cdot \bv_{bb'\bk}=-|\beps_m\cdot \bv_{bb'\bk}|\nabk |\beps_m\cdot \bv_{bb'\bk}|  }
is an odd function of $\bk$. To appreciate this, apply that $\beps_m$ is real, the velocity operator inverts sign under time reversal, and the time-reversal symmetry of cell-periodic wave functions [\q{timereversecell}]:
\e{ \beps_m\cdot\bv_{bb',-\bk}= -e^{i\phi_{b\bk}-i\phi_{b'\bk}}\overline{\beps_m\cdot \bv_{bb'\bk}}.}\\

(b) is related to the photonic shift vector [\q{photonicshift2}] by the following identity:
\e{\real \beps_m\cdot \bv_{b'b\bk}\big[-i\nabk  + (\bA_{b'b'\bk}-\bA_{bb\bk})\big]\beps_m\cdot \bv_{bb'\bk} =|\beps_m\cdot \bv_{bb'\bk}|^2 \bS^m_{b'\bk \lea b\bk}.}
Plugging the above equation and
\q{defineXk} into \q{jb}, one finally derives \q{bismine2} with \q{photonicshift2} and \q{photonrate}. 

\subsubsection{Sum rule for the phononic shift current}\la{app:sumrulephononic} 

Substituting \qq{rbb}{pbb} into the right-hand side of 
\e{
0\eq \braopket{B'}{[\hbr,\hat{V}^{-m}]}{B}_{1}=\sum_{B''}\big( \br_{B'B''}V^{-m}_{B''B}-V^{-m}_{B'B''}\hbr_{B''B}\big), \la{sumrule2}
}
applying the standard identity $f(x,x')\partial_x\delta(x-x')=\delta(x-x') \partial_{x'}f(x,x')$, and separating the band-diagonal and band-off-diagonal matrix elements of the position operator, one obtains:
\e{
0\eq \big( \;i\nabkp +i\nabk +\bA_{b'b'\bkp}-\bA_{bb\bk} \;\big) V^{-m}_{B'B}+\delta_{\bkp,\bk-\bq}\bigg(\bA^{\text{off}}_{\bk-\bq}\breve{V}^{-m}_{\bk-\bq}-\breve{V}^{-m}_{\bk-\bq} \bA^{\text{off}}_{\bk}\bigg)_{b'b}.
}
Plugging this into our expression for the phononic shift current [\q{eq6bisphonon}],
\e{\bj_{phon}
\eq -\imag  \frac{2 |e|}{\hbar\calv} \sum_m^{phonon}\sum_{BB'}\bigg\{{N_m}{f_{B'B}}-{f_{B}}\big(1-{f_{B'}}\big)\bigg\}\f{V^m_{BB'}\big(i\nabkp+i\nabk + \bA_{b'b'\bk}-\bA_{bb\bk}\big)V^{-m}_{B'B}}{E_{B'B}+\hbar\omega_m+i0^+},  \la{eq6bis2}
}
with $B=(b\bk), B'=(b\bkp), f_{B'B}=f_{B'}-f_B$ and $E_{B'B}=E_{B'}-E_B$. By applying the  Sokhotski–Plemelj theorem, one can decompose $\bj_{phon}=(a)+(b)$, with
\e{
(a) \propto&\; \sum_{\bk\bkp\bq}\bigg\{{N_m}{f_{B'B}}-{f_{B}}\big(1-{f_{B'}}\big)\bigg\}CPV \f{\imag V^m_{BB'}\big(i\nabkp+i\nabk + \bA_{b'b'\bk}-\bA_{bb\bk}\big)V^{-m}_{B'B}}{E_{B'B}+\hbar\omega_m},\la{jphona}\\
(b) \eq \frac{2\pi |e|}{\hbar\calv} \sum_m^{phonon}\sum_{BB'}\bigg\{{N_m}{f_{B'B}}-{f_{B}}\big(1-{f_{B'}}\big)\bigg\}\delta(E_{BB'}-\hbar\omega_m)\real V^m_{BB'}\big(i\nabkp+i\nabk + \bA_{b'b'\bk}-\bA_{bb\bk}\big)V^{-m}_{B'B}.\la{jphonb}
}
To simplify the above expressions, 
it is worth recalling $\overline{V^{m}_{BB'}}= V^{-m}_{B'B}$ from \q{Vnotselfadj}.\\

 (a) vanishes by time-reversal symmetry, which imposes that
\e{\imag V^m_{BB'}\big(i\nabkp+i\nabk + \bA_{b'b'\bk}-\bA_{bb\bk}\big)V^{-m}_{B'B}=|V^m_{BB'}|( \nabkp+\nabk) |V^m_{BB'}| \la{imv} }
is odd under simultaneusly inverting $(\bk,\bkp,\bq)\ri (-\bk,-\bkp,-\bq)$, and the rest of the integrand in \q{jphona} is even. (Certainly all energies are even functions, and we have argued for $f_{B}\approx f_{-B}
$; we suppose further that $N_m\approx N_{-m}$, i.e., that any time-reversal-breaking of the phonon occupations is proportional to the light intensity, and does not affect the second-order shift current.) To prove oddness of \q{imv}, it suffices to show that $|V^m_{BB'}|$ is even, i.e., $|V^m_{BB'}|=|V^{-m}_{-B,-B'}|$ with the minus signs denoting a reversal in wavevectors. Recalling how time reversal acts on $\hat{V}^m$ [\q{Vnotselfadj}] and on Bloch waves [\q{timereversecell}], 
\e{V^m_{-B,-B'}= e^{i(\phi_B-\phi_{B'})}V^m_{B'B} \imp |V^{-m}_{-B,-B'}|^2= V^{-m}_{-B,-B'}V^{{m}}_{-B',-B} =V^{-m}_{{B}'{B}}V^{{m}}_{{B}{B}'}= |V^m_{BB'}|^2. } 

Plugging 
\e{\real V^m_{BB'}\big(i\nabkp+i\nabk + \bA_{b'b'\bk}-\bA_{bb\bk}\big)V^{-m}_{B'B}=-|V^{{m}}_{BB'}|^2\bigg\{ -(\nabkp+\nabk)\arg V^{{m}}_{BB'}+ \bA_{bb\bk}-\bA_{b'b'\bk}\bigg\} \la{rev} }
into \q{jphonb} and interchanging $B\leri B'$, one finally derives \q{bismine2}, \q{phononicshift2} and \q{phononrate}. \\

It is worth justifying our interpretation of \q{phononrate} as a difference between absorption and emission rates:\\

\noi{i} Suppose a Bloch state transits from $B \ri B'$ while absorbing a phonon of mode $m$; this is implemented by the electron-phonon interaction $U^e_m(a_m+\dg{a}_{-m})$ [cf.\ \q{defineU}], or more specifically by $V^m_{B'B}\dg{c}_{B'}\pdg{c}_B a_m$ [cf.\ \q{defineUem}].  Thus one expects the associated shift vector for this process to be $-\nabk \arg V^m_{B'B} +\bA_{b'b'\bkp}-\bA_{bb\bk}=\bS^m_{B'\lea B}$ [cf.\ \q{phononicshift2}]. 
By the golden rule, one expects a transition probability that is proportional to $|V^m_{B'B}|^2$ and given by the first term in \q{phononrate}, namely $\cala_{B'\lea B}^m$. The associated contribution to the current is then $-(|e|/\calv) \;\bS^m_{B'\lea B}\;\cala_{B'\lea B}^m$, which is the first term in \q{bismine2}. \\

\noi{ii} Suppose a Bloch state transits from $B' \ri B$ while emitting a phonon of mode $m$; this is implemented by the electron-phonon interaction $U^e_{-m}(a_{-m}+\dg{a}_{{m}})$ [cf.\ \q{defineU}], or more specifically by $V^{-m}_{BB'}\dg{c}_{B}\pdg{c}_{B'} \dg{a}_m$ [cf.\ \q{defineUem}].  Thus one expects the associated shift vector for this process to be $-\nabk \arg V^{-m}_{BB'} +\bA_{bb\bk}-\bA_{b'b'\bkp}=\bS^{-m}_{B\lea B'}$ [cf.\ \q{phononicshift2}]. By the golden rule, one expects a transition probability that is proportional to $|V^{-m}_{BB'}|^2=|V^m_{B'B}|^2$ [cf.\ \q{Vnotselfadj}] and given by (negative of) the second term in \q{phononrate}, namely $\cale_{B\lea B'}^m$. Why the minus sign in \q{phononrate}; equivalently, why the minus sign in \q{bismine2}? The reason is that the current contributed by this transition is 
\e{ -\f{|e|}{\calv}\bS^{-m}_{B\lea B'}\cale_{B\lea B'}^m =-\f{|e|}{\calv}(-\bS^{{m}}_{B'\lea B})\cale_{B\lea B'}^m,  }
which is the second term in \q{bismine2}. Note that $\bS^{-m}_{B\lea B'}=-S^{{m}}_{B'\lea B}$ follows from $\overline{V^{m}_{BB'}}= V^{-m}_{B'B}$ [cf.\ \q{Vnotselfadj}].

\subsubsection{The zeroth-order quasiparticle distribution is not thermal}\la{app:zerothrho}

Let us define the non-perturbative quasiparticle distribution as 
\e{
f^{stat}_B \eq \tr[n_B^e\rho]; \as n_B^e = \dg{c}_Bc_B.
}
In the Schr\"odinger representation (indicated by $\substack{S\\ =}$ below), density matrices can be time-dependent but operators (such as $n_B^e$) are time-independent:
\e{
\partial_t f^{stat}_B\condeq{S} \tr[n_B^e\partial_t\rho].\la{ssch}
}
Because $\rho$ is stationary, the non-perturbative quasiparticle distribution is steady:
\e{
 0\eq \partial_t\rho=-\f{i}{\hbar}[H,\rho] \imp 0= \partial_t f^{stat}_B.
}
In the Heisenberg representation, density matrices are generally time-independent, but operators (like $n_B^e$) satisfy Heisenberg's equation of motion:
\e{
\partial_t f^{stat}_B = \tr[(\partial_t n_B^e)_H\rho_H] = \f{i}{\hbar}\tr\{ [U_H,(n_B^e)_H]\rho_H \} ,
}
with  $O_H$ denoting an operator $O$ in the Heisenberg representation;\footnote{This may be verified by substituting $\rho= e^{-iHt/\hbar}\rho_H e^{iHt/\hbar}$ and $n_B^e=e^{iHt/\hbar}(n_B^e)_H e^{-iHt/\hbar}$ into \q{ssch}.} here, it should be recalled that $H=H_H= (H_0)_H+U_H$ and $[H_0,n_B^e]=0 \imp [(H_0)_H,(n_B^e)_H]=0$.
Since traces are independent of the representation:
\e{
\tr\{ [U_H,(n_B^e)_H]\rho_H \}=\tr\{ [U,n_B^e]\rho \},
}
and we may insert the perturbative expansion for $\rho$ in  \q{rhoperturb}. A term in this perturbative expansion that is even in powers of $U$   has a vanishing contribution to $\tr\{ [U,n_B^e]\rho \}$, because one traces over an odd multiple of the bosonic creation/annilation operator. In particular, $\tr\{ [U,n_B^e]\rho^{(0)} \}=0$ because $\rho^{(0)}= \sum_{\mu}F_{\mu}\ketbra{\mu}{\mu}$ [cf.\ \q{rhoperturb}] and $\braopket{\mu}{a_m+\dg{a}_{-m}}{\mu}=(a_m+\dg{a}_{-m})_{\mu\mu}=0$. Let us therefore evaluate $\tr\{ [U,n_B^e]\rho^{(1)} \}$, using our expression for $\rho^{(1)}$ in \q{rho1}:
\e{
-i\hbar \partial_t f_B^{stat} \eq \sum_{m\mu\nu} \big\{ [U^e_m,n_B^e](a_m+\dg{a}_{-m})\big\}_{\nu\mu} \f{F_{\mu\nu}U_{\mu\nu}}{E_{\mu\nu}-i0^+}+O(U^4).
}
Each photon/phonon that is created must be subsequently annihilated, and vice versa:
\e{
-i\hbar \partial_t f_B^{stat} \eq \sum_{m\mu\nu}  [U^e_m,n_B^e]_{\nu\mu}(U_{-m}^e)_{\mu\nu} \f{F_{\mu\nu}U_{\mu\nu}}{E_{\mu\nu}-i0^+} \big\{ (a_m)_{\nu\mu}(\dg{a}_m)_{\mu\nu}+(\dg{a}_{-m})_{\nu\mu}(a_{-m})_{\mu\nu}\big\}.
}
Switching $m\ri -m$ in the second term, and applying the standard matrix elements for bosonic operators,
\e{
-i\hbar \partial_t f_B^{stat} 
\eq  \sum_{m\mu\nu}  [U^e_m,n_B^e]_{\nu\mu}(U_{-m}^e)_{\mu\nu} \f{F_{\mu\nu}}{E_{\mu\nu}-i0^+}  N_m^{\mu}\delta_{N^{\mu},N^{\nu}+m}+\sum_{m\mu\nu}  [U^e_{-m},n_B^e]_{\nu\mu}(U_{m}^e)_{\mu\nu} \f{F_{\mu\nu}}{E_{\mu\nu}-i0^+} (N_m^{\mu}+1)\delta_{N^{\mu}+m,N^{\nu}}.
}
$\delta_{N^{\mu},N^{\nu}+m}$ is a Kronecker delta function enforcing $N_{m'}^{\mu}=N_{m'}^{\nu}$ for all $m'$, except for $N^{\nu}_m+1=N^{\mu}_m$; for $\delta_{N^{\mu}+m,N^{\nu}}$, it is $N^{\mu}_m+1=N^{\nu}_m$ that is the exception. We use this delta function to kill the summation over $N^{\nu}$:
\e{
-i\hbar \partial_t f_B^{stat} \eq
  \sum_{m n^\mu n^\nu N^{\mu}}  [U^e_m,n_B^e]_{\nu\mu}(U_{-m}^e)_{\mu\nu} \f{p_{n^{\mu}}P_{N^{\mu}}-p_{n^{\nu}} P_{\ldots N_m^{\mu}-1\ldots }}{E^e_{\mu\nu}+\hbar\omega_m-i0^+}  N_m^{\mu}\lin
  &+\;    \sum_{m n^\mu n^\nu N^{\mu}}   [U^e_{-m},n_B^e]_{\nu\mu}(U_{m}^e)_{\mu\nu} \f{p_{n^{\mu}}P_{N^{\mu}}-p_{n^{\nu}} P_{\ldots N_m^{\mu}+1\ldots }}{E_{\mu\nu}^e-\hbar\omega_m-i0^+} (N_m^{\mu}+1).
}
$E^e$ is the electronic component of $E$. Carrying out the sum over $N^{\mu}$,
\e{
-i\hbar \partial_t f_B^{stat} \eq
  \sum_{m n^\mu n^\nu}   [U^e_m,n_B^e]_{\nu\mu}(U_{-m}^e)_{\mu\nu} \f{p_{n^{\mu}}\langle  N_m\rangle -p_{n^{\nu}} \langle  N_m+1\rangle}{E^e_{\mu\nu}+\hbar\omega_m-i0^+} 
  +    \sum_{m n^\mu n^\nu}   [U^e_{-m},n_B^e]_{\nu\mu}(U_{m}^e)_{\mu\nu} \f{p_{n^{\mu}}\langle  N_m+1\rangle-p_{n^{\nu}} \langle  N_m\rangle}{E_{\mu\nu}^e-\hbar\omega_m-i0^+} .
}
Interchanging summation variables $n^{\mu} \leftrightarrow n^{\nu}$ for the second term,
\e{
-i\hbar \partial_t f_B^{stat} \eq
  \sum_{m n^\mu n^\nu} \bigg( - \f{[n_B^e,U^e_m]_{\nu\mu}(U_{-m}^e)_{\mu\nu}}{E^e_{\mu\nu}+\hbar\omega_m-i0^+} +\f{[U^e_{-m},n_B^e]_{\mu\nu}(U_{m}^e)_{\nu\mu}}{E_{\mu\nu}^e+\hbar\omega_m+i0^+} \bigg) \bigg(  p_{n^{\mu}}\langle  N_m\rangle -p_{n^{\nu}} \langle  N_m+1\rangle \bigg).
}
By applying that $n_B^e$ is self-adjoint and $U^e_m=\dg{(U^e_{-m})}$ [cf.\ \q{defineU}], one recognizes one fraction to be the complex conjugate of the other:
\e{
-i\hbar \partial_t f_B^{stat} \eq
  \sum_{m n^\mu n^\nu} \bigg( 2i \imag \f{[U^e_{-m},n_B^e]_{\mu\nu}(U_{m}^e)_{\nu\mu}}{E_{\mu\nu}^e+\hbar\omega_m+i0^+}\bigg) \bigg(  p_{n^{\mu}}\langle  N_m\rangle -p_{n^{\nu}} \langle  N_m+1\rangle \bigg).\la{numin}
}
At this point we split the photonic and phononic contributions:
\e{
\partial_t f_B^{stat} \eq (\partial_t f_B^{stat})^{phot}+ (\partial_t f_B^{stat})^{phon}
}
by splitting the sum over the bosonic modes: $\sum_m=\sum_m^{phot} +\sum_m^{phon}.$ Focusing first on the photonic contribution,
we evaluate the numerator in \q{numin} with help from \q{commutatorelecop}, \q{defineUem} and $\eps_m=\eps_{-m}\in \R$, 
\e{
[U^e_{-m},n_B^e]\eq \sum_{b'}\sharp \eps_{-m}\cdot \big( v^{off}_{b'b\bk}\dg{c}_{b'\bk}c_{b\bk}-  (b\leftrightarrow b')\big); \as \sharp = \sqrt{\tf{he^2}{\omega_{m}\calv}},
}
\e{
[U^e_{-m},n_B^e]_{\mu\nu}(U_{m}^e)_{\nu\mu} \eq \sum_{b'}|\sharp \eps_{m}\cdot  v^{off}_{bb'\bk}|^2\bigg\{ (1-n_{B'}^{\nu})n^{\nu}_B(1-n_B^{\mu})n_{B'}^{\mu} \delta_{n^{\mu},n^{\nu}-B+B'} -  (b\leftrightarrow b')\bigg\}.
}
Since  the numerator in \q{numin} is manifestly real,  it suffices to evaluate the imaginary part of the denominator:
\e{
\delta_{n^{\mu},n^{\nu}-B+B'}\imag \f1{E_{\mu\nu}^e+\hbar\omega_m+i0^+}=-\pi \delta_{n^{\mu},n^{\nu}-B+B'} \delta(E_{\mu\nu}^e+\hbar\omega_m)=-\pi \delta_{n^{\mu},n^{\nu}-B+B'}\delta(\var_{B'B}+\hbar\omega_m),
}
with $\var_B$ being a one-electron energy.
Summing over electron occupancies,
\e{
 \sum_{ n^\mu n^\nu}  p^{n^\mu} (1-n_{B'}^{\nu})n^{\nu}_B(1-n_B^{\mu})n_{B'}^{\mu} \delta_{n^{\mu},n^{\nu}-B+B'}= (1-f_B)f_{B'}; \as f_B=\langle n_B \rangle.
}
Combining it all, we arrive at a steady-state condition on the quasiparticle occupancies:
\e{
0= \partial_t f_B^{stat} \eq  (\partial_t f^{stat}_B)^{phot}_{gain} -  (\partial_t f^{stat}_B)^{phot}_{loss} + (\partial_t f^{stat}_B)^{phon} +O(U^4)\lin
 (\partial_t f^{stat}_B)^{phot}_{gain} \eq \f{2\pi}{\hbar} \sum_{mb'}|\sharp \eps_{m}\cdot  v^{off}_{bb'\bk}|^2(1-f_B)f_{B'}\bigg\{ \langle  N_m\rangle \delta(\var_{BB'}-\hbar\omega_m) +\langle  N_m+1\rangle \delta(\var_{BB'}+\hbar\omega_m)\bigg\}\lin
  (\partial_t f^{stat}_B)^{phot}_{loss} \eq \f{2\pi}{\hbar} \sum_{mb'}|\sharp \eps_{m}\cdot  v^{off}_{bb'\bk}|^2(1-f_{B'})f_{B}\bigg\{ \langle  N_m\rangle \delta(\var_{B'B}-\hbar\omega_m)  + \langle  N_m+1\rangle \delta(\var_{B'B}+\hbar\omega_m) \bigg\}.
}
It may be seen that the gain and loss rates are of the form expected from Dirac's time-dependent perturbation theory, i.e., Fermi's golden rule.
The phononic contribution may be evaluated analogously and also has the form expected from Dirac's time-dependent perturbation theory. \\

In conclusion, for the non-perturbative quasipartice distribution $f_B^{stat}$ to be steady (up to $U^4$ corrections), the zeroth-order quasiparticle distribution $f_B$ is the steady solution of $I_{coll}[f_B]=0$, where $I_{coll}$ is the collisional integral (evaluated  by Fermi's golden rule) in the presence of the light source. In particular, $f_B$ is \underline{not} the thermal quasiparticle distribution in the absence of the light source, contrary to the way in which most authors approach perturbation theory in nonlinear optical response.

\subsection{Numerical implementation of the BIS shift-current formula}\la{app:benchmark}

This appendix explains how to simulate an iso-energy-averaged quasiparticle distribution $f_E$ that is a steady solution to  the kinetic equation derived in \app{app:kineticmodel}, and how $f_E$ is subsequently inputted  to the BIS formula [Eq.\eqref{bismine2}] to determine the shift conductivity and its threefold decomposition. The conductivity will be determined for the model Hamiltonian [\q{modelham} with $\tilde{Q}=1$ and $\tilde{P}=4$] that is characterized by large time-reversal-symmetric Berry curvature; in particular, we would like the reader to be able to reproduce the conductivity plot in \fig{fig:Q05_maintext}(e).   \\

In \app{app:nonequil}, we have motivated  the momentum-resolved collisional integral in \qq{ick}{longi}, and derived  the corresponding iso-energy-averaged collisional integral in \qq{esipovmodel2}{belowcut}, having assumed that the quasiparticle distribution is iso-energy symmetric: $f_{\bk}\approx f_E$; cf.\  \q{isoenergy}. This assumption is justified to the extent that the collisional integral is  iso-energy symmetric, meaning that \qq{ick}{longi} is well-approximated by \qq{esipovmodel2}{belowcut}. Whether this is a good approximation depends on the parameters chosen in our model Hamiltonian [\q{modelham}] as well as the source radiation frequency $\omega_s$. We have checked that the e-isotropy condition approximately holds with our chosen parameters ($\tilde{Q}=1$ and $\tilde{P}=4$) in the frequency range $\hbar\omega/E_o \in [0.8,1.5]$.\footnote{The dipole matrix element becomes  iso-energy asymmetric at higher frequencies, as explained in \s{sec:vortex}.} The iso-energy symmetric assumption was made to save computational simulation time, but one may do without this assumption if one is numerically sophisticated.\\

\begin{figure}[H]
\centering
\includegraphics[width=1\textwidth]{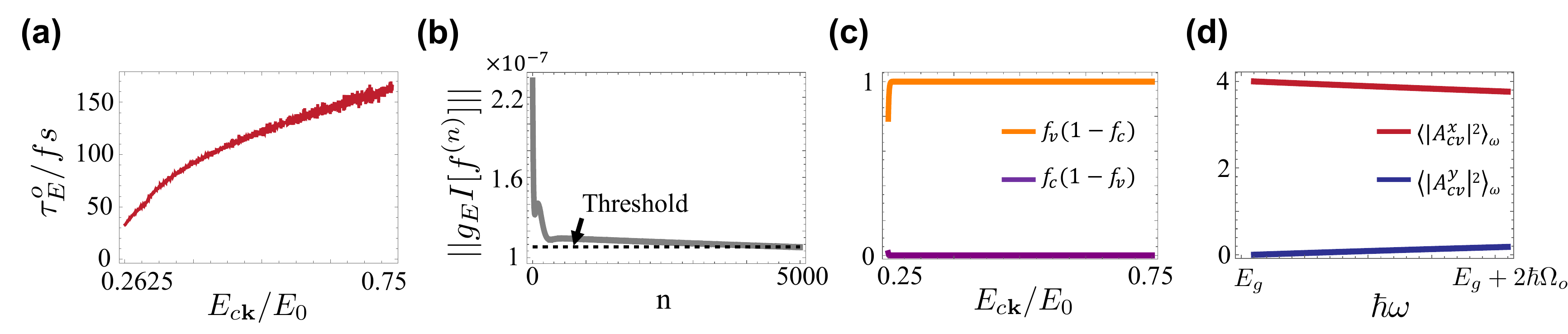}
\caption{(a) shows the plot of $\tau_{E}^{o}$ versus $E$ in the active region. (b) shows the plot of $||g_{E}I[f^{(n)}]||$ versus the evolution step $n$. (c) compares $f_{c}(1-f_{v})$ and $f_{v}(1-f_{c})$. (d) compares $\langle |A_{cv}^{x}|^2\rangle_{\omega}$ and $\langle |A_{cv}^{y}|^2\rangle_{\omega}$ in the passive region, i.e., $\hbar\omega \in [E_{g},E_{g}+2\hbar\Omega_{o}]$.}
\label{fig:appc}
\end{figure}

There remains some work in fixing the parameters in both sets of collisional integrals, chief among them being the electron-optical-phonon coupling constant $\zeta$ in \q{longi}, as well as the time scale $\tau^o_E$ for spontaneous emission of optical phonons in \q{esipovmodel2}.  The two parameters are related through \q{zeta}, which can be simplified as:
\begin{equation}
\label{zeta2}
\begin{aligned}
\zeta \;\frac{ a}{\mathcal{V}^2 g_E}\sum^{cut}_{\mathbf{k}\mathbf{k}^{\prime}}\frac{\left|\braket{u_{\bkp}}{u_{\bk}}_{\sma{\text{cell}}}\right|^2}{|\mathbf{k}-\mathbf{k}^{\prime}|^2} \delta(E_{c\bk}-E)\delta(E_{c\mathbf{k}\mathbf{k}^{\prime}}-\hbar\Omega_{o})
=\frac{1}{\tau^{o}_{E}}.
\end{aligned}
\end{equation}
The summation is restricted by the condition $\delta k=|\mathbf{k}-\mathbf{k}^{\prime}| \leq \mathscr{G}/10$, with $\mathscr{G}=2\pi/a$ being the reciprocal lattice period. A typical scale for $\tau^o_E$ is $0.1ps$,\cite{lundstrom_book,na_damascelli_electronphonon}
hence we set $\tau^o_{E^*}=0.1ps$ for a reference energy $E^*= 0.41375E_0$ in the active region; this fixes $\zeta=6.329 \frac{E_{0}}{0.1ps}$ and causes $\tau_{E}^{o}$ to vary from value 33.3 fs to value 165.2fs in the active region, as illustrated in  Fig.~\ref{fig:appc} (a). To be clear, all plotted energies are defined to equal zero in the middle of the gap, in contrast to the carrier energies defined with respect to the band extrema. The other parameters in the kinetic model are fixed to be:  $\tau_{\text{rec}}=1 ns$ (a typical interband recombination time\cite{sturmanfridkin_book}); $\tau_{E}^{s}=1 ns$ for all $E$ (a typical energy relaxation time due to spontaneous emission of acoustic phonons\cite{zakharchenya_photoluminescence}); \footnote{$\tau_{E}^{s}\gg \tau_{E}^{o}$ and the optical phonon scattering explicitly dominates in the active region.}  and $a=5\ang$ (a typical lattice period). All calculation in this appendix are presented for a linearly polarized source: $\beps_{s}=\vec{x}$.  \\

Our first step is to simulate $f_E$ which sets the iso-energy-averaged collisional integral [\qq{esipovmodel2}{belowcut}] to zero. We begin by discretizing the energy: $\ldots, E_j,E_{j+1},E_{j+2},\ldots$ such that  adjacent energy levels are separated by $\Delta E=E_{j+1}-E_j$. Conduction-band Bloch states are binned according to the following rule: if $E_{j}-\Delta E/2 \leq  E_{c\bk} < E_j +\Delta E/2$, then the Bloch state $(c\bk)$ belongs in the $j$'th bin. By choosing $\hbar \omega=n_{\omega}\Delta E$ and $\hbar\Omega_o=n_{\Omega} \Delta E$ to be integer multiples of $\Delta E$, one can translate Dirac delta functions  to Kronecker delta functions: $\delta(E_{c\mathbf{k}\mathbf{k}^{\prime}}-\hbar\Omega_{o})\ri \delta_{E_{c\mathbf{k}\mathbf{k}^{\prime}},\hbar\Omega_{o}}/\Delta E$ and $\delta(E_{cv\mathbf{k}}-\hbar\omega)\ri \delta_{E_{cv\mathbf{k}},\hbar\omega}/\Delta E$.
For instance, $\delta_{E_{c\mathbf{k}\mathbf{k}^{\prime}},\hbar\Omega_{o}}=1$ if and only if $E_{c\bk}\ri E_j$ and $E_{c\bkp}\ri E_{j-n_{\Omega}}$ for some bin index $j$. 
Then the photon-absorption term is  discretized  as 
\begin{equation}
\label{eq:Iexc}
G_{\uparrow}[f]\,\delta(E-E_{c,exc})\ri g_{E}\sum_{m_s} I^{m_s}_{{\text{exc}},E};\as I^m_{{\text{exc}},E} =2\pi^2\alpha_{fs}c \frac{\Delta N_{m}}{\mathcal{V}}\,2E\,\left(1-2f_E \right)\langle|\epsilon_{m}\cdot \mathbf{A}_{cv\mathbf{k}}|^ 2  \rangle_{\omega_m}\frac{\delta_{2E,\hbar \omega_m}}{\Delta E},
\end{equation}
with  $\alpha_{fs}=|e|^2/(\hbar c)$ being the fine-structure constant, and $I^{m}_{\text{exc},E}$ being the iso-energy average of $I_{exc\bk}^{\omega\epsilon}$; cf.\ Eq.~\eqref{exciterate}. In practice, we have chosen $\Delta E=E_{0}/1600$ and $\hbar\Omega_{o}=20\Delta E$. To avoid certain artifacts of our energy discretization scheme,  we introduced   a small frequency bandwidth  ($4\Delta E/\hbar$) for the source-generated photons; this means that the source produces an equal number of photons in each of four modes ($m_s=1,2,3,4$), with differing frequencies $\omega_{s}-2\Delta E/\hbar$, $\omega_{s}-\Delta E/\hbar$, $\omega_{s}$, $\omega_{s}+\Delta E/\hbar$ but identical polarization $\beps_{s}$. \\

We initialize the distribution as a Boltzmann-Maxwell distribution: $f^{(0)}_{E}=f_{E,BM}^{T}=\exp(-(E-\mu_e)/k_BT)$,
which is the steady distribution favored by the Fokker-Planck term:  $(1+k_BT\partial_E)f^T_E=0$.\footnote{ hot-carrier photoluminescence spectra support the hypothesis that most photoexcited carriers are distributed \`a la Maxwell-Boltzmann; cf. \app{app:relaxationmechanisms}.} 
$\mu_e$ is generically not the chemical potential in thermal equilibrium; instead, it   is determined by balancing recombination and excitation rates for the conduction band as a whole: $\sum_{E} g_{E }f_{E}^{T}\Delta E/\tau_{\text{\text{rec}}}=G_{\sma{\uparrow}}$, with $\sum_E\Xi(E)$ our shorthand for $\sum_j\Xi(E_j)$.\\

Beginning from our ansatz Maxwellian distribution, we evolve the system over  a  discrete time interval $\delta t$ to obtain a new distribution:
\begin{equation}
f_E^{(n+1)}=f_E^{(n)}+I_E[f^{(n)}]\delta t,
\end{equation}
for $n=0,1,2,....$, with  the collisional integral $I_E[f]$ defined in \qq{esipovmodel2}{belowcut}. This is a numerical procedure to obtain a steady state, and in no way reflects the actual time evolution of quasiparticle distribution in an experiment.
We stop this iterative process  when the norm  
\begin{equation}
    ||g_{E}I[f^{(n)}]||=\sqrt{\sum_{E} (g_{E}I[f^{(n)}(E)])^2},
\end{equation}
decays below a certain threshold, i.e., $0.05\%$ of $\sum_{E,m_{s}}g_{E}I^{m_{s}}_{\text{exc}}$.
Supposing the threshold is crossed when $n=n_0$, then we say $f^{(n_{0})}$ is a \textit{numerically steady} solution of the kinetic equation. \\

For illustration, Fig.~\ref{fig:activeregion} (c) represents a numerically steady distribution calculated using the above scheme, with $\Delta N_{m_s}/\mathcal{V}=(1ns)^{-1}\frac{\Delta E \mathscr{G}^2}{440\pi^2 c \alpha_{fs} E_{0}}\approx  10^{10} cm^{-3}$ for each of the four source modes, $E_{\text{exc}}=5\hbar\Omega_{o}$, $n_0=5000$ steps, and a time step $\delta t=1 fs$.  Fig.~\ref{fig:appc} (b) illustrates a decay of $||g_E \,I[f_{n}||$ below our threshold of $(5\times 10^{-4})\sum_{E}g_{E}I_{\text{exc}}\approx 1.08\times 10^{-7}$.\\

To calculate the shift current, we input the numerically steady $f^{(n_0)}_E$ to the threefold-decomposed current formulas in \q{definejexc}, \q{jintraoptical} and \q{approxjrec}. The discrete analogs of these formulas are:
\e{
\label{BISexc}
\mathbf{j}_{\text{exc}}\eq -2_{\uparrow\downarrow}\frac{|e|}{\mathcal{V}}\sum_{m_s}\sum_{\mathbf{k}}S^{\eps_s}_{c\mathbf{k}\leftarrow v\mathbf{k}}I_{\text{exc},E_{\bk}}^{m},\\
\label{eq:BIS3}
\mathbf{j}_{\text{intra}}\eq 2_{\uparrow\downarrow} 2_{cv}\frac{|e|}{\mathcal{V}}\sum_{\mathbf{k}\mathbf{k}^{\prime}}^{cut}\frac{\zeta a}{\mathcal{V}}\frac{1}{|\mathbf{k}^{\prime}-\mathbf{k}|^2} f_{c\bkp}
(1-f_{c\bk}) \frac{\delta_{E_{\mathbf{k}\mathbf{k}^{\prime}},\hbar\Omega_{o}}}{\Delta  E} \bOmega_{c,(\bk+\bkp)/2}\times (\bkp-\bk),\\
\label{eq:ptcurrentem}
\mathbf{j}_{\text{\text{rec}}}\eq 2_{\uparrow\downarrow}\frac{|e|}{\mathcal{V}}\sum_{\mathbf{k}\in \text{pass}}\mathbf{S}^{\hat{x}}_{c\kext \leftarrow v\kext }\frac{f_{c\mathbf{k}}}{\tau_{\text{\text{rec}}}}.
}
We will explain each equation in turn:\\

\noi{i-excitation} $I_{\text{exc}}^{m}$ was defined in \q{eq:Iexc} and $\sum_{m_s}$ sums over the aforementioned source modes.\\

\noi{ii-intra} \q{eq:BIS3} is derived by substituting the electron-phonon matrix element  [Eq.~\eqref{longi}] and the anomalous shift vector [\q{anomalousshift}] into Eq.~\eqref{jintraoptical}, and then summing over both conduction and valence bands. In this sum, each band contributes equally due to the presumed electron-hole symmetry [$f_{c\bk}=1-f_{v\bk}$; cf.\ \app{app:ehsymmetry}], hence the factor of $2_{cv}=2$ in \q{eq:BIS3}. To see why, note
for any two-band model that $\bOmega_{c\bk}=-\bOmega_{v\bk}$, hence $\bS^{ano}_{c;\bkp \lea \bk}=-\bS^{ano}_{v;\bkp \lea \bk}$ and  $f_{c\bkp}(1-f_{c\bk})\bS^{ano}_{c;\bkp \lea \bk}=(1-f_{v\bkp})f_{v\bk}(-\bS^{ano}_{v;\bkp \lea \bk})$. Recognizing from \q{anomalousshift} that $\bS^{ano}_{v;\bkp \lea \bk}=-\bS^{ano}_{v;\bk \lea \bkp}$, we find that \q{jintraoptical} is  identical for valence and conduction bands.\\

\noi{iii-recombination} $\sum_{\mathbf{k}\in \text{pass}}$ in \q{eq:ptcurrentem} integrates over the passive $\bk$-volume, based on a previous argument [cf.\ \s{sec:kinetictheory}] that the majority of photo-excited carriers are steadily distributed within the passive region; this argument is corroborated by our numerical simulation in Fig.~\ref{fig:activeregion}(c), bearing in mind that $g_{E}$ is constant in our quasi-2D model. Because a dipole selection rule fixes $A^{y}_{cv \bk}=0$ for $k_x=0$, and $A^{y}_{cv \bk}$ cannot vary substantially in the small passive region [assuming the band gap is not anomalously small], it may be deduced that
$|A^{y}_{cv,\mathbf{k}}|^2\ll |A^{x}_{cv,\mathbf{k}}|^2$ everywhere in the passive region; cf. Fig.~\ref{fig:appc} (d). Therefore, one may as well approximate all recombination transitions as being mediated by $x$-polarized photons, with the spontaneous emission rate $\mathcal{E}^{sp,\hat{x}}_{v\mathbf{k}\leftarrow c\mathbf{k}}=f_{c\mathbf{k}}/\tau_{\text{\text{rec}}}$. The corresponding photonic shift vector $\mathbf{S}^{\hat{x}}_{v\mathbf{k}\leftarrow 
c\mathbf{k}}$ is also approximated as $\mathbf{S}^{\hat{x}}_{v\kext\leftarrow c\kext}$, because the variation of the photonic shift vector within the passive region is small. \\

The threefold-decomposed  conductivities are obtained by diving each of $\jexc,\jintra$ and $\jrec$ by $|\mathcal{E}_{\omega}|^2$; cf.\ \q{sigmathreefold}. It is advantageous to express the  squared electric amplitude $|\mathcal{E}_{\omega}|^2$ in terms of the discrete $I^m_{\text{exc}}$ [\q{eq:Iexc}]: 
\begin{equation}
\label{efieldIexc}
|\mathcal{E}_{\omega}|^2=\left(\sum_{E,m_s}g_{E}\Delta E I^{m_s}_{\text{exc},E}\right)/[2 \pi c \alpha_{fs}(1-2f_{E})\langle|\epsilon_{m_{s}}\cdot \mathbf{A}_{cv\mathbf{k}}|^ 2  \rangle_{\omega} JDOS_{\uparrow}],
\end{equation} 
in accordance with $\sum_{m_s}\Delta N_{m_s}\hbar\omega_{s}/\mathcal{V}=|\mathcal{E}_{\omega}|^2/(2\pi)$; c.f. Eq.~\eqref{elecfield}. For   the conductivity plot in \fig{fig:Q05_maintext}(e), we had chosen $\sum_{m_s}\Delta N_{m_s}/\mathcal{V}\approx 10^{10} cm^{-3}$.
For comparison, in a typical argon-ion-laser experiment with a radiation intensity of $40Wcm^{-2}$,\cite{koch_BaTiO3} the number density of source photons  is approximately $1/3 \times 10^{10}\,cm^{-3}$.

\subsection{Comparison with the  Kraut-Baltz-Sipe-Shkrebtii formula and dissipative Floquet methods}\la{sec:krautbaltz}

The  Kraut-Baltz-Sipe-Shkrebtii formula (KBSS) for the shift current is\cite{kraut_anomalousbulkPV,baltz_bulkPV,sipe_secondorderoptical} 
\e{\bj_{KBSS} = \bsigma^{KBSS}_{\beps,\omega}|\cale_{\omega}|^2, \as \bsigma^{KBSS}_{\beps,\omega} = -2\pi \f{|e|^3}{\hbar}\sum_{bb'}\int \f{d^3k}{(2\pi)^3}f^{T}_{bb'\bk}|\beps\cdot\bA_{b'b\bk}|^2 \bS_{b'\bk \lea b\bk}\delta(E_{b'b\bk}- \hbar \omega), \la{kbssformula}}
with $E_{b'b\bk}=E_{b'\bk}-E_{b\bk}$ and $f^{T}_{bb'\bk}=f^{T}_{b\bk}-f^{T}_{b'\bk}$.
One can convert \q{kbssformula} to a proportionality relation with the radiation intensity (within the dielectric medium) by $\cali_{rad}=(c/2\pi)n_{\omega}|\cale_{\omega}|^2$, assuming the medium is non-magnetic with a frequency-dependent refractive index $n_{\omega}$ that is spatially uniform and isotropic.\footnote{The time-averaged  Poynting  vector (within the dielectric medium) has the form $\cali_{rad}\hbq$, with $\cali_{rad}=(c/2\pi)n_{\omega}|\cale_{\omega}|^2$ having dimensions of energy per unit area per unit time, and  $\hbq$ being  the unit directional vector of the electromagnetic wave propagation. We adopt the same, real-valued definition of the refractive index as in \ocite{landaulifshitz_electrodynamics}. In an absorptive medium, $\cali_{rad}$ should be multiplied by a coordinate-dependent, exponential damping factor;\cite{landaulifshitz_electrodynamics} however  this factor is negligible if the attenuation length greatly exceeds the thickness of the medium. }\\

The KBSS formula has been derived in a variety of models and methods,\cite{kraut_anomalousbulkPV,baltz_bulkPV,sipe_secondorderoptical,morimoto_nonlinearoptic,barik_nonequilibriumnature,matsyshyn_rabiregime,parker_diagrammatic,ahn_riemanniangeometry,holder_trsbreaking,hikaru_chiralphotocurrent} which may have created an impression that the KBSS formula is universally truthful.  The actual reason for the universality is a largely unjustifiable and often implicit assumption shared by  all these models, namely that the electronic quasiparticle distribution retains its equilibrium value under continuous-wave irradiation. It is an experimental fact that this assumption does not hold, as is most vividly demonstrated by hot-carrier photoluminescence spectroscopy.\cite{zakharchenya_photoluminescence,esipov_temperatureenergy}\\

This formula was originally derived by Kraut and Baltz\cite{kraut_anomalousbulkPV,baltz_bulkPV}
and subsequently rederived by Sipe and Shkrebtii\cite{sipe_secondorderoptical} using more-or-less standard perturbation theory. In the Kraut-Baltz derivation, relaxation was accounted for in a crude relaxation time approximation, with the relaxation time eventually taken to be arbitrarily small compared to the Rabi oscillation period at resonance; in other words, relaxation to equilibrium is assumed to be such a strong effect (relative to the optical excitation) that the electronic quasiparticle distribution never deviates from the equilibrium value. (Similar perturbative derivations\cite{hikaru_chiralphotocurrent,ahn_riemanniangeometry,holder_trsbreaking} have proposed without rigorous justification to view the imaginary infinitesimals in the energy denominator as an inverse relaxation time.) In the Sipe-Shkrebtii derivation\cite{sipe_secondorderoptical} (and similar diagrammatic methods\cite{parker_diagrammatic}), relaxation was omitted entirely; because their method is based on perturbing an equilibrium state in the lowest orders for the electric field, it is not surprising that their final formula is expressed in terms of the equilibrium quasiparticle distribution. The KBSS formula has been alternatively derived from dissipative Floquet methods\cite{morimoto_nonlinearoptic,barik_nonequilibriumnature,matsyshyn_rabiregime} in the regime of strong dissipation: relaxation rate $\gg$ Rabi frequency. This is another model where relaxation to equilibrium is assumed to be overwhelmingly strong.\\

The rest of this appendix will be used to demonstrate that the BIS formula \textit{also} reduces to the KBSS formula if the electronic quasiparticle distribution is thermal. On one hand, this planned demonstration can be viewed as a consistency check of the BIS formula. On the other hand, the BIS-to-KBSS reduction  crystallizes what is missing from the KBSS formula: namely, the missed photocurrent can be precisely attributed to  the deviation of the steady quasiparticle distribution  from its equilibrium value, given a  realistic model of relaxation in which relaxation also causes shifts.\\

Without further ado, the KBSS formula in \q{kbssformula} is related to the BIS formula in \q{bismine2} by
\e{\bj_{KBSS}=\bj[f^{T}_{B},N^{T;phot}_{m}+\Delta N_{s}\delta_{m,m_s},N^{T;phon}_{m}]. \la{definejtran23}}
That the BIS formula is a functional of the quasiparticle, photon and phonon occupancies has been explained in \s{app:occupancies}. The KBSS formula is thus the BIS formula with a very specific input for occupancies: $f^T_B$ is a Fermi-Dirac distribution  [\q{fermidirac} ], $N_m^{T;phon}$ is a Planck distribution [\q{planck}] with the same temperature, and the photon occupancy is a sum of thermal and non-thermal contributions; the non-thermal photons  are generated by a mono-modal source with mode index $m_s$.  \\

All bosonic modes  with a thermal occupancy cannot contribute to the shift current, owing to detailed balance; cf.\ \q{detailedbalance}. For the source mode $m_s$, the net transition rate [\q{photonrate}] can be decomposed as 
\e{\big(\cala^{m_s}_{C\lea V}-\cale^{m_s}_{V \lea C}\big)_{f^{T}_{B},N^{T}_{m_s}+\Delta N_{s}}\eq \big(\cala^{m_s}_{C\lea V}-\cale^{m_s}_{V \lea C}\big)_{f^{T}_{B},N^{T}_{m_s}}+ \f{( 2\pi e)^2\omega_{s}}{\calv} \;\big|\beps_s\cdot \bA_{cv\bk}\big|^2\delta(E_{cv\bk}- \hbar \omega_{s})f^{T}_{vc\bk}\Delta N_s, \la{sourcems}}
with $C=(c\bk)$ and $V=(v\bk)$.
The first term on the right-hand side of \q{sourcems} vanishes by detailed balance [\q{detailedbalance}], hence the right-hand side of \q{definejtran23} reduces to \q{kbssformula}.\\

We will say a few words about what is missed from the KBSS formula, how the BIS formula does better, and why  dissipative Floquet models (in their present formulation) do not. 
As explained in \s{sec:kinetictheory} and elaborated in \app{app:jtranjexc}, the KBSS current is approximately the  transient photocurrent, or equivalently the excitation component of the steady photocurrent:
\e{ \bsigma^{KBSS}_{\beps,\omega} \approx \bsigma^{\text{exc}}_{\beps,\omega};   \as \bsigma_{\beps,\omega}=\bsigma^{\text{exc}}_{\beps,\omega}+\bsigma^{\text{intra}}_{\beps,\omega}+\bsigma^{\text{rec}}_{\beps,\omega}.\la{kbsscomp} }
As defined through the BIS formula, the shift conductivity $\bsigma_{\beps,\omega}$  [\q{shiftconductivity}] has a threefold decomposition  explained in   \q{decomposesigma}; apparently, the KBSS formula misses out on current contributions by intraband relaxation and interband recombination.\\

Because the KBSS formula is derived by dissipative Floquet methods in the strongly dissipative regime,\cite{morimoto_nonlinearoptic,barik_nonequilibriumnature,matsyshyn_rabiregime} it is evident that these methods  also miss out on the effects of intraband relaxation and interband recombination. The present formulation of Floquet methods are inadequate for the following reasons: (a) The premise of time-periodic Hamiltonians  relies on a classical approximation of the radiation field, and precludes the quantum effect of radiative recombination by spontaneous emission. (b) In \ocite{morimoto_nonlinearoptic} and \ocite{matsyshyn_rabiregime}, the use of experimentally-unrealizable `fermionic baths' as a relaxation mechanism precludes the phonon-induced shift [\q{anomalousshift}] responsible for $\bsigma^{\text{intra}}_{\beps,\omega}$. (c) In \ocite{barik_nonequilibriumnature}, Barik and Sau considered  electron-phonon scattering as a relaxation mechanism; however, they also missed the phonon-induced shift  [\ \q{phononicshift2}] due to an unjustifiable assumption that the electron-phonon matrix element is momentum-independent.

\section{Loop formulation of the steady shift current}\la{app:loop}

\begin{figure}[H]
\centering
\includegraphics[width=16 cm]{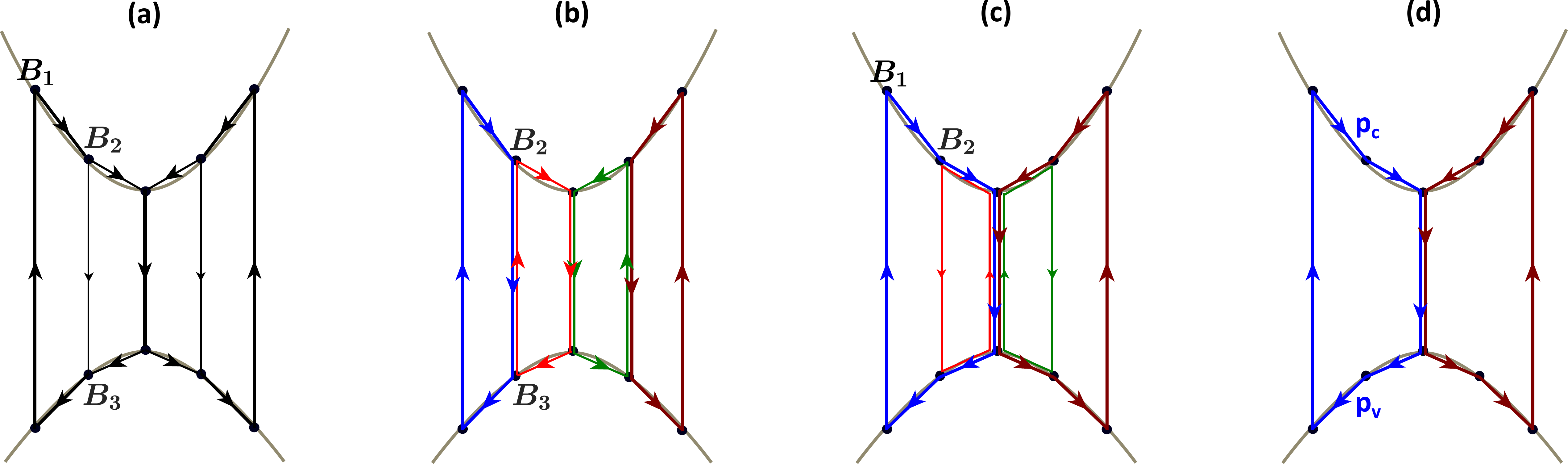}
\caption{(a) Caricature of a probability flow network. Panels (b) and (c) illustrate two distinct loop decompositions of the network in panel (a). Panel (d) illustrates a geodesic approximation of the network in (a). Part of the approximation amounts to neglecting the radiative recombination transition between the two Bloch states labelled $B_2$ and $B_3$, as justified in \s{sec:anomalous}. }
\label{fig:flownetwork}
\end{figure}

We present an equivalent formulation of the steady shift current, namely that the BIS formula in \q{bismine2} is equivalent to a sum of loop currents:
\e{\text{Loop current theorem:}\as \bj=-\f{|e|}{\calv}\sum_{B,B',m } \bS^{m}_{B' \lea B} \bigg( \cala^{m}_{B' \lea B}-\cale^{m}_{B \lea B'} \bigg)=\sum_{\text{loop}}\jloop, \la{loopcurrentthm}}
with $\jloop$ being the current contributed by a closed flow line (in energy-momentum space) of one-electron probability, as illustrated in \fig{fig:flownetwork}(b-c).  The precise definition of $\jloop$ is given in \q{jloop} after some preliminary preparations.\\

As a first step to reformulating the shift current in terms of loop currents, \app{app:flownetwork} shows how to interpret the flow of one-electron probabilities in energy-momentum space as an oriented graph with nodes corresponding to Bloch states, as illustrated in \fig{fig:flownetwork}(a); it will be shown that this graph can be decomposed into loops, and for each loop one can associate a net shift vector [\q{sloop}] and a current [\q{jloop}]. We will then prove the loop current theorem in \app{app:loopthm}, and subsequently discuss two applications: \\

\noi{i} The loop-current formula manifests that the intraband-Berry connection terms: $(\bA_{b'b'\bkp}-\bA_{bb\bk})$ in the shift vector [\qq{phononicshift}{photonicshift}] always cancel out, when all transitions in the steady state are accounted for. From this follows a revision of a purported  relation  between the shift current and interband polarizaton differences,\cite{fregoso_opticalzero} as discussed in \app{app:loopimp}.\\

\noi{ii} The loop formulation naturally leads to equitable approximations of the shift current, which treat excitation, relaxation and recombination on equal footing. The approximation lies in identifying  a reduced family of loops which contribute most substantially to the shift current. Once a reduced family of loops is identified, calculating the shift current via \q{loopcurrentthm} requires far less computational resources than a direct calculation of the BIS formula [cf.\ \app{app:benchmark}]. This work focuses on the geodesic loops [\fig{fig:flownetwork}(d)] which predominate the shift current in direct-gap semiconductors.  For 3D semiconductors, the geodesic approximation to the shift conductivity [\q{shiftconduct}] is derived from the loop-current formula [\q{loopcurrentthm}] in \app{app:geodesic}, and as a small-angle-scattering limit of the BIS formula [\q{bismine2}] in \app{app:smallanglelimit}.  
Finally, the geodesic approximation is extended to quasi-2D semiconductors in \app{app:geodesicquasi2D}.

\subsection{The shift loop and the loop current} \la{app:flownetwork}

 Let us define a \textit{link} as a pair of Bloch labels. A link is said to be ordered if the band energy of the first label is larger than the band energy of the second:
\e{\text{ordered link} \;\equiv\; \link; \as E_{\bgt\blt}=E_{\bgt}-E_{\blt}>0.}
 A \textit{general link} written as $(B',B)$ admits any possible ordering of $E_{B'}$ and $E_B$. For instance, given the three Bloch labels in \fig{fig:flownetwork}(a), one may write $(B_1>B_2)$ and $(B_3,B_2)$ but not $(B_3>B_2)$. \\

For every ordered link, we define  the \textit{ordered transition rate} as the sum of one-electron transition rates over all possible bosonic modes indexed by $m$:
\e{\text{ordered transition rate}\;=R_{\link} \eq \sum_m\big( \cala^{m}_{\bgt \lea \blt}-\cale^{m}_{\blt \lea \bgt} \big), \la{rlink} }
and the  \textit{ordered shift vector} as a weighted average of the shift vector [cf.\ \qq{phononicshift2}{photonicshift2}] over all bosonic modes:
\e{\text{ordered shift vector}\;\bS_{\link} \eq \sum_m \bS^m_{\bgt \lea \blt}\f{\cala^{m}_{\bgt \lea \blt}-\cale^{m}_{\blt \lea \bgt}}{R_{\link}}.\la{slink}}
As a reminder, $\cala$ is the absorption rate and $\cale$ the emission rate defined in \qq{phononrate}{photonrate}. Because  $\cala^m_{\bgt \lea \blt}-\cale^{m}_{\blt \lea \bgt}\propto \delta(E_{\bgt\blt}-\hbar\omega_m)$ with a bosonic energy $\hbar\omega_m$ that is strictly positive,\footnote{As remarked earlier in \s{sec:algebra}, quantized phonons/photons are not well-defined for zero $\omega_m$.} we wrote $\link$ in \q{slink} rather than $(B>B')$. Let us discuss two classes of ordered shift vectors: \\

\begin{tcolorbox}
\noindent \textit{ Ex-1: Phononic ordered shift vector} \\

If $\bgt$ and $\blt$ differ in electronic wavevectors, then, within the dipole approximation for the electron-photon coupling, one can restrict $\sum_m$ in \qq{rlink}{slink} to phononic modes.\\

\noi{a} If the difference in wavevectors ($\bk$ and $\bkp$) is small and the transition is intraband ($b=b'$), $\bS^{\bq p}_{\tran}$ in \q{slink} is well-approximated by the anomalous shift $\bS^{ano}_{b;\bkp \lea \bk}$ [\q{anomalousshift}], which does not depend on the phonon branch $p$. It should also be recalled from \q{phononrate} that $\cala^{m}_{\bgt \lea \blt}-\cale^{m}_{\blt \lea \bgt}\propto \delta_{\bq,\bkp-\bk}$. Altogether, these imply that   \q{slink} reduces to  $\bS_{(b\bkp > b\bk)}=\bS^{ano}_{b;\bkp \lea \bk}$. \\

\noi{b} If the difference in wavevectors ($\bk$ and $\bkp$) is not necessarily small, but the  phonon energy $\hbar\omega_m=E_{B'B}$ is nondegenerate, then $\sum_m$ in \qq{rlink}{slink} is restricted to one phonon branch (say $m$), and  $\bS_{\link}=\bS^{m}_{\tran}$ as defined in \q{phononicshift2}.
\end{tcolorbox}

\begin{tcolorbox}
\noindent \textit{ Ex-2: Photonic ordered shift vector}\\

If $\bgt$ and $\blt$ are identical in electron wavevectors ($\bk=\bkp$), with $E_{b'b\bk}=E_{b'\bk}-E_{b\bk}$ exceeding the optical phonon energies, than one may restrict $\sum_m$ in \qq{rlink}{slink} to photonic modes. \\

\noi{a} If $\bk$ does not lie on the excitation surface, $\sum_m$ in \qq{rlink}{slink} is restricted (by energy conservation) to photonic modes whose occupations are thermal, i.e., $N_m=N_{m}^{T_l}$ has the Planck form and does not depend on the orientation $\hbq$ of the photon wavevector. In fact, the only quantities in \q{slink} that depend on $\hbq$ is the photonic shift vector [\q{photonicshift2}] and the square of the interband Berry connection [\q{photonrate}]. Thus \q{slink} simplifies to:
 \e{\bS_{(b'\bk >b\bk)}=\f{\int d\lambda_{\hbq} \sum_{p=1}^2|\beps_{\bq p}\cdot \bA_{b'b\bk}|^2\;\bS^{\bq p}_{b'\bk\lea b\bk}}{\int d\lambda_{\hbq} \sum_{p=1}^2|\beps_{\bq p}\cdot \bA_{b'b\bk}|^2},\la{orderedshiftphoton}}
where we integrate over $\hbq$ (parametrized by solid angle $\lambda_{\hbq}$) and sum over both transverse polarizations.\\

\noi{b} If $\bk$ lies on the excitation surface, $\sum_m$ in \qq{rlink}{slink} sums over all photonic modes with the same frequency $\omega_s$ as the source-generated photons.
For a bright source, an argument in \app{app:threefold} conveys that $\sum_m$ in \qq{rlink}{slink} may as well be restricted to  the single source mode $m_s$, so that \q{slink} simplifies to   $\bS_{(b'\bk >b\bk)}=\bS^{m_s}_{b'\bk \lea b\bk}$ [\q{photonicshift2}].
\end{tcolorbox}

It would also be useful to discuss the net transition rate for  $B'\lea B$, with $E_{B'}$ not necessarily greater than $E_B$. For this purpose, we define the:
\e{ \text{oriented transition rate}\; R_{\tran}=\sgn [E_{B'B}] R_{\link}=-R_{B\lea B'}, \la{orientedtranrate}}
such that $R_{\tran}>0$ represents a net probability flow from $B$ to $B'$, independent of the ordering of band energies.\\

 We may draw a cartoon to visualize the flow of probability in energy-momentum space. In \fig{fig:flownetwork}(a), we represent every link by an arrow; the thickness of the arrow shaft is proportional to $|R_{\link}|$; the arrowhead points from $\bgt \lea \blt$  if   $R_{\link}>0$, and vice versa. Our cartoon is thus an oriented graph/network, with each node/vertex corresponding to a Bloch state, and with each link/edge oriented according to direction of the probability flow. We will use node${=}{B}$ interchangeably. \\

By comparing the BIS formula [\q{bismine2}] with the definitions of $R_{\link}$ and $\bS_{\link}$ in \qq{rlink}{slink}, one deduces that the shift current  is essentially the sum of $R_{\link}\bS_{\link}$ over all ordered links in the probability-flow network:
\e{\bj \eq -\f{|e|}{\calv}\sum_{\link} \bS_{\link} R_{\link}. \la{bismine222}
}

 	In the steady state, the time-independence of the occupancy of each Bloch state implies that for each node (say, $B$) in the graph, incoming transition rates must exactly balance outgoing transition rates: $\sum_{B'}R_{\tran}=0$. The probability-flow network can therefore be viewed as a discrete analog of a divergence-free/solenoidal vector field. 
This discrete solenoidal condition allows  to decompose the probability-flow network into  loops, as illustrated in \fig{fig:flownetwork}(b).\footnote{Analogously, a divergence-free vector field can be approximated by a superposition of elementary solenoids, which includes the case of finite-length loops.\cite{smirnov_solenoidalvector}} Each loop represents the closed flow line of an electron's probability in energy-momentum space, with the perspective that forward-moving holes are backward-moving electrons. \\

More precisely, here are three defining properties of a  loop:\\

\noi{a} The first property of a loop is that it is a closed concatenation of general links: 
\e{\text{loop with}\;N\;\text{links}\;\eq  (B_N,B_{N-1})(B_{N-1},B_{N-2})\ldots(B_2,B_1)(B_1,B_N).\la{loop}
}
If $(B',B)$ is one of the $N$ links appearing above, then we say the link is contained in the loop: $(B',B)\in loop$; if $E_{B'B}>0$ [resp. $<0$], we would further say that  $(B'>B)\in loop$ [resp.\ $(B>B')\in loop$].\\

\noi{b} To each loop, we associate a positive-valued \textit{loop rate} $|\delta R_{\text{loop}}|$ which is the magnitude of the probability flow rate along the loop.\\

\noi{c} Each loop has a $\Z_2$-valued orientation ($Or_{\text{loop}}$) which  determines the direction of probability flow:
\e{
Or_{\text{loop}}\eq+1: \as B_1 \ri B_2 \ri \ldots \ri B_N \ri B_1 \la{looporientation1}\\
Or_{\text{loop}}\eq -1:\as B_1 \lea B_2 \lea \ldots \lea B_N \lea B_1.\la{looporientation}
}\\

 It follows from (a-c) that one can assign an \textit{oriented loop rate} to each link in the loop:
\e{\delta R^{\text{loop}}_{B_{n+1}\lea B_{n}}=-\delta R^{\text{loop}}_{B_{n}\lea B_{n+1}}=Or_{\text{loop}}|\delta R_{\text{loop}}|, \iwith B_{N+1}\equiv B_1. \la{looporientedrate}}
The sense in which the probability-flow network is decomposed to loops is that for each link in the network,
\e{R_{\tran}=\sum_{loop \ni (B',B)}\delta R^{\text{loop}}_{\tran},\la{eq:deltaRloop}} 
where the summation is over all loops that contain the link $(B',B)$; $R_{\tran}$  is given by the Fermi's golden rule [c.f. Eqs.~\eqref{phononrate} and ~\eqref{photonrate}] and depends on the carrier distribution. Equivalently, for every ordered link in the network,
\e{R_{\link}=\sum_{loop \ni \link}\delta R^{\text{loop}}_{\tran}.\la{decomposeorderedrate}}
Consider the cartoon of \fig{fig:flownetwork}(b) for illustration: $(B_3,B_2)$ is contained in two loops colored red and blue, hence $R_{B_3\lea B_2}$ is given by a sum of two $\delta R$'s. The loop decomposition is not unique, meaning that a different set of loops may satisfy \q{decomposeorderedrate} for the same network, as illustrated in \fig{fig:flownetwork}(c).  \\

For each loop, the shift loop  is defined by summing the ordered shift vector over all ordered links in the loop, weighted by a sign that encodes the direction of probability flow in that loop:
\e{\text{Shift loop}\as \sloop = \sum_{\link \in loop} \sgn[\delta R^{\text{loop}}_{B'\lea B}]\;\bS_{\link}.  \la{sloop}}
Because the summation is over ordered links, $E_{B'B}>0$, and $\sgn[\delta R^{\text{loop}}_{B'\lea B}]=+1$ (resp. $-1$) if the loop-decomposed probability flow is toward increasing band energies (resp. decreasing band energies).

\begin{tcolorbox}
To motivate this sign factor, consider an example which elaborates on case (b) of \textit{Ex-1} [cf. box under \q{slink}].  For the conduction-band link $(B_1>B_2)$ illustrated in \fig{fig:flownetwork}(c), the probability flow is toward decreasing band energies, which reflects the predominance of phonon emission over absorption. Then the link's contribution to $\sloop$ is simply 
$ \sgn[\delta R^{\text{loop}}_{B_1\lea B_2}]\;\bS_{(B_1{>}B_2)}=-\bS^{\bq p}_{B_1\lea B_2}$, which equals $\bS^{-\bq,p}_{B_2\lea B_1}$ by the inversion symmetry of the phonon shift vector [\q{phononicshift2}]. As explained at the end of \app{app:sumrulephononic},  $\bS^{-m}_{B_2\lea B_1}$ is precisely the shift vector associated to emitting a phonon of mode $m$.  
\end{tcolorbox}

With $\sloop$ and $|\delta R_{\text{loop}}|$  in hand, we can now define the \textit{loop current} 
\e{
 \jloop = -\f{|e|}{\calv}\bS_{\text{loop}}|\delta R_{\text{loop}}|, \la{jloop}
}
 that enters our loop current theorem in \q{loopcurrentthm}. It may be seen that the loop current depends implicitly on the carrier population through $|\delta R_{loop}|$, as per Eq.~\eqref{eq:deltaRloop} and with identifying $R_{\tran}$ as the golden-rule transition rates in Eqs.~\eqref{phononrate} and \eqref{photonrate}.

\subsection{Derivation of loop current theorem}\la{app:loopthm}

Beginning from the right-hand side of \q{loopcurrentthm}, we input   the definitions of the loop current in \q{jloop} and the shift loop in \q{sloop}, 
\e{
\sum_{\text{loop}}\jloop\eq  -\f{|e|}{\calv}\sum_{\text{loop}}\bS_{\text{loop}}|\delta R_{\text{loop}}|=  -\f{|e|}{\calv}\sum_{\text{loop}}\sum_{\link \in loop} \sgn[\delta R^{\text{loop}}_{B'\lea B}]|\delta R_{\text{loop}}|\;\bS_{\link}.
} 
Utilizing the definition of the oriented loop rate in \q{looporientedrate},
 \e{
\sum_{\text{loop}}\jloop\eq   -\f{|e|}{\calv}\sum_{\text{loop}}\;\sum_{\link \in loop} \delta R^{\text{loop}}_{B'\lea B}\;\bS_{\link}= -\f{|e|}{\calv}\sum_{\link }\;\sum_{loop\ni \link}\delta R^{\text{loop}}_{B'\lea B}\;\bS_{\link}.\la{proof23}
} 
In the last step, we have applied that summing over all ordered links in a given loop and subsequently summing over all loops is equivalent to summing over all loops which contain a given ordered link and subsequently summing over all ordered links. Carrying out the restricted summation over loops on the right-hand side of \q{proof23} and utilizing the rate decomposition condition in \q{decomposeorderedrate}, we obtain the BIS formula [\q{bismine2}], which completes the proof.

\subsection{Gauge invariance of the reduced shift loop}\la{app:loopimp}

The theorem allows to simply derive general properties of the steady shift current. We focus on one such property, namely that the $(\bA_{b'b'\bkp}-\bA_{bb\bk})$ terms in both phononic and photonic shift vectors [\qq{phononicshift2}{photonicshift2}] cancel out, when all transitions are accounted for.  This cancellation was pointed out by BIS without an explicit demonstration,\cite{belinicher_kinetictheory} but is a simple consequence of the loop current theorem.\\ 

Recall that the shift vector (in either the photonic or phononic case)  may be decomposed into a term that depends on the bosonic mode and terms that do not:
\e{\bS^m_{\tran}=\delta S^m_{\tran} + \bA_{B'}-\bA_{B}; \as \bA_{B'}=\bA_{b'b'\bkp}; \as \bA_B= \bA_{bb\bk}. \la{reducedshift}}
The mode-dependent term is  the negative gradient of an argument of a certain transition matrix element [\qq{phononicshift2}{photonicshift2}]; we will refer to $\delta S^m_{\tran}$ as the \textit{reduced shift vector}. It follows that the ordered shift vector [\q{slink}] decomposes similarly as
\e{\bS_{\link} \eq \bA_{B'}-\bA_{B} + \sum_m \delta \bS^m_{\bgt \lea \blt}\f{\cala^{m}_{\bgt \lea \blt}-\cale^{m}_{\blt \lea \bgt}}{R_{\link}}.\la{slink2}}
One may verify that the intraband connection terms cancel out in the shift loop $\sloop$, for any loop. Indeed, in the case of the first orientation in \q{looporientation1}, the shift loop decomposes as
\e{\sloop =\delta \sloop + (\bA_{B_2}-\bA_{B_1}) + (\bA_{B_3}-\bA_{B_2})+\ldots +(\bA_{B_N}-\bA_{B_{N-1}}) + (\bA_{B_1}-\bA_{B_N})=\delta \sloop.\la{reducedsloop}}
$\delta \sloop$, the \textit{reduced shift loop}, is defined by replacing all shift vectors by reduced shift vectors [cf.\ \q{reducedshift}] in \q{sloop}. Thus it follows that each loop current, being proportional to $\sloop$, is invariant if the intraband connection terms are dropped. Finally, the steady shift current, being a sum of loop currents, also satisfies the same invariance property.  \\

\q{reducedsloop} implies that the reduced shift loop is a well-defined, gauge-invariant quantity, despite the fact that the reduced shift vector [\q{reducedshift}] of a single transition is not gauge-invariant. By `gauge-invariance', we mean being invariant under redefining one-electron Bloch wave functions by a Bloch-label-dependent phase $\phi_B$ that is differentiable with respect to $\bk$: $\ket{u_B}_{\sma{\text{cell}}}\ri e^{i\phi_B}\ket{u_B}_{\sma{\text{cell}}}$. \\

The cancellation in \q{reducedsloop} calls into question a  claim made by Fregoso-Morimoto-Moore,\cite{fregoso_opticalzero} namely that large polarization differences $[-|e|\int_{BZ}(\bA_{cc\bk}-\bA_{vv\bk})d^3k/(2\pi)^3]$ between the conduction and valence bands imply a large shift current, in the absence of optical vortices. At best, the Fregoso-Morimoto-Moore claim holds for the frequency-integrated transient shift conductivity [\app{app:jtranjexc}], but not the steady shift conductivity.

\subsection{Geodesic approximation of loop currents for 3D semiconductors}\la{app:geodesic}

The steady shift current is well approximated well by keeping the most relevant loops in \q{loopcurrentthm}. 
This appendix focuses on the reduced family of \textit{geodesic loops}, which predominate the shift current in an intrinsic, direct-gap semiconductor with a single minimum for  $E_{c\bk}-E_{v\bk}$, namely $E_{c\kext}-E_{v\kext}=E_g$, and conditioned on (i) carrier-optical-phonon scattering being the dominant mechanism for energy relaxation in the active region, (ii) small optical phonon energies (relative to $E_g$ and the largest energy of a photoexcited carrier), (iii) low temperature $k_BT_l\ll E_g,\hbar\Omega_o$ ($\Omega_{o}$ the optical phonon threshold frequency). The goal of this section is to derive the geodesic approximation to the shift conductivity [\q{shiftconduct}] from the loop-current formula [\q{loopcurrentthm}].\\

\begin{figure}[H]
\centering
\includegraphics[width=12 cm]{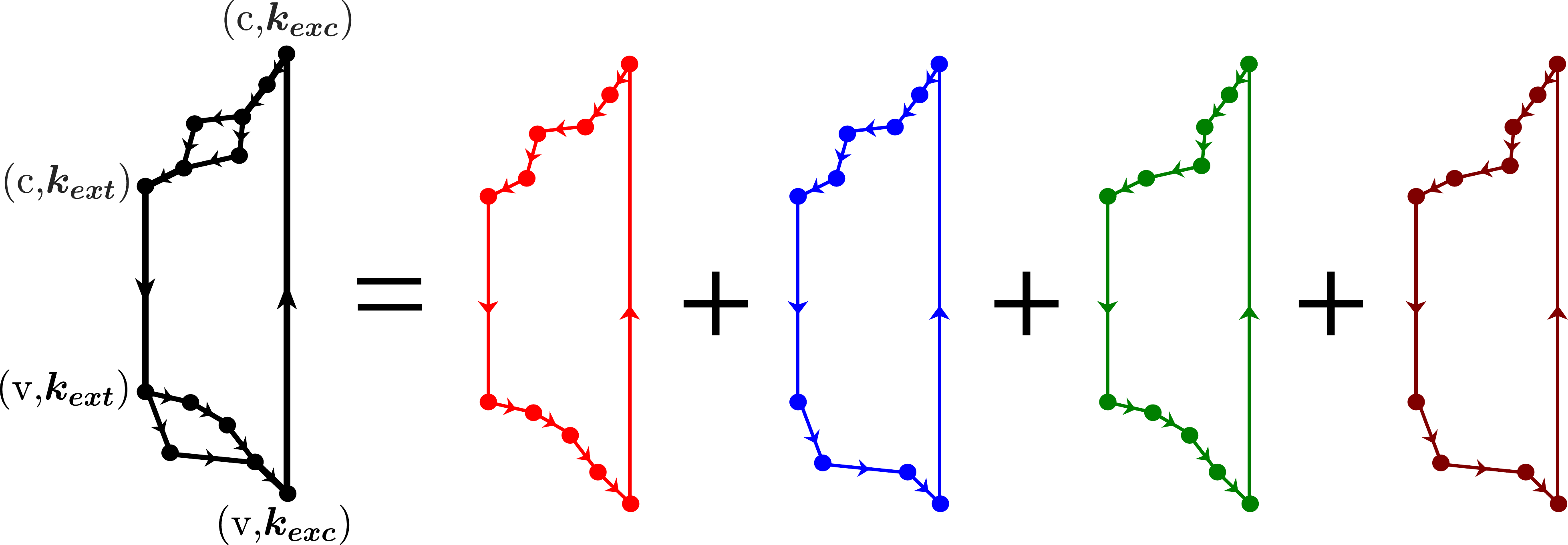}
\caption{Loop decomposition of a probability flow subgraph. A similar subgraph was considered in \fig{fig:loops}(c).  }
\label{fig:subgraph}
\end{figure}

To motivate the geodesic loop, let us first consider  a pair of Bloch states with Bloch labels $V=(v,\kexc)$ and $C=(c,\kexc)$; $\kexc$ lies on the optical surface, $v$ denotes the valence band, and $c$ the conduction band. The  oriented transition rate $R_{C\lea V}$ [cf.\ \q{orientedtranrate}] is assumed to be dominated by the absorption of  non-thermal, source-created photons. The probability-flow subgraph that includes the link $(C,V)$ is caricatured in \fig{fig:subgraph}. We will not repeat the arguments [detailed in \s{sec:kinetictheory} and \s{sec:anomalous}] that explain why such a subgraph is predominant; our goal here is to explain how such a subgraph can be approximated by a geodesic loop.\\ 

Granted some poetic license, one may view the subgraph as a cyclic probability river that rises in elevation, then splits into tributaries which eventually merge into a waterfall. The splitting reflects the multiple possible intraband relaxation pathways in the conduction band; the merging reflects the existence of a band-energy extremum that causes relaxation pathways to converge toward the extremal wavevector $\kext$. The cyclic river may be decomposed into $N$ cyclic streams, such that for each stream, the flow rate  is constant along the stream. ($N=4$ for our caricature in \fig{fig:subgraph}.) This constant flow rate is identified with $|\delta R_{loop_n}|$, with $\{loop_n\}_{n=1}^N$ being labels for the $N$ streams.   All streams merge at $(C,V)$, such that  the  sum of the stream flow rates [\q{looporientedrate}]  equals the river flow rate: 
\e{\sum_{n=1}^{N} |\delta R_{loop_n}|=R_{C\lea V}; \as C=(c\kexc); \as V=(v\kexc). \la{streamsadd}}
We have chosen a stream decomposition such that all the streams flow with the same orientation as the river, and this is always possible to choose. In principle, one may choose a stream decomposition in which some of the streams flow against the river along $(C,V)$; then for those counter-flowing streams, one would replace $|\delta R_{loop_n}|\ri -|\delta R_{loop_n}|$ in \q{streamsadd}.\\

We proceed without further use of metaphors. The contribution of the above subgraph to the shift current\footnote{Bear in mind that `current' has nothing to do with the metaphorical river current.} is a sum of $N$ loop currents: 
\e{\bj_{subgraph[\kexc]} = -\f{|e|}{\calv}\sum_{n=1}^N \bS_{loop_n} |\delta R_{loop_n}|.\la{subgraphcurrent}} \\

As argued in \s{sec:anomalous}, the predominant intraband-relaxation pathways do not deviate far from geodesic paths connecting $\kexc$ to $\kext$; we remind the reader that the geodesic path is orthogonal to all iso-energy contours.  Let us define the \textit{geodesic loop} as  combining  an  excitation transition at $\kexc$, geodesic-path relaxation to $\kext$ through the conduction band, recombination at $\kext$, and geodesic-path relaxation back to $\kexc$ through the valence band, as caricatured in \fig{fig:flownetwork}(d). We denote the geodesic, oriented $\bk$-paths by $p_c$ and $p_v$ respectively, and  the geodesic loop by $loop[\kexc]$; the associated shift loop is denoted $\bS_{loop[\kexc]}$, with $\sloop$ generally defined in \q{sloop}. \\

It is of interest to show how $\bS_{loop[\kexc]}$ simplifies to an expression for the shift loop [\qq{decomposeshiftloop}{srec}] that we have used in the main text:\\

\noi{i} For the recombination transition associated to $\link=(c\kext > v\kext)$ [left-most link in \fig{fig:subgraph}], one  applies \q{orderedshiftphoton} to show that  $\sgn[\delta R^{\text{loop}}_{B'\lea B}]\;\bS_{\link}=\bS_{\text{rec}}$, as defined in \q{srec}. $\sgn[\delta R^{\text{loop}}_{B'\lea B}]$ being $-1$ accounts for the reversed orientation in a recombination transition, but this minus sign can be absorbed by $-\bS^m_{c\bk\lea v\bk}= \bS^m_{v\bk \lea c\bk}$. \\

\noi{ii} For the excitation transition associated to $\link=(c\kexc > v\kexc)$, $\sgn[\delta R^{\text{loop}}_{B'\lea B}]\;\bS_{\link}=\bS^{m_s}_{c\kexc\lea v\kexc}$ assuming that the source is mono-modal and bright [cf.\ argument in \app{app:threefold}].\\

\noi{iii} For an  intraband transition associated to $\link=(c\bkp > c\bk)$, it is assumed small-angle scattering predominates  ($||\bkp-\bk||\ll$ Brillouin-zone dimension), such that  $\sgn[\delta R^{\text{loop}}_{B'\lea B}]\;\bS_{\link}$ reduces to the asymptotic expression  $-\bOmega_{c\kave}\times \delta \bk$ [cf.\ case (a) in \textit{Ex-1} of \app{app:flownetwork}]. By approximating a discrete sum over intraband links as a line integral, one obtains the first line integral in \q{decomposeshiftloop}. The second line integral is obtained in an analogous manner.\\

Because $loop_n$ does not deviate far from $loop[\kexc]$, we approximate $\bS_{loop_n}\approx  \bS_{loop[\kexc]}$ for all loops that make up the subgraph; this is the geodesic approximation. The approximation is justified to the extent that small-angle scattering predominates over large-angle scattering, as elaborated in \app{app:smallanglelimit}.  Applying the geodesic approximation to \qq{streamsadd}{subgraphcurrent},
\e{\bj_{subgraph[\kexc]} \approx  -\f{|e|}{\calv}\bS_{loop[\kexc]}R_{C\lea V}. \la{approxjsubgraph}}\\

The shift current is a sum of loop currents over loops that constitute the full probability-flow network, and not just the subgraph containing $(C,V)$. [For the one-dimensional caricature in \fig{fig:flownetwork}(d), the full network is composed of two subgraphs.] In other words, one should sum $\bj_{subgraph[\kexc]}$ over all $\kexc$ on the excitation surface:
\e{\bj\approx -\twoupdo\f{|e|}{\calv}\sum_{\bk}\bS_{loop[\bk]}R_{(c\bk)\lea (v\bk)},\la{sumsubgraph}}
with the understanding that $R_{(c\bk)\lea (v\bk)}\propto\delta(E_{cv\bk}-\hbar\omega)$ [\q{photonrate}] constrains $\sum_{\bk}$ to the excitation surface; we have also included a factor of $2$ to account for spin.   Assuming a bright, mono-modal light source, we may follow the argument in \app{app:threefold} to derive that  $R_{(c\bk)\lea (v\bk)}\approx I_{exc\bk}^{\beps\omega}$ with $I_{\text{exc}}$ defined in 
 \q{exciterate}. Converting the source mode occupancy $\Delta N_s$ to an electric-field amplitude $\cale_{\omega}$ through \q{elecfield}, one obtains:
\e{\bj \approx -2\pi \f{|e|^3}{\hbar} \langle f_{vc\bk}|\beps\cdot \bA_{cv\bk}|^2\bS_{loop[\bk]} \rangle_{\omega} (2\jdosup ) |\cale_{\omega}|^2, \la{geodesicj} }
with $f_{vc}=f_{v\bk}-f_{c\bk}$ a difference in the steady-state quasiparticle distribution functions, and $\langle\ldots \rangle_{\omega}$ and $\jdosup$ defined in \q{averageES}. When expressed in terms of a nonlinear conductivity:  $\bj=\bsigma_{\beps,\omega} |\cale_{\omega}|^2$, \q{geodesicj} is equivalent to  \q{shiftconduct}.

\subsection{Geodesic approximation as a small-angle-scattering limit of BIS formula }\la{app:smallanglelimit}

Because the geodesic approximation of the shift conductivity [\q{shiftconduct}] has been used in all model calculations, it is of interest to clarify the regime of validity of the approximation. Here, we will demonstrate that \q{shiftconduct} derives as a \textit{small-angle-scattering limit} of the BIS formula \q{bismine2}:
$\limit{s\ri \infty} \bsigma_{BIS}=\bsigma_{geo};$ with  $s$
a parameter  that controls the angle of scattering. The choice of $s$ is not unique. One possible choice is to increase the power  in the square of the electron-phonon matrix element: $|V^m_{b\bkp,b\bk}|^2 \propto 1/||\bkp-\bk||^s$  [\q{defineUem}], bearing in mind that this is a theoretical exercise to elucidate the essence of the geodesic approximation; the physical value of   $s$ is two, for polarization scattering with optical phonons; cf.\ \s{sec:anomalous}.\\

Implementing the threefold decomposition of both $\bsigma_{geo}$ [\q{decomposesigma}] and $\bsigma_{BIS}$ [\app{app:threefold}], one can straightforwardly verify that the excitation components match exactly, while the recombination components match to a  good approximation, bearing in mind that recombination transitions predominantly occur at $\bk$ near the extremal wavevector.\footnote{As described in  \s{sec:kinetictheory} and elaborated here, $\sigma_{BIS}^{\text{rec}}$  reduces to $\sigma_{geo}^{\text{rec}}$  if the photonic shift vector  $\bS^{m}_{c\bk \lea v\bk}$  [in \q{approxjrec}, with $\bk$ in the passive $\bk$-volume] is   approximated to be $\bS^{m}_{c\kext \lea v\kext}$. This approximation leads to a relative error of order $\hbar \Omega_o/E_g$,   assuming that the band gap $E_g$ is the energy scale for significant variation of the shift vector. If the photo-excited carriers within the passive region follow a Maxwellian distribution, with electron temperature $k_BT_e<\hbar\Omega_o$ and hole temperature $k_BT_h<\hbar\Omega_o$ [cf.\ \app{app:relaxationmechanisms}],  then  the relative error is reducible to $\sigma_{BIS}^{\text{rec}}=\sigma_{geo}^{\text{rec}} +O_r(k_B T/E_g)$, with $T$ being the smaller of $\{T_e,T_h\}$.} This appendix will demonstrate  for the intraband components that $\limit{s\ri \infty} \bsigma^{\text{intra}}_{BIS}=\bsigma^{\text{intra}}_{geo}$.\\

Assuming only two bands are optically excited, the intraband conductivity decomposes into contributions by individual bands: $\bsigma^{\text{intra}}=\bsigma^{\text{intra}}_c+\bsigma^{\text{intra}}_v$, and we will prove for the  conduction band that
\e{\text{BIS-geodesic reduction:} \as \limit{s\ri \infty} \bsigma^{\text{intra}}_{c,BIS}=\bsigma^{\text{intra}}_{c,geo} +O_r\big(\surd{{\tfrac{\cramped{\hbar\Omega_o}}{\cramped{E_{\text{exc}}}}}},\surd \tfrac{\hbar\Omega_o}{E_g}\big),\la{limit}}
with $\Omega_o$ the optical phonon threshold [\q{opticalphononocc}] and $E_{\text{exc}}$ the excitation energy measured from the conduction-band minimum [\q{excenergy}]; the meaning of $O_r$ is relative error, i.e., $a+O_r(b,c)$ means $O_r(b,c)$ has a magnitude less than or comparable to  $maximum\{|ba|,|ca|\}$, assuming $b$ and $c$ to be dimensionless.
The BIS-geodesic reduction  for the valence band [\q{limit} with $c\ri v$] also holds true, but is a straightforward extension requiring no further substantiation. \\

To clarify, $\bsigma^{\text{intra}}_{c,geo}$ is given by \qq{decomposeshiftloop}{shiftconduct} with  the shift loop reduced to the line integral over the geodesic path $p_{c\bk}$ connecting $\bk$ to $\kext$:   
\e{\bsigma^{\text{intra}}_{c,geo} \eq - 2\pi \f{|e|^3}{\hbar} \left\langle f_{vc\bk}|\beps\cdot A_{cv\bk}|^2\int_{p_{c\bk}}\bOmega_c \times d\bk \right\rangle_{\omega} \twoupdo \jdosup, \la{sigmageo} }
while $\bsigma^{\text{intra}}_{c,BIS}$ is taken from \qq{jintraoptical}{sigmathreefold}:  
\e{
\bsigma^{\text{intra}}_{c,BIS}=-\twoupdo\f{|e|}{\calv|\cale_{\omega}|^2}\sum_{\bk,\bkp } \bS_{c\bk \lea c\bkp} \cale^{sp}_{c\bk \lea c\bkp}.   \la{sigmaintra2}
}
We have omitted the phonon mode $m=(\bq p)$ superscript on the phononic shift $\bS^m_{c\bkp\lea c\bk}=-\bS^{-m}_{c\bk\lea c\bkp}$ and spontaneous emission rate $\cale^{sp,m}$, with the understanding that $p$ is fixed to a single branch of optical phonons and $\bq =\bk-\bkp$ is fully determined by momentum conservation; cf.\ \q{phononrate}. Henceforth, we will  simplify notation by  omitting the  $c$ subscript  on all quantities, except in instances  where such omission may lead to confusion. \\

In addition to certain assumptions that justify the predominance of geodesic loops [summarized in the beginning of \app{app:geodesic}], we will make additional model assumptions which simplifies the demonstration of the BIS-geodesic reduction [\q{limit}], though we do not believe
these additional assumptions are ultimately necessary for the reduction: \\

\noi{i} The optical phonon frequency is roughly a constant equal to $\hbar\Omega_o$ for the small phonon wavevectors we consider.  \\

\noi{ii} Both conduction and valence bands have isotropic dispersions, i.e., $E_{c\bk}$ and $E_{v\bk}$ depend on $\bk$ through $||\bk||$, as may be expected near band extrema with cubic symmetry.\\

\noi{iii} In the active region,  electron-optical-phonon scattering overwhelmingly dominates over electron-acoustic-phonon scattering as the primary mechanism for energy relaxation. One way to formalize this is to take $\eta_E$ defined in \q{defineeta} to zero. \\

\noindent Some implications of (i-iii) will hereby be elucidated, in preparation to prove the BIS-geodesic reduction [\q{limit}]. \\

\noindent\underline{Excitation rate}\\

(i-ii) imply that the excitation energy $E_{\text{exc}}$ [cf. \q{excenergy}] of conduction-band states is degenerate, i.e., the excitation rate [\q{exciterate}]  is nonzero only if $E_{\bk}=E_{\text{exc}}$:
\e{ I_{exc\bk}=\tilde{I}_{exc\bk}\delta_{\bk,E_{\text{exc}}}; \as \tilde{I}_{exc\bk}= \frac{2\pi|e|^2}{\hbar}f_{vc\bk}|\epsilon \cdot \mathbf{A}_{cv\bk}|^2\f{|\nabk E_c|}{|\nabk E_{cv}|}\calv g_{E_{\text{exc}}}|\mathcal{E}_{\omega}|^2.\la{itildeexck}}
We collect here a few useful properties of \textit{surface projectors}:
\e{\delta_{\bk,E} =\f{\delta(E_{\bk}-E)}{\calv g_E};\as \sum_{\mathbf{k}}\delta_{\mathbf{k},E}=\sum_{E}\delta_{\mathbf{k},E}=1; \as \delta_{\bk,E}\delta_{\bk,E}=\delta_{\bk,E};\as \delta_{\bk,E_{j}} \delta_{\bk,E_{j^{\prime}}}=\delta_{\bk,E_{j}}\delta_{jj^{\prime}},\la{surfaceproj}
}
which encode their completeness (with $\sum_E$ meaning $\int \calv g_E dE$), idempotence and orthogonality.\footnote{If the reader is bothered by $(\delta_{\bk,E})^2$  being a product of two Dirac delta functions, one may regularize the surface projector as:
\begin{align}
    \delta^{\calv}_{\bk,E}\eq \begin{cases}
        1, & |E_{\bk}-E|< 1/2\calv g_E\\
        0, & \text{otherwise},
    \end{cases} \la{kronv}
\end{align}
multiply two regularized projectors and then subsequently take $\calv\ri \infty$.}
Integrating a surface-projected test function is equivalent to averaging the test function over a two-sphere parametrized by the solid angle $\lambda$:
\begin{equation}
\label{intsa}
\aven{\Xi(\bk)}{\bk E}=\sum_{\bk}\delta_{\bk,E}\;\Xi(\bk)=\int\frac{d\lambda}{4\pi} \;\Xi(\bk)\big |_{\bk=(k_{E},\lambda)}; \as \sum_{\bk}=\int_0^{cutoff} \f{\calv k^2 dk}{(2\pi)^3} \int d\lambda,
\end{equation}
with $k_{E}$ being the inverse of the isotropic band dispersion $E_k$. $\aven{\Xi(\bk)}{\bk E}$ is referred to as the \textit{iso-energy average} of $\Xi(\bk)$. \\

\noindent\underline{Quasiparticle distribution}\\

(i-iii) imply that the non-equilibrium quasiparticle distribution within the active region is singularly peaked at periodic intervals:\cite{esipov_oscillatoryeffects,esipov_temperatureenergy}
\e{ f_{\bk}  =\sum_{j=0}^{j_{max}} \tilde{f}_{\bk,j} \delta_{\bk,E_j}; \as E_j=E_{\text{exc}}-j\hbar\Omega_o. \la{fsingular}}
$f_{\bk}$ thus has a ladder-like structure,
with the top rung of the ladder corresponding to the excitation energy ($E_0=E_{\text{exc}}$), and the lowest rung $E_{j_{max}}$ lying just above the passive region. The singular nature of $f_{\bk}$ originates from the source being monochromatic and the predominant phonons being dispersionless. One may verify that the regular function $f_E$ in \q{highestpeak} becomes proportional to a Dirac delta function as $\eta_E \ri 0$.\\

\noindent\underline{Spontaneous emission rate}\\

A related implication of (i-iii) is that the spontaneous emission rate is a sum of terms that connect adjacent rungs of the ladder:
\begin{equation}
\label{semission}
\mathcal{E}^{sp}_{\bk \leftarrow \bkp}= \sum_{j=0}^{j_{max}-1}\frac{\tilde{\tilde{\mathcal{E}}}^{sp}_{\bk \leftarrow \bkp}}{||\bk-\bkp||^s}\delta_{\bk,E_{j+1}}\delta_{\bkp,E_{j}}.
\end{equation}
We have extracted $1/||\bk-\bkp||^s$ and the singular delta functions such  that $\tilde{\tilde{\mathcal{E}}}^{sp}_{\bk \leftarrow \bkp}$ is regular as $\bk$ approaches $\bkp$.  To derive the surface projector $\delta_{\bkp,E_j}$ in  \q{semission}, apply that $\mathcal{E}^{sp}_{\bk \leftarrow \bkp}$ [\q{sponemirate1}] is proportional to the singular distribution $f_{\bkp}$ [\q{fsingular}]; the second surface projector $\delta_{\bk,E_{j+1}}$ in \q{semission} originates from energy conservation: $E_{\bkp\bk}=\hbar\Omega_o$ [\q{sponemirate1}]. The ladder structure in \q{semission} implies that the operator $\sum_{j=0}^{j_{max}-1}\delta_{\bkp,E_j}$  acts trivially on the emission rate:
\e{\sum_{j=0}^{j_{max}-1}\delta_{\bkp,E_j}\mathcal{E}^{sp}_{\bk \leftarrow \bkp} = \mathcal{E}^{sp}_{\bk \leftarrow \bkp}, \la{donothing}}
owing to the idempotence of surface projectors; cf.\ \q{surfaceproj}.\\

\noindent\underline{Kinetic equation}\\

The kinetic equation for the steady quasiparticle distribution [\qq{ick}{longi}] simplifies to
\begin{equation}
\label{anisokin}
I_{\text{exc},\bkp}-\sum_{\bk}\mathcal{E}^{sp}_{\bk\leftarrow\bkp}+\sum_{\bk^{\prime\prime}}\mathcal{E}^{sp}_{\bkp\leftarrow\bk^{\prime\prime}}=0,
\end{equation}
for $\bkp$ in the active region [\q{activepassive}]; $I_{\text{exc},\bkp}$ is given in \q{itildeexck} and $\mathcal{E}^{sp}_{\bk\leftarrow\bkp}$ in \q{semission}. We have dropped the recombination component [\q{iphot}] of the kinetic equation because the loss rate due to spontaneous emission of optical phonons greatly outweighs the loss rate due to interband recombination; cf.\ the discussion under \q{esipovmodel}.\\ 

The last preparation for the BIS-geodesic reduction [\q{limit}] will be to relate the excitation  and spontaneous emission rates as:
\begin{equation}
\label{IE2}
\tilde{I}_{\text{exc},(k_{0},\lambda)}=\tilde{\tilde{\mathcal{E}}}^{sp}_{(k_{j+1},\lambda) \leftarrow (k_{j},\lambda)}\lim_{s\rightarrow\infty}\left\langle1/q^{s}\right\rangle^{j+1}_{j}; \as \left\langle 1/q^{s}\right\rangle_{j}^{j+1}\equiv\int\frac{d\lambda'}{4\pi}\frac{1}{||\bk-\bkp||^s} \bigg|_{\mathbf{k}=(k_{j+1},\lambda');\bkp=(k_j,\lambda)}. 
\end{equation}
$k_j$ is short for $k_{E_j}$, meaning it  is the radius of the spherical iso-energy surface with energy $E_{j}$. The term on the right-hand side of $\tilde{I}_{\text{exc}}$  can be interpreted as the rate at which a quasiparticle on the $j$'th iso-energy surface drops to the $(j+1)$'th surface by spontaneously emitting an optical phonon. \\

\q{IE2} is ultimately a consequence of the conservation of probability flow in energy-momentum space. Proving \q{IE2} takes three steps: (A) we first relate the excitation rate to  the rate of phonon-mediated transitions between the $0$'th/excitation surface to the $1$'th iso-energy surface. (B) We then relate the rate of phonon-mediated transitions between the $(j-1)$'th and $j$'th surfaces  to the rate of phonon-mediated transitions between the   $j$'th and $(j+1)$'th surfaces. (C) Combining our relations from (A) and (B) and taking the small-angle-scattering limit  gives us \q{IE2}.\\

\noi{Step (A)} Projecting the kinetic equation  [Eq.~\eqref{anisokin}] onto the excitation surface tells us
\begin{equation}
0=\delta_{\bkp,E_{0}}\left(I_{\text{exc},\bkp}-\sum_{\bk}\mathcal{E}^{sp}_{\bk\leftarrow\bkp}+\sum_{\bk^{\prime\prime}}\mathcal{E}^{sp}_{\bkp\leftarrow\bk^{\prime\prime}}\right)=I_{\text{exc},\bkp}-\delta_{\bkp,E_{0}} \sum_{\bk}\mathcal{E}^{sp}_{\bk\leftarrow\bkp},\la{above}
\end{equation}
with the last term dropping out because there are no quasiparticles with energies exceeding $E_{\text{exc}}$ that can drop to the excitation surface by emitting a phonon; cf.\ \q{semission}. Let us substitute the ladder formula for the emission rate [Eq.~\eqref{semission}] into \q{above} and  apply the orthogonality of surface projectors [\q{surfaceproj}] to reduce $\sum_j$ to the $j=0$ term. We then  convert $\sum_{\bk}\delta_{\bk E_1}$ to a solid-angular integral via Eq.~\eqref{intsa} to obtain
\begin{equation}
\label{conservationES}
\begin{aligned}
\delta_{\bkp,E_{0}}\tilde{I}_{\text{exc},\bkp}
&=\delta_{\bkp,E_{0}}\int\frac{d\lambda}{4\pi}\frac{\tilde{\tilde{\mathcal{E}}}^{sp}_{\bk \leftarrow \bkp}}{||\bk-\bkp||^s} \bigg|_{\mathbf{k}=(k_{1},\lambda)},
\end{aligned}
\end{equation}

\noi{Step (B)} If we project the kinetic equation  [Eq.~\eqref{anisokin}] to the $j$'th iso-energy surface with $j\neq 0$ and $\neq j_{max}$, then it is the excitation term that drops out:
\begin{equation}
\delta_{\bkp,E_{j}} \left(\sum_{\bk^{\prime\prime}}\mathcal{E}^{sp}_{\bkp\leftarrow\bk^{\prime\prime}}-\sum_{\bk}\mathcal{E}^{sp}_{\bk\leftarrow\bkp}\right)=0. \la{likehow}
\end{equation}
Like how we derived the right-hand side of Eq.~\eqref{conservationES}, \q{likehow} can be massaged to the form:
\begin{equation}
\label{conservationact}
\begin{aligned}
\int\frac{d\lambda^{\prime\prime}}{4\pi}\;\frac{\tilde{\tilde{\mathcal{E}}}^{sp}_{\bkp \leftarrow \bk^{\prime\prime}}}{||\bkp-\bk^{\prime\prime}||^s} \bigg|_{\bk^{\prime\prime}=(k_{j-1},\lambda^{\prime\prime})}=\int\frac{d\lambda}{4\pi}\;\frac{\tilde{\tilde{\mathcal{E}}}^{sp}_{\bk \leftarrow \bkp}}{||\bk-\bkp||^s} \bigg|_{\mathbf{k}=(k_{j+1},\lambda)}.
\end{aligned}
\end{equation}
for any $\bkp$ on the  $j$'th iso-energy surface.  With $\bkp$ as a reference point, \q{conservationact} encodes that the incoming probability flow from the $(j-1)$'th surface matches the outgoing probability flow to the $(j+1)$'th surface. \\

\noi{Step (C)} 
Both Eq.~\eqref{conservationES} and Eq.~\eqref{conservationact} involve solid-angular integrals  which simplify in the small-angle-scattering limit:  fixing $\bkp=(k_{j},\lambda^{\prime})$, 
\begin{equation}
\label{smallangle}
\begin{aligned}
\lim_{s\rightarrow\infty}\int\frac{d\lambda}{4\pi}&\;\frac{\tilde{\tilde{\mathcal{E}}}^{sp}_{\bk \leftarrow \bkp}}{||\bk-\bkp||^s}\bigg|_{\mathbf{k}=(k_{j+1},\lambda)}
=\tilde{\tilde{\mathcal{E}}}^{sp}_{(k_{j+1},\lambda^{\prime}) \leftarrow \bkp}\lim_{s\rightarrow\infty}\left\langle 1/q^{s}\right\rangle_{j}^{j+1},
\end{aligned}
\end{equation}
with $\langle 1/q^s\rangle$ defined in \q{IE2}. 
The crucial step taken here is to replace $\tilde{\tilde{\mathcal{E}}}^{sp}_{\bk \leftarrow \bkp}$ in the integral by its value when  $||\bk-\bkp||^{-s}$ is maximized, or equivalently when $||\bk-\bkp||$ is minimized. This replacement is justified asymptotically as $s\ri \infty$, and may be seen as an application of Laplace's method.\cite{olver_asymptotics} To manifest the usual form of the integral seen in Laplace's method, we momentarily adopt spherical-angular coordinates $\lambda=(\cos\theta,\phi)$ such that $\bk=(k\sin\theta \cos \phi,k\sin \theta \sin \phi,k\cos\theta)$ and $\bkp=(0,0,k')$; then for any smooth function $f_{\bk}=f(k,\lambda)$,
\e{ L_s= \int \frac{d\lambda}{4\pi} \;\f{f(k,\lambda)}{||\bk-\bkp||^s} = \int^1_{-1}  e^{s R(x)} \bigg[\int_0^{2\pi}\f{f(k,x,\phi)}{4\pi}d\phi\bigg]dx; \as R(x)=-\half \text{ln}\; [k^2+{k'}^2-2kk'x ], \la{laplace1}
}
with $x=\cos \theta$. $R(x)$ has a unique global maximum at $x=1$, which is an end point of the interval of integration. Applying a standard formula from asymptotic analysis,\cite{olver_asymptotics}
\e{ \limit{s\ri \infty}L_s = \f{f(k,1,\phi)}{2} \f{e^{sR(1)}}{sR'(1)} +O_r(s^{-1}) = f_{0,0,k} \f{1}{2skk'|k-k'|^{s-2}} +O_r(s^{-1}),  \la{laplace2}}
with $R'=dR/dx$. 
In our application, $f_{0,0,k}$ corresponds to $\tilde{\tilde{\mathcal{E}}}^{sp}_{(k_{j+1},\lambda^{\prime}) \leftarrow (k_j,\lambda')}$ in \q{smallangle}.\\

Substituting \q{smallangle} into Eq.~\eqref{conservationact}, we relate the transition rates between two adjacent pairs of iso-energy surfaces as:
\begin{equation}
\label{EE}
\frac{\tilde{\tilde{\mathcal{E}}}^{sp}_{(k_{j},\lambda) \leftarrow (k_{j-1},\lambda)}}{\tilde{\tilde{\mathcal{E}}}^{sp}_{(k_{j+1},\lambda) \leftarrow (k_{j},\lambda)}}=\lim_{s\rightarrow\infty}\frac{\left\langle1/q^{s}\right\rangle^{j+1}_{j}}{\left\langle1/q^{s}\right\rangle^{j-1}_{j}}.
\end{equation}
Combining Eq.~\eqref{conservationES} and Eq.~\eqref{EE}, we relate the excitation rate to the transition rate between a pair of iso-energy surfaces:
\begin{equation}
\tilde{I}_{\text{exc},(k_{0},\lambda)}=\tilde{\tilde{\mathcal{E}}}^{sp}_{(k_{j+1},\lambda) \leftarrow (k_{j},\lambda)}\lim_{s\rightarrow\infty}\left\langle1/q^{s}\right\rangle_{0}^{1}\prod_{i=1}^{j}\frac{\left\langle1/q^{s}\right\rangle^{i+1}_{i}}{\left\langle1/q^{s}\right\rangle^{i-1}_{i}}.\la{because}
\end{equation}
The solid-angular integral in  \q{IE2} is evaluated to be
\begin{equation}
\label{symmint}
\left\langle 1/q^{s}\right\rangle_{j}^{j+1}=\f1{2(s-2)k_jk_{j+1}}\bigg[ \f1{|k_j-k_{j+1}|^{s-2}}- \f1{(k_j+k_{j+1})^{s-2}} \bigg],
\end{equation}
which manifests that $\left\langle 1/q^{s}\right\rangle_{j}^{j+1}$ is symmetric under interchanging $j$ and $j+1$, hence
\q{because} simplifies to \q{IE2}, as desired.\\

\noindent\underline{Proof of BIS-geodesic reduction [\q{limit}]}\\

Let us begin the proof by demonstrating that the ratio between Eq.~\eqref{sigmageo} and Eq.~\eqref{sigmaintra2}
reduces to 
\begin{equation}
\label{geoBISratio1}
\begin{aligned}
\frac{\bsigma_{geo}^{\text{intra}}}{\bsigma_{BIS}^{\text{intra}}}&=\frac{
\big\langle \big\langle \tilde{I}_{\text{exc},\mathbf{k}} \int_{p_{\bk }}\mathbf{\Omega}\times d\mathbf{k}\big\rangle \big\rangle_{\bk E_{0}}}{\sum_{j=0}^{j_{max}-1}\big\langle \big\langle\sum_{\mathbf{k}}\bS_{\mathbf{k} \leftarrow \mathbf{k}^{\prime}} \mathcal{E}^{sp}_{\mathbf{k} \leftarrow \mathbf{k}^{\prime}}\big\rangle\big\rangle_{\bkp E_{j}}},
\end{aligned}
\end{equation}
Beginning with the geodesic expression in Eq.~\eqref{sigmageo}, we insert the integral expression of $\langle \ldots \rangle_{\omega}$ from  \q{averageES} and decompose the excitation rate  ${I}_{exc\bk}$ according to \q{itildeexck}:
\e{- \f{\calv|\cale_{\omega}|^2}{\twoupdo|e|}\bsigma_{geo}^{\text{intra}} \eq \sum_{\bk}\delta_{\bk,E_{\text{exc}}}\tilde{I}_{exc\bk}\int_{p_{\bk}}\bOmega \times d\bk.\la{lala}  }
The right-hand side of the above equation is simply the numerator of \q{geoBISratio1},  per our definition of iso-energy averaging in \q{intsa}.
Working now on the BIS formula [Eq.~\eqref{sigmaintra2}], we insert the trivial  operator $\sum_{j=0}^{j_{max}-1}\delta_{\bkp,E_j}$ [\q{donothing}] and apply again the definition of iso-averaging in \q{intsa}:
\begin{equation}
\label{equivalencenum}
\begin{aligned}
- \f{\calv|\cale_{\omega}|^2}{\twoupdo|e|}\bsigma^{\text{intra}}_{BIS}
&=\sum_{j=0}^{j_{max}-1}\big\langle \big\langle\sum_{\mathbf{k}}\bS_{\mathbf{k} \leftarrow \mathbf{k}^{\prime}} \mathcal{E}^{sp}_{\mathbf{k} \leftarrow \mathbf{k}^{\prime}}\big\rangle\big\rangle_{\bkp E_{j}}.
\end{aligned}
\end{equation}
Taking the ratio of \q{lala} and \q{equivalencenum} gives \q{geoBISratio1}, as desired.\\

Focusing on a summand of fixed $j$ and taking the small-angle-scattering limit,
\begin{equation}
\label{geoBISratio3}
\begin{aligned}
S\cale_j=\lim_{s\rightarrow\infty}\big\langle \big\langle\sum_{\mathbf{k}}\bS_{\mathbf{k} \leftarrow \mathbf{k}^{\prime}} \mathcal{E}^{sp}_{\mathbf{k} \leftarrow \mathbf{k}^{\prime}}\big\rangle\big\rangle_{\bkp E_{j}}&=\int \f{d\lambda}{4\pi}\;\sum_{\bk}\bS_{\mathbf{k} \leftarrow (k_j,\lambda)} \lim_{s\rightarrow \infty}\mathcal{E}^{sp}_{\mathbf{k} \leftarrow (k_j,\lambda)}.
\end{aligned}
\end{equation}
We then apply Laplace's method [\qq{laplace1}{laplace2}] and replace $\bS_{\mathbf{k} \leftarrow \mathbf{k}^{\prime}} \tilde{\tilde{\mathcal{E}}}^{sp}_{\mathbf{k} \leftarrow \mathbf{k}^{\prime}}$ by its value when $||\bk-\bkp||$ is minimized:
\e{
S\cale_j
&=\int \frac{d\lambda}{4\pi}\;\bS_{(k_{j+1},\lambda) \leftarrow (k_{j},\lambda) } \;\tilde{\tilde{\mathcal{E}}}^{sp}_{(k_{j+1},\lambda) \leftarrow (k_{j},\lambda)}\lim_{s\rightarrow\infty}\left\langle 1/q^{s}\right\rangle_{j}^{j+1}
=\int \frac{d\lambda}{4\pi}\; \mathbf{\Omega}_{\bk_{ave}^j }\times\delta\bk^j\;\tilde{I}_{\text{exc},(k_{0},\lambda)},
}
with $\bk_{ave}^j=(\bk^j+\bkp^j)/2$, $\delta\bk^j=\bkp^j-\bk^j$, $\bkp^j=(k_{j+1},\lambda)$ and $\bk^j=(k_j,\lambda)$.
In the last step, we substituted the spontaneous emission rate with the excitation rate in accordance with \q{IE2}, and replaced the phonon-mediated shift vector with its asymptotic small-angle limit [\q{anomalousshift}]. The sum of $S\cale_j$ over $j$ may be regarded as a  Riemann sum which approximates a line integral over the geodesic path:
\e{
\sum_j\mathbf{\Omega}_{\bk_{ave}^j }\times\delta\bk^j = \int_{p_{\bk}}\bOmega \times d\bk + O_r\big(\surd{\tfrac{\hbar\Omega_o}{E_{\text{exc}}}},\surd \tfrac{\hbar\Omega_o}{E_g}\big).\la{riemann}
}
Indeed, it may be seen that the discrete transitions between iso-energy surfaces:
\e{ (k_{j_{max}},\lambda)\lea (k_{j_{max}-1},\lambda)\lea \ldots \lea (k_{1},\lambda) \lea (k_0,\lambda)=\bk }
concatenate into a straight path $p'_{\bk}$ of fixed solid-angular orientation; the geodesic path $p_{\bk}$ similarly connects  $\kext \lea \bk$ in a straight path. One caveat is that  $(k_{j_{max}},\lambda)$ is not $\kext$ and lies just outside the passive $\bk$-volume [\q{activepassive}], thus $|p'_{\bk}|$ is shorter than $|p_{\bk}|$ by about $k_{\hbar\Omega_o}$. In the parabolic-band approximation, $k_{\hbar\Omega_o}/k_{E_{\text{exc}}} =(\hbar \Omega_0/E_{\text{exc}})^{1/2}$, which is the reason for the relative error in  \q{riemann}. This estimate presumes the band gap $E_g$ is comparable to $E_{\text{exc}}$. For semiconductors with anomalously small band gaps, the Berry curvature may be concentrated in an energy interval comparable to $E_g$, hence the relative error is modified to
$(\hbar \Omega_0/E_g)^{1/2}$.
Altogether,
\e{
\lim_{s\rightarrow\infty}\sum_{j=0}^{j_{max}-1}\big\langle \big\langle\sum_{\mathbf{k}}\bS_{\mathbf{k} \leftarrow \mathbf{k}^{\prime}} \mathcal{E}^{sp}_{\mathbf{k} \leftarrow \mathbf{k}^{\prime}}\big\rangle\big\rangle_{\bkp E_{j}}= \big\langle \big\langle \tilde{I}_{\text{exc},\mathbf{k}} \int_{p_{\bk }}\mathbf{\Omega}\times d\mathbf{k}\big\rangle \big\rangle_{\bk E_{0}} +O_r\big(\surd{{\tfrac{\cramped{\hbar\Omega_o}}{\cramped{E_{\text{exc}}}}}},\surd \tfrac{\hbar\Omega_o}{E_g}\big),
}
which combines with \q{geoBISratio1} to give the BIS-geodesic reduction [\q{limit}].\\

For finite $s=2$ which is appropriate to polarization scattering, one should  expect the ratio  $\bsigma^{\text{intra}}_{geo}/\bsigma^{\text{intra}}_{BIS}$ to deviate from unity. In practice,   we find this deviation to be small:  for the model calculation in \s{sec:anomalous},  the ratio turns out to be   $1.08$  for a source photon energy of $\hbar\Omega=0.8 E_{0}$;  cf.\ Fig.~\ref{fig:Q05_maintext}(e).

\subsection{Geodesic approximation for quasi-2D semiconductors}\la{app:geodesicquasi2D}

Having formulated the geodesic approximation for 3D direct-gap semiconductors, we would like to extend the notion to quasi-2D direct-gap semiconductors, as exemplified by the model Hamiltonian in \q{modelham}.\\

\begin{figure}[H]
\centering
\includegraphics[width=8 cm]{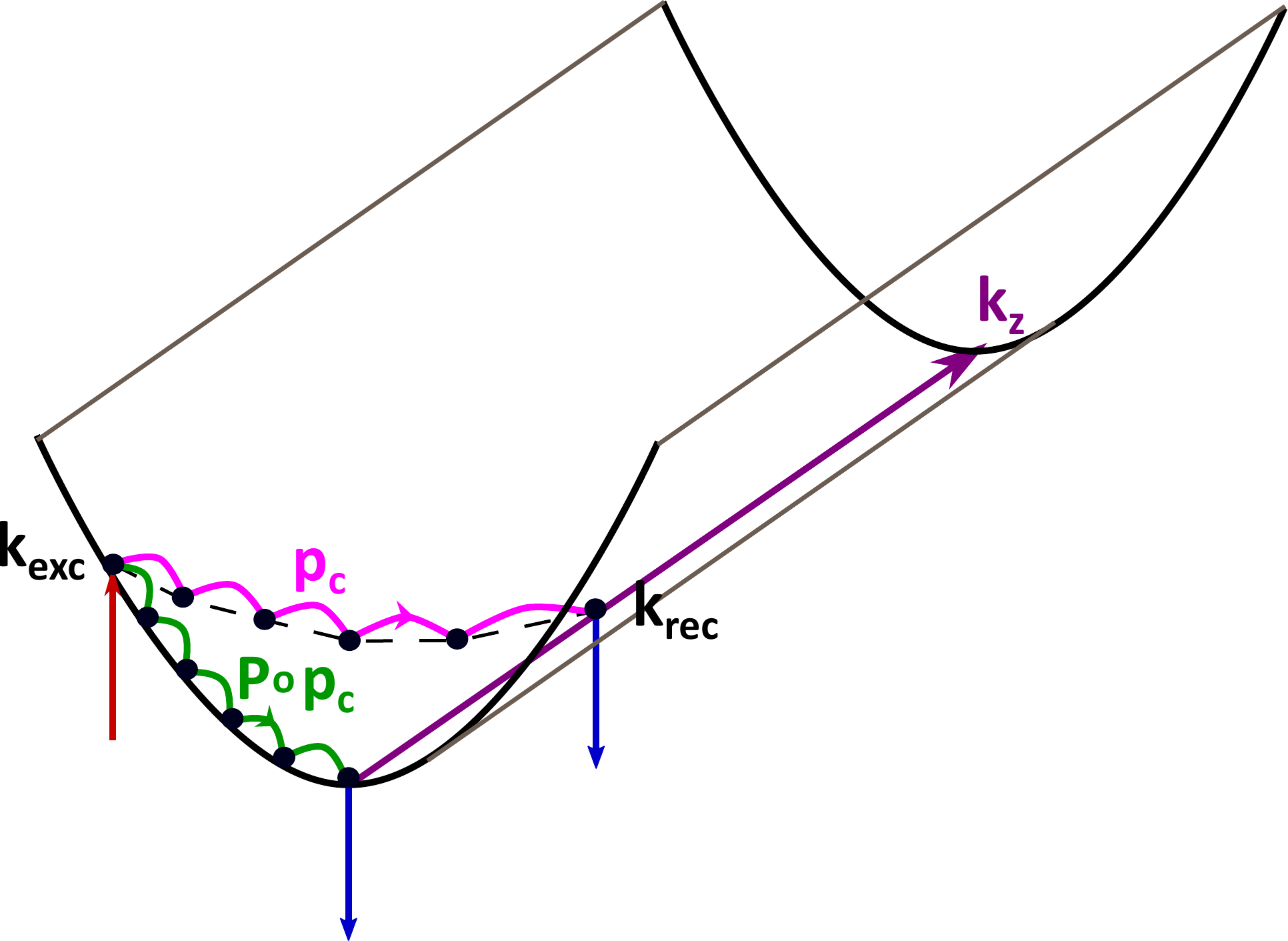}
\caption{Quasi-two-dimensional conduction band plotted over $(E,k_x,k_z)$, with $E$ parametrizing an implicit vertical axis.  }
\label{fig:quasi2D}
\end{figure}

By quasi-2D, we mean that the electronic band energies $E_{b\bk}$ and cell-periodic wave functions $\ket{u_{b\bk}}_{\sma{\text{cell}}}$ are approximately independent of one wavevector coordinate, say, $k_z$. The former condition implies that the band gap is minimized not at a single $\bk$-point but along a $\bk$-line. In our model [\q{modelham}], this $\bk$-line is parametrized by $\bk=(0,0,k_z)$, as illustrated by the purple line in \fig{fig:quasi2D}. 
The latter condition on the wave function implies that the intraband Berry curvature vector is collinear with the z unit directional vector: $\bOmega_{b\bk}=\Omega^z_{b\bk}\vec{z}$, and that the shift current vanishes in the z direction. Indeed, a nonzero z-component of the photonic/phononic shift vector requires that $\ket{u_{b\bk}}_{\sma{\text{cell}}}$ nontrivially depend on $k_z$, as deducible from \qq{phononicshift2}{photonicshift2} and \q{defineUem}.\\

Let us then consider the shift current $\bj_{\per}$ orthogonal to $\vec{z}$. We would like to demonstrate that $\bj_{\per}$ is well approximated by \q{geodesicj}, with $loop[\bk]$ reinterpreted as a \textit{planar geodesic loop}: a geodesic loop confined to the $k_x-k_y$ plane that contains $\bk$. Precisely, we mean that all nodes in $loop[\bk]$ have identical values for $k_z$, and $\bk$ is connected by a geodesic path to  the extremal wavevector that lies closest to $\bk$, as illustrated by the green trajectory in \fig{fig:quasi2D}. A consequence of $loop[\bk]$ being planar is that the affinity shift loop in  \q{geodesicj} simplifies to a planar integral:
\e{
\langle f_{vc\bk}|\beps\cdot \bA_{cv\bk}|^2\bS_{loop[\bk]} \rangle_{\omega} =  \int \f{dk_xdk_y}{(2\pi)^2a_z} \f{\delta(E_{cv\bk}-\hbar\omega)}{\jdosup}f_{vc\bk}|\beps\cdot \bA_{cv\bk}|^2\bS_{loop[\bk]}.\la{planarsim}
} 
We have introduced a lattice constant $a_z$ such that $\int dk_z = 2\pi/a_z$. The z component of $\bk$ in the integrand can be arbitrarily chosen, and the integrand only depends on  band energies and wave functions within the arbitrarily chosen $k_x-k_y$ plane. This justifies our use of the planar model Hamiltonian  in \q{modelham}, which explicitly depends on $k_x$ and $k_y$ but not $k_z$. \\

To recapitulate, being quasi-two-dimensional allows to  simplify  the loop analysis to planar loops, as if the problem were strictly two-dimensional. Such a simplification is not a priori obvious, since a hot photo-excited electron with initial wavevector $\kexc$ (on the excitation surface) may relax to any point along the conduction-band minimum, including points which differ from $\kexc$ in the z component [cf.\ pink trajectory in \fig{fig:quasi2D}]. If $subgraph[\kexc]$,   the probability-flow subgraph that includes the link $(c\kexc > v\kexc)$ [cf.\ \fig{fig:subgraph}], is decomposed into loops, one expects to find loops which are extended in the $k_z$ direction.\\

Let us denote the pink-colored trajectory by $p_c$;
the lower-energy boundary point of $p_c$ corresponds to the wavevector $\krec$ of recombination; it is assumed the electron traces a path $p_v$ from Bloch label $(v,\kexc)$ to $(v,\krec)$, which is not illustrated in \fig{fig:quasi2D}. Altogether, $p_c,p_v$  and the vertical links at $\kexc$ and $\krec$ combine to form $loop(\kexc,\krec)$; its associated shift loop $\bS_{loop(\kexc,\krec)}$ is defined through \q{sloop}. We define $P\circ loop(\kexc,\krec)$ as the projection of $loop(\kexc,\krec)$ onto the $k_x-k_y$ plane containing $\kexc$. If $k_{\text{exc}}^z=0$, then this projection amounts to setting $k_z=0$ for all nodes along the loop, such that the pink trajectory collapses to the green trajectory in \fig{fig:quasi2D}.   \\

We would prove that the shift loop is invariant under such a projection: $\bS_{loop(\kexc,\krec)}=\bS_{P\circ loop(\kexc,\krec)}$. Points (i-iii) in \app{app:geodesic} can be used to show that $\bS_{loop(\kexc,\krec)}$ has the same form as the right-hand side of \q{decomposeshiftloop}, with $\bS_{\text{rec}}$ defined as in \q{srec} but with $\kext$ replaced by $\krec$.   Because the photonic shift vector [\q{photonicshift2}] and the interband Berry connection is purely a function of $\ket{u_{b\bk}}$ which is $k_z$-independent, the two photonic terms in  \q{decomposeshiftloop} are invariant under changing the z component of $\krec$. What remains is to demonstrate a similar invariance for the anomalous component of the shift loop, which is given by a sum of the two line integrals in \q{decomposeshiftloop}.  $p_c$ differs from $P\circ p_c$ only in that the $\bk$-path of integration is extended in the $k_z$ direction [\fig{fig:quasi2D}]. Since $\bOmega_{c\bk}=\Omega^z_{c\bk}\vec{z}$ is $k_z$-independent,  it follows that $\bOmega_{c\bk}\times d\bk =\bOmega_{c,P\circ \bk}\times P\circ d\bk$, meaning $\int_{p_c}\bOmega_{c\bk}\times d\bk$ is invariant under projecting $p_c \ri P\circ p_c$. The same argument and conclusion holds for $v\ri c$. This completes our proof of invariance for the shift loop.\\

It is a straightforward generalization to demonstrate that the invariance property: $\bS_{\text{loop}}=\bS_{P\circ loop}$ holds for any loop, not just the simple loop we considered above.   Thus for the purpose of evaluating the loop current contribution by $subgraph[\kexc]$  [\q{subgraphcurrent}], one may as well project the entire subgraph to the $k_x-k_y$ plane containing $\kexc$. \\

At this point, one may apply essentially the same arguments that led to approximating \q{subgraphcurrent} by \q{geodesicj}, with the only modification being that  all loops are now planar, and in particular,  $loop(\bk)$ in \qq{approxjsubgraph}{sumsubgraph} is a planar geodesic loop. This completes the proof of \q{planarsim}.

\section{The transient current approximates the excitation-induced current}\la{app:jtranjexc}

\app{app:nonequil} establishes concepts and notations which are  prerequisite to understanding this section. \\

We focus  on the photo-excited carrier density regime: $n\lesssim n_h$, where energy relaxation in the active region is dominated by optical phonons.  Assuming that the excitation energy [\q{excenergy}] of photo-excited carriers lies in the active region,  we would  demonstrate that the transient current $\jtran$ is well approximated by the excitation-induced component $\jexc$ [cf.\ \q{definejexc}] of the steady current.\\

Before the tackling the transient and non-equilibrium currents, let us take a step back to consider an equilibrated mix of electrons, photons and phonons in the absence of the light source. The  quasiparticle occupancy then follows  the  Fermi-Dirac distribution: $f^{T_0}_{b\bk}$ [\q{fermidirac} ], while the occupancy of photons and phonons follow the Planck distribution: 
$N_m^{T_0}$ [\q{planck}] with the same equilibrium temperature. The shift current, viewed as a functional of the quasiparticle, photon and phonon occupancies [\q{definejsteady}], vanishes: 
\e{ \text{Equilibrium:} \as \bj[ f^{T_0}_{B},N^{T_0;phot}_{m},N^{T_0;phon}_{m} ]=0,\la{detailed}
}
owing to detailed balance; cf.\ \q{detailedbalance}.\\ 

At the onset of  turning on a light source (with frequency $\omega_s$, mode $m_s$, polarization $\beps_s$), the quasiparticles and phonons retain their equilibrium distributions, but the photon occupancy is modified to $N^{T_0;phot}_{m}+\Delta N_{s}\delta_{m,m_s}$. We define
the transient current as the current at the onset of radiation:
\e{\text{Onset:} \as\jtran=\bj[f^{T_0}_{B},N^{T_0;phot}_{m}+\Delta N_{s}\delta_{m,m_s},N^{T_0;phon}_{m}]. \la{definejtran2}}
All bosonic modes  with a thermal occupancy cannot contribute to the shift current, owing to detailed balance; cf.\ \q{detailedbalance}. For the source mode $m_s$, the net transition rate [\q{photonrate}] can be decomposed just as in \q{sourcems}, with $T$ replaced by $T_0$. Because the first term on the right-hand side of \q{sourcems} vanishes by detailed balance [\q{detailedbalance}], the transient current is simply proportional to the source-generated photon occupancy:
\e{\jtran = -\f{|e|}{\calv}\sum_{\bk} \bS^{\beps_s}_{C \lea V} \f{( 2\pi e)^2\omega_s}{\calv} \;\big|\beps_s\cdot \bA_{cv\bk}\big|^2\delta(E_{cv\bk}- \hbar \omega)f^{T_0}_{cv\bk}\Delta N_s. \la{jtranexp}}
The formula here assumes a two-band semiconducting model [\app{app:smcmodel}], but more generally one would just sum over contributions from all resonant interband transitions.
It should be borne in mind that $f^{T_0}_{C}$ is exponentially suppressed with exponent $E_g/k_BT_e \gg 1$ for an intrinsic semiconductor:
 \e{ f^{T_0}_{C}\ll 1,\as \text{and}\as 1-f^{T_0}_{V}\ll 1. 	\la{fermid}}
\q{jtranexp} manifests that $\jtran \neq 0$  must originate solely from the disruption of detailed balance between pairs of Bloch states  that are resonantly coupled by the light source, i.e., pairs   labelled $(c\bk)$ and $(v\bk)$,  with $\bk$ on the excitation surface $ES$; cf.\ \q{excsurf}.  It follows that in evaluating $\jtran=\bj[f^{T_0},\ldots]$, one may as well restrict the wavevector summations $\sum_{\bk\bkp}$ in \q{bismine2} with the condition $\bk=\bkp \in ES$:
\e{\jtran=\bj[f^{T_0}_{B},N^{T_0;phot}_{m}+\Delta N_{s}\delta_{m,m_s},N^{T_0;phon}_{m}]_{\bk=\bkp\in ES}.\la{jtrann}}

As derived in \app{app:threefold}, the excitation-induced component of the steady shift current [cf.\ \q{definejexc}] differs from \q{jtranexp} only in that $f_{cv\bk}^{T_0}$ is replaced by the non-equilibrium $f_{cv\bk}=f_{C}-f_{V}$.  If one accepts that the non-equilibrium quasiparticle distribution over the excitation surface  satisfies:
 \e{\forall \bk \in ES: \as f_{C}\ll 1,\as \text{and}\as 1-f_{V}\ll 1, 	\la{outof}} 
then $f_{cv\bk}\approx f^{T_0}_{cv\bk}$ (on the excitation surface), and therefore the  excitation-induced current approximates the transient current: 
\e{\jtran =\bj[f^{T_0}]|_{\bk=\bkp\in ES}\approx \bj[f]|_{\bk=\bkp\in ES}=	\jexc, \la{definejtran}}
in accordance with \qq{jtranexp}{jtrann}.\\

For $n\ll n_h$, we believe the inequalities in \q{outof}  hold generally, due to an argument presented in the main text and reproduced here: the smallness of $f_{C}$ and $(1-f_{V})$  originates from the slowness in optical excitations compared to the fastness of energy relaxation  by carrier-carrier and carrier-phonon scatterings.\\

We will flesh out this argument by deriving an explicit expression of $f_C$ for the  kinetic model  set up in \app{app:kineticmodel}. This model 
encodes certain assumptions which caricature reality,  as detailed in \app{app:ehsymmetry} and \app{app:eisotropy}. Thus our explicit expression for $f_{C}$ should be understood as an order-of-magnitude estimate for more realistic distributions; this is fine because the advertised inequality  [\q{outof}] is a statement about orders of magnitude.\\

With this caveat in mind, let us reproduce from \q{esipovmodel2} the kinetic equation for the iso-energy-averaged quasiparticle distribution $f_E$ in the conduction band:  
	\e{E>\hbar\Omega_o:\as G_{\sma{\uparrow}}\delta(E-E_{c,exc}) -\f{g_Ef_E}{\tau^o_{E}} +\f{g_{E_+}f_{E_+}}{\tau^o_{E_+}} +\partial_E \bigg[\f{g_EE}{\tau^{s}_E}\bigg(1 + k_BT_e\partial_E\bigg)f_E\bigg]=0; \as E_+=E+\hbar\Omega_o. \la{esipovmodel}}
We assume the reader has read the discussion leading to \q{esipovmodel2}, and we will not repeat the definitions and descriptions of each term in the kinetic equation. However, we will mention two slight differences between the above equation and \q{esipovmodel2}: \\ 

\noi{i} We have dropped the interband recombination term $[-g_Ef_E/\tau_{\text{rec}}]$ that was present in \q{esipovmodel2}. This is alright for $E>\hbar\Omega_o$ (the active region), because electron-optical-phonon scattering results in a substantially larger loss rate: $-{g_Ef_E}/{\tau^o_{E}}$, given that $\tau_{\text{rec}}\sim 1 \,ns$ and $\tau^o_E \sim 100 fs$.\cite{lundstrom_book,na_damascelli_electronphonon,sturmanfridkin_book}\\

\noi{ii} The diffusive Fokker-Planck term in \q{esipovmodel} carries a more general meaning than the corresponding term in \q{esipovmodel2}:\\

\noi{ii-a} For $n\ll n_l$, the diffusive term encodes electron-acoustic-phonon scattering, and
$\tau^s\equiv \tau_A$ is the energy relaxation time  due to spontaneous emission of acoustic phonons; a typical value is $\tau_A \sim 1\, ns$.\cite{esipov_temperatureenergy,zakharchenya_photoluminescence}\\

\noi{ii-b} For $n_h\gg n\gg n_l$, the diffusive term encodes  electron-electron scattering, and $\tau^s=\tau_{ee}$ is  the time taken   for a hot `test electron' (with initial energy $\gg k_BT_e$) to cool down to an energy comparable to $k_BT_e$.\cite{esipov_temperatureenergy} By assumption for this density regime, electron-electron scattering is more efficient in relaxing an electron's energy than   electron-acoustic-phonon scattering, meaning $\tau^{ee}\ll \tau^A \sim 1 ns$. It is also possible for electron-electron collisions to establish an electron temperature $T_e$ that exceeds the lattice temperature $T_l$.\cite{esipov_temperatureenergy} \\

The solution to the differential equation  [\q{esipovmodel}] has been derived in \ocite{esipov_oscillatoryeffects} and \ocite{esipov_temperatureenergy}.  Here, we extract a few salient facts from these references that help to prove \q{outof}: In the absence of the secondary scattering process ($\tau^{s}_E\ri\infty$), the distribution is a sum of Dirac-delta functions centered at $E_k:=E_{c;exc}-k\hbar\Omega_o$.  The effects of the secondary scatterers is that each peak shifts as:
$E_k\ri E_k-\eta_{E_k}(k+1)\hbar\Omega_o$,
as well as broadens to a regular function.  Assuming $\hbar\Omega_o/k_B{T_e}\gtrsim 1$ and $k\sim 1$, the width of each peak remains small compared to $\hbar\Omega_o$. The highest peak $(k=0)$ has the functional form:
\e{ f_E\eq \f{G_{\sma{\uparrow}}\tau^o}{g} \f{\varrho-1}{4k_BT_e\sqrt{\varrho}}\exp\bigg[-x-\sqrt{\varrho}|x|\bigg]\bigg|_{x=(E-E_{c;exc})/2k_BT_e};\as \varrho= 1+4\f{k_BT_e}{\hbar\Omega_o}\eta^{-1},\la{highestpeak}}
with $g,\tau^o$ and $\eta$ evaluated at $E_{c;exc}$.\footnote{The solution presented in \ocite{esipov_temperatureenergy} is missing a factor of $1/k_BT_e$, which we presume is a minor typographical oversight.} In particular, 
\e{
f_{E_{c;exc}}\eq \f{G_{\sma{\uparrow}}\tau^o}{g\hbar\Omega_o \eta \sqrt{\varrho}} \as \substack{\eta \hbar\Omega_o/k_BT_e \ll 1 \\ \approx} \as \half \f{G_{\sma{\uparrow}}}{g}\sqrt{\f{\tau^{s}\tau^o}{E_{c;exc} k_BT_e}}.\la{analog}
}

Let us estimate $G_{\sma{\uparrow}}/g$ under realistic experimental conditions. Recalling $G_{\sma{\uparrow}}=\alpha_{\sma{\uparrow}}\cali_{rad}/\hbar\omega$ from \q{defineGup}, and assuming typical values for the lattice period $a\sim 5\ang$, density of states $g\sim eV/a^3$, absorption coefficient $\alpha_{\sma{\uparrow}} \sim 10^3 cm^{-1}$,\cite{didomenico_oxide}  and continuous-wave laser intensity $\cali_{rad} \sim 40 Wcm^{-2}$,\cite{koch_batio3expt} one finds a modest value for $G_{\sma{\uparrow}}/g \sim 10eV/s$. \\

$f_{E_{c;exc}}$ is the product of $G_{\sma{\uparrow}}/g$ with a quantity which has dimensions of time over energy. This quantity encodes the microscopic energy relaxation processes, which occur at much shorter time scales than $1s$: as a reminder, $\tau^o\sim 100 fs$ and $\tau^s \lesssim 1\, ns$.\cite{lundstrom_book,na_damascelli_electronphonon,esipov_temperatureenergy,zakharchenya_photoluminescence} Thus, $f_{E_{c;exc}}\ll 1$ even at the low temperature of  $T_e\sim 1\,K$. Given that $f_C\ll 1$ for $\bk \in ES$, $1-f_V=f_C \ll 1$ immediately follows from  the electron-hole symmetry of our model; cf.\ \app{app:ehsymmetry}. This completes our demonstration of \q{outof}.

\section{Model calculations with optical vortices} \la{app:vortexmodel}

This appendix details the model calculations that support certain claims  stated in \s{sec:vortex}, which  we reproduce here for easy reference:

\noi{I} $\bsigma_{\vec{x},\omega}$ is dominated by the recombination-induced current; 

\noi{II} $\bsigma_{\vec{y},\omega}$ is dominated by the excitation-induced and intraband currents; 

\noi{III} The signs of $\bsigma^y_{\vec{x},\omega}$ and $\bsigma^y_{\vec{y},\omega}$ differ over a broad range of frequencies. 

\noi{IV} The linear disparity in the conductivity is large:  $|\bsigma^y_{\vec{x},\omega}-\bsigma^y_{\vec{y},\omega}| \sim mAV^{-2}$. 

\noi{V} The current response to unpolarized light is given by: $|\bsigma^y_{\vec{x},\omega}+\bsigma^y_{\vec{y},\omega}|/2 \sim 0.1 mAV^{-2}$.

\noindent Some aspects of the following demonstration will be a more quantitative elaboration of qualitative arguments made in \s{sec:vortex}.\\

\begin{figure}[H]
\centering
\includegraphics[width=10 cm]{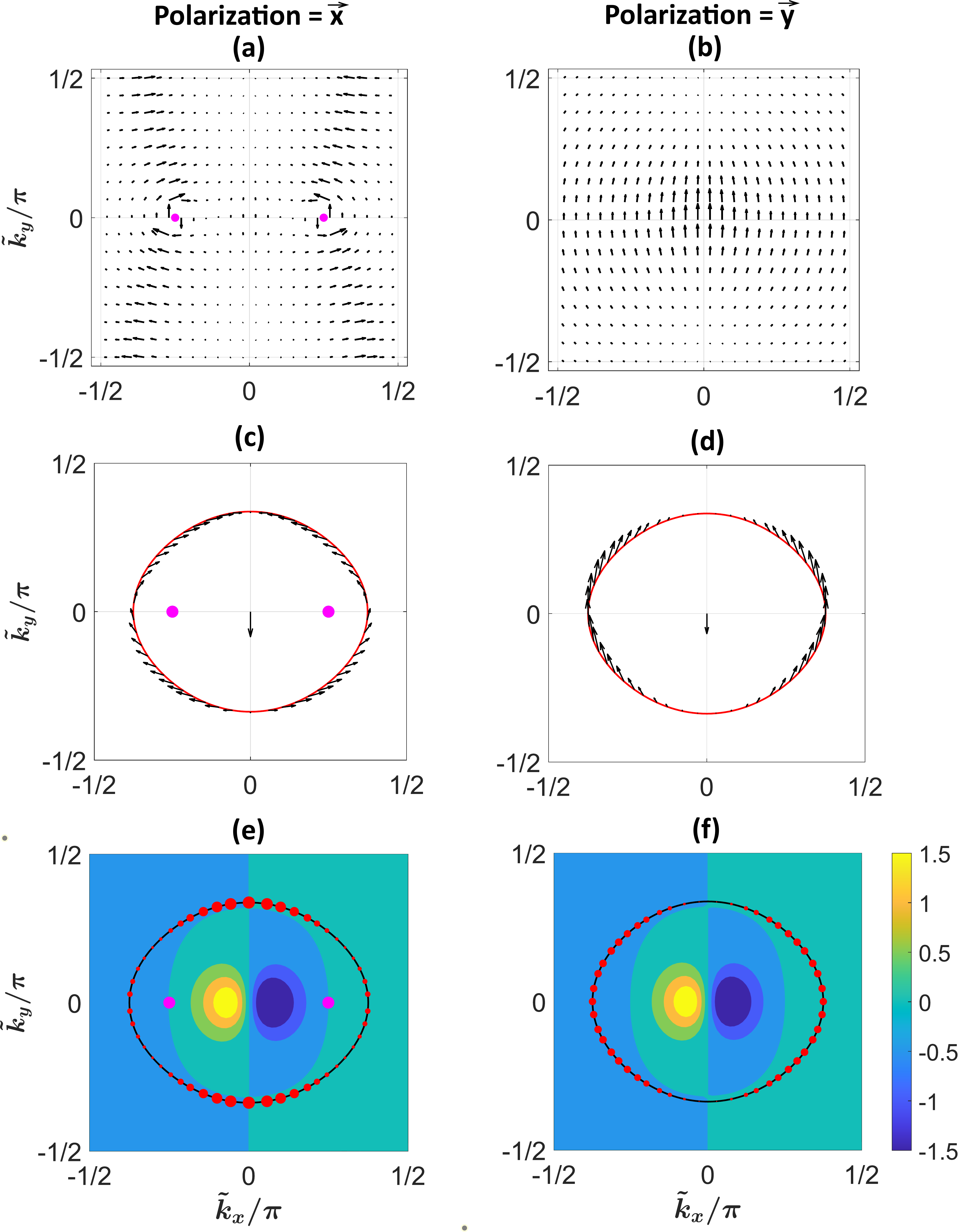}
\caption{Characterization of the model Hamiltonian in \q{modelham} with $\tilde{Q}=-1$. Panels (a,c,e) are characterizations for a light source with linear polarization vector $\beps_s=\vec{x}$, and (b,d,f) for  $\beps_s=\vec{y}$. The pink dots in (a,c,e) represent the $\bk$-locations of $\vec{x}$-vortices. Panels (a) and (b) depict the photonic shift vector field $\bS^{\beps_s}_{c\bk\lea v\bk}$, with $\beps_s=\vec{x}$ and $\vec{y}$ respectively. For panels (c) and (d), the red ellipse represents the excitation surface for a photon frequency $\hbar\omega=4.5E_0$; arrows on the ellipse represent the vectors $|\beps_s\cdot \bA_{cv\bk}|^2\bS^{\beps_s}_{c\bk\lea v\bk}$ for $\bk$ on the excitation surface; the central arrow represents the recombination component of the affinity shift loop: $ASL^{\text{rec}}_{\beps,4.5E_0}$; cf.\ \q{asl}. For panels (e) and (f), the size of the red dots indicates the magnitude of $|\beps_s\cdot \bA_{cv\bk}|^2$ for $\bk$ on the excitation surface; the colored background represents the Berry curvature scalar field $\Omega^z_{c\bk}$ in units of $a^2=(\calv_{\sma{\text{cell}}})^{2/3}$, with a color legend on the right.  }
\label{fig:Qm05_menagerie}
\end{figure}

The form of our model Hamiltonian is identical to the one studied in the context of the anomalous shift; cf.\ \q{modelham}. Having studied the case of $\tilde{Q}=1$, we now tune $\tilde{Q}$  from positive to negative values. The conduction and valence bands touch (at $\bk=0$) when $\tilde{Q}=0$ and subsequently untouch for negative $\tilde{Q}$. This untouching  is  accompanied by the  nucleation of two time-reversal-related $\vec{x}$-vortex lines at $(\tilde{k}_x,\tilde{k}_y)\approx (\pm \sqrt{-\tilde{Q}/(1-\tilde{Q}/2)},0)$, as illustrated in \fig{fig:Qm05_menagerie}(a); there are no $\vec{y}$-vortices in this model [\fig{fig:Qm05_menagerie}(b)]. Henceforth, we fix $\tilde{Q}=-1$. \\

\begin{figure}[H]
\centering
\includegraphics[width=16 cm]{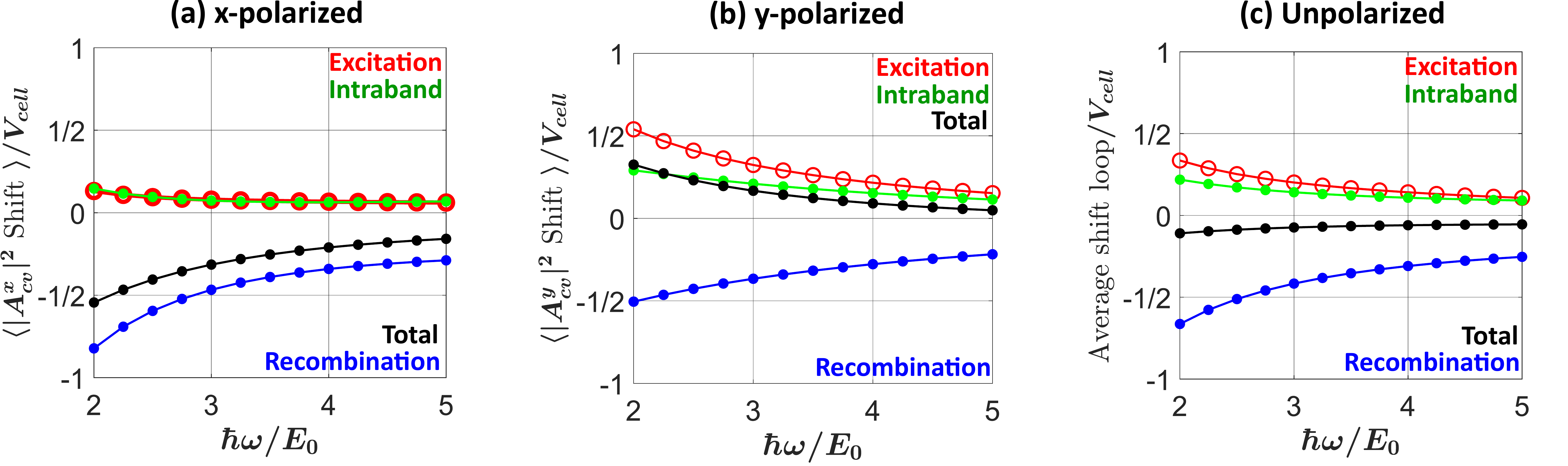}
\caption{ The black curves in panels (a) and (b) represents the affinity shift loop $\langle |\beps_s\cdot \bA_{cv}|^2\bS^{\beps_s}_{y,loop}\rangle_{\omega}$ (in units of $\calv_{\sma{\text{cell}}}$) vs the photon frequency $\omega$, for polarization $\beps_s=\vec{x}$ and $\vec{y}$ respectively. Non-black curves represent  the three components of the affinity shift loop: excitation (red curve), intraband relaxation (green curve), and recombination (blue curve). Panel (c) averages  the affinity shift loop over two orthogonal light polarizations.    }
\label{fig:Qm05_shiftloop_unpol}
\end{figure}

Firstly, let us consider the case of an $\vec{x}$-polarized light source and make the case that the excitation-induced current is outweighed by the recombination-induced current: $||\bj_{\text{exc}}[\vec{x}]|| \ll ||\bj_{\text{rec}}[\vec{x}]||$, owing to the vortex-induced orientational disorder of the photonic shift vector field. This inequality simplifies to  $||\bsigma^{\text{exc},y}_{\vec{x},\omega}|| \ll ||\bsigma^{rec,y}_{\vec{x},\omega}||$ for the $y$-component of the shift conductivity [\qq{decomposeshiftloop}{decomposesigma}], because a  mirror symmetry ($x\ri -x$) of the model Hamiltonian\footnote{$M_xH(\bk)M_x^{-1}=H(-k_x,k_y)$ with $M_x=\bsigma_3$.} constrains the $x$ component of the shift current to vanish, while the $z$ component vanishes owing to the quasi-two-dimensionality of the model; cf.\ \app{app:geodesicquasi2D}. Because the shift conductivity is essentially the product of the joint density of states ($\jdosup$) [cf.\ \q{definejdos}] with  the affinity shift loop [cf.\ \qq{aveshiftloop}{shiftconduct}], one may as well compare the excitation and recombination components of the affinity shift loop: 
\e{ASL^{\text{exc}}_{\beps,\omega} \equiv \langle |\beps\cdot \bA_{cv\bk}|^2 S^{\beps}_{y,c\bk\lea v\bk} \rangle_{\omega}
\as \text{vs}\as ASL^{\text{rec}}_{\beps,\omega} \equiv \langle|\beps\cdot \bA_{cv\bk}|^2\rangle_{\omega} S_{y,rec}, \la{asl}}
for $\beps=\vec{x}$ and $\langle \ldots \rangle_{\omega}$ denoting an average over the excitation surface; cf.\ \q{averageES}. $S_y$ means the $y$ component of $\bS$, and $S^{\beps}_{y,c\bk\lea v\bk}$ is the photonic shift vector defined in \q{photonicshift2}. The recombination shift $\bS_{\text{rec}}$ is defined in \q{srec} but simplifies in the present context to   $\bS^{\vec{x}}_{v\kext\lea c\kext}$, owing to a mirror-symmetry-imposed dipole selection rule.\footnote{Conduction- and valence-band states with $k_x=0$ transform under different representations of mirror symmetry $M_x$, hence $A^y_{cv\kext}=A^z_{cv\kext}=0$. This implies for any $\beps$ that is not orthogonal to $\vec{x}$ that $\nabk \arg \beps\cdot \bA_{cv\kext} = \nabk \arg A^x_{cv\kext}$ and $\bS^{\beps}_{v\kext \lea c\kext}=\bS^{\vec{x}}_{v\kext \lea c\kext}$.  } A numerical calculation of \q{asl} reveals for a wide range of photon frequencies that $ASL_{\vec{x}}^{\text{exc}}$ and $ASL_{\vec{x}}^{\text{rec}}$ have opposite signs, and that  $|ASL_{\vec{x}}^{\text{exc}}| \ll |ASL_{\vec{x}}^{\text{rec}}|$ by a multiplicative factor ranging from 1/5 to 1/8, as illustrated in \fig{fig:Qm05_shiftloop_unpol}(a).\\

To rationalize this multiplicative factor, we illustrate $|A^x_{cv\bk}|^2\bS^{\vec{x}}_{c\bk\lea v\bk}$ as  arrows in
\fig{fig:Qm05_menagerie}(c), for $\bk$ along a representative excitation surface  encircling the $\vec{x}$-vortices.  The central arrow in \fig{fig:Qm05_menagerie}(c) represents $\langle|A^x_{cv\bk}|^2\rangle_{\omega} \bS^{\vec{x}}_{v\kext\lea c\kext}$. All arrows are drawn with a common scale to allow for mutual comparison. It is evident that proximity to the $\vec{x}$-vortex causes the direction of $|A^x_{cv\bk}|^2\bS^{\vec{x}}_{c\bk\lea v\bk}$ to rotate along the excitation surface; the average of $|A^x_{cv\bk}|^2\bS^{\vec{x}}_{c\bk\lea v\bk}$ over the excitation surface is therefore diminished; this average just equals $ASL^{\text{exc}}_{\beps,\omega}$; cf.\ \q{asl}. In contrast, recombination occurs in the vicinity of the extremal wavevector $\kext$, where the photonic shift vector is roughly constant. Thus follows a general principle: ceteris paribus, the orientational disorder induced by $\vec{x}$-vorticity reduces $\bj_{\text{exc}}[\vec{x}]$ relative to $\bj_{\text{rec}}[\vec{x}]$, for an $\vec{x}$-polarized source. \\

Ceteris paribus, the same orientational disorder reduces $\bj_{\text{exc}}[\vec{x}]$ relative to $\bj_{\text{exc}}[\vec{y}]$, for reasons explained in \s{sec:vortex}. This implies a linear disparity of the excitation-induced current $\jexc$, which applies to a broad range of photon frequencies; compare red curves of \fig{fig:Qm05_shiftloop_unpol}(a) and (b).\\

To understand the linear disparity of the intraband current $\jintra$, we have indicated the $\bk$-dependent magnitude of $|A^{x}_{cv}|^2$ (resp. $|A^{y}_{cv}|^2$) by the size of dots imprinted over the excitation surface in \fig{fig:Qm05_menagerie}(e) [resp. \fig{fig:Qm05_menagerie}(f)]; in both figures, the same Berry curvature scalar field ($\Omega^z_{c\bk}$) is represented by a color plot. It may be seen that $|A^{x}_{cv}|^2$ and $|A^{y}_{cv}|^2$ are both anisotropic over the excitation surface, but each favors a different segment of the excitation surface, for reasons explained in \s{sec:vortex}. We deduce for the $\vec{y}$-polarized source that the predominant relaxation pathways are roughly parallel to $k_x$ [cf.\ \fig{fig:Qm05_maintext}(f)]  and intersect the Berry-curvature hot spots, leading to a larger anomalous shift than the case of the $\vec{x}$-polarized source.  Once again, this effect is not limited to a fine-tuned photon frequency; compare green curves of \fig{fig:Qm05_shiftloop_unpol}(a) and (b).\\

Altogether, the linear disparity of $\jexc$ and $\jintra$ results in the net shift current being dominated by $\jexc+\jintra$ for a $\vec{y}$-polarized  source [cf.\ black curve in \fig{fig:Qm05_shiftloop_unpol}(b) and claim (II)], and by $\jrec$ for a $\vec{x}$-polarized source [black curve in \fig{fig:Qm05_shiftloop_unpol}(a) and claim (I)]; the net current changes sign if the polarization is flipped [claim (III)]. The linear disparity of the affinity shift loop [i.e., the difference of the two black curves in \fig{fig:Qm05_shiftloop_unpol}(a) vs (b)] is comparable to $-1$ (in units of $\calv_{\sma{\text{cell}}}$, the real-space volume of the  primitive unit cell) over a broad range of frequencies; this corresponds to a linear disparity of the conductivity: $\bsigma^y_{\vec{x},\omega}-\bsigma^y_{\vec{y},\omega}\approx 2 mAV^{-2}$ [cf.\ \q{shiftconduct}, \fig{fig:coverpic2}(d) and claim (IV)], assuming a generic value for  $\jdosup\approx  (\calv_{\sma{\text{cell}}}eV)^{-1}$.\footnote{ We choose $\tilde{Q}=-1$ and $\tilde{P}=12$ such that $\jdosup\approx (\calv_{\sma{\text{cell}}}eV)^{-1}$.} \\

The response to an unpolarized light source is given by  $(\bsigma^y_{\vec{x},\omega}+\bsigma^y_{\vec{y},\omega})/2$, which $\approx 0.2 mAV^{-2}$ over a broad range of frequencies  [cf.\ black curve in \fig{fig:Qm05_shiftloop_unpol}(c) and claim (V)].\\

We end this appendix with a caveat: the calculated values of $\bsigma$ should be taken with a grain of salt. A reliable calculation of $\bsigma$ should also account for the dependence of Bloch wave functions over continuous space,\cite{aa_topologicalprinciple} but such dependence is discarded when the Hilbert space is  reduced to a two-dimensional vector space at each $\bk$ point, as was done for all model Hamiltonians in this work. A more realistic model would incorporate ab-initio-derived wave functions as additional model parameters.\cite{julen_wannierinterpolation,julen_assessingrole} Reassuringly, our qualitative arguments for vortex-induced shifts do not rely on the two-band approximation and  are equally applicable to realistic, continuous-space Hamiltonians.

\section{Chern-vorticity theorem}\la{app:chernvorticity}

The Chern-vorticity theorem in \q{eq:chernvortex} relates  the Chern numbers ($C_v,C_c$) of the  valence and conduction states (over any  closed 2D $\bk$-manifold $\bSigma$) to the net optical vorticity $(Vort)$ within $\bSigma$. \\

\begin{figure}[h]
\centering
\includegraphics[width=10 cm]{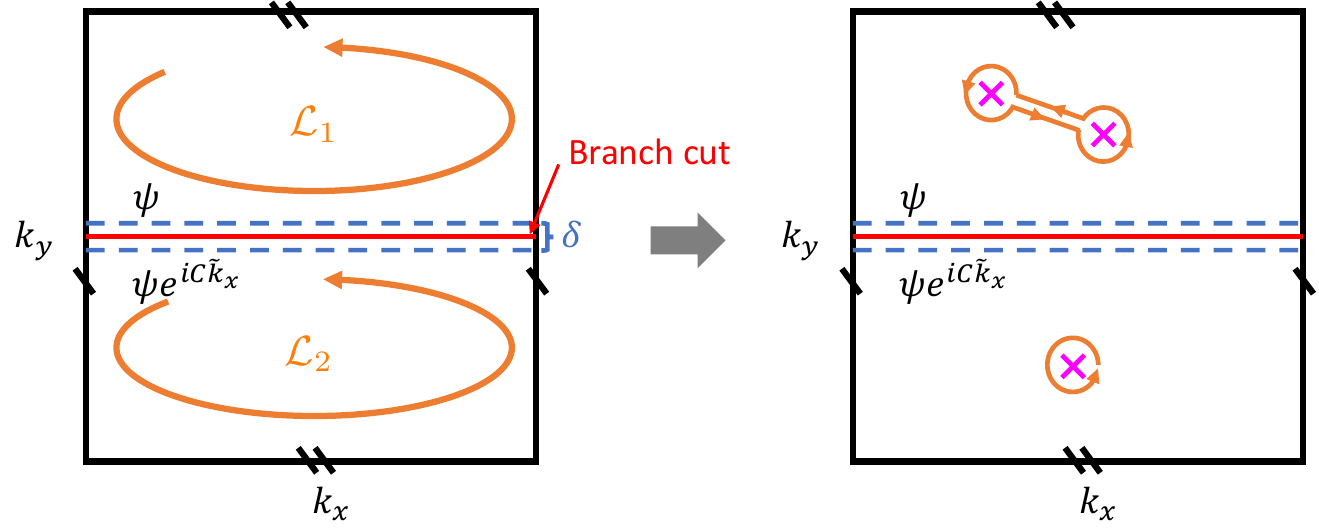}
\caption{Illustration of the two-patch decomposition of $\bSigma$,  as well as the paths for the line integrals in Eqs.~\eqref{eq:pathint} and ~\eqref{eq:circulation}. The magenta crosses represent the optical vortices.}
\label{fig:chernvortex}
\end{figure}

To prove the theorem, we first recall that if $C_v ({\rm resp.} \ C_v) \neq 0$, the wave function cannot be made continuous and periodic over $\bSigma$, i.e., $\bA_{vv\bk}({\rm resp.} \ \bA_{cc\bk})$ must be singular somewhere on $\bSigma$. To be concrete, supposing $\bSigma$ were a two-torus; Fig.~\ref {fig:chernvortex} illustrates how $\bSigma$  is decomposed into two patches, such that the wave function in the interior of each patch is analytic in $\bk$,  but $\bA_{vv\bk}({\rm resp.} \ \bA_{cc\bk})$ is singular at the patch boundary:
\begin{equation}
\label{eq:pathint}
\lim_{\delta\rightarrow 0}\frac{1}{2\pi}\left(\int_{\mathcal{L}_{1}}+\int_{\mathcal{L}_{2}}\right)\bA_{bb\bk}\cdot d\bk
=C_{b}, \ \operatorname{with} \ b=v,c.
\end{equation}
Here, $C_{b}$ is the Chern number of the band labelled by $b$, and $\delta$ is an infinitesimal parameter illustrated in Fig.~\ref {fig:chernvortex}. \\

Performing  the same line integral with the Berry connection replaced by the photonic shift vector gives zero for any linear polarization vector $\beps$:
\e{
\lim_{\delta\rightarrow 0}\frac{1}{2\pi}\left(\int_{\mathcal{L}_{1}}+\int_{\mathcal{L}_{2}}\right)\bS^{\beps}_{c\bk \leftarrow v\bk}\cdot d\bk=0,
}
because the shift vector is gauge invariant and smoothly defined except at optical vortices, and one can always choose the patch boundary to avoid those vortex points.  \\

Comparing the last two equations with the definition of the photonic shift vector in \q{photonicshift}, one infers that there must be nonzero circulations in $\arg[\beps\cdot \bA_{cv\bk}]$ to compensate for the singularity of $\bA_{vv\bk}({\rm resp.} \ \bA_{cc\bk})$. Specifically,
\begin{equation}
\label{eq:circulation}
\lim_{\delta\rightarrow 0}\frac{1}{2\pi}\left(\int_{\mathcal{L}_{1}}+\int_{\mathcal{L}_{2}}\right)\nabk \arg[\beps\cdot \bA_{cv\bk}]\cdot d\bk=C_{c}-C_{v},
\end{equation}
which means that the net phase vorticity of $\beps\cdot \bA_{cv\bk}$ over $\bSigma$ is $C_{c}-C_{v}$.  The line integral over $\mathcal{L}_{1}$ (resp. $\mathcal{L}_{2}$) equals the winding number of $\operatorname{arg}[\beps\cdot\bA_{cv\bk}]$ in patch 1 (resp. patch 2), and is thus topologically invariant upon contracting the $\mathcal{L}_1$ (resp. $\mathcal{L}_2$) to infinitesimally encircle any optical vortex in patch 1 (resp. patch 2); this contraction is illustrated in Fig.~\ref{fig:chernvortex}. This invariance implies that Equation~\eqref{eq:circulation} is equivalent to Equation~\eqref{eq:chernvortex}.\\

\section{Supporting our case study of BiTeI}\label{app:sbandedge}

Our case study of BiTeI is based on a four-band Hamiltonian $H_{\text{BiTeI}}(\bk)$ [cf.\ \q{eq: Hamiltonian}] with energies ordered as $E_{1}\leq E_{2}<E_{3}\leq E_{4}$. We focus on photon frequencies which resonantly excite quasiparticles from the highest-energy valence band to the lowest-energy conduction band: $b'=3$ and $b=2$. Minimizing $E_{3\bk}$ with respect to $\bk$ defines a circular ring contained in the zero-$k_z$ plane; maximizing $E_{2\bk}$ with respect to $\bk$ also defines a circular ring contained in the zero-$k_z$ plane; actually, the two rings coincide, as suggested pictorially in \fig{fig:BiTeI_illustration}. This coincidence may be rationalized: the O(2) symmetry about the z axis, combined with time reversal symmetry ($T$), imply the existence of a $C_{2z}T$ symmetry (two-fold rotation composed with time reversal) which maps $\bk \ri (k_x,k_y,-k_z)$; thus if $E_{3\bk}$ is minimized on a single O(2)-symmetric ring, this ring  must lie on the $C_{2z}T$-symmetric plane with $k_z=0$. Within this plane, the Hamiltonian has a chiral symmetry that relates positive to negative energies:
\e{
\tau_{2}\sigma_{3} H_{\text{BiTeI}}(k_{x},k_{y},0)\tau_{2}\sigma_{3}=-H_{\text{BiTeI}}(k_{x},k_{y},0),\la{chiralsymm}
}
which implies that $E_{2k_xk_y0}$ is maximized wherever  $E_{3k_xk_y0}$ is minimized. We will refer to this ring as the \underline{band-edge ring}.

\subsection{Effective description by a massive Dirac fermion}

Near the topological phase transition between  a trivial insulator and a $\Z_2$ topological insulator, the two bands that touch are effectively described by a massive Dirac fermion in two momentum dimensions. Here, we provide a detailed derivation of the massive Dirac Hamiltonian in the $k_{x^{+}}-k_{z}$ half-plane [c.f. Eq.~\eqref{eq:massivedirac}]. \\

We start by restricting $H_{BiTeI}$ [Eq.~\eqref{eq: Hamiltonian}] to the $k_{x^{+}}-k_{z}$ half-plane and  Taylor-expanding the Hamiltonian around   $\mathbf{k}_{0}=(\lambda,0,0)$:
\e{
\label{eq:expand}
H_{BiTeI}\eq H_{0}+H_{1}; \lin
H_{0}\eq \hbar v\lambda (\tau_{1}\sigma_{3}-\tau_{2}\sigma_{1}),\lin
H_{1}\eq (m'-2A\lambda q_{x})\tau_{3}\sigma_{0}-\hbar v(q_{x}\tau_{2}\sigma_{1}+q_{z}\tau_{2}\sigma_{3})+O(q^2).
}
We have introduced wavenumbers $q_{x}$ and $q_{z}$ which are the deviations from $\mathbf{k}_{0}$ in the $k_{x^{+}}-k_{z}$ half-plane; $\tau_{i}\sigma_{j}$ is the Kronecker product of $\tau_{i}$ and $\sigma_{j}$, i.e., $\tau_{i}\otimes\sigma_{j}$~\cite{enwiki:1237563175}. $\mathbf{k}_{0}$ is a point where the bands touch during the topological phase transition; by construction, $H_1$ vanishes at $\bk_0$ when $m^{\prime}=0$, and the touching bands correspond to the two zero-energy eigenstates of $H_0$, which 
we label as $\ket{1}=(i,1,0,0)^{T}/\sqrt{2}$ and $\ket{2}=(0,0,-i,1)^{T}/\sqrt{2}$. \\

In the low-energy subspace spanned by $\ket{1}$ and $\ket{2}$, the effective Hamiltonian is given by degenerate perturbation theory as:
\begin{equation}
\label{eq:perturbation}
H^{\prime}_{i,j}=\bra{i}H_{1}\ket{j}, \ i,j=1,2.
\end{equation}
Given that 
\begin{equation}
\label{eq:perturbation}
\bra{i}\tau_{3}\sigma_{0}\ket{j}=(\gamma_{3})_{ij}, \bra{i}\tau_{2}\sigma_{1}\ket{j}=-(\gamma_{1})_{ij}, \bra{i}\tau_{2}\sigma_{3}\ket{j}=-(\gamma_{2})_{ij} \ i,j=1,2,
\end{equation}
with $\gamma_{1,2,3}$ being Pauli matrices of the Hilbert space spanned by $\ket{1}$ and $\ket{2}$,
Eq.~\eqref{eq:perturbation} directly gives Eq.~\eqref{eq:massivedirac} in the main text.

\subsection{Vanishing shift at the band edge, for $x$- and $y$-polarized light}

This section aims to explain why  $\vec{z}\cdot \jexc$ dominates over $\vec{z}\cdot \jrec$ in the low-frequency regime of \fig{fig:BiTeI}(c).
This reduces to explaining the smallness of
the recombination shift vector $\vec{z}\cdot \bS_{\text{rec}}$ [\q{srec}] relative to the excitation shift vector [\q{sexc}], according to the average-shift-loop formula in \qq{decomposeshiftloop}{shiftconduct}.  Given that $\vec{z}\cdot \bS_{\text{rec}}$ is an affinity-weighted average of $\vec{z}\cdot \bS^{\beps}_{v\kext \lea c\kext}$ over all polarization vectors of the spontaneously emitted photon [\q{srec}], it may be argued that $\vec{z}\cdot \bS_{\text{rec}}$ is small because of the vanishing of the band-edge shift vectors 
\e{\text{On the band-edge ring:}\as \vec{z}\cdot \bS^{\vec{x}}_{v\kext \lea c\kext}=\vec{z}\cdot \bS^{\vec{y}}_{v\kext \lea c\kext}=0.\la{vanishshiftedge}} 
This is equivalent to the vanishing of the band-edge shift connections,
\e{\text{On the band-edge ring:}\as \vec{z}\cdot \bS^{\vec{x}}_{v\kext \lea c\kext} |\vec{x}\cdot \bA_{cv\kext}|^2=\vec{z}\cdot \bS^{\vec{y}}_{v\kext \lea c\kext} |\vec{y}\cdot \bA_{cv\kext}|^2=0,\la{vanishshiftconnedge}} 
because the optical affinity is non-vanishing throughout the band-edge ring; after all, there are no optical vortex loops intersecting the band-edge ring, as illustrated in \fig{fig:BiTeI_illustration}. \\

For any tight-binding Hamiltonian $H(\bk)$, the photonic shift connection can be expressed as~\cite{cook_designprinciples}
\begin{equation}
\label{eq:shiftconnection}
\vec{z}\cdot \bS^{\vec{x}}_{b'\leftarrow b}|\vec{x}\cdot \bA_{b'b}|^2 =\operatorname{Im}\left\{\frac{\overline{{v}^{x}_{bb'}}}{(\omega_{bb'})^2}\left[\bra{u_{b\bk}}\partial_{z}\partial_{x}H\ket{u_{b'\bk}}-\frac{v^{z}_{bb'}\Delta^{x}_{bb'}+v^{x}_{bb'}\Delta^{z}_{bb'}}{\omega_{bb'}}+\sum_{b'' \neq b, b'}\left(\frac{v_{bb''}^z v_{b''b'}^x}{\omega_{b b''}}-\frac{v_{bb''}^x v_{b''b'}^z}{\omega_{b'' b'}}\right)\right]\right\}.
\end{equation}
Here, $\partial_{z}\equiv \partial_{k_{z}}$, $v^{z}_{bb'}=\bra{u_{b\bk}}\frac{1}{\hbar}\partial_{z}H\ket{u_{b'\bk}}$, $\Delta^{z}_{bb'}=\partial_{z}E_{b}-\partial_{z}E_{b'}$,  $\omega_{bb'}=(E_{b}-E_{b'})/\hbar$ and $\sum_{b'' \neq b, b'}$ means to sum over all band indices $b''$ which are neither $b$ nor $b'$. \\

Let us show that Eq.~\eqref{eq:shiftconnection} vanishes for $H=H_{BiTeI}$ throughout the band-edge ring:\\

\noi{i} The first term in the square bracket of Eq.~\ref{eq:shiftconnection} vanishes, because $H_{\text{BiTeI}}$ depends quadratically on $\bk$ as $k_x^2+k_y^2+k_z^2$.  \\

\noi{ii} The second term in the square bracket  vanishes, because band energy functions are extremized  at the band edge: $\partial_{z}E_{3}=\partial_{z}E_{2}=0$. \\

\noi{iii} The third term also vanishes, but the argument is longer: firstly, observe from \q{eq: Hamiltonian} that $\partial_{z}H_{\text{BiTeI}}\big|_{k_z=0}=\tau_{2}\sigma_{3}$ is simply the chirality operator in \q{chiralsymm}, meaning that $\partial_{z}H_{\text{BiTeI}}$  maps between energy eigenstate with inverted energies. This implies $\hbar v_{bb''}^z=\bra{u_{b\bk}}\partial_{z}H_{\text{BiTeI}}\ket{u_{b''\bk}}\big|_{k_z=0}$ is only nonzero if $E_{b''}=-E_{b}$, but $E_{b''}=-E_{b}$ cannot be satisfied because of the constraint $\sum_{b'' \neq b, b'}$ in \q{eq:shiftconnection}. A similar argument proves that $v_{b''b'}^z=0$, hence altogether the third term in the square bracket vanishes.\\

The above demonstration holds if one replaces $x\ri y$, meaning that the z-component of the shift connection also vanishes for $\vec{y}$-polarized light, throughout the band-edge ring. This completes the proof of \q{vanishshiftconnedge}.

\subsection{Asymptotic behavior of anomalous-shift integrals}\la{app:anomalousshiftintegrals}

One result in \s{sec:casestudy} was the {$(1/E_{g})$-divergence of the intraband shift conductivity across the topological phase transition, with $|E_g|$ the band gap and $\sgn(E_g)=-1$ on the trivial side of the transition. This divergence relied on an inequality between two anomalous-shift integrals:
\begin{equation}
\label{eq:affweightedas2}
\left|\left(\int_{-\mathcal{P}}+\int_{\mathcal{P}}\right)\vec{z}\cdot \bOmega_{c\bk}\times d\bk \right| \ll \left|\left(\int_{-\mathcal{P}}-\int_{\mathcal{P}}\right)\vec{z}\cdot \bOmega_{c\bk}\times d\bk \right|, 
\end{equation}
which is asymptotically valid as $|E_{g}|$ approaches zero; $\pm\calp$ are any pair of diametrically-opposite geodesic paths, as representatively illustrated in \fig{fig:pathillu}.
\begin{figure}[H]
\centering
\includegraphics[width=6 cm]{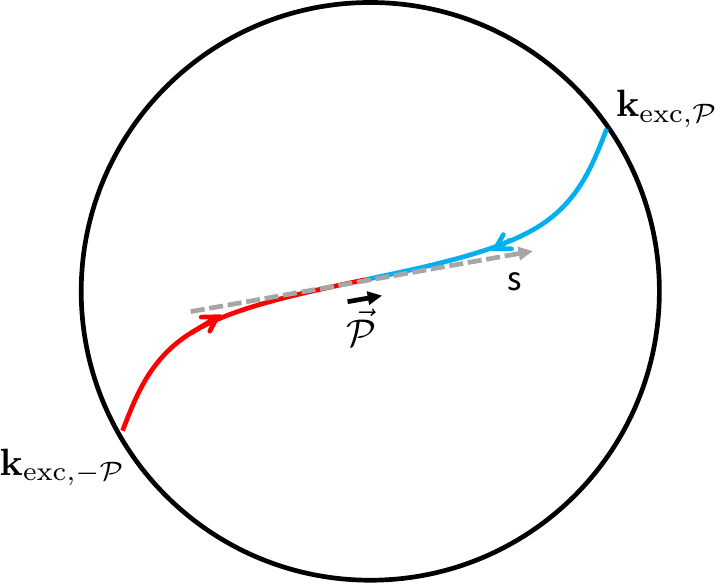}
\caption{Representative pair of diametrically-opposite geodesic paths, with $-\calp$ in red and $\calp$ in blue. The precise meaning of `diametrically-opposite' is that both paths approach the band-extremal wavevector ($\bk_{ext,\calp}$) along the same tangent line, which we illustrate as a grey dashed line with tangent vector $\vec{\calp}$.}
\label{fig:pathillu}
\end{figure}
\noindent We have demonstrated in \s{sec:casestudy} that the right-hand side of \q{eq:affweightedas2} diverges as $1/E_g$, thus to prove \q{eq:affweightedas2} it suffices to show that the magnitude of the left integral is decreasing as $|E_g|\ri 0$. This is the main result of this subsection.\\

To begin,  consider the Berry curvature  within one cross-section of the torus enclosed by the excitation surface, as exemplified by the $k_x^+$-$k_x$ half plane. As $|E_g|\ri 0$, $\vec{z}\cdot \bOmega_{c\bk}\times d\bk/|d\bk|=\Omega_{c\bk}^y$ becomes localized to a `hot spot' centered at the band-extremal wavevector $\bk_{ext,\calp}$, with a spot width comparable to $|m'|/\hbar v \propto |E_g|$. Indeed, writing the massive Dirac Hamiltonian as a dot product of three-vectors: 
\e{
H'=\bd \cdot \boldsymbol{\gamma},\as \bd=(d_1,d_2,d_3)=(\hbar v q_x, \hbar v q_z,m'-2A\lambda q_x), \as \boldsymbol{\gamma}= (\gamma_1,\gamma_2,\gamma_3),
}
the conduction-band Berry curvature can be expressed as 
\e{
\Omega_{c\bk}^h= -\f1{4d^3}\eps_{hij}\bd\cdot \nabla_{k_i}\bd \times \nabla_{k_j}\bd \imp \Omega_{c\bk}^y=\f{1}{2}\f{m'}{D^{3/2}}; \as
D\eq  (d_1)^2+(d_2)^1+(d_3)^2; \as d= \sqrt{D}, 
 \la{omegahot}
}
The $\bk$-location  of the Berry-curvature maximum can be identified by
\e{
0=\partial_s \Omega_{c\bk}^y = \f{3}{4} m'\f{\partial_s D}{D^{5/2}},
}
with $(\partial_{s})^m$ being the $m$'th-order derivative in the direction that is tangential to $\mathcal{P}$ at the band extremum. The Berry-curvature maximum (of the hot spot) coincides (in $\bk$-location) with the band-extremal wavevector $\bk_{ext,\calp}$; this is because the energy spectrum of $H'$ has an $E\ri -E$ symmetry  at each $\bk$,  which implies that extremizing the $\bk$-dependent energy gap $(E_{c\bk}-E_{v\bk})$ is equivalent to extremizing the conduction-band energy:
\e{
E_{c\bk} = \sqrt{D_{\bk}} \imp 0=\partial_s E_{c}\bigg|_{\bk_{ext,\calp}} = \f{\partial_sD}{\sqrt{D}}\bigg|_{\bk_{ext,\calp}}. \la{partialD}
}
The band gap is defined as the extremal value of $(E_{c\bk}-E_{v\bk})$:
\e{
|E_g|=2d\bigg|_{\bk_{ext,\calp}}=\f{m'}{\sqrt{1+u^2}}; \as u=\f{2A\lambda}{\hbar v},
}
and the extremal value of the Berry curvature can be expressed in terms of the signed band gap ($E_g =\sgn[m']|E_g|$) as
\e{
\Omega_{c\bk_{ext,\calp}}^y=\f{1}{2}\f{m'}{D^{3/2}}\bigg|_{\bk_{ext,\calp}}=4\sqrt{1+u^2}\f{E_g}{|E_g|^3}.\la{omegaeg}
}
That the width of the Berry-curvature hot spot is of order $|m'|/\hbar v$ can be deduced from dimensional analysis of \q{omegahot}, assuming that $|A\lambda|$ is less than or comparable to $|\hbar v|$. \\

The localization of Berry curvature in momentum space allows to express the anomalous-shift integral as
\e{
\left(\int_{-\mathcal{P}}+\int_{\mathcal{P}}\right)\vec{z}\cdot \bOmega_{c\bk}\times d\bk = \int_{-\Lambda}^{\Lambda} \Omega^{(0)}_s \Theta_s ds + \text{correction}, \la{onlyodd}
}
with a cutoff $\Lambda>0$ for the integration variable $s$ along the grey dashed tangent line in Fig.~\ref{fig:pathillu}; we have introduced $\Omega_{s}^{(0)}\equiv \Omega^{y}_{c\bk_{ext,\calp}+s\vec{\calp}}$, with $\vec{\mathcal{P}}$ being the  unit-norm vector parallel to the tangent line. 
$\Theta_s$ is the symmetric step function that equals $-1$ for positive $s$, and $+1$ for negative $s$. This step function arises because  $\pm \calp$ are oriented paths beginning on diametrically-opposed points on the excitation surface and ending at the same point: $\bk_{ext,\calp}$. For fixed $\Lambda$, it is evident that the magnitude of the correction in \q{onlyodd} decreases as $E_g\ri 0$, owing to the increasing localization of the hot spot (which has a width $|m'|/\hbar v \propto |E_g|$).\\

To prove the main result of this section, what remains is to show that $\int_{-\Lambda}^{\Lambda} \Omega^{(0)}_s \Theta_s ds$ is also decreasing; actually, we will prove a stronger statement that this integral just vanishes.  Indeed, so long as  the massive-Dirac Hamiltonian has an energy gap (i.e., $m'\neq 0$), $\bOmega_{c\bk}$ is an analytic function of $\bk$ and hence $\Omega^{(0)}_s$ is an analytic function of $s$, meaning it admits a convergent Taylor expansion $\Omega^{(0)}_s= \sum_{n \in \mathbb{N}} \Omega^{(n)}_0 s^n/n!$, with the $n$'th-order derivative $\Omega^{(n)}_0$ to be  $(\partial_s)^n\Omega^{(0)}_{s}$ evaluated at the band extremum ($s=0$).  
Only the odd-order derivatives contribute to \q{onlyodd}, owing to the symmetric step function $\Theta_s$ being an odd function of $s$, hence
\e{
\left(\int_{-\mathcal{P}}+\int_{\mathcal{P}}\right)\vec{z}\cdot \bOmega_{c\bk}\times d\bk \approx 2\sum_{n\in 2\mathbb{N}+1} \f{\Omega^{(n)}_0}{n!}  \int_{0}^{\Lambda} s^n ds. \la{onlyodd2}
}
The following discussion proves that $\Omega^{(n)}_0$ vanishes for any odd $n$. 
It follows from a dimensional analysis of \q{omegahot} that the $n$'th-order derivative can be expanded as 
\e{
\Omega^{(n)}_s =\f{1}{D^{3/2+n}}\sum_{m_1=0}^2\ldots\sum_{m_n=0}^2 c_{m_1\ldots m_n} (\partial_{s_1})^{m_1}\ldots (\partial_{s_n})^{m_n} D_{s_1}\ldots D_{s_n} \bigg|^{m_1+\ldots m_n=n}_{s_j\ri s},
}
with linear coefficients $c_{m_1\ldots m_n}$ that depend on $m_j$.
The meaning of the subscript $s_j\ri s$ is that after performing all the differentiations $[(\partial_{s_1})^{m_1}\ldots (\partial_{s_n})^{m_n}]$, the resultant function of $(s_1,\ldots, s_n)$ is to be replaced by a function of $(s,\ldots,s)$. The summations over $m_j$ are restricted such that $m_1+m_2+\ldots +m_n=n$, and each $\sum_{m_j=0}^2$ is capped at two, because $D$ is a second-order polynomial of momenta variables, which follows from the linearization of the massive-Dirac Hamiltonian. If $\Omega^{(n)}_s$ is evaluated at the band extremum $(s=0)$, then one can further drop $m_j=1$ (in each of the summations over $m_j$) because $\partial_s D\big|_{s=0}=0$ [cf.\ \q{partialD}]. It becomes apparent that the condition  $m_1+\ldots m_n=n$ cannot be satisfied for odd $n$ and $m_j\in \{0,2\}$, implying that $\Omega^{(n)}_0=0$ for odd $n$. This completes the proof.

\section{Alternative derivation of the anomalous shift}\la{app:alternative}	

We provide an alternative derivation of the anomalous shift vector [\q{anomalousshift}] that aims to demystify the appearance of the Berry curvature. 
Beginning from an expression for the intraband phononic shift that was derived in \s{sec:anomalous} and is valid for small momentum transfer:
\e{Phonon:\as \bS^m_{\bkp \lea \bk}\approx -\nabla_{\kave}(\bA_{\kave}\cdot\delbk) +\bA_{\bkp}-\bA_{\bk},}
with $\bA_{\bk}$ the intraband Berry connection, $\kave=(\bk+\bkp)/2$ and $\delbk=\bkp-\bk$. We have omitted the band index to simplify  notation. We consider the $y$-component of the above shift vector, and express the derivative as the limiting value of a difference: 
\e{\vec{y}\cdot\nabla_{\kave}(\bA_{\kave}\cdot\delbk) = \limit{\eps \ri 0} \f{(\bA_{\kave+\eps \vec{y}/2}-\bA_{\kave-\eps \vec{y}/2})\cdot\delbk}{\eps},}
such that the shift vector component can be expressed as a line integral of the connection:
\e{ (\bS^m_{\bkp \lea \bk})^y \approx \limit{\eps\ri 0}\f1{\eps}\oint \bA_{\bk}\cdot d\bk \approx (\bOmega_{\kave}\times \delbk)^y} 
along an infinitesimally-thin parallelogram drawn in \fig{fig:anomalous_shift}. Finally, one converts the line integral to an area integral of the curvature by Stokes' theorem. This proof is easily generalized for the $x$ and $z$ components. 
\begin{figure}[H]
\centering
\includegraphics[width=4 cm]{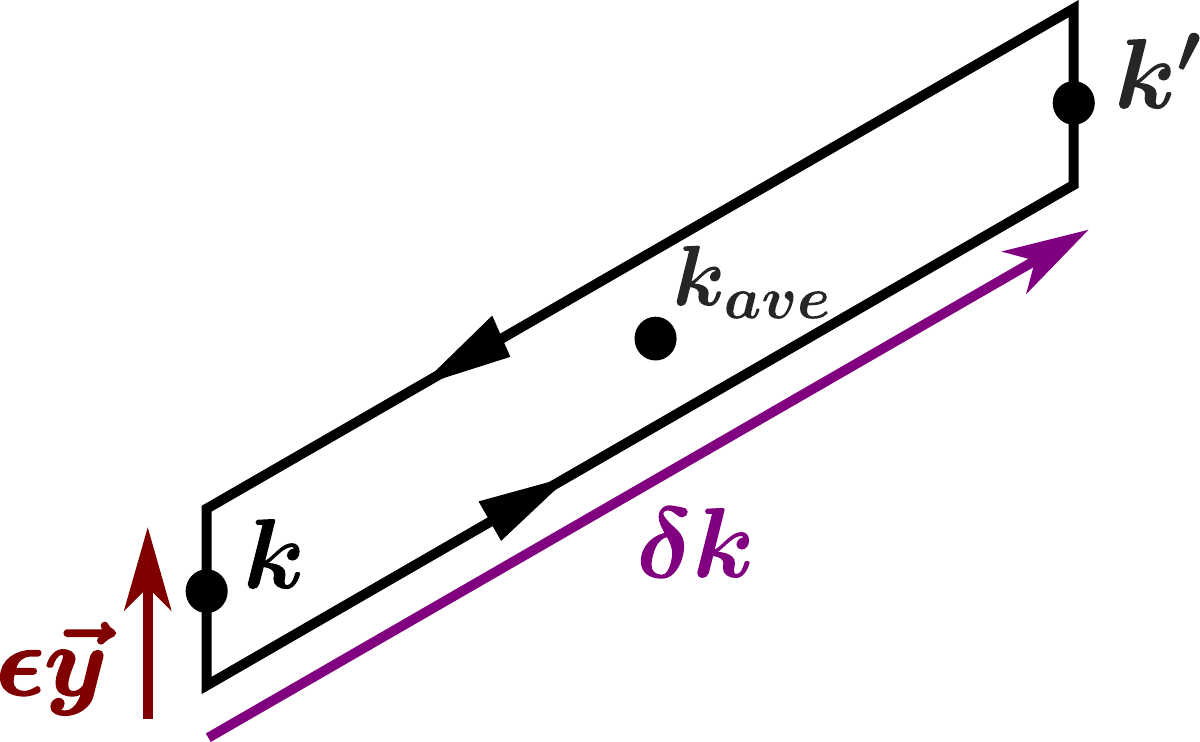}
\caption{Infinitesimally-thin parallelogram.}
\label{fig:anomalous_shift}
\end{figure}

\section{Difficulties of the parallel-transport gauge}\la{app:deceptive}

It has been claimed in the literature that $\braket{u_{b\bk p}}{u_{b\bk}}_{\sma{\text{cell}}}=1+O(\delta k^2)$ can be chosen as a gauge choice for the wave function.\cite{vogl_electronphonon,antoncik_overlapintegrals} This gauge corresponds to a parallel transport condition [$\delbk\cdot \bA_{bb\bk}=0$] in the direction of $\delbk=\bkp-\bk$.\cite{alexey_smoothgauge,AA_glideresolved} 
It is not uncommon to find textbooks which ignore the wave-function dependence of the electron-phonon scattering rate.\cite{lundstrom_book} All Berry-curvature effects (including the anomalous shift [\q{anomalousshift}]) are missed if one blithely adopts the parallel-transport gauge.\\

It is therefore of interest to expose the fallacies inherent in $\braket{u_{b\bkp}}{u_{b\bk}}_{\sma{\text{cell}}}=1+O(\delta k^2)$,\footnote{There is, of course, no controversy in the claim that $|\braket{u_{b\bkp}}{u_{b\bk}}_{\sma{\text{cell}}}|^2=1+O(\delta k^2)$; cf.\ \q{metric}.} of which there are two related kinds:  \\

\noi{i} For fixed $\delbk$, it is generically impossible to set $\delbk\cdot \bA_{bb\bk}=0$ for all $\bk$ in the Brillouin zone; this is tantamount to assuming that the single-band Berry phase vanishes for all momentum loops parallel to $\delbk$. This assumption may hold if $\delbk$ is orthogonal to a mirror plane, in which case the Berry phase of a single spinless band (in the absence of spin-orbit coupling) is indeed quantized to $0$ or $\pi$, but one cannot rule out the case of $\pi$ a priori. \\

\noi{ii} For fixed $\delbk$, it \textit{is} possible to impose $\delbk\cdot \bA_{bb\bk}=0$  for all $\bk$ in a ball-shaped subregion of the Brillouin zone. However, it is generically impossible to simultaneously impose  $\delbk'\cdot \bA_{bb\bk}=0$ within the same ball, for $\delbk'$ that is not collinear with $\delbk$. The simultaneous imposition is equivalent to assuming a vanishing Berry phase for an infinitesimal loop encircling $\bk$, i.e., that the Berry curvature $\delbk\times\delbk' \cdot \bOmega_{b}$ vanishes at $\bk$. Certainly, one must allow for  phonons of all possible wavevectors ($\delbk,\delbk' \in BZ$) to completely describe the electron-phonon interaction. Without finetuning, the Berry curvature $\bOmega_{b}$ vanishes at a generic $\bk$-point only in $PT$-symmetric materials with negligible spin-orbit coupling.\cite{100page} $PT$ is certainly not a symmetry in the present case study of noncentric (meaning no $P$), non-magnetic (meaning $T$-symmetric) materials. 

\section{Energy conversion efficiency} \la{app:efficiency}

We will derive an ideal expression for the energy conversion efficiency  for Pusch et al.'s model\cite{pusch_conversionefficiency} of a shift-current photovoltaic cell. Our derivation closely follows that in Sec II of Ref.\ \cite{pusch_conversionefficiency}, which we recommend as prerequisite reading. However, our final expression for the efficiency  [\q{efficiency}] is less heuristic than Eq. (11) of Ref.\ \cite{pusch_conversionefficiency}, in that ours is wholly expressed in terms of kinetic and band-structure parameters which can be extracted from ab-initio calculations. \\

We adopt the same device geometry that is illustrated in Fig. 1 of Ref.\ \cite{pusch_conversionefficiency}: light falls onto a semiconductor facet with illuminated area $A_{illum}=dw$; $d$ is the separation between two electrodes and $w$ the width of each electrode. For concreteness, we will fix the facet's normal vector  to be parallel to the unit directional vector $\vec{z}$ ;
the photovoltaic current flows between the electrodes in the $x$ direction, and $w$ is the linear dimension of the electrode in the $y$ direction. Assuming that the radiation falls onto the facet with normal incidence, the Poynting vector within the semiconductor decays exponentially as
\e{ \text{Poynting vector}\;\eq \cali_{rad}(z)\vec{z}; \as \cali_{rad}(z)=\cali_{rad}(0)e^{-\alpha_{abs} z},   }
with an attenuation length given by the inverse of  the absorption coefficient:
\e{ \alpha_{abs} \eq \f{4\pi^2}{\alpha_{fs}} \f{\hbar\omega}{n_{\omega}} \langle f_{vc\bk}|\beps_s\cdot \bA_{cv\bk}|^2\rangle_{\omega} \,\jdosupdo. \la{abscoeff}  }
Here, $\alpha_{fs}\approx 137$ is the fine structure, $n_{\omega}$ is the refractive index, $f_{vc\bk}=f_{V}-f_C$ is a difference of the steady quasiparticle distributions, and $\jdosupdo$ is the spin-doubled joint density of states. Our semiclassical expression for the absorption coefficient presumes that $\alpha_{abs}^{-1}$ greatly exceeds the lattice period; the same type of semiclassical approximation implies that the shift current density has the same exponential decay owing to being proportional to $\cali_{rad}(z)$: 
\e{  
j_x(z)= -|e| \fraks \,\text{Abs}(z); \as \text{Abs}(z)= \f{\alpha_{abs}\cali_{rad}(z)}{\hbar\omega}.\la{semiclassicalj}
}
$\fraks$, the average shift per photo-excited electron-hole pair, has been defined in \q{fraks};
$\text{Abs}(z)$ is understood as the photon absorption rate per unit volume, at a distance $z$ from the illuminated facet. 
\q{semiclassicalj} is equivalent to $j_x=\sigma_{\beps_s,\omega}|\cale_{\omega}|^2$ with $\sigma_{\beps_s,\omega}$ the geodesic-approximated shift conductivity in \q{shiftconduct}; to derive the equivalence, revert to Gaussian units and replace $|\cale_{\omega}(z)|^2=2\pi \cali_{rad}(z)/cn_{\omega}$ [cf.\ footnote in \app{sec:krautbaltz}] and $e^2/\hbar c \approx 1/137.$\\

Assuming ideally that the contacts with the electrodes do not introduce additional resistance,    the energy conversion efficiency is given by 
\e{ 
\text{Eff} \eq \f1{4}\f{V_{oc}I_{sc}}{\cali_{rad}A_{illum}},
}
with $V_{oc}$ the open-circuit photovoltage  and $I_{sc}$ the short-circuit shift current. The latter quantity is obtained by integrating 
\e{
I_{sc}= w\int^t_0 j_x(z)dz \approx wj_x(0)/\alpha_{abs},\la{isc}
}
with $t$ the thickness of the semiconductor in the $z$ direction. In the last step of \q{isc}, we assumed $t\gg \alpha_{abs}^{-1}$. The open-circuit photovoltage is determined by the condition that the  shift and drift currents cancel out at each $z$:
\e{j_x(z) \eq \sigma_{ph}(z)\f{V_{oc}}{d}.}
Assuming ideally that the temperature is sufficiently low ($k_BT\ll E_g$) for the dark conductivity to be negligible,\footnote{See Sec V of Ref.\ \cite{pusch_conversionefficiency}
}
the drift current is simply proportional to the linear conductivity of photo-excited carriers; this conductivity is assumed to have the Drude form:
\e{\sigma_{ph}(z) \eq e^2\tau_{tr}\bigg( \f{n(z)}{m_e}+\f{p(z)}{m_h} \bigg), }
with $\tau_{tr}$ the transport lifetime, $n$ (resp. $p$) the photo-excited electron density (resp. hole density), and $m_e$ (resp. $m_h$) the effective mass for electrons (resp. holes).  Assuming that the semiconductor is intrinsic, 
\e{ n(z)=p(z) = \tau_{rec} \text{Abs}(z); \as \sigma_{ph}(z) = \f{e^2\tau_{tr}\tau_{rec}}{m_r}\text{Abs}(z),  }
with $\tau_{rec}$ the recombination time [cf.\ \q{esipovmodel2}] and $m_r^{-1}=m_e^{-1}+m_h^{-1}$. Combining all the above equations,
\e{ 
\text{Eff} \eq \f1{4}\f{.511\,MeV}{\hbar\omega}\f{m_r}{m_{f}} \f{\fraks^2/\tau_{tr}\tau_{rec}}{c^2}, \la{efficiency}
}
which is equivalent to \q{efficiency1}.

\bibliography{bib_Apr2018}

\end{document}